\documentclass[b5paper,10 pt,twoside]{book}

\usepackage{setspace}
\usepackage{verbatim}
\usepackage[english]{babel}
\usepackage{amsmath,amsthm}
\usepackage{makeidx}
\usepackage{epsfig}
\usepackage{graphicx}
\usepackage{subfigure}
\usepackage{float}
\usepackage{fancyhdr}
\usepackage{appendix}
\usepackage{pdflscape}
\usepackage{longtable}
\usepackage{hyperref}
\usepackage[latin1]{inputenc}
\usepackage{color}
\usepackage[labelfont=bf,small]{caption}  
\usepackage[top=4 cm, bottom=5 cm, left=3.5 cm, right=2.5 cm]{geometry}
\usepackage{changepage}

\makeindex
\begin{document}

\begin{titlepage}
\begin{center}
\includegraphics[width=3 cm]{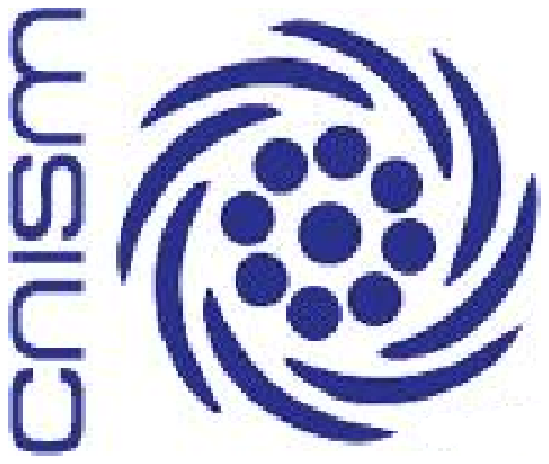}
\\[0.2 cm]
\linespread{0.3}
\textbf{Consorzio Nazionale Interuniversitario \\ per le Scienze dei Materiali} \\
\vspace{0.2 cm}
\rule{\linewidth}{0.5mm}
\\[0.1 cm]

\linespread{1.2}
\Large
Challenges for first--principles methods in theoretical and computational physics: 
multiple excitations in many--electrons systems
and the Aharonov--Bohm effect in carbon nanotubes
\rule{\linewidth}{0.5mm}
\vspace{-0.5 cm}

\linespread{1.}
\normalsize
\begin{tabular}{c c c}
  \textbf{Universit\`{a} degli}  & &  \textbf{Universit\`{a} degli} \\ 
  \textbf{studi di Milano}       & &  \textbf{studi di Roma Tre}    \\ 
  \includegraphics[width=3 cm]{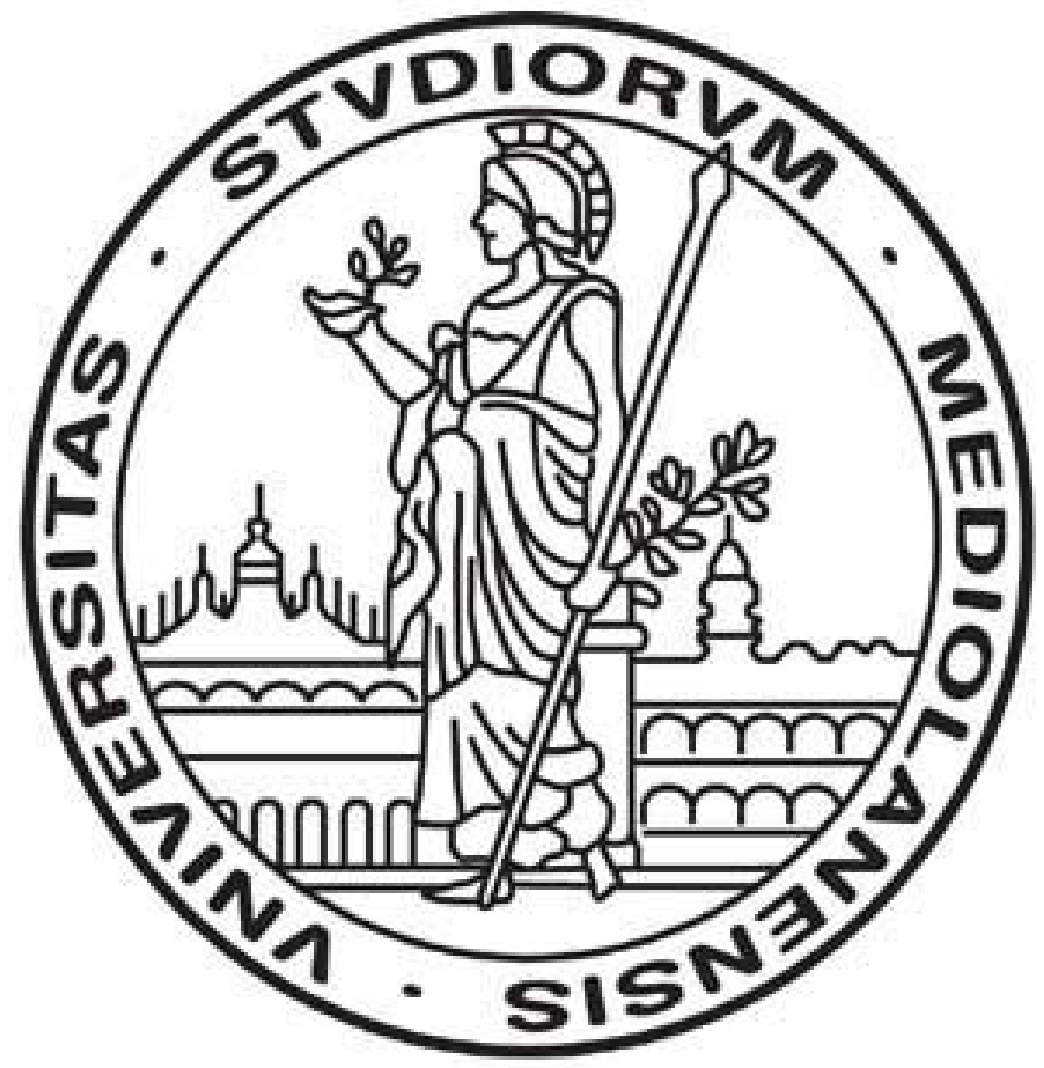}
  &\ \ \ \ \ \ \ \ \ \ \ \ \ \ \ \ \ \ \ \ \ \ \ \ \ \ \  &
  \includegraphics[width=3 cm]{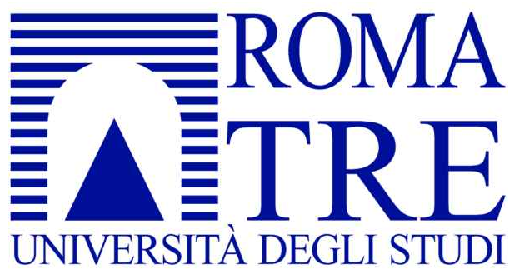}            \\
\end{tabular}
\vspace{0.5 cm}
\end{center}

\begin{flushleft} \large
\emph{Supervisor:}  Giovanni ONIDA     \\
\emph{Cosupervisor:}  Andrea MARINI  \\
\end{flushleft}

\vspace{0.5 cm}
\vfill
\begin{center}
\large
\textbf{\emph{Davide Sangalli}}
\end{center}

\end{titlepage}
\thispagestyle{empty}

\begin{titlepage}

\changepage{3.cm}{1.5cm}{1.cm}{-1.cm}{}{-1.cm}{}{}{}
\begin{center}
\includegraphics[width=0.99\textwidth]{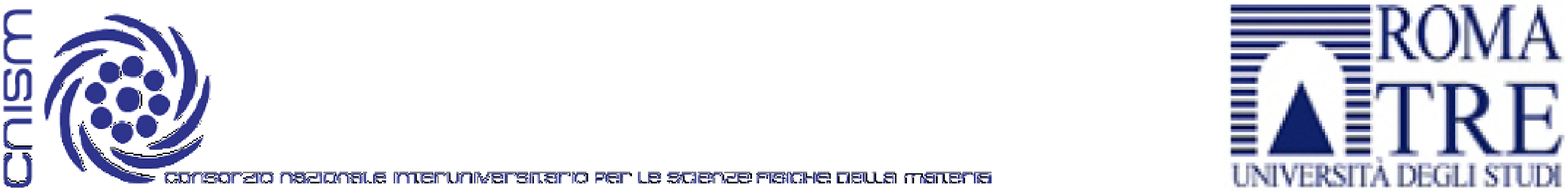}    \\
\vspace{1.cm}
\Large
\textbf{Universit\`a degli Studi Roma TRE \\[0.3 cm] e \\[0.3 cm]
Consorzio Nazionale Interuniversitario per le Scienze Fisiche della Materia\\}
\vspace{1.cm}
\large
\textbf{Dottorato di Ricerca in Scienze Fisiche della Materia \\
XXIII ciclo \\}
\vspace{1.cm}
\normalsize
\textbf{Challenges for first--principles methods in theoretical and \\
computational physics: multiple excitations in many--electrons  \\
systems and the Aharonov--Bohm effect in carbon nanotubes \\}
\vspace{1.cm}
\large
\textbf{Tesi di dottorato del dott. Davide Sangalli\\}
\end{center}

\vspace{2.5cm}
\normalsize
\begin{minipage}{0.45\textwidth}
\begin{flushleft} 
Relatore:  Prof. Giovanni Onida     \\
Correlatore:  Dott. Andrea Marini  \\
\end{flushleft}
\end{minipage}
\begin{minipage}{0.45\textwidth}
\begin{flushright}
Coordinatore Dottorato: \\
Prof. Settimio Mobilio
\end{flushright}
\end{minipage}

\vspace{1. cm}
\vfill
\begin{center}
\normalsize
a.a. 2010 / 2011
\end{center}

\end{titlepage}
\thispagestyle{empty}

\setcounter{secnumdepth}{1}  
\frontmatter
\renewcommand{\[}{\left[}
\renewcommand{\]}{\right]}
\renewcommand{\(}{\left(}
\renewcommand{\)}{\right)}

\makeatletter
\def\overbracket{\@ifnextchar [ {\@overbracket} {\@overbracket
[\@bracketheight]}}
\def\@overbracket[#1]{\@ifnextchar [ {\@over@bracket[#1]}
{\@over@bracket[#1][0.3em]}}
\def\@over@bracket[#1][#2]#3{
\mathop {\vbox {\m@th \ialign {##\crcr \noalign {\kern 3\p@
\nointerlineskip }\downbracketfill {#1}{#2}
                              \crcr \noalign {\kern 3\p@ }
                              \crcr  $\hfil \displaystyle {#3}\hfil $%
                              \crcr} }}\limits}
\def\downbracketfill#1#2{$\m@th \setbox \z@ \hbox {$\braceld$}
                  \edef\@bracketheight{\the\ht\z@}\downbracketend{#1}{#2}
                  \leaders \vrule \@height #1 \@depth \z@ \hfill
                  \leaders \vrule \@height #1 \@depth \z@ \hfill
\downbracketend{#1}{#2}$}
\def\downbracketend#1#2{\vrule depth #2 width #1\relax}
\makeatother
\pagestyle{fancy}
\renewcommand{\chaptermark}[1]
             {\markboth{#1}{}}
\renewcommand{\sectionmark}[1]
             {\markright{\thesection\ #1}}
\fancyhf{}
\fancyhead[LE,RO]{\bfseries\thepage}
\fancyhead[LO]{\bfseries\rightmark}
\fancyhead[RE]{\bfseries\leftmark}
\renewcommand{\headrulewidth}{0.5pt}
\renewcommand{\footrulewidth}{0pt}
\addtolength{\headheight}{0.5pt}
\fancypagestyle{plain}{%
              \fancyhead{}
              \renewcommand{\headrulewidth}{0pt}}

\def\ai         {{\it ab initio}\;}
\def\Ai         {{\it Ab initio}\;}

\tableofcontents
\chapter{Introduction}
Many--body physics is a branch of physics whose scope is to understand physical phenomena where a number of interacting bodies is present. The presence of the interaction is what makes the description of such systems challenging but at the same time exciting. Interacting particles can give birth to new physical process which cannot be simply described as the sum of the behaviour of each single element. The superconducting phase at low temperatures, plasmon peaks in the absorption spectrum, Mott transitions are only a few examples. New physics emerge as a result of the coherent behaviour of the many--body system.

The description of interacting particles requires sophisticated many--body techniques and the exact mathematical solution to the problem is almost never available: approximations are needed. To construct a practical approximation one need to have some clue as to which are the most relevant phenomena, which are the physical aspects that can be discarded and which can be treated in an approximate way as perturbations. Often even the approximate equation cannot be solved analytically, a computational approach is needed. Computational Physics can be seen as an approach which stands in the middle between Theoretical Physics and Experimental Physics. Some of the results presented in the present work have been obtained through a numerical approach.

In the present work we will describe some of these techniques with a focus on a specific phenomenon: the description of double excitations\index{double excitations} in the absorption spectrum. Double excitations are a peculiar effect of interacting systems which does not have a counterpart in non--interacting ones. The optical absorption spectrum of a system is obtained by shining light on it. At the microscopic level photons hit the electrons which sit in the ground state and change their configuration. If the light source is not too intense this can be described in linear response; that is only ``one photon'' processes are involved, only one electron per time can be influenced. Here is where the interaction comes in. The hit electron is linked to the others and so other process take place, one of these is the appearance of multiple excitations. These are, strictly speaking, virtual processes as the real time evolution of the system is different from the one described. Nevertheless the physical effect is there and can be measured as extra peaks in the absorption spectrum.

Double excitations is not the only subject of the present work and understanding is not the only scope of many--body physics. The same techniques can be used to make accurate quantitative predictions of the behaviour of a material. These allow us to control physical phenomena and possibly to use them in technological applications. In the second part of the Thesis we focus on the application of more standard techniques to the description of carbon nanotubes (CNTs). In particular we focus on the effects of magnetic fields on CNTs. 

CNTs are quasi 1D-systems composed by carbon atoms which have been discovered in 1952\footnote{A large percentage of academic and popular literature attributes their discovery to Sumio Iijima of NEC in 1991~\cite{Iijima_1991,Charlier2007}, however already in 1952 L. V. Radushkevich and V. M. Lukyanovich published clear images of 50 nanometer diameter tubes made of carbon in the Soviet Journal of Physical Chemistry, the publication however was in Russian. For a detailed review on the discovery of CNTs we address the reader to Ref.~\cite{CNT_discovery}.}. They have the shape of a hollow cylinder with a nanometric diameter ($10^{-9}\ m$), a micrometric length ($10^{-6}\ m$) and the thickness of a single atomic layer\footnote{Here we refer to single--walled CNTs. Multi--walled CNTs, which are composed by concentric single--walled CNTs exist too.}. What makes such objects so interesting is that they are mechanically very strong and stable. These properties makes them ideal system both for many possible technological approaches and for testing the physical behavior of electrons in 1D system as well as in cylindrical topologies. In this work we are in particular interested in the effect of magnetic fields related to topology.

Under the effect of a magnetic field electrons delocalized on a cylindrical surface display a peculiar behaviour, known as Aharonov--Bohm\index{Aharonov--Bohm} effect. The Aharonov--Bohm\index{Aharonov--Bohm} is a pure quantum mechanical effect which does not have any counterpart in classical physics. In CNTs the Aharonov--Bohm\index{Aharonov--Bohm} modify the electronic gap and so can be used to tune the electronic properties. Though a model able to account for such process is available in the literature, in the present work we will describe the effect of magnetic fields ``\ai''. \Ai is any approach which describes the physics starting from first principles and without the use of any external parameter. As pointed out in the first part of the introduction the exact solution to the many--body problem is in practice never available and approximations are needed. In the description of CNTs we will use standard approximations which are by far much more accurate and general than any approximation introduced in phenomenological descriptions based on model systems.

In part I the general many--body problem is introduced. In particular Density--Functional Theory (DFT\index{DFT}) and Many--Body Perturbation Theory (MBPT) are described according to our needs for the forthcoming parts. In part II the problem of double excitations\index{double excitations} is presented together with experimental evidence and the state of the art. In this part we will propose a new approximation which could be used in the standard approach for the description of absorption spectra in both the MBPT and DFT\index{DFT} framework. This approximation is able to describe double excitations\index{double excitations}. Finally in part III the general problem of CNTs in magnetic fields will be considered. After a brief overview on the main experimental evidence of the Aharonov--Bohm\index{Aharonov--Bohm} effect, the Zone Folding Approach (and the Tight Binding model) will be introduced. Then we will describe how magnetic field effect are included in our \ai approach and finally we will compare our results with the predictions of the models.

\chapter*{Notation and conventions}
This is a brief overview of the conventions used to express operators and their related physical quantities. The same conventions are introduced in Ch.~\ref{chap:Many Body Systems}. Atomic units are used in Part I of the thesis, while in Part II the international system (SI) of units is used.

The one body operators are written in second quantization according to the following expression:
\begin{equation*}
\hat{A} = \sum_{\sigma_1,\sigma_2} \int d^3\mathbf{x}_1  d^3\mathbf{x}_2  dt_1 dt_2\
                       A_{\sigma_1\sigma_2}(\mathbf{x}_1t_1,\mathbf{x}_2t_2)
                       \hat{\psi}_{\sigma_1}^{\dag}(\mathbf{x}_1,t_1)\hat{\psi}_{\sigma_2}(\mathbf{x}_2,t_2) 
\text{.}
\end{equation*}
Here $\sigma$ is a spin variable, while $\mathbf{x}$ and $t$ are space and time variables respectively and $A_{\sigma_1\sigma_2}(\mathbf{x}_1t_1,\mathbf{x}_2t_2)$ is the kernel of the operator. A compact notation is also often used
\begin{eqnarray*}
\hat{A} &=&\int d1 d2\ A(1,2)\hat{\psi}^{\dag}(1)\hat{\psi}(2) \text{ ,} \\ 
\hat{A} &=& \int d\mathbf{1} d\mathbf{2}\
                     A(\mathbf{1}t_1,\mathbf{2}t_2)\hat{\psi}^{\dag}(\mathbf{1}t_1)\hat{\psi}(\mathbf{2}t_2)
\text{.}
\end{eqnarray*}

The compact notation will be preferred wherever it will not be source of confusion. In this notation repeated primed variables are supposed to be integrated, i.e.
\begin{equation*}
\Sigma^{\star}(1,2) = i G(2',1) W(1,2') \Gamma^{\star}(2,2';1')
\text{ ,}
\end{equation*}
means
\begin{equation*}
\Sigma^{\star}(1,2) = \int d1' d2'\ i G(2',1) W(1,2') \Gamma^{\star}(2,2';1')
\text{.}
\end{equation*}
The symbol $\int d1'$ stands for $\sum_\sigma' \int d\mathbf{x'} dt'$. However, when only partial integration will be performed (i.e. only on space, time or spin variables), the sums / integrals will be written explicitly.
The notation $G(1,2^+)$ will be used for
\begin{equation*}
\lim_{\epsilon\rightarrow 0^+} G(\mathbf{x_1},t;\mathbf{x_2},t+\epsilon)
\text{,}
\end{equation*}
and the notation $\langle\hat{A} \rangle$ for the expectation value on the ground state of an operator:
\begin{equation}
\langle\Psi_0 |\hat{A}|\Psi_0 \rangle
\text{.}
\end{equation}

For the coulomb interaction we will use
\begin{equation*}
w(1,2)=\frac{1}{|\mathbf{x}_1-\mathbf{x}_2|}\delta(t_2-t_1)\delta_{\sigma_1\sigma_2}
\text{;}
\end{equation*}
for the one particle part of the Hamiltonian
\begin{equation*}
\hat{H}_0=-\frac{1}{2}\nabla^2+V_I(\mathbf{x})
\text{.}
\end{equation*}

A functional will be expressed with the following notation $E[\rho]$ which means that the energy $E$ is a functional of the density $\rho(\mathbf{x},t)$.

Finally we list here some (but not all) of the symbols used in the thesis:
\begin{itemize}
\item $\mathbf{B}$ is the magnetic field,
\item $\mathbf{A}$ is the vector potential,
\item $\Phi$ is the magnetic flux
\item $\mathbf{H}$ is the magnetic induction field,
\item $\mathbf{E}$ is the electric field,
\item $\mathbf{D}$ is the electric displacement field,
\item $\mathbf{P}$ is the total polarization,
\item $\mathbf{M}$ is the total magnetization,
\item $\mathbf{j}$ is the current--density, with $\mathbf{j}^{(p)}$ the paramagnetic and $\mathbf{j}^{(A)}$ the diamagnetic component
\item  $\rho$ is the density
\item $\epsilon$ is the dielectric constant,
\item $\alpha$ is the polarizability,
\item $\chi$ is the response function, $\chi_0$ the independent--particle one, and $\chi_{KS}$ the Kohn--Sham one,
\item $L$ is a four--point response function, while $\tilde{L}$ is four--point in space and two--point in time,
\item $T$ is the time--ordering operator while,
\item $\hat{T}$ and $T[\rho]$ are the kinetic energy operator and the kinetic energy of the system respectively,
\item $\Sigma$ is the self--energy,
\item $\Sigma_H=\Sigma+v_H$, that is the sum of the self--energy and the Hartree potential ($v_H$),
\item $\Sigma^\star$ is the reduced self--energy,
\item $\Sigma_s$ is a static self--energy while $\Sigma_d$ is a dynamical self--energy,
\item $\Pi$ is the polarizability, which correspond to the density--density response function\footnote{The two quantities differ only because $\Pi$ is $T$--ordered, while $\chi_{\rho\rho}$ is a retarded quantity},
\item $\Pi^\star$ is the reduced polarizability,
\item $\Gamma$ is the vertex function and $\Gamma^\star$ is the reduced vertex function;
\item $W$ is the screened coulomb interaction,
\item $g$ is the independent--particle Green's function\index{Green's function}, $g_H$ the Hartree Green's function\index{Green's function}, and $G$ the many--body Green's function\index{Green's function},
\item $Z$ is the renormalization factor,
\item $E_{xc}$ is the exchange--correlation energy, $v_{xc}$ the exchange--correlation potential and $f_{xc}$ the exchange--correlation kernel,
\item $\Xi$ is the kernel of the Bethe--Salpeter equation (BSE) while $K$ the kernel of a generalized equation (including the dynamical BSE) for the four--point response function,
\item $\mu$ is the magnetic susceptibility or the chemical potential, will be clear from the context 
\item $\Psi$ the many--body wave--function and $\Psi_0$ the many--body ground state, $\Psi_s$ the Kohn--Sham ground state.
\item $\Phi_0$ will be used for the non--interacting many--body ground state or for the magnetic flux quantum $\Phi_0=\frac{h}{e}$, with $h$ the Planck's constant and $e$ the electron charge,
\item when explicitly specified $\Psi_S$ will be the wave--function in the Schr\"odinger's picture, $\Psi_H$ in the Heisenberg's picture, and $\Psi_I$ in the interaction picture,
\item $\delta A$ stands for a variation of the quantity $A$, while $\delta(1,2)$ is the Dirac's delta function,
\end{itemize}

\mainmatter
\part{Theoretical background}          \label{part:Theoretical background}
\chapter{Many--Body Systems}            \label{chap:Many Body Systems}
Any known many--body system is constituted of interacting particles and, at least in principle, any physical aspect can be understood describing their dynamics. Elementary particles are in general ruled by the equations of quantum mechanics and special relativity (or general relativity) and four possible kind of interactions are known to exist: the Electromagnetic interaction, the Nuclear Weak interaction, the Nuclear Strong interaction and the Gravitational interaction. For this reason even the description of a single atom, where in principle all forces have to be taken into account, appear an almost impossible problem. Moreover any macroscopic body is constituted of an enormous number of interacting particles, for a reference the Avogadro's number; $N_A=6.0221415\ \times\ 10^{23}\ [\text{particles/moles}]$. So any attempt to solve the many--body problems seems doomed to fail. Despite this discouraging scenario there are two factors which in fact make possible to tackle the many--body problem from a microscopic point of view. First the majority of the processes which happen in everyday life involve a thin energy window such that only the electromagnetic interaction plays a role and, quite often, only the dynamics of the electrons needs to be described. Secondly the majority of the macroscopic objects are constituted by fundamental building blocks which almost completely determine their properties: these are the molecules which constitutes the gases and the liquids and the unitary cells which are repeated an infinite number of times in many solid systems\footnote{For gases and liquids only the properties related to the electronic dynamics can be studied from a microscopic point of view. Other properties require a statistical description of the system.}.

A crucial role is, then, played by the equation which describes the dynamics of few interacting electrons immersed in the Coulomb potential of the nuclei that, in a first approximation can be considered frozen in their instantaneous positions: the Scr\"{o}dinger Equation (SE) in the Born-Oppenheimer (BO) approximation. Possibly the electrons can interact with external fields which can be used either to explore or to tune the properties of the materials. The first part of this thesis will be dedicated to the description of light absorption experiments where an external light source is used to investigate the optical properties of a many--electrons system. The second part will be focused on the study of magnetic field effect on Carbon Nano-Tubes (CNTs) and how the external field can be used to tune the electronic properties of the CNTs.

The SE  can be obtained applying an Hamiltonian operator to the electronic wave--function. The operator can be divided into two terms $\hat{H}^{int}+\hat{H}^{ext}$. The first term describes the electrons--nuclei interaction, while the second term is due to the presence if an external perturbation that, in this work, is the electromagnetic field. The SE in atomic units reads:
\begin{equation}\label{ManyBody_Hamiltonian}
\left(i\frac{\partial}{\partial t}-\hat{V}^{ext}\right) \Psi=
  \frac{1}{2}\left[ \left(i\mathbf{\nabla}-\hat{\mathbf{A}}^{ext}\right)^2
                     +\hat{\mathbf{B}}^{ext} \boldsymbol{\sigma}
                     +\hat{w}+\hat{V}_{I}
             \right] \Psi
\text{.}
\end{equation}
$\hat{V}^{ext}$ and $\hat{\mathbf{A}}^{ext}$ accounts for the external potentials and the term $\mathbf{B}^{ext} \boldsymbol{\sigma}$ accounts for the interaction of the spin with a possible applied external field. $\boldsymbol{\sigma}$ is a vector constituted by the Pauli matrices, $\Psi$ is the electronic wave--function; $\hat{w}$ is the Coulomb interaction while $\hat{V}_{I}$ accounts for the ionic potential. Finally the $\mathbf{\nabla}$ operator takes into account the kinetic energy of the electrons.

It is important to observe that some of the terms discarded in Eq. (\ref{ManyBody_Hamiltonian}) does not have in practice a role in the physical process we are interested in. The reason is that the energy scales involved are so different that the corresponding dynamics can be neglected. This is the case of the nuclear forces and of the gravitational force. Some other terms instead are discarded as they are usually very small although they could be needed for the description of some physical phenomena. This is the case of relativistic corrections\footnote{The terms due to the presence of a magnetic field included in Eq. (\ref{ManyBody_Hamiltonian}) are already relativistic corrections to $\hat{H}^{ext}$. Here we consider these terms as in this thesis we are interested in the description of the Aharonov--Bohm\index{Aharonov--Bohm} effect in CNTs. The relativistic corrections to the $\hat{H}^{int}$ part of the Hamiltonian, that is the magnetic field generated by the electronic current (and spin) and the magnetic field due to the nuclei (the spin orbit interaction mainly) are neglected here.} which are needed in the description of materials composed of heavy nuclei; the dynamics of the nuclei, which is neglected in the BO approximation, is relevant for example for the description of the superconducting phase of some materials. Many development of the state--of--the--art are devoted to overcome such approximations.

On the other hand the idea of describing solid state devices as an infinite repetition of the same fundamental building blocks can be applied to the description of specific kind of materials only. As predicted by Feynman times ago, objects at the nanometric scale can display very peculiar properties which are completely different from the case of molecules or of bulk systems. In these direction the state-of-the-art tools need to be pushed beyond their present limits. From one side the common approximations involved, which are often based on physical intuition, cease to be valid and new approximations are needed. From the other side the lack of a repetitive structure calls for the need of instruments able to describe systems with hundred, thousand and even more interacting electron. Only thanks to the very recent increase of computational power and, in the same time, to recent developments of the techniques, it has become possible to tackle, at least in some cases, the description of such system ``ab-initio'' and so to test the prediction of more simple theoretical models.

\section{Looking for the ground--state}
The solution to Eq. (\ref{ManyBody_Hamiltonian}) is the main goal of the many--body physics. The first objective is to solve such equation with $\hat{H}^{ext}=0$. That is to find out the ground--state of the system.

An exact analytical solution can be obtained only in oversimplified systems such as the hydrogen atom. For any realistic system this is far beyond our possibilities. A computational solution can be obtained at the price of a computational time, that grows exponentially with the number of electrons, only for very small systems. This is why approximations are needed. The problematic part of the Hamiltonian is the interaction term $w(|\mathbf{r_1-r_2}|)$ for which different possible strategies are available. In this work we will tackle the problem using Many--Body Perturbations Theory (MBPT) and the Density--Functional Theory (DFT). Both methods start from the consideration that the many--body wave--function contains much more information than really needed. So instead of the exact wave--function, which is a function of $3N$ variables, where $N$ is the number of electrons, one can look for simpler quantities which contains only the informations needed to give a quantitative description of experiments. In quantum mechanics any measurable quantity is related to an Hermitian operator. In particular one body operators can be written in second quantization as:
\begin{equation}\label{one_body_operator}
\hat{A} = \sum_{\sigma_1,\sigma_2} \int d^3\mathbf{x}_1  d^3\mathbf{x}_2  dt_1 dt_2\
                       A_{\sigma_1\sigma_2}(\mathbf{x}_1t_1,\mathbf{x}_2t_2)
                       \hat{\psi}_{\sigma_1}^{\dag}(\mathbf{x}_1,t_1)\hat{\psi}_{\sigma_2}(\mathbf{x}_2,t_2) 
\text{,}
\end{equation}
We introduce here the notations $1=(\mathbf{1},t)=(\mathbf{x},t,\sigma)$ which will be used from now on. So for example Eq. (\ref{one_body_operator}) can be written in the two forms:
\begin{eqnarray}
\hat{A} &=&\int d1 d2\ A(1,2)\hat{\psi}^{\dag}(1)\hat{\psi}(2) \text{ ,} \\ 
\hat{A} &=& \int d\mathbf{1} d\mathbf{2}\
                     A(\mathbf{1}t_1,\mathbf{2}t_2)\hat{\psi}^{\dag}(\mathbf{1}t_1)\hat{\psi}(\mathbf{2}t_2)
\text{.}
\end{eqnarray}
Within this notation we finally define $1^+=lim_{\epsilon\rightarrow 0^+} (\mathbf{1},t_1+\epsilon)$.

\begin{figure}[t]
 \centering
 \includegraphics[width=0.8\textwidth]{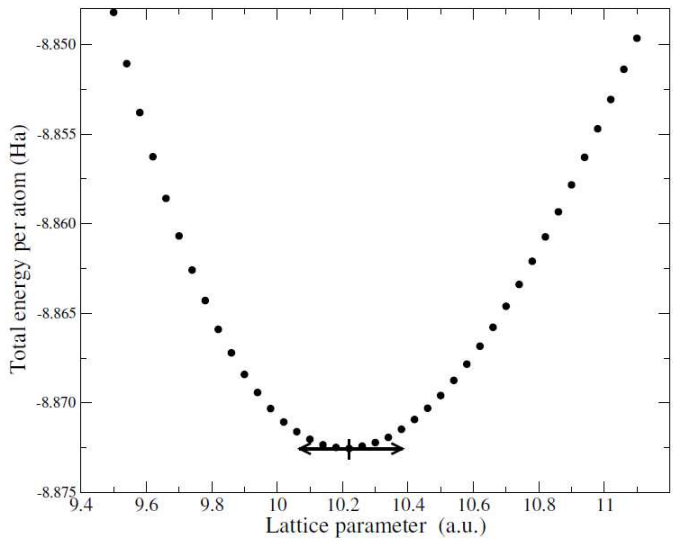}
 \caption{Total energy minimization: by looking at the minimum of the total energy is
          possible to obtain the ground--state properties of the system. Here, as an example, the 
          total energy of Silicon Bulk as a function of the lattice parameter is shown. From ref.~\cite{PhD_Thesis_Fabien}. }
\label{fig:Energy_minimization}
\end{figure}
Eq. (\ref{one_body_operator}) defines a ``one body operator'', that, in the jargon of the second quantization, acts only on one particle states. The field operator $\hat{\psi}(2)$ destroys a particle at the space-time coordinate $2$ and his Hermitian counterpart $\hat{\psi}^{\dag}(1)$ creates a particle at $1$. If we set $x_2=x_1$, $t_2=t_1$ or $\sigma_2=\sigma_1$ the operator is said to be local in space, time or spin respectively.

As observables are related to the expectation value $\langle\hat{A}\rangle$, it is then natural to construct a theory which is able to evaluate, instead of the many--body wave--function, the expectation value of any given one body operator. This can be done introducing the one body Green's function\index{Green's function} (GF)
\begin{equation}
G(1,2)=-i\langle \Phi_0 | T \left[  \hat{\psi}(1)\hat{\psi}^{\dag}(2) \right] | \Phi_0 \rangle
\text{.}
\end{equation}
$\Phi_0$ is the interacting ground--state many--body wave--function of the system and $T$ is the time ordering operator\footnote{The $T$ operator order any couple of operators. $T[\hat{A}(t_1)\hat{B}(t_2)]=\theta(t_1-t_2)\hat{A}(t_1)\hat{B}(t_2)\pm \theta(t_2-t_1)\hat{B}(t_2)\hat{A}(t_1)$ where the sign is $+$ or $-$ according to whether the operators are bosonic or fermionic.}.

A particularly meaningful physical quantity is the total energy of the system, as its knowledge enables to obtain many informations on the system. An example is shown in Fig. (\ref{fig:Energy_minimization}), where the total energy is used to determine the equilibrium lattice parameter of silicon bulk. Unfortunately the total energy operators involves a two body operator, the coulomb interaction
\begin{equation}
w(1,2)=\frac{1}{|\mathbf{x}_1-\mathbf{x}_2|}\delta(t_2-t_1)\delta_{\sigma_1\sigma_2}
\end{equation}
whose average cannot be easily obtained in terms of the GF\index{Green's function}s. Nevertheless the Galitiskii-Migdal equation~\cite{Fetter_Walecka} ensures that the total energy of any system can be expressed in terms of the one particle GF\index{Green's function}. We do not give the proof here but we just observe that, from the SE it's possible to obtain the identity
\begin{equation}\label{exp_int}
\langle i\frac{\partial}{\partial t}-\hat{H}_0\rangle=\langle \hat{w} \rangle
\end{equation}
where we have introduced $\hat{H}_0(1)$ for the one particle part of the Hamiltonian. All the operators on the right hand side of Eq.~(\ref{exp_int}) are one particle operators and so the average can be expressed in terms of the one particle GF\index{Green's function}:
\begin{equation}
\langle \hat{w} \rangle = -\frac{i}{2} \sum_{\sigma} \int d^3\mathbf{x_1}
            \lim_{2\rightarrow 1^+}
            \left[ i\frac{\partial}{\partial t} - H_0(1) \right] G(1,2)
\hspace{0.5 cm} \text{.}
\end{equation}

In the next chapter we will briefly review how the GF\index{Green's function} can be used in practice to get approximate expressions for several observables.  

\section{Perturbing the ground--state}
The solution of the many--body problem with $H^{ext}=0$ provides informations on the system such as its mechanical stability, the electric character (metal / semiconductor / insulator), etc. As long as the external Hamiltonian is zero the body will remain in is ground--state and will not do anything special: a very boring condition. What is much more interesting, and at the same time closer to what happens in an experimental situation, is to see how a system react to an external perturbation. A system is never isolated and always interact with the enviroment. Experimentalists have to probe with an external field to see if a system is a conductor or an insulator, have to shine light on it to see which frequency the system absorbs (roughly speaking to see the color of the system) or they have to try to break it too see how much it is mechanically resistant. As a first guess one could imagine that knowing the ground--state we can understand how it will react to an external perturbation, at list to first order. This is not always the case, the interaction is still there and things are much more complicated (and much more exciting!).

\begin{figure}[t]
 \centering
 \includegraphics[width=0.8\textwidth]{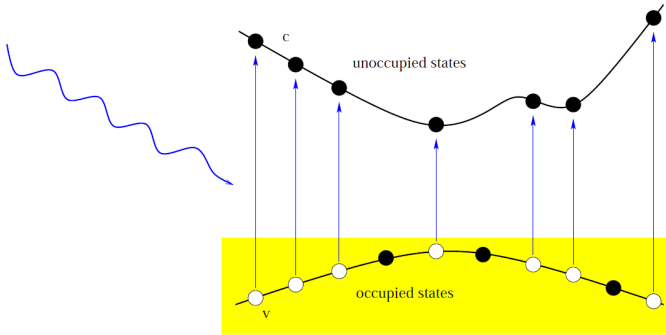}
 \caption{Perturbing the ground--state: a schematic illustration of an external electromagnetic field which
          perturbs the ground--state of a semi--conductor. Electrons are excited to valence states. This is an
          independent particle representation. From ref.~\cite{PhD_Thesis_Sottile}.}
\end{figure}

In this work we will only deal with electromagnetic external perturbations that we will artificially distinguish in two classes: the static perturbations and the time dependent ones. Static perturbations can be embodied in the ground--state Hamiltonian and the same techniques we will describe for the  $H^{ext}=0$ case will apply. This is what we will to study on CNTs immersed in external magnetic fields. On the other hand time dependent perturbations are more problematic to deal because, when the Hamiltonian is time dependent, it's not possible to speak of a ground--state of the system. New techniques are needed to follow the time evolution of the system; anyway often experimentalist just uses small external perturbations to probe the system. In these situations linear response calculations give us all the information we need to describe the experiment. For this reason other key quantities we will encounter are the response functions.

The equation of motion in the presence of an electromagnetic field can be expressed in terms of the density, the magnetization density and the current density
\begin{equation}
\begin{split}
\hat{\rho}(\mathbf{x},t)&=\sum_{\sigma_1,\sigma_2}\delta_{\sigma_1,\sigma_2}\hat{\psi}^\dag(1)\hat{\psi}(1^+)             \text{,}    \\
\hat{\mathbf{m}}(\mathbf{x},t)&=\sum_{\sigma_1,\sigma_2}\boldsymbol{\sigma}_{\sigma_1,\sigma_2}\hat{\psi}^\dag(1)\hat{\psi}(1^+)  \text{,}\\
\hat{\mathbf{j}}(\mathbf{x},t)&=\sum_{\sigma_1,\sigma_2}\delta_{\sigma_1,\sigma_2}j(\mathbf{x},t)\hat{\psi}^\dag(1)\hat{\psi}(1^+)
\text{,}
\end{split}
\end{equation} 
where
\begin{equation}
j(\mathbf{x},t)=\frac{i}{2}\int d^3\mathbf{x}'
            \left(\delta(\mathbf{x}-\mathbf{x}')\frac{\partial}{\partial\mathbf{x}}+
                                       \frac{\partial}{\partial\mathbf{x}}\delta(\mathbf{x}-\mathbf{x}')\right)
                +A(\mathbf{x},t)  \text{.}
\end{equation}
The expression of the external Hamiltonian is then
\begin{equation}\label{external_Hamiltonian}
\hat{H}^{ext}(\mathbf{x},t)=V^{ext}(\mathbf{x},t)\hat{\rho}(\mathbf{x},t)
              -\mathbf{A^{ext}}(\mathbf{x},t)\hat{\mathbf{j}}(\mathbf{x},t)
              +\mathbf{B}^{ext}(\mathbf{x},t)\hat{\mathbf{m}}(\mathbf{x},t)
\text{.}
\end{equation}
Note that Eq. (\ref{external_Hamiltonian}) defines a local one particle operator. Its effect on the electronic wave--function can be highly non linear. This is the case, for example, of multi-photon excitations where the one particle operator $\hat{H}^{ext}$ act $n$-times on the ground--state wave--function, thus inducing, at least, an $n$-particles effect. In the linear regime however, only terms to first order in $\hat{H}^{ext}$ need to be considered; one could be tempted to say that, then, only one particle process are involved. We will see that this is not the case due to the effect of the particle--particle interactions which induce many--particles process, such as double excitations. There is then an overlap between linear and non linear processes. This happens because the electrons react to the total electromagnetic field, that we have artificially divided into $V^{ext}$ and $w$ (and $V_I$); even when $V^{ext}$ is small the changes induced in $w$ could be big enough to induce effects beyond the linear order. We will describe double excitations more in details in the Part II of the present work.

As our perturbation couples to the density, the current and the spin it is convenient to look at their variation induced by weak external fields\footnote{Here the expression $\delta A= \chi_{A,B} \delta B$ stands for $\delta A(\mathbf{x},t)= \chi_{A,B}(\mathbf{x},t;\mathbf{x'},t') \delta B(\mathbf{x'},t')$, where primed variables are integrated.}:
\begin{eqnarray} \label{response_general}
\delta\rho       &= \chi_{\rho\rho}\ \delta V^{ext}
               + \chi_{\rho\mathbf{j}}\  \delta \mathbf{A}^{ext}
               + \chi_{\rho\mathbf{m}}\  \delta \mathbf{B}^{ext}& \text{,}           \\
\delta \mathbf{j}&= \chi_{\mathbf{j}\rho}\ \delta V^{ext}
               +  \chi_{\mathbf{j}\mathbf{j}}\ \delta \mathbf{A}^{ext}
               +  \chi_{\mathbf{j}\mathbf{m}}\ \delta \mathbf{B}^{ext}&  \text{,}   \\  
\delta \mathbf{m}&= \chi_{\mathbf{m}\rho}\  \delta V^{ext}
               + \chi_{\mathbf{m}\mathbf{j}}\ \delta \mathbf{A}^{ext}
               + \chi_{\mathbf{m}\mathbf{m}}\ \delta \mathbf{B}^{ext}&  \text{.}
\end{eqnarray}
$\delta\rho$, $\delta\mathbf{j}$ and $\delta\mathbf{m}$ represent the variation of the density, the magnetization--density and the current--density. For example $\delta\rho=\langle \hat\rho(\mathbf{x},t) \rangle-\rho_0(\mathbf{x},t)$, however we drop the spatial and the time dependence in order to have a compact notation and focus on the relation between the physical quantities and external perturbations. $\chi_{\rho\rho}$, $\chi_{\mathbf{jj}}$ and $\chi_{\mathbf{mm}}$ are respectively the density--density, current--current and spin--spin response functions. Similarly $\chi_{\rho\mathbf{j}}$, $\chi_{\rho\mathbf{m}}$ and $\chi_{\mathbf{jm}}$ are ``mixed'' response functions. $\delta V$, $\delta \mathbf{A}$ $\delta \mathbf{B}$ represents small external perturbations. Eq. (\ref{response_general}) can be seen as the non relativistic limit of the general equation which couples the four potentials $(V,\mathbf{A})$ to the four current $(\rho,\mathbf{j})$. The magnetization enter as a result of the reduction, in the non relativistic limit, of the Dirac equation to the Schr\"{o}dinger equation\footnote{In the relativistic formulation the spin is naturally included in the four--current density. In particular the spin is always related to a spatial current which, inserted in the Maxwell equations, generates the magnetic field usually associated with a magnetic spin moment. This magnetic field is very small and, consequently, it is generally neglected in the non--reativistic limit.}.

In this thesis we will study the response to electric fields, neglecting much smaller dynamic magnetic field. We will work in the coulomb gauge to remove the coupling of the external potentials with the currents. We are then reduced to consider only the density response function in presence of a scalar potential.
\begin{equation}
\delta\rho(1) = \chi(1,1') \delta V(1')
\text{.}
\end{equation}
Here repeated primed variables are integrated. From now on we will keep this notation.
The density response function is obtained by summing over the spin variables: $\delta\rho(\mathbf{x},t)=\sum_\sigma\delta\rho(1)$; $\chi_{\rho\rho}(\mathbf{x_1},t_1;\mathbf{x_2},t_2)=\sum_{\sigma_1\sigma_2}\chi(1,2)$ and $\ \delta V^{ext}(\mathbf{x},t)=1/2\sum_{\sigma}V(1)$. The response function $\chi(1,2)$ can be related to the Green's functions observing that it involves the expectation value of four field operator, two coming from the density and two from the external potential. In the next chapter we will derive this relation. As we can argue from the number of operators the relation will involve a four point, that is a two particle, GF\index{Green's function}.

\section{The macroscopic and the microscopic world} \label{Sec:micro_macro}
While for the ground--state we have a receipt to directly obtain the expectation value of any physical quantity, for the linear response regime there is no a direct equation which relates the response function to any measurable quantity.

\begin{figure}[t]
 \centering
 \includegraphics[width=0.8\textwidth]{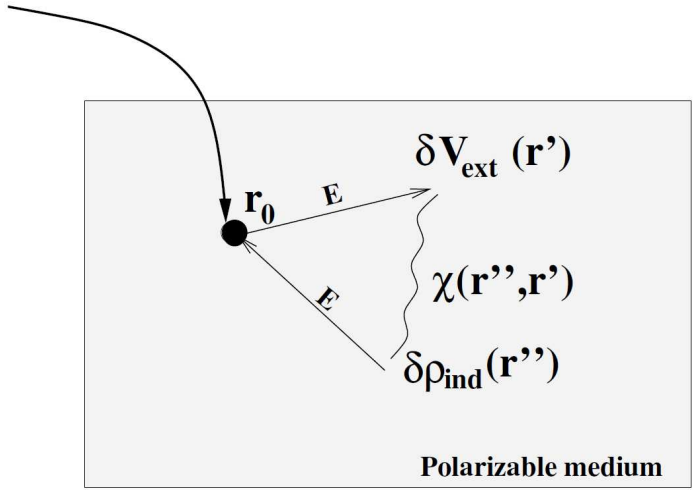}
 \caption{Polarization of a bulk system. An external charge placed in $r_0$ perturbes the system throught an external potential. This induce a variation of the density which creates an induced Electric field. From ref.~\cite{PhD_Thesis_Fabien}. }
\end{figure}

For example in an absorption experiment, the macroscopic electromagnetic fields are measured. Therefore we need relations between the microscopic and macroscopic quantities. We star from the macroscopic Maxwell equations
\begin{equation} \label{Maxwell_equations}
\begin{split}
\mathbf{\nabla}\times\mathbf{E}=-\frac{\partial \mathbf{B}}{\partial t}
\text{,}\ \ \ \ \ \ \
&\mathbf{\nabla} B= 0
\text{,}\\
\mathbf{\nabla} D= \rho^{mac}
\text{,}\ \ \ \ \ \ \
&\mathbf{\nabla}\times\mathbf{H}=\mathbf{j}^{mac}+\frac{\partial \mathbf{D}}{\partial t}
\text{.}
\end{split}
\end{equation}
Eqs. (\ref{Maxwell_equations}) together with the relations
\begin{equation}
\mathbf{D}=\epsilon_0\mathbf{E}+\mathbf{P}
\ \ \ \ \ \ \ 
\mathbf{H}=\mu_0(\mathbf{B}+\mathbf{M})
\end{equation}
completely define the macroscopic fields acting in a medium. Here $(\mathbf{E},\mathbf{B})$ are the electric and magnetic field, $(\mathbf{D},\mathbf{H})$ are the electric displacement and magnetizing field, $(\mathbf{P},\mathbf{M})$ are the polarization and magnetization of the system. In our work we will only consider linear systems with $\mu\simeq\mu_0$. This means that:
\begin{equation}
\begin{split} \label{linear_media}
\mathbf{P}&=\chi_e E
\text{,}\ \ \ \ \ \ \ \ \
\mathbf{E}=\epsilon^{-1}\mathbf{D}
\text{,}\\
\mathbf{M}&=\chi_m B
\text{,}\ \ \ \ \ \ \ \
\mathbf{B}=\mu_0\mathbf{H}
\text{.}
\end{split}
\end{equation}
Finally we consider isotropic systems where both $\chi_e$, $\chi_m$ and also $\epsilon^{-1}$ are diagonal tensors and behave as simple costants.

$\mathbf{D}$ does not depend on the internal charges of the system and it can be identified with the external electric field, $\mathbf{D}=\epsilon_0\mathbf{E}^{ext}$. On the other hand $E$ is the macroscopic total electric field, obtained as an average of the microscopic field over a region of space $V$ which is big enough to smooth the strong oscillations of the atomic fields but small enough
so that $\lambda\gg V^{1/3}$, where for periodic systems $V$ correspond to the volume of the unit cell\footnote{For absorption experiments typical energy range are around $1 / 20 eV$ so that $\lambda$ ranges from $1 \mu m$ to $50 nm$}. The macroscopic field is then defined as a function of the position of the unit--cell, let's call it $\mathbf{R}$, and if the unit cell is small enough we can consider it a function of a continuous variable $\mathbf{E}(\mathbf{R})\simeq\mathbf{E}(\mathbf{x})$.

In the long wave--length regime we can just take the microscopic version of eq. (\ref{linear_media}) and average over the volume $V$. As the external field is slowly varying over the volume we obtain
\begin{eqnarray} \label{epsilon_Mac}
\langle\mathbf{E_{micro}}\rangle&=&\langle\epsilon_{micro}^{-1}\rangle \mathbf{D}/\epsilon_0 \\
\mathbf{E}&=&\epsilon_M^{-1}\ \ \mathbf{D}/\epsilon_0 \\
\end{eqnarray}
So all we need to simulate realistic experiments is to calculate a microscopic dielectric function $\epsilon^{-1}_{micro}$ which will be averaged over the volume $V$. To be precise in absorption experiments the ratio between the total field and the applied external field is measured, that is 
\begin{equation}
\epsilon_M=\mathbf{\frac{D}{E}}=\frac{1}{\langle\epsilon_{micro}^{-1}\rangle}
\text{.}
\end{equation}
The quantity $\epsilon_M^{-1}$ is instead measured in Electron Energy Loss Spectroscopy (EELS) experiments~\cite{Schafer_Wegener} where the energy loss by a fast electron, approximated as a classical particle, while traveling through a solid is measured. The difference between the two experiments is related to the long range term of the interaction as explained in App. \ref{App:EELS vs absorption}.

For linear isotropic materials the dielectric function is a number and, therfore, it can be easely manipulated in the Maxwell equations as well as in the equation for the electromagnetic potentials. In particular, in the coulomb gauge $\mathbf{E}=-\mathbf\nabla V$, and the microspopic screening can be found relating the external and the total potential: $V^{tot}=\epsilon^{-1}V^{ext}$. We will define $\epsilon_{micro}$ in the next chapter and we will see that it is related to response function through the relation
\begin{equation}
\epsilon^{-1}_{micro}=1+w\chi
\text{.}
\end{equation}

In the case of isolated systems it is not possible to define a meaningful and finite volume $V$. Strictly speaking $V\rightarrow\infty$ and all averaged quantities goes to zero. Consequently, instead of the dielectric function, we introduce the polarizability $\alpha$ which relates the change of the dipole moment to the external field $\mathbf{D}$\footnote{Microscopic fields do not appear in the macroscopic equations as their macroscopic average goes to zero. However they are still present. Here for example microscopic fields modify the dipole.}.
\begin{equation} \label{polarizability}
\delta\mathbf{p}=\alpha\frac{\mathbf{\delta D}}{\epsilon_0}
\end{equation}

Here the long wave--length limit appear as a dipole expansion for the external applied field. As $k=\frac{2\pi}{\lambda}<<1$ the external field and the external potential can be written as:
\begin{equation}
\mathbf{E}^{ext}(\mathbf{x},t)=\mathbf{E_0}e^{i\mathbf{k}\mathbf{x}-\omega t}
                        \simeq \mathbf{E_0} e^{-i\omega t} \\
\text{,}
\end{equation}
from which 
\begin{equation}
V^{ext}(\mathbf{x},t) \simeq -\mathbf{E_0}\mathbf{x}e^{-i\omega t}
\end{equation}
Observing now that $\delta\mathbf{p}=-\langle\mathbf{x}\delta\rho\rangle$ we can construct the microscopic equivalent of eq. (\ref{polarizability}) starting from equation $\delta\rho=\chi\delta V^{ext}$.
\begin{equation}
\langle\mathbf{x}\delta\rho\rangle=\langle\mathbf{x}\chi\mathbf{x}\rangle \mathbf{E}_0  \\
\end{equation}
with $\alpha=\langle\mathbf{x}\chi\mathbf{x}\rangle$ the polarizability tensor. Although isolated systems are almost never isotropic, experimentally the absorption spectrum of molecules is often obtained in the gas phase\footnote{Many measurements for molecules are in condensed phase, either in solution or in molecular crystals. For these statistical analysis is carried out to relate molecular response properties to macroscopic response properties. Solution and solid effects are more and more frequently included in calculations of molecular absorption spectra.}. The gas by itself is isotropic and the experiment can be interpreted by using the spatial average of the polarizability: $Tr(\alpha)=(\alpha_{xx}+\alpha_{yy}+\alpha_{zz})/3$.

\chapter{Green's Function approach}    \label{chap:Green's Function approach}
In the previous chapter we have introduced the one particle Green's Function (GF\index{Green's function}) $G(1,2)$ as the natural quantity to describe in order to evaluate the ground state properties of the system. In contrast to the Many--Body wave--function, which is an $3N+1$ variables function, with $N$ the number of electrons, $G$ is a function of only two space-time coordinates, i.e. $8$ variables. Nevertheless the GF\index{Green's function} of a fully interacting many--body system is as complicate to calculate as the ground state wave--function. This can be easily seen writing down the equation of motion for $G$, starting from the relation
\begin{equation}\label{operator_time_hamiltonian}
\frac{\partial \hat{A}(t)}{\partial t}= -i [\hat{H},\hat{A}(t)]
\text{,}
\end{equation}
where $\hat{A}(t)$ is any operator. In the case of the field operators $\hat{\psi}(1)$, $\hat{\psi}^{\dag}(1)$ the Eq. (\ref{operator_time_hamiltonian}) yields an equation of motion for the GF\index{Green's function}
\begin{equation} \label{Green_EOM}
\left[i\frac{\partial}{\delta t} - H_0(1)\right]G(1,2)=\delta(1,2)
 -i U(1,3)G(1,3;2,3^+)
\text{.}
\end{equation}
Eq. (\ref{Green_EOM}) shows that the one particle GF\index{Green's function} depends on the two particles GF\index{Green's function}. In the same way the two particles GF\index{Green's function} introduces the three particles and so on. This defines an hierarchy of equations that cannot be closed exactly. Approximations are required.

There is an advantage in computing the GF\index{Green's function} instead of the many--body wave--function: the GF\index{Green's function}  is the minimal object needed to compute the expectation value of any one particle operator, while the many--body wave--function contains much more informations which often are not needed\footnote{Only for small molecules wave--function methods remain dominant as they give more accurate answers than Green's function methods. However for infinite system explicitly--correlated wave--function methods are impractical.}. In this chapter we will introduce the techniques to evaluate $G(1,2)$ and we will illustrate how the response function, the key quantity of linear response theory, can be written in terms of the GF\index{Green's function}.

\section{The time--evolution operator and the role of the interaction}
The GF\index{Green's function} approach allows, through the second quantization formalism, to isolate the complicated part of the Hamiltonian, the interaction. This can be done by writing the field operators in the ``interaction picture''. In the Schr\"{o}dinger picture the SE reads:
\begin{equation}
i \frac{\partial}{\partial t} |\Psi_S(t)\rangle = \hat{H} 
   | \Psi_S(t)\rangle
\text{.}
\end{equation}
The wave--functions are the key quantities of the theory and it is possible to define a time evolution operator $U(t)=e^{-i\hat{H}t}$. In second quantization the key quantities are the operators, so the idea of the Heisenberg picture is to describe the system evolution in terms of the operators. The wave--functions is imposed to be static:
\begin{equation} \label{Heisenberg-def}
|\Psi_H\rangle = e^{i\hat{H}t} |\Psi_S(t)\rangle 
\text{,}
\end{equation}
and accordingly the time evolution operator correspond to the identity. By demanding that 
\begin{equation}
\langle \Psi_H | \hat{O}_H | \Psi_H \rangle = \langle \Psi_S | \hat{O} | \Psi_S \rangle
\text{,}
\end{equation}
the time dependent operators are, then, constructed:
\begin{equation}
\hat{O}_H = e^{i\hat{H}t} \hat{O} e^{-i\hat{H}t}
\text{.}
\end{equation}
The interaction picture is obtained in a similar way by moving only the complicated part of the time evolution on the operators. So the Hamiltonian is split in two parts:
\begin{equation}
\hat{H}=\hat{H}_0 + \hat{H}_1
\text{,}
\end{equation}
and the wave--function is defined as
\begin{equation} \label{interazione-def}
| \Psi_I(t) \rangle = e^{i\hat{H}_0 t} | \Psi_S(t) \rangle 
\text{.}
\end{equation}
As a consequence the wave--functions are solutions of the following Schr\"{o}dinger equation
\begin{equation} \label{eq-interazione}
i \frac{\partial}{\partial t} |\Psi_I(t)\rangle 
              =\ e^{i\hat{H}_0 t} \hat{H}_1 e^{-i\hat{H}_0 t}
                  |\Psi_S(t)\rangle  
\text{.}
\end{equation}
Similarly operators are written as
\begin{equation}
\hat{O}_I = e^{i\hat{H}_0t} \hat{O} e^{-i\hat{H}_0t}
\text{.}
\end{equation}
In the interaction picture the time evolution of the wave--function can be cast in a generalized time evolution operator
\begin{equation} \label{ev-interazione}
\begin{split}
\hat{U}(t,t_0)=& \sum_{n=0}^{+\infty} \left(i\right)^n \frac{1}{n!}
                \int_{t_0}^{t}dt_1 \ldots \int_{t_0}^{t}dt_n\ T
                [\hat{H}_1(t_1)] \ldots \hat{H}_1(t_n)]   \\
              =&\ T \left[ e^{-i\int_{t_0}^{t}\hat{H}_1(t')dt' } \right]
\end{split}
\end{equation}
defined in such a way that
\begin{equation}\label{Psi_evolution}
|\Psi_I(t)\rangle=U_I(t,t_0)|\Psi(t_0)\rangle
\text{.}
\end{equation}

\section{Equilibrium properties}
\subsection*{The one particle propagator}
The interaction picture is a practical starting point thanks to the Gell-Mann and Low theorem\cite{Fetter_Walecka}. The Hamiltonian
\begin{equation}
\hat{H}_{\epsilon}(t)=\hat{H}_0+e^{-\epsilon|t|}\hat{H}_1
\end{equation}
is considered
By using Eq. (\ref{Psi_evolution}) the time evolution of any eigenstate can be written as
\begin{equation} \label{GellMann_Low}
| \Psi_I(t) \rangle = 
\frac{\hat{U}_{\varepsilon}(t,t_0)|\Phi^\epsilon_I\rangle}
    {\langle \Phi^\epsilon_I | \hat{U}_{\varepsilon}(t,t_0)|\Phi^\epsilon_I \rangle}
\text{,}
\end{equation}
where $|\Phi^\epsilon_I\rangle$ is the eigenstate at time $t=t_0$. The Gell-Mann and Low Theorem states that in the limit $t_0\rightarrow - \infty$ and $\epsilon\rightarrow 0$~\footnote{It is important to take the limit in this order in order to obtain a meaninful result.}\ \ $\Phi^\epsilon_I$ reduce to an eigenstate of $\hat{H}_0+\hat{H}_1$. A relation among the eigenstates of $H_0$ and the eigenstate of $\hat{H}_0+\hat{H}_1$ is then established. Finally assuming that the not interacting ground state slowly evolves to the ground state of the interacting system we obtain~\cite{Fetter_Walecka} $<<$\emph{The most useful result of quantum field theory}$>>$ :
\begin{equation} \label{Green_series}
\begin{split}
iG(1,2) =& 
 \sum_{n=0}^{+\infty} \frac{(-i)^n}{n!}
 \int_{-\infty}^{+\infty}dt_{1'} \ldots dt_{n'}   \\
 &  \phantom{\sum_{n=0}^{+\infty} -i^n}
        \frac{\langle \Phi_0 | \hat{T} \left[\hat{H}_1(t_{1'}) \ldots \hat{H}_1(t_{n'})
        \hat{\psi}^{\dag}_I(1) \hat{\psi}_I(2) \right] | \Phi_0 \rangle}
      {\langle \Phi_0 | \hat{S} | \Phi_0 \rangle}    \\
       =& \frac{\langle \Phi_0 | \hat{T} \left[\hat{S}
        \hat{\psi}^{\dag}_I(1) \hat{\psi}_I(2) \right] | \Phi_0 \rangle}
      {\langle \Phi_0 | \hat{S} | \Phi_0 \rangle}
\text{,}
\end{split}
\end{equation}
with
\begin{equation}
\hat{S}=\sum_{n=0}^{+\infty}\frac{(-i)^n}{n!}\int_{-\infty}^{+\infty}dt_{1'} \ldots dt_{n'}
        \hat{T} \left[\hat{H}_1(t_{1'}) \ldots \hat{H}_1(t_{n'}) \right]
\text{.}
\end{equation}
Eq. (\ref{Green_series}) is a perturbation expansion in the interaction which involve the expectation values of many field operators. Any term of order $n$ involves $n$-times the interaction operator $\hat{H}_1=\hat{w}(1,2)$. Thanks to the Wick theorem \cite{Fetter_Walecka} we can express each term in the series in terms of the not interacting GF\index{Green's function}
\begin{equation}
ig(1,2)=\frac{\langle \Phi_0 | \hat{T} \left[
        \hat{\psi}^{\dag}_I(1) \hat{\psi}_I(2) \right] | \Phi_0 \rangle}
      {\langle \Phi_0 | \Phi_0 \rangle}
\text{.}
\end{equation}
Any term in the expansion starts contains a not interacting GF\index{Green's function} which start from point $2$, $g(\ ,2)$, and a not interacting GF\index{Green's function} which ends up at point $1$, $g(1,\ )$, while the presence of the interaction affect the propagation in between. So the interacting GF\index{Green's function} can be expressed as
\begin{equation} \label{Hedin0_Green}
G(1,2) = g(1,2) + g(1,1') \Sigma_H(1',2') g(2',2)   
\text{,}
\end{equation}
where we have introduced the $\Sigma_H(1,2)$ self--energy\index{self--energy} function. This functions includes terms which are repeated an infinite number of times. By introducing a reduced self--energy\index{self--energy} $\Sigma_H^{\star}(1,2)$ which satisfy the equation
\begin{equation}
\Sigma_H(1,2)= \Sigma_H^{\star}(1,2)+\Sigma_H^{\star}(1,1')g(2',2')\Sigma_H(2',2)
\text{,}
\end{equation}
Eq. (\ref{Hedin0_Green}) then becomes
\begin{equation} \label{Hedin_Green}
G(1,2) = g(1,2) + g(1,1') \Sigma_H^{\star}(1',2') G(2',2)   
\text{.}
\end{equation}
Eq. (\ref{Green_series}) and Eq. (\ref{Hedin_Green}) represent the starting point for the development of approximations to the many--body problem. In particular Eq. (\ref{Green_series}) can be approximated by truncating the series expansion to finite order. This is a reasonable approximation for systems where correlations effects are less important and the interaction can be treated as a small perturbation. For more correlated system, and in general for solids, the expansion in the bare interaction in not meaningful as screening effects are dominant. As discussed in the next sections in this regime the electrons are screened and the interaction is considerably weaker than in isolated systems. Eq. (\ref{Hedin_Green}) can be used, choosing approximations for the self-energy $\Sigma_H^{\star}$, one of the most common being the $GW$ approximation, which is the first order term in a possible expansion for the GF\index{Green's function} in powers of the screened interaction $W$.

Finally inserting Eq. (\ref{Hedin_Green}) in Eq. (\ref{Green_EOM}) the EOM for the GF\index{Green's function} reads
\begin{equation} \label{Green_EOM_Sigma}
\left[i\frac{\partial}{\delta t} - \hat{H}_0(1)\right]G(1,2)=\delta(1,2)
 + \Sigma_H^{\star}(1,1') G(1',2)
\text{.}
\end{equation}

\subsection*{The Hedin's equations (I): bare interaction}
To device useful approximations for the self--energy\index{self--energy} we follow the method developed by Lars Hedin \cite{Hedin}. An expression for the self--energy\index{self--energy} can be obtained by studying how the full GF\index{Green's function} react to a fictitious external potential $\delta\varphi(1)$. This potential is then set to zero at the end of the derivation to recover the ground state GF\index{Green's function}.

In the following we include the interaction in the first part of the Hamiltonian $\hat{H}=\hat{H}_0+\hat{w}(1,2)$, while the potential $\hat{H}_1(1)=\hat{\psi}^\dag(1)\hat{\psi}(1)\delta\varphi(1)$ is included in the $\hat{S}$ operator:
\begin{equation}
iG_{\varphi}(1,2) = \frac{\langle \Psi_0 | T \left[\hat{S}
        \hat{\psi}^{\dag}(1) \hat{\psi}(2) \right] | \Psi_0 \rangle}
               {\langle \Psi_0 | \hat{S} | \Psi_0 \rangle}
\text{.}
\end{equation}
The variation of the GF\index{Green's function} with respect to the potential $\delta\varphi$ reads
\begin{equation} \label{Green_1p_and_2p}
\frac{\delta G_{\varphi}(1,2)}{\delta \varphi(3)} = -G_{\varphi}(1,3;2,3^+)
+G_{\varphi}(1,2)G_{\varphi}(3,3^+)
\text{.}
\end{equation}
Eq. (\ref{Green_1p_and_2p}) can be used in the Eq. (\ref{Green_EOM}), to express the two particle GF in terms of $G(1,2)$:
\begin{multline}
\left[i\frac{\delta}{\delta t}-\hat{H}_0(1)\right] G(1,2)
      +i\ w(1,1')G(1',1'^+)G(1,2) \\
 -i\ w(1,1')\frac{\delta G(1,2)}{\delta \varphi(1')}= \delta(1,2)
\text{.}
\end{multline}
Using the equality
\begin{equation} \label{der_identity}
\frac{\delta G(1,2)}{\delta \varphi(3)} =
-G(1,1')\frac{\delta G^{-1}(1',2')}{\delta \varphi(3)}G(2',2)
\text{,}
\end{equation}
we define the vertex\index{vertex} function
\begin{equation} \label{Hedin0_Vertex}
\Gamma(1,2;3)=-\frac{\delta G^{-1}(1,2)}{\delta \varphi(3)}
\text{.}
\end{equation}
By using Eq. (\ref{Green_EOM_Sigma}) and Eq. (\ref{Hedin0_Vertex}) we obtain the following expression for the self--energy\index{self--energy}:
\begin{equation}
\begin{split}
&\Sigma_H^{\star}(1,2)= -i\delta(1,2)\ w(1,1')G(1',1'^+)+ \\
&\phantom{\Sigma^{\star}(1,2)= -i\delta(1,2)}
-i\ w(1,2')G(1,1')\frac{\delta G^{-1}(1',2)}{\delta \varphi(3')} \\
&\phantom{\Sigma^{\star}(1,2)} =\delta(1,2)v_H(1) + i\ w(1,3')G(1,1')\Gamma(1',2;3')
\text{.}
\end{split}
\end{equation}
Using the Dyson equation for $G(1,2)$ to compute the functional derivative of $G^{-1}$ we get
\begin{flalign}
\label{Hedin bare equations}
&G(1,2) = g_H(1,2) + g_H(1,1') \Sigma^{\star}(1',2') G(2',2)\text{,}                         \\
&\Sigma^{\star}(1,2) = i G(2',1) w(1,2') \Gamma(2,2';1')\text{,}\label{Hedin0_SelfEnergy}    \\
&\Gamma(1,2;3)=\delta(1,2) \delta(1,3)+   \nonumber                                      \\
&\phantom{\Gamma(1,2;3)=\delta(1,2)}
              \frac{\delta \Sigma^{\star}(1,2)}{\delta G(1',2')} 
               G(2',3') G(4',1') \Gamma(3',4';3)
\text{.}
\end{flalign}
The Hartree potential is included in the definition of the bare GF\index{Green's function}
\begin{equation}
g_H^{-1}(1,2)=g^{-1}(1,2)+v_H(1)\delta(1,2)
\text{.}
\end{equation}

\subsection*{The Hedin equations (II): screened interaction}

Lars Hedin introduced a fundamental breakthrough by realizing that, instead of the bare Coulomb interaction $w(1,2)$, one should consider a perturbative expansion in the screened potential $W(1,2)$. In many--body systems, indeed, the interaction is always screened. Formally we introduce the potential
\begin{equation} \label{pot-eff}
\delta V(1)=\delta\varphi(1)+\delta v_H(1)
\end{equation}
defined as the sum of the fictitious potential $\varphi$ and its classical screening. This is the classical screened potential because it is obtained by neglecting the changes in the quantistic self-energy. Considering linear and isotropic systems the microscopic dielectric function can be defined as the relation between the total potential $\varphi$ and the external potential (see Ch. (\ref{chap:Many Body Systems}), Sec. (\ref{Sec:micro_macro})):
\begin{equation} \label{epsilon_micro}
\epsilon^{-1}(1,2)=\ \frac{\delta V(1)}{\delta \varphi(2)}=\ \delta(1,2)+ \ w(1,1')\Pi(1',2)
\text{,}
\end{equation}
where $\Pi(1,2)$ is the response function
\begin{equation} \label{pi-def}
\Pi(1,2)=-i\frac{\delta G(1,1^+)}{\delta \varphi(2)}
\text{.}
\end{equation}
The screened interaction is then
\begin{equation} \label{Hedin0_Interaction}
\begin{split}
W(1,2) =& \ \epsilon^{-1}(1,1') w(1',2)       \\
       =&\ w(1,2) + \ w(1,1')\Pi(1',2')w(2',2)
\text{.}
\end{split}
\end{equation}
The polarization function can be interpreted as the self-energy of the screened interaction $W$ and, similarly to the case of $\Sigma$, it can be reduced
\begin{equation} \label{Reduced_polarization}
\begin{split}
\Pi(1,2)=&\ -i\ \frac{\delta G(1,1^+)}{\delta V(2')} \frac{\delta V(2')}{\delta \varphi(2)} \\
        =&\ \Pi^{\star}(1,2)+ \Pi^{\star}(1,1')w(1',2')\Pi(2',2)
\hspace{0.5 cm} \text{,}
\end{split}
\end{equation}
to obtain
\begin{equation} \label{Hedin_Interaction}
W(1,2)= w(1,2)+ \ w(1,1')\Pi^{\star}(1',2')W(2',2)
\end{equation}
The equation for the reduced polarization can be obtained From Eq. (\ref{der_identity}) 
\begin{equation} \label{Hedin-polarizzazione}
\Pi^{\star}(1,2) = -i\ G(1',1)G(1,2')\Gamma^{\star}(1',2';2)
\text{.}
\end{equation}
where we have defined a reduced vertex\index{vertex} function
\begin{equation} \label{Hedin_Vertex}
\begin{split}
\Gamma^{\star}(1,2;3)&=\frac{G^{-1}(1,2)}{\delta V(3)}   \\
                     &=\delta(1,2)\delta(1,3)+\frac{\Sigma^{\star}(1,2)}{\delta V(3)}   \\
                     &=\delta(1,2)\delta(1,3)+                                          \\
                     &\phantom{\delta(1,2)\delta(1,3)}
                      \frac{\delta\Sigma^{\star}(1,2)}{\delta G(1',2')}
                      G(1',3')G(4',2')\Gamma^{\star}(3',4';3)               
\text{.}
\end{split}
\end{equation}
The self-energy can be expressed in terms of the new quantities in order to obtain a closed set of equations, the Hedin equations:
\begin{figure}[t]
 \centering
 \includegraphics[width=0.95\textwidth]{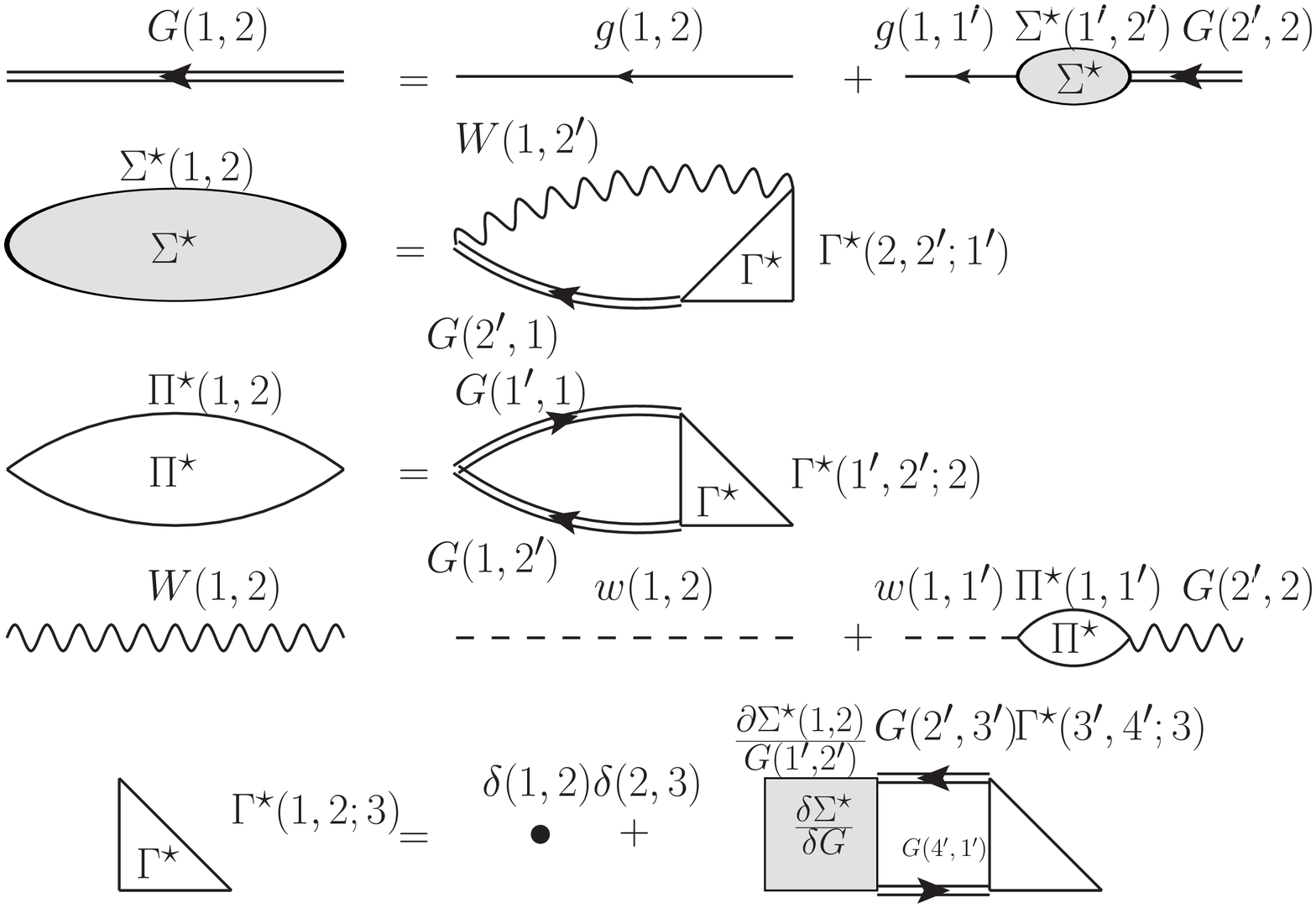}
 \caption{Diagrammatic representation of the Hedin Equations.}
\end{figure}
\begin{flalign}
\label{Hedin's equations}
&G(1,2) = g_H(1,2) + g_H(1,1') \Sigma^{\star}(1',2') G(2',2)      \text{,} \\
&W(1,2) = w(1,2) + w(1,1') \Pi^{\star}(1',2') W(2',2)             \text{,} \\
&\Pi^{\star}(1,2) = -i\ G(1,1')G(2',1)\Gamma^{\star}(1',2';2)     \text{,} \\
&\Sigma^{\star}(1,2) = i G(2',1) W(1,2') \Gamma^{\star}(2,2';1')  \text{,} \\
&\Gamma^{\star}(1,2;3)=\delta(1,2) \delta(1,3)+   \nonumber       \\
&\phantom{\Gamma(1,2;3)=\delta(1,2)}
              \frac{\delta \Sigma^{\star}(1,2)}{\delta G(1',2')} 
               G(2',3') G(4',1') \Gamma^{\star}(3',4';3)
\text{.}
\end{flalign}
This set of equations is exact. Indeed their solution is not easier than the solution of the many--body problem. Nevertheless the Hedin's equation offer a convenient starting point to develop approximations.
\begin{figure}[t]
 \centering
 \includegraphics[width=0.4\textwidth]{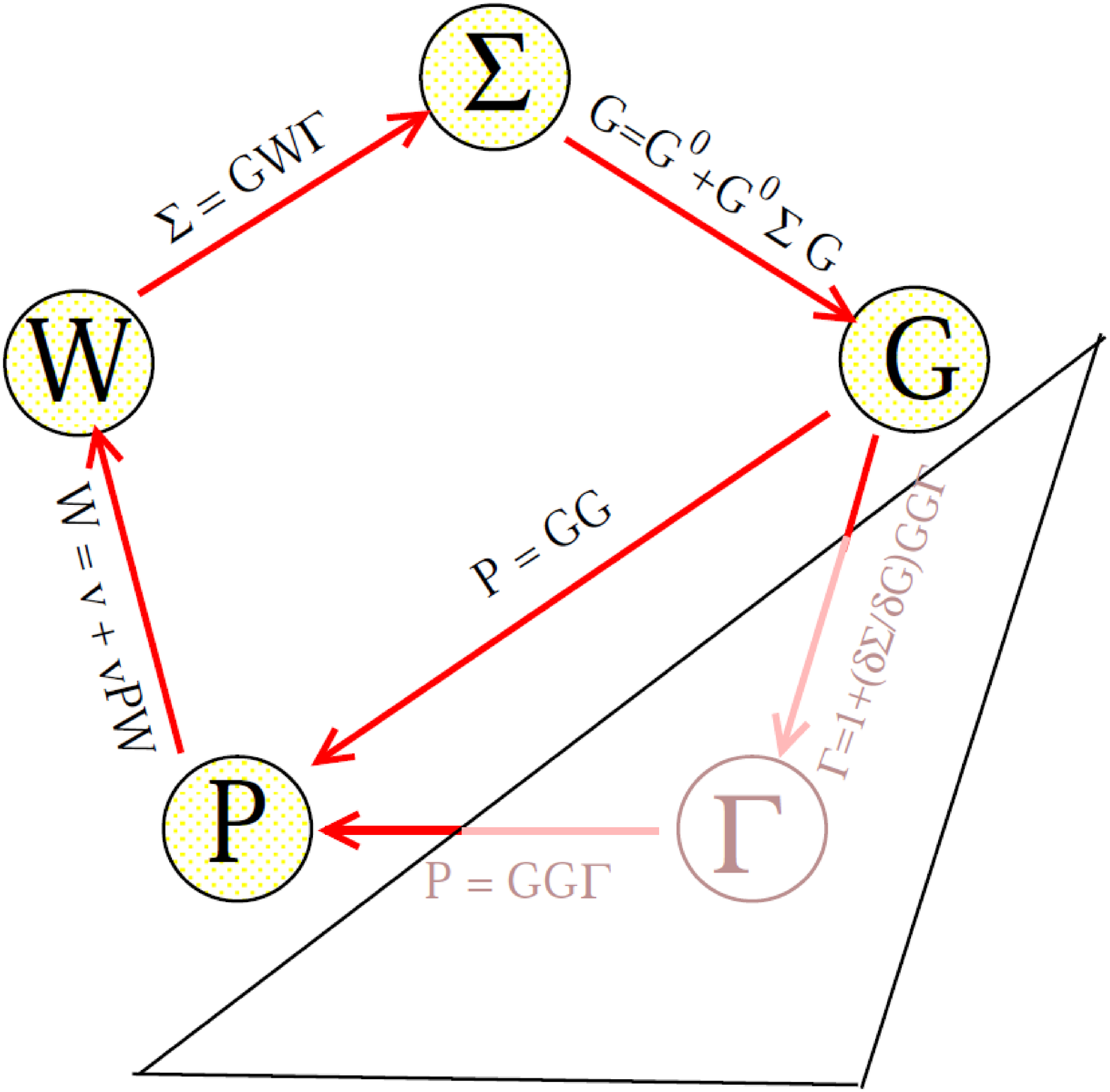}
 \caption{The Hedin Pentagon and the GW approximation. From ref.~\cite{PhD_Thesis_Sottile}.}
\end{figure}
One of the most common approximation to this set of equation is to set $\Gamma^{\star}(1;2,3)=\delta(1,2) \delta(1,3)$, that is to drop the complicate functional derivative $\frac{\delta \Sigma^{\star}(2,3)}{\delta G(6,7)}$. This approximation is known as $GW$ approximation. The $GW$ approximation is simply the first order approximation in the expansion of the self--energy\index{self--energy} in powers of the screened interaction\cite{Hedin} and goes one step beyond the Hartree and the Hartree Fock approximations. The infinite order resummation of Feynman diagrams is already included in $\Sigma$ which is then iterated to construct the GF\index{Green's function}. Indeed to Dyson equation are solved, one for $w(1,2)$ and another for $G(1,2)$.

\section{The quasiparticle concept}

Starting from Eq. (\ref{Green_EOM_Sigma}) it is possible to introduce the quasiparticle (QP) concept, which was first introduced by Landau in the description of Fermi liquids\cite{Landau_Fermi_Liquid}. First we Fourier transform eq. (\ref{Green_EOM_Sigma}) 
\begin{equation}
\left( \omega-H_0(\mathbf{1})-v_H(\mathbf{1}) \right) G(\mathbf{1},\mathbf{2};\omega)-\Sigma(\mathbf{1},\mathbf{1}';\omega)G(\mathbf{1}',\mathbf{2};\omega)=\delta(\mathbf{1},\mathbf{2})
\text{,}
\end{equation}
then perform an analytic continuation of the frequency variable to the complex plane, $\omega\rightarrow z$. A formal solution to the equation can be obtained by using a biorthonormal representation of the GF\index{Green's function}
\begin{equation}
G(\mathbf{1},\mathbf{2};z)=\sum_\lambda \frac{\Psi_\lambda(\mathbf{1},z)\tilde{\Psi}_\lambda(\mathbf{2},z)}{z-E_\lambda(z)}
\text{.}
\end{equation}
The right and left wave--functions $\Psi_\lambda(\mathbf{1},z)$ and $\tilde{\Psi}_\lambda(\mathbf{1},z)$ satisfy the equations
\begin{equation}
\begin{split}
\left(H_0(\mathbf{1})+v_H(\mathbf{1}) \right) \Psi_\lambda(\mathbf{1},z)
        +\Sigma^{\star}(\mathbf{1},\mathbf{1}';z)\Psi_\lambda(\mathbf{1},z)=E_\lambda(z)\Psi_\lambda(\mathbf{1},z) \text{,}   \\
\left(H_0(\mathbf{1})+v_H(\mathbf{1}) \right) \tilde{\Psi}_\lambda(\mathbf{1},z)
        +\Sigma^{\star\dag}(\mathbf{1},\mathbf{1}';z)\tilde{\Psi}_\lambda(\mathbf{1},z)=E_\lambda(z)\tilde{\Psi}_\lambda(\mathbf{1},z)
\text{.}
\end{split}
\end{equation}

The QP concept can be introduced assuming that the dominant contribution to the GF\index{Green's function} comes from the fixed points of $E_\lambda(z)$:
\begin{equation}\label{QP_definition}
E_i^{QP}=E_{\lambda}(E_i^{QP})
\text{.}
\end{equation}
The QP equation is then defined as
\begin{equation}\label{QP_equation}
\left(H_0(\mathbf{1})+v_H(\mathbf{1}) \right) \Psi_i(\mathbf{1})
        +\Sigma^{\star}(\mathbf{1},\mathbf{1}';E_i^{QP})\Psi_i(\mathbf{1})=E_i^{QP}\Psi_i(\mathbf{1})   \\
\text{.}
\end{equation}

\subsection*{The renormalization factor}
In practice the solution of eq. (\ref{QP_equation}) is usually obtained as a first order correction to some mean field theory, like the Hartree approximation, that is eq. (\ref{QP_equation}) with $\Sigma=0$. The solution of Eq. (\ref{QP_definition}) is the approximated by
\begin{equation}\label{First_order_QP}
E_i^{QP}\simeq\epsilon^H_j+\langle j | \Sigma(E_i^{QP}) | j \rangle
\text{.}
\end{equation}
Eq. (\ref{First_order_QP}) is solved by linearizing the frequency dependence of the Self-Energy around the Hartree poles $\epsilon_H$
\begin{equation} \label{linear_freq_depend}
\langle \Sigma(E^{QP}) \rangle \simeq
    \langle \Sigma(\epsilon^H) \rangle +
    \langle \frac{\partial\Sigma(\omega)}{\partial\omega}\bigg|_{\omega=\epsilon} \rangle (E^{QP}-\epsilon^{H})
\text{.}
\end{equation} 
Defining the renormalization factor
\begin{equation}
Z=\left(1-\langle\frac{\partial\Sigma(\omega)}{\partial\omega}\bigg|_{\omega=\epsilon}\rangle\right)^{-1}
\text{,}
\end{equation}
we obtain
\begin{equation}\label{First_order_QP_linear}
E_j^{QP}\simeq\epsilon^H_j+Z\langle \Sigma(E_j^{QP}) \rangle
\text{.}
\end{equation}
\begin{figure}[t] 
 \centering
 \includegraphics[width=0.8\textwidth]{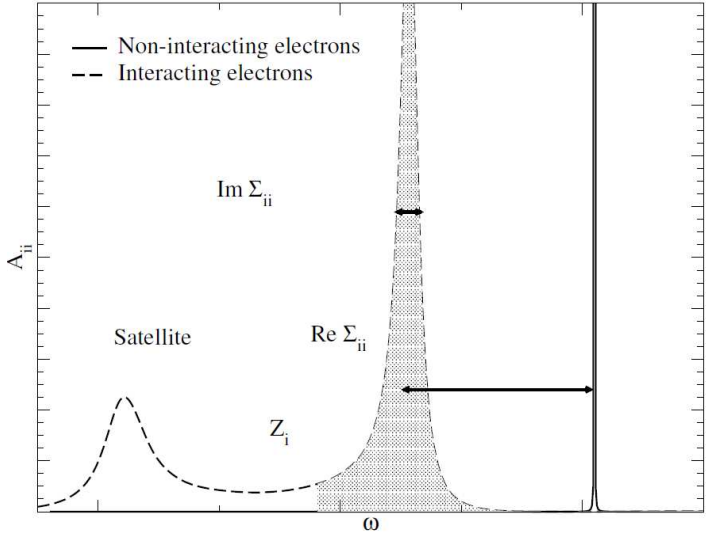}
 \caption{Quasi Particle representation of the Green's Function. The pole of the not-interacting GF\index{Green's function} is shifted and broadened. QP are connected to the coherent part of $\Sigma$. Additional peaks in the GF\index{Green's function} can appear due to the incoherent contributions (plasmons, polarons, resonances,...). These additional peaks cannot be interpreted in a IP theory. From ref.~\cite{PhD_Thesis_Sottile}}
 \label{fig:QP_renormalization}
\end{figure}

In general the number of GF\index{Green's function}'s poles $E_i^{QP}$ is larger then the number of IP states. Indeed only if the self--energy\index{self--energy} frequency dependence is linearized according to Eq. (\ref{linear_freq_depend}) the GF\index{Green's function} poles coincide with the QP states. Such an approximation fails in the description of satellites, as the one shown in Fig. (\ref{fig:QP_renormalization}), which appear as extra poles in the spectral function $A(\omega)$ defined as
\begin{equation}
\begin{split}
\langle A(\omega)\rangle &=\frac{1}{\pi} sign(\mu-\omega)\langle Img[G(\omega)] \rangle \\
                    &\simeq\frac{1}{\pi} \frac{|Img[\Sigma_i(\omega)]|}
                         {(\omega-\epsilon_i-Re[\Sigma_i(\omega)])^2+(Im[\Sigma_i(\omega)])^2}
\text{.}
\end{split}
\end{equation}
In the second line the non diagonal part of the self--energy\index{self--energy} is neglected. The extra poles, in the quantum chemistry language, can be described as ``multiple excitations''\cite{Cederbaum1}. Eq. (\ref{First_order_QP}) is a good approximation only if the starting wave--functions are closed enough to the QP wave--functions. This is usually not the case for the Hartree theory and for this reason in practice QP corrections are calculated on top of Density--Functional Theory (DFT\index{DFT}) calculations. We will introduce DFT\index{DFT} in the next chapter.

We will find a similar distinction between QPs and satellites (extra-poles) in the case of neutral-excitations. In Sec. (\ref{Sec:BSE}) the Bethe Salpeter Equation (BSE), a Dyson--like equation to compute neutral excitations, will be introduced. Like the self--energy\index{self--energy}, $\Sigma$, the kernel of the BSE, $\Xi$, will be able to produce extra poles; however while $\Sigma$ is usually almost diagonal, $\Xi$ is not. This means that IP poles of the GF\index{Green's function} are shifted to QP and all incoherent peaks are easily identified as satellites, while in the BSE IP electrons-holes transition are strongly mixed and the resulting spectrum is usually broaden, making difficult to identify possible satellites. This argument will be subject of the Part. II of the present thesis on double excitations.

\subsection*{The Lehmann representation}
The biorthonormal representation of the GF\index{Green's function} in the complex plane is linked to the Lehmann representation of the GF\index{Green's function} on the real frequencies axis.
\begin{figure}[t]
 \centering
 \includegraphics[width=0.8\textwidth]{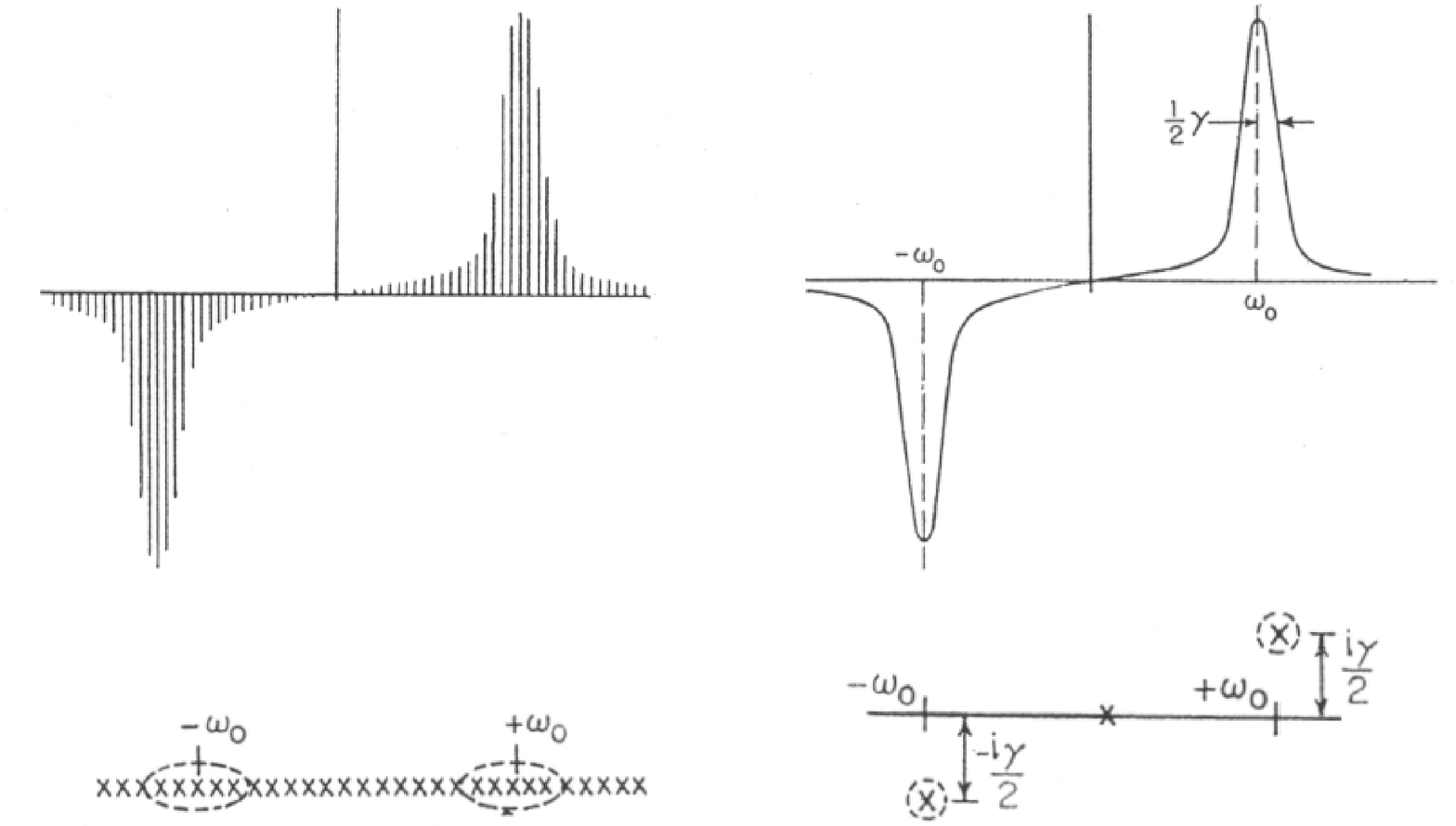}
 \caption{Lehmann representation, on the left, compared with quasiparticle representation, on the right. From ref.~\cite{PhD_Thesis_Gatti}}
 \label{fig:Lehmann_Vs_QP}
\end{figure}
This can be obtained using the full many--body wave--function and the identity
\begin{equation}\label{ManyBody_Identity}
\hat{I}=\sum_{J} |\Psi_J \rangle \langle \Psi_J|
\text{,}
\end{equation}
with $\Psi_J$ the exact many--body eigenstate. By inserting Eq. (\ref{ManyBody_Identity}) in the GF\index{Green's function} definition, the Fourier transform of $G$ yields
\begin{multline} \label{Lehmann_Green}
G(\mathbf{1},\mathbf{2};\omega)=\sum_J
                   \frac{\langle\Psi_0|
                          \hat{\psi}^{\dag}(\mathbf{1})|\Psi_J\rangle
                          \langle\Psi_J|\hat{\psi}(\mathbf{2})|\Psi_0\rangle}
                        {\omega-\omega_J+i\eta}+    \\
                   -\frac{\langle\Psi_0|\hat{\psi}(\mathbf{2})|\Psi_J\rangle
                          \langle\Psi_J|\hat{\psi}^{\dag}(\mathbf{1})|\Psi_0\rangle}
                        {\omega-\omega_J-i\eta}
\text{.}
\end{multline}
Eq. (\ref{Lehmann_Green}) defined the Lehmann representation. This is an exact representation expressed on the real axis instead of the complex plane. The Lehmann and the biorthonormal representations are closely linked, as exemplified in Fig. (\ref{fig:Lehmann_Vs_QP}). The concept represented in the figure emerge if we consider the thermodynamical limit of the Lehmann representation
\begin{eqnarray} \label{Lehmann_Green_term_limit}
G(\mathbf{1},\mathbf{2};\omega)&=&\sum_J
                   \frac{f_J(\mathbf{1})f^*_J(\mathbf{2})}{\omega-\omega_J+i\eta}+    
                  -\frac{g_J(\mathbf{1})g^*_J(\mathbf{2})}{\omega+\omega_J-i\eta}   \\
                               &\simeq& \int dx \frac{A(\mathbf{1,2}|x)}{\omega-x}
\end{eqnarray}
where $f_J(\mathbf{1})=\langle\Psi_0|\hat{\psi}^{\dag}(\mathbf{1})|\Psi_J\rangle$,
$g_J(\mathbf{1})=\langle\Psi_0|\hat{\psi}(\mathbf{2})|\Psi_J\rangle$ and
\begin{equation}
A(x)=\sum_J \bigg( f_J(\mathbf{1})f^*_J(\mathbf{2})\theta(\mu-E_J) +
                  g_J(\mathbf{1})g^*_J(\mathbf{2})\theta(E_J-\mu)\bigg)\delta(E_J-x)
\end{equation}
is another expression for the spectral function previously defined. Eq. (\ref{Lehmann_Green_term_limit}) is an integral representation of the GF\index{Green's function} which can be analytically continued: $\omega\rightarrow z$. In complex analysis a branch cut, here the series of poles in the Lehmann representation of the GF\index{Green's function} in the thermodynamical limit, can be expressed as the integral of a complex pole, here the poles of the spectral function.

In this respect the QP concept well describes the power of MBPT\index{MBPT}. Indeed the QP poles are not eigenstate of the Hamiltonian being formed by a macroscopically large number of almost degenerate eigenstates of $\hat{H}$. QPs have a finite lifetime. 

An infinite series of closely lying poles which appear as a branch-cut in the Lehmann representation is replaced with a single complex pole in the QP picture. Nevertheless the QP representation fails to describe the continuous part of the GF\index{Green's function}. For other details on the QP concept see App.~\ref{App:On the quasiparticle concept}

\section{Neutral excitations: the Bethe--Salpeter equation} \label{Sec:BSE}
Once we have computed the one particle GF\index{Green's function} we can evaluate any physical proprieties of the system at rest. Still we have no information about how the system would react if disturbed by a time dependent external potential. In principle it is possible to calculate the perturbed time dependent one particle GF\index{Green's function}, but in the case of weak perturbations we can use the linear response. In this approach the key quantity is the linear response function which can be calculated in terms of the GF\index{Green's function}. This is the subject of the present section. The Lehmann representation shows that the linear response function describes the neutral excitations of the system where two or more particles are involved. Therefore the two particle GF\index{Green's function} will naturally emerge as the starting point to compute linear response properties.

We start from the wave--function evolution induced by a time dependent perturbation
\begin{equation}
\begin{split}
|\Psi_S(t)\rangle =&\ e^{-i\hat{H}t/\hbar}\hat{U}(t,t_0) |\Psi(t_0)\rangle   \\
                  =&\ e^{-i\hat{H}t/\hbar}|\Psi(t_0)\rangle +
                      e^{-i\hat{H}t/\hbar}\int_{t_0}^t dt'\ \delta\hat{V}_{\hat{H}}^{ext}(t')
                        |\Psi(t_0)\rangle + ...
\text{.}
\end{split}
\end{equation}
Then the expectation value of any operator can be written as
\begin{equation} \label{linear_variation_operator}
\langle \hat{O}(t) \rangle = \langle \Psi(0) | \hat{O}_{\hat{H}}(t) | \Psi(0) \rangle +
                             \int_{t_0}^t dt' \langle \Psi(0) |[\hat{O}_{\hat{H}}(t),
                             \delta\hat{V}_{\hat{H}}^{ext}(t') | \Psi(0) \rangle + ...
\text{.}
\end{equation}
We consider perturbations of the form
\begin{equation}\label{Eqn:V_ext}
\delta \hat{V}^{ext}= \delta V^{ext}(1) \hat{\rho}(1)
\hspace{0.5 cm} \text{,}
\end{equation}
and by using the density operator Eq. (\ref{linear_variation_operator}) becomes:
\begin{equation}
\rho(1)=\rho_0(1)+ \Theta(t-t') \delta V^{ext}(1')
\langle \Psi(0) |\left[\hat{\rho}(1),
                         \hat{\rho}(1')\right]|\Psi(0) \rangle
\text{.}
\end{equation}
The retarded response function is thus defined as
\begin{equation}
i\chi^R(1,2)=\langle \Psi(0) |\left[\hat{\rho}(1),
             \hat{\rho}(2)\right] |\Psi(0) \rangle \Theta(t_1-t_2)
\text{.}
\end{equation}
Formally this is different from the polarization function $\Pi$ that is a $T$-ordered quantity while $\chi$ is a retarded quantity. However it can be proven through the Lehmann representation that the two are equivalent a part from a small shift of the poles along the imaginary axis. The $T$-ordered response function is:
\begin{equation}
i\Pi(1,2)=\langle
\Psi(0) |T\left[\delta\hat{\rho}(1)
                \delta\hat{\rho}(2)\right]|\Psi(0) \rangle
\hspace{0.5 cm} \text{,}
\end{equation}
where the operator $\delta\hat{\rho}=\hat{\rho}-\langle\hat{\rho}\rangle_0$ is used to ensure that the terms which involves the expectation value on the ground state are canceled, as for the retarded function.

We have already evaluated $\Pi$ by solving the Hedin equations. However in contrast to the GF\index{Green's function}, the $GW$ approximation, that is $\Gamma=1$, is not a good approximation for the response function. The reason is that we are looking to different physical processes. In order to find a better approximation for $\Pi$ we use the Hedin equation for the vertex\index{vertex}
\begin{multline}
\Gamma(1,2;3)=\delta(1,2)\delta(1,3)+
                \frac{\delta}{\delta G(1',2')}
                \big[\delta(1,2)v_H(2)+\Sigma^{\star}(1,2)\big]
                \\
                G(2',3')G(4',1')\Gamma(3',4';3)
\text{,}
\end{multline}
to obtain an equation for the Polarization
\begin{multline} \label{Dyson_try_Pi}
\Pi(1,2)= -iG(1,2)G(2,1) +
          -iG(1,1')G(2',1) \\
          \frac{\delta}{\delta G(3',4')}
          \big[\delta(1',2')v_H(2')+\Sigma^{\star}(1',2')\big]
          G(4',5')G(6',3')\Gamma(5',6';2)
\text{.}
\end{multline}
Eq. (\ref{Dyson_try_Pi}) has the structure of a Dyson equation of the form,
\begin{equation}\label{almost_Dyson_Pi}
\Pi= \Pi_0 + \Pi_0 \Xi \Pi
\text{,}
\end{equation}
with $\Pi_0=-iGG$. However the combination $-iGG\Gamma$ factor appearing in r.h.s of Eq. (\ref{Dyson_try_Pi}) is not a two point quantity. The reason is that the kernel, 
\begin{equation}
\Xi(1,2;3,4)=i\frac{\delta}{\delta G(3,4)}
           \big[\delta(1,2)v_H(2)+\Sigma^{\star}(1,2)\big]
\text{,}
\end{equation}
is a four point quantity. Eq. (\ref{almost_Dyson_Pi}) can be closed using the extended space of two--particles GF\index{Green's function}s, defined as 
\begin{equation}
iL(1,2;3,4)=-G_2(1,2;3,4) + G(1,3)G(2,4)
\text{.}
\end{equation}
The polarization is written in terms of $L$ as
\begin{equation}
L(1,2;1,2)=\Pi(1,2)
\text{,}
\end{equation}

Looking at the derivation of the Hedin equations we see that
\begin{equation}
iL(1,2;3,4)=\frac{\delta G(1,3)}{\delta \varphi(4,2)}
\end{equation}
and using Eq. (\ref{der_identity}) and the definition of the vertex\index{vertex} function we finally obtain
\begin{equation} \label{Bethe-Salpeter}
\begin{split}
L(1,2;3,4)&=-iG(1,4)G(2,3)-iG(1,1')G(2',2) \\
&\phantom{G(1,4)G(2)}
\ i\frac{\delta \big(\delta(1',2')v_H(1')+\Sigma^{\star}(1',2')\big)}{\delta G(3',4')}
(-i)\frac{\delta G(3',3)}{\delta \varphi(4,4')}  \\
&=L_0(1,2;3,4)+L_0(1,2;1',2')\Xi(1',2';3',4')L(3',4';3,4)
\text{.}
\end{split}
\end{equation}
Eq. (\ref{Bethe-Salpeter}) is the Bethe Salpeter Equation (BSE). It's then possible to construct approximations to the response function, via the kernel $\Xi$. The most common approximation is derived from the $GW$ approximation to the self--energy\index{self--energy}, $\Sigma=iGW$, by neglecting terms second order in $W$. The kernel reads:
\begin{equation}\label{common_BSE_kernel}
\Xi(12,34)=\delta(1,2)\delta(3,4)w(1,3)-\delta(1,3)\delta(2,4)W(1,2)
\end{equation}
The screened interaction appearing in $\Xi$ is usually taken as static. In the case of double excitations we will see that this static approximation cannot describe multiple neutral excitations.

\chapter{Density--Functional Theory}    \label{chap:Density Functional Theory}
In the previous chapter we have introduced the MBPT formalism. The main advantage of this approach, compared to the solution of the full many--body Schr\"{o}dinger equation, is that it makes possible to develop efficient approximations. The reason is that the electron--electron interaction, which constitutes the complicated part of the many--body Hamiltonian, can be treated in a perturbative manner. The solution of the MBPT equations is anyway a demanding task and simpler approaches are desiderable. In this chapter we will present the Density--Functional Theory (DFT\index{DFT}). The key quantity of this approach is the density of the system which is a considerably simpler quantity than the GF. The density  completely describes a system of classical interacting particles and is the only quantity needed within the Hartree approximation. To get a more realistic description, terms approximating the quantum mechanical effects need to be expressed as functional of the density.

The first attempt in this direction goes back to the Thomas-Fermi (TF) model\cite{Thomas_1927,Fermi_DFT} in the 1927 and its extension due to Dirac in 1928~\cite{Dirac_DFT}. However Tomas-Fermi-Dirac theory remained rather inaccurate for most applications. Only some time later, in 1964, the Hohenberg-Kohn (HK) theorem\cite{HK} put the theory on a firm theoretical footing. The HK states that the ground--state of a is an exact functional of the density. The proof of the existence of an exact functional gave a boost in the developments of density--based approach. However this was mainly thanks to the surprising success of the Kohn-Sham (KS) scheme (1965)\cite{KS}, within the Local--Density Approximation (LDA), that DFT\index{DFT} became popular. In this chapter we will start from the demonstration of the HK theorem and then we will introduce the KS scheme and the LDA approximation. In particular we will show why the LDA approximation, which had been initially designed for uniform systems, has been found to be accurate even for strongly inhomogeneous systems such as isolated atoms and molecules.

While MBPT offers a direct recipe to evaluate all ground--state properties starting from the Green function, DFT\index{DFT} gives only the total density and the total energy. As stated in the introduction of the present thesis, by using the total energy many properties of the system can be obtained but others, as for example the electronic gap, are not directly accessible. Anyway it is common practice to interpret the KS band gap and to look at the KS wave--functions to obtain more informations on the system. In practice DFT\index{DFT} provides a ``zero--order Hamiltonian'' for a first understanding of the physical properties of the system and also for MBPT calculations.

\section{The Hohenberg--Kohn theorem}
The HK theorem states that: [HK1] for any system constituted by interacting particles, for a fixed the interaction, there exists a bijective relation among the external potential (up to an arbitrary additive constant), the ground--state many--body wave--function and the ground--state density of the system,
\begin{equation}
\hat{V}^{ext} \Longleftrightarrow 
\Psi_0 \Longleftrightarrow
\rho_0
\nonumber \hspace{0.5 cm} \text{;}
\end{equation}
[HK2] the ground--state energy and density can be determined by minimizing a functional of the charge--density.
As a consequence of the HK theorem the knowledge of the ground--state wave--function gives access to all physical observables. Moreover the external potential fixes the Hamiltonian and knowing the Hamiltonian we can in principle access the excited states of the system\cite{DFT_book}. It is surprising that given the ground--state density we can access all this physical information. However DFT\index{DFT} is in practice used only for ground--state properties, as there is no practical scheme able to describe excited stated starting from the ground--state density. Excited state properties can be obtained from the extension to the time domain of DFT\index{DFT}: this is Time--Dependent DFT (TDDFT\index{TDDFT}).

The proof of the [HK1] theorem is straightforward. First we show that given a ground--state wave--function there exists a unique external potential which determines it. The relation in the opposite direction is trivial, at least for systems which does not present a degenerate ground--state, as the Schr\"{o}dinger equation has a unique solution. The second step is to show that given a ground--state density there exists a unique wave--function which determines such a density. Again the reverse relation is trivial as the density is an observable of the system and any observable can be obtained from the wave--function:
\begin{equation}
\rho_0=\langle \Psi_0| \hat{\rho} |\Psi_0\rangle \text{.}
\end{equation}

\textbf{Step one.} Let's suppose that there exist two different external potentials which have the same ground--state wave--function: 
\begin{eqnarray}
(\hat{T}+\hat{w}+\hat{V}^{ext}) |\Psi_0 \rangle = E_0 |\Psi_0 \rangle   \nonumber \\
(\hat{T}+\hat{w}+\hat{V}'^{ext}) |\Psi_0 \rangle = E'_0 |\Psi_0 \rangle
\text{.}
\end{eqnarray}
If we subtract the two equations assuming that the wave--functions are different from zero only in a subspace of null measure we obtain
\begin{equation}
V^{ext}(\mathbf{x})-V'^{ext}(\mathbf{x})= (E_0 - E'_0)
\text{.}
\end{equation}
This means that the two potentials can differ only up to a constant which anyway is irrelevant as the total energy of a system is always determinated up to an additive constant.

\textbf{Step two.} Suppose that $\Psi_0$ and $\Psi_0'$ are the ground--state wave--function of two different Hamiltonians $H$ and $H'$. From the definition of ground--state it follows that
\begin{equation}\label{Minimum_principle}
E_0 = \langle \Psi_0  | \hat{H}  | \Psi_0  \rangle <
\langle \Psi'_0 | \hat{H} | \Psi'_0 \rangle
\text{.}
\end{equation}
The r.h.s. of Eq. (\ref{Minimum_principle}) can be written as
\begin{equation}
\langle \Psi'_0 | \hat{H}'-\hat{V}'^{ext}+\hat{V}^{ext} | \Psi'_0 \rangle =
E'_0 + \int d^3 \mathbf{r}\ \rho'(\mathbf{r}) \big[V^{ext}(\mathbf{r})-V'^{ext}(\mathbf{r})\big] 
\text{,}
\end{equation}
which means
\begin{equation}\label{inequality1}
E_0 < E'_0 + \int d^3 \mathbf{r}\ \rho'(\mathbf{r})
\big[V^{ext}(\mathbf{r})-V'^{ext}(\mathbf{r})\big]
\text{.}
\end{equation}
By following the same procedure but inverting the roles of $\Psi_0$ and $\Psi_0'$ we obtain 
\begin{equation}\label{inequality2}
E_0 < E'_0 + \int d^3 \mathbf{r}\ \rho'(\mathbf{r})
\big[V^{ext}(\mathbf{r})-V'^{ext}(\mathbf{r})\big]
\text{.}
\end{equation}
If we assume $\rho_0=\rho_0'$, summing up Eq. (\ref{inequality1}) and (\ref{inequality2}) we obtain
\begin{equation}
E_0 + E'_0 < E_0 + E'_0
\text{.}
\end{equation}
This is clearly impossible and so the assumption $\rho_0=\rho_0'$ is wrong.

Once [HK1] has been proven\footnote{The proof of HK1 assumes v--representability of the density: i.e. that given a ``reasonably well behaved'' non negative function $\rho(\mathbf{r})$ one can always find a local external potential $V^{ext}(\mathbf{r})$, so that $\rho(\mathbf{r})$ is the ground state density of the Hamiltonian $\hat{H}=\hat{T}+\hat{w}+\hat{V}^{ext}$. Unfortunately this is not always the case, for a detailed discussion see Ref. \cite{DFT_book}.}
the functional of the [HK2] part of the theorem can be written as:
\begin{equation}
E[\rho]=\langle \Psi[\rho] | \hat{H} | \Psi[\rho] \rangle
\text{.}
\end{equation}
This functional has a minimum when $\rho=\rho_0$ is the ground state density of the system. Moreover it can be written as $E[\rho]=F[\rho]+\int d^3\mathbf{x}\rho(\mathbf{x})V^{ext}(\mathbf{x})$, where $F[\rho]=\langle \Psi[\rho] | \hat{T}+\hat{w} | \Psi[\rho] \rangle$ is an universal functional for any many--electrons system.

The basic assumption in the proof of the theorem is that the coupling of the system with the environment is given by a term of the form $V^{ext}\rho$. This is not always the case. If an external magnetic field is present then the terms $\mathbf{A}\mathbf{j}$ or $\mathbf{B}\boldsymbol{\sigma}$ or both must be included. In this case the modified HK theorem leads to Spin Density--Functional Theory (SDFT\index{SDFT}) for density and magnetization~\cite{DFT_book}, Current--Density--Functional Theory (CDFT\index{CDFT}) for density and current~\cite{CDFT,CDFT_book} and to SCDFT\index{SCDFT} when density, magnetization and current are considered.

\section{The Kohn--Sham scheme}
The HK theorem states that the density determines the ground--state of an interacting system. Still, it does not provide any recipe to use it. The prescription is introduced by the KS scheme. The idea is to use the HK theorem for an auxiliary system of non-interacting particles whose ground--state density is assumed to be the same as that of the interacting system: $\rho_s=\rho_0$. Then we look for the external potential $v_s[\rho]$ and for the non-interacting wave--function $\Psi_s[\rho]$ related to such density by the HK theorem.

We start from the energy functional of the real system
\begin{equation}
E[\rho]=\langle\Psi[\rho]|\hat{T}+\hat{V}^{ext}+\hat{w}|\Psi[\rho]\rangle 
       = T[\rho]+U^{ext}[\rho]+U_w[\rho]
\text{,}
\end{equation}
then we rewrite it, by using the relation $\rho\Longleftrightarrow\Psi_s$, starting from the non interacting many--body wave--function
\begin{equation}
\begin{split}
E_s[\rho] =& \langle \Psi_s[\rho]
                     | \hat{T} + \hat{V}^{ext} | \Psi_s[\rho] \rangle
                     +\frac{1}{2} \int \int 
                     \frac{\rho(\mathbf{r}) \rho(\mathbf{r}')}
                                       {\left| \mathbf{r}-\mathbf{r}' \right|}
                      d^3\mathbf{r}\ d^3 \mathbf{r}'
                     + E_{xc}[\rho]  \\    
                  =& T_s[\rho] + U^{ext}[\rho] + E_H[\rho]
                     +E_{xc}[\rho]
\text{.}
\end{split}
\end{equation}
The non-interacting wave--function enters the kinetic energy term, while the external potential energy is the same as that of the interacting system. We have also introduced a term which describes the energy of an interacting system of classical particle $E_H[\rho]$ and a last term $E_{xc}[\rho]$ so that $E[\rho]=E_s[\rho]$. This means that
\begin{equation}
E_{xc}[\rho]= \left( T[\rho] - T_s[\rho] \right) +
\left( E_w[\rho] - E_H[\rho] \right)
\text{.}
\end{equation}
$E_{xc}[\rho]$ is the xc--energy. Now we look for the ground--state by minimizing the functional $E_s[\rho]$. In a non-interacting system $\Psi_s=\Pi_i | \psi_i |$\footnote{Here the notation $\Pi_i | *_i |$ means one have to perform an ``antisymmetrized'' product} and $\rho=\sum_i |\psi_i|^2$ so that we can minimize the energy functional with respect to $\psi_i$. This is a constrained minimization as we want the $\psi_i$ to be orthonormal. That is, using the theory of Lagrangian multiplier, we have to minimize the functional
\begin{equation}
E[\rho(\psi_1,...,\psi_n)] +
             \sum_{h,k} \left( \delta_{h,k} - \lambda_{h,k} \int \psi_h(\mathbf{r})
             \psi_k(\mathbf{r}) d^3\mathbf{r} \right)
\text{.}
\end{equation}
Using the relation
\begin{equation}
\frac{\delta}{\delta \psi_i^{\ast}} = \frac{\delta \rho}{\delta \psi_i^{\ast}}
       \frac{\delta}{\delta \rho}
        = \psi_i \frac{\delta}{\delta \rho}
\end{equation}
we find
\begin{equation} 
\frac{\delta}{\delta \psi_i^{\ast}} T_s + \psi_i\frac{\delta}{\delta \rho}
    \left( E_H + U^{ext} + E_{xc} \right) = \lambda_i \psi_i
\text{,}
\end{equation}
that leads to
\begin{equation} \label{KS-equation}
\left( \hat{t}+ \hat{v}_H[\rho]+ \hat{v}^{ext}+
       \hat{v}_{xc}[\rho] \right) \psi_i = \lambda_i \psi_i
\text{.}
\end{equation}
Eq. (\ref{KS-equation}) is the KS equation, $v_{xc}[\rho]=\delta E_{xc}[\rho] / \delta \rho$ is the unknown xc--potential\footnote{Here we assumed that the functional derivative exists; however this must be verified for any given energy functional.} which encloses all the difficulties of the interacting many body system beyond the classical Hartree potential. Though Eq. (\ref{KS-equation}) has to be solved self--consistently with respect to the density, the scheme offers a very appealing starting point as the xc--potential appears in the equations in the form of a \emph{local} multiplicative operator. If compared to the exchange term of the HF approach or to the Self-Energy of the MBPT, which are non local operators, the advantage is evident.  Unfortunately no hint to approximate the xc--potential or the xc--energy functionals is given.

\section{The local--density approximation}

One of the reasons beyond the success of DFT\index{DFT}, despite its simplicity, is that the simplest approximation proposed for $E_{xc}$, the Local--Density Approximation (LDA) successfully describes both extended and isolated systems. Such an approximation was designed to work for systems where the density is almost spatially uniform or slowly varying. The idea is to compute the exchange correlation energy of the uniform electron gas as a function of the constant density, $\rho_0$, $E^{hom}_xc(\rho_0)$ and then to express the general energy functional of the density as
\begin{equation}
E_{xc}[\rho]
       \simeq \int d^3\mathbf{r}\ \rho(\mathbf{r})\epsilon^{hom}_{xc}(\rho(\mathbf{r}))
\text{,}
\end{equation}
where 
\begin{equation}
\epsilon^{hom}_{xc}(\rho)=
\frac{E^{hom}_{xc}(\rho)}{V\rho}
\end{equation}
is the energy per electron.

The total energy of the homogeneous electron gas can be obtained accurately from Quantum Monte-Carlo calculations.

\subsection*{The success of the local--density approximation}

To explain the success of the LDA we need to introduce the exchange--correlation hole--density which is linked to the success of the LDA approximation.

First we introduce the expectation value of the pair correlation function
\begin{equation} \label{correlazione-coppie}
g[\rho](\mathbf{r,r'})=\mathbf{\frac{\langle\hat{\rho}(r)\hat{\rho}(r')\rangle}
{\rho(r)\rho(r')}-\frac{\delta(r-r')}{\rho(r)}}
\text{,}
\end{equation}
which can be used to write the energy correlation due to the interaction $\langle \hat{w} \rangle$ as
\begin{equation}
E_{w}[\rho]=\frac{1}{2}\int d^3\mathbf{r}\int d^3\mathbf{r'}\ \frac{\rho(\mathbf{r})\rho(\mathbf{r'})}
{|\mathbf{r-r'}|}g[\rho](\mathbf{r,r'})
\text{.}
\end{equation}
Than we make use of the adiabatic connection and of the HK theorems. We define a group of Hamiltonians $\hat{H}_{\lambda}$ such that
\begin{multline}
\hat{H}_{\lambda}=\hat{T}+\sum_{\sigma}\int d^3\mathbf{r}\ v_{\lambda}(\mathbf{r})
\hat{\psi}^{\dag}_{\sigma}(\mathbf{r})\hat{\psi}_{\sigma}(\mathbf{r})  \\
+\frac{\lambda}{2}\sum_{\sigma\sigma'}\int\int d^3\mathbf{r}d^3\mathbf{r}'\
w(\mathbf{r},\mathbf{r}')
\hat{\psi}^{\dag}_{\sigma}(\mathbf{r})\hat{\psi}^{\dag}_{\sigma'}(\mathbf{r'})
\hat{\psi}_{\sigma}(\mathbf{r})\hat{\psi}_{\sigma'}(\mathbf{r}')
\text{,}
\end{multline}
with $v_\lambda$ such that $\rho_\lambda=\rho_1$ for each $\lambda$. For $\lambda=0$ we have the KS system.
\begin{figure}[H] 
\centering
\subfigure[Exchange--correlation hole]{\includegraphics[width=0.95\textwidth]{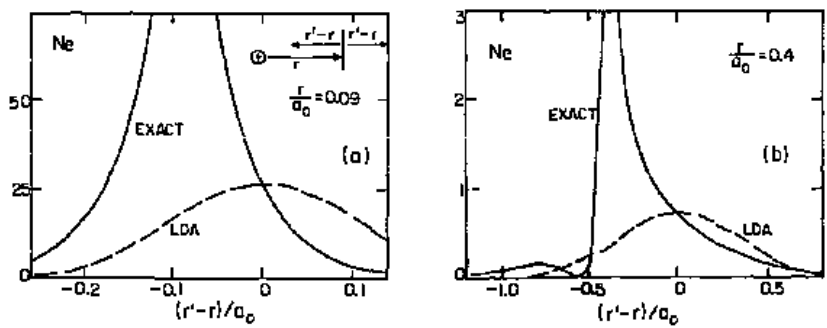}}
\subfigure[Spherical average of the Exchange--correlation hole]{\includegraphics[width=0.95\textwidth]{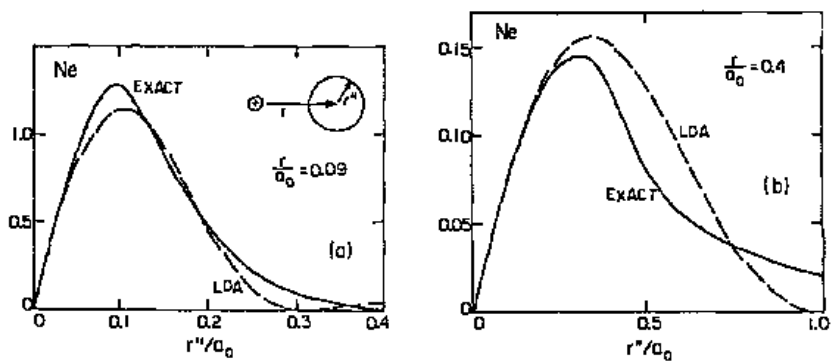}}
\caption{Exchange-correlation hole and its spherical average within LDA are compared with the exact solution at two different densities $r/a_0=0.09$ and $r/a_0=0.4$. From ref.~\cite{LDA_figure}.}
\label{fig:LDA_h_rho_xc}
\end{figure}
\newpage

Now we compute, using the Hellmann-Feynman theorem
\begin{equation} \label{Hellmann-Feynman}
\begin{split}
\frac{dE(\lambda)}{d \lambda}=&\ \left\langle \Psi^0_{\lambda} \bigg|
\frac{\partial \hat{H}_{\lambda}}{\partial \lambda} \bigg| \Psi^0_{\lambda} \right\rangle \\
       =&\ \frac{1}{2}\int d^3\mathbf{r}\int d^3\mathbf{r'}\ \frac{\rho(\mathbf{r})\rho(\mathbf{r'})}
{|\mathbf{r-r'}|}g_\lambda[\rho](\mathbf{r,r'})
+\frac{\partial}{\partial \lambda}\int d^3\mathbf{r}\ v_{\lambda}(\mathbf{r})\rho(\mathbf{r})
\text{.}
\end{split}
\end{equation}
where $g_\lambda[\rho](\mathbf{r,r'})$ is the pair correlation function for the Hamiltonian $\hat{H}_\lambda$ and $\Psi^0_\lambda$ its ground--state. Finally we express the total energy for $\lambda=1$ as
\begin{equation}
\begin{split}
E(1)&=E(0)+\int_0^1 d\lambda \frac{dE(\lambda)}{d\lambda}  \\
    &=T_{s}[\rho]+E^{ext}[\rho]+E_H[\rho]\\
    &\phantom{T_{s}[\rho]+E^{ext}[\rho]}+\frac{1}{2} \int\int d^3\mathbf{r}d^3\mathbf{r}'\
      w(\mathbf{r,r'})\rho(\mathbf{r})\rho(\mathbf{r'})
      \int_0^1 d\lambda \left(g_{\lambda}[\rho](\mathbf{r,r'})-1\right)
\text{,}
\end{split}
\end{equation}
where, in the last line, we have subtracted the Hartree energy from the expression for $\langle\hat{w}\rangle_\lambda$. This last term is clearly the exchange correlation energy. Defining the exchange--correlation hole--density 
\begin{equation}
\varrho_{xc}(\mathbf{r,r'})=\rho(\mathbf{r'})
\left[\int_0^1 d\lambda \left(g_{\lambda}[\rho](\mathbf{r,r'})-1\right)\right]
\text{,}
\end{equation}
we can therefore study the behavior of the $\varrho_{xc}$ to check the LDA performance. In particular Fig. (\ref{fig:LDA_h_rho_xc}) shows how $\varrho_{xc}$ behaves for the neon atom. The exact solution is compared with the LDA one. In the two upper frames, $(a)$, $\varrho_{xc}$ is plotted for a fixed $\mathbf{r}$ coordinate, as a function of $\mathbf{r'}$ along a direction parallel to $\mathbf{r}$. In the two bottom panels $(b)$ its spherical average is plotted. We see that compared to the exact xc-hole the LDA performs quite badly, while the $\varrho_{xc}$ spherical average is well described. For an interaction which depends only on the modulus of distance among the particle $E_{xc}[\rho]$ depends only on the spherical average and this is one of the reasons why LDA well performs even for not homogeneous systems. Another reason is that the LDA xc-hole satisfy the sum rule
\begin{equation}
\int d^3\mathbf{r'}\varrho_{xc}(\mathbf{r,r'})=-1
\text{,}
\end{equation}
which helps to guarantee error cancellations.

\subsection*{Spin dependent local--density approximation}
For spin polarized systems, that is systems whose ground--state present a magnetization $m_z\neq 0$, the DFT\index{DFT} approach could be ideally used. However in these approach the magnetization would be an unknown functional of the density and the LDA approximation is not guaranteed to work. For these reason it is preferable to work within the SDFT formalism where the magnetization can be obtained directly from the KS wave--functions and its possible to design the Spin LDA (SLDA) as a direct extension of the LDA. To this end the quantity
\begin{equation}
\xi =\frac{ \rho_\uparrow - \rho_\downarrow}{\rho_\uparrow + \rho_\downarrow}
\text{,}
\end{equation}
which measures the polarization of the system, is defined. Then the approximation
\begin{equation}
\epsilon_{xc}(\rho,\xi)=\epsilon_{xc}(\rho,\xi=0)+\(\epsilon_{xc}(\rho,\xi=1)-\epsilon_{xc}(\rho,\xi=0)\)g(\xi)
\end{equation}
is used, where $\epsilon_{xc}(\rho,\xi=0)=\epsilon^{LDA}_{xc}(\rho)$, $\epsilon_{xc}(\rho,\xi=1)$ can be compute from the energy of the homogeneous electron gas in the configuration $\xi=1$ and $g(\xi)$ is an interpolation function, which is usually chosen as
\begin{equation}
g(\xi)=\frac{ (1+\xi)^{4/3} + (1-\xi)^{4/3} -2 }{2\ (2^{1/3}-1)}
\text{.}
\end{equation}

\section{Time--dependent density--functional theory}
DFT\index{DFT} is a convenient theory to obtain the ground--state of a system through minimization of the energy as a functional of the density. However within DFT\index{DFT} it is not possible to describe the evolution of a system perturbed by an external time dependent potential. The extension of DFT\index{DFT} to the time domain, the Time Dependent DFT (TDDFT\index{TDDFT}) relies on a formal extension of the HK theorem given by Runge and Gross (RG). In 1984 RG~\cite{RG} showed that, given an initial time $t_0$ where the system is in a state $\Psi_0$, is possible to establish a bijective correspondence between the time dependent external potential, the evolution of the density and the evolution of the wave--function. On the basis of this theorem they derived three schemes to calculate the time dependent density, one of which, a stationary action principle, can be seen as the extension of the second HK theorem\footnote{The RG action functional has been found out to be wrong. A corrected functional has been provided some years later by Van Leeuwen~\cite{TDDFT_Leeuwen}.}. The theory provides a formalism where, at least in principle, it is possible to describe the evolution of a system even in the presence of strong perturbations. In this work we are interested in small perturbations and so we can restrict the analysis to the linear regime. The response function within TDDFT can be obtained through a single particle Schr\"odinger Equation, or Time--Dependent KS (TDKS) equation (that is the third scheme proposed in RG paper), by observing
\begin{equation}
\delta\rho(1)=\chi_{KS}(1,1')\delta v_{KS}(1')
\end{equation}
where $\chi_{KS}(1,2)$ is the response function of the KS non interacting system, while $\delta v_{KS}(1)$ is the variation of the TDKS potential needed to follow the evolution of the density when an external potential $\delta V^{ext}(1)$ is applied. From the previous section we know  $v_{KS}(1)=V^{ext}(1)+v_h[\rho](1)+v_{xc}[\rho](1)$. Assuming that the TDKS potential is identical to the potential of the static theory. Its variation is, to linear order in the density,
\begin{equation}
\delta v_{KS}(1)=\delta V^{ext}(1)+\frac{\delta v_h[\rho](1)}   {\delta\rho(1')}\delta\rho(1')
                                  +\frac{\delta v_{xc}[\rho](1)}{\delta\rho(1')}\delta\rho(1')
\text{.}
\end{equation}
If we Compute the variation of the Hartree potential, defining the exchange correlation kernel $f_{xc}(1,2)=\delta v_{xc}[\rho](1) /\delta\rho(2)$ and remembering that $\delta\rho(1)=\chi(1,1')\delta V^{ext}(1')$ we obtain a Dyson equation for the response function:
\begin{equation} \label{Dyson-tddft}
\chi^R(1,2)= \chi_{KS}(1,2) + \ \chi_{KS}(1,1') f_{Hxc}(1',2') \chi^{R}(2',2)
\text{.}
\end{equation}
Here $f_{Hxc}(1,2)$ is the sum of the Hartree plus the $xc$ kernel. The most common approximations adopted for $v_{xc}$ and so for $f_{xc}$ are the adiabatic extensions of the ground--state approximations. However within TDDFT excited states, that can have different spin configurations, are described. For this reason it is convenient to derive the approximation from the SDFT energy functionals:
\begin{equation}
f^{xc}_{\sigma_1\sigma_2}[\rho,\xi](\mathbf{x}_1t_1,\mathbf{x}_2t_2) =
\frac{\delta v^{xc}_{\sigma_1}[\rho,\xi](\mathbf{x}_1,t_1)}{\delta \rho_{\sigma_2}(\mathbf{x}_2,t_2)}
\simeq \delta(t_1-t_2)\frac{\delta v^{xc,A}_{\sigma_1}[\rho_{t_1},\xi_{t_1}](\mathbf{x}_1)}{\delta \rho_{\sigma_2}^{t_1}(\mathbf{x}_2)}
\text{;}
\end{equation}
the variable $\xi$ is eventually set to zero after the derivatives have been computed, for spin unpolarized systems to have a functionals of only the density. As for the ground--state, in the case of spin polarized systems TD-Spin-DFT (TDSDFT) can be used used, that is $\xi$ is not set to zero.
In particular starting from the LSDA we have the Adiabatic LDA/LSDA (ALDA/ALSDA) approximation
\begin{multline} \label{K_alda}
f^{ALDA}_{xc}(1,2)= \delta(\mathbf{x_1,x_2})\delta(t_1,t_2)
 \bigg( \frac{\partial \epsilon^{hom}_{xc}(\rho,\xi)}{\partial \rho_{\sigma_1}} + \\
      \frac{\partial \epsilon^{hom}_{xc}(\rho,\xi)}{\partial \rho_{\sigma_2}} +  
  \rho\frac{\partial^2 \epsilon^{hom}_{xc}(\rho,\xi)}{\partial \rho_{\sigma_1} \partial \rho_{\sigma_2}}  \bigg)
\text{.}
\end{multline}
Similarly to the case of DFT\index{DFT}, TDDFT\index{TDDFT} must be formally extended to TD-Current-DFT and/or TDSDFT if external magnetic fields are considered.
However practical calculations indicate that this makes little difference in practice.

\section{The electron--electron interaction}

\begin{figure}[t]
 \centering
 \includegraphics[width=0.8\textwidth]{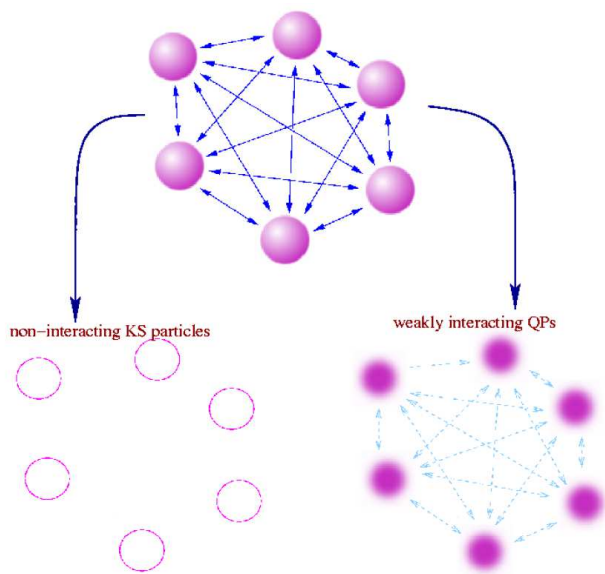}
 \caption{Dealing with the interaction. Form ref.~\cite{PhD_Thesis_Sottile}.}
\end{figure}

The electron--electron interaction is the main ingredient of Many Body systems. Both DFT\index{DFT} and MBPT offer a practical approach to deal with it. In DFT\index{DFT}, and in particular in the KS scheme, the starting point is a system of not-interacting particles where the effect of the interaction is included through an external potential. In MBPT the classical part of the interaction, the Hartree potential only, is a local potential. Correlation is included in the self-energy $\Sigma$ which is an highly non local frequency dependent and, in the quasi particle picture, complex function. Compared to KS electrons QP can be seen as weakly interacting dressed electrons characterized by a finite lifetime.
 
The main advantage of DFT\index{DFT} compared to MBPT is that it's much less demanding from a computational point of view. This is the reason why nowadays  the majority of codes solve the MBPT equations starting from DFT\index{DFT} calculations. Still the main drawback of DFT\index{DFT} is that it does not offer a recipe to construct systematic approximations. For this reason attempts to improve the LDA are not always satisfactory. For example the Generalized Gradient Approximation (GGA) does not always improve upon LDA in lattice constants and can even predict less accurate results~\cite{GGA_worse,GGA_worse2}. Improvements to the LDA are also derived from other approaches such as hybrid functionals defined from the Hartree Fock (HF) approximation. Within these approaches DFT\index{DFT} can include more physics at the price of loosing, at least in part, its main advantage. As a matter of fact the more physics is included beyond the LDA the more complicated are the functionals constructed. Following this path, though many important results have been obtained, DFT\index{DFT} is bound to become at a certain point more expensive than other competing approaches.

In the present thesis we prefer to focus on MBPT in the first part where we will look for development of new approximations, in particular for the description of double excitations, and to look at LDA/DFT\index{DFT} as the zero--order approach. For this same reason we will use DFT\index{DFT} in the second part where we will explore the predictions of ab-initio calculations on carbon nanotubes in a magnetic field.

\part{Double Excitations}                               \label{part:Double Excitations}
\chapter{Introduction to the problem}                   \label{chap:Introduction to the problem}
To observe an object with our eyes we need to shine light on it. An absorption experiment is nothing but a detailed study of what happen when the light hit the object. In this thesis we focus on the theoretical description of spectra in the visible energy range: the goal of theoretical spectroscopy is to give an accurate description of the microscopic process involved. The theory which describes the evolution of the many body electronic wave--function interacting with the electromagnetic field is the quantum electrodynamics. Unfortunately it is not possible to solve the exact equations and, as we have explained in the first part of the present work, approximations need to be introduced. Besides mathematical rigour, physical intuition can give essential guidelines in the development of approximations. Relying on concepts emerging in the macroscopic world, obtained as the classical limit of quantum mechanics, it is often possible to give an intuitive interpretation of the microscopic world. As an example we can consider the idea of quasiparticles, which can capture many of the features observed in the absorption experiments describing the main peaks as excitons. Similarly we have the idea of collective excitations, which is used to explain other possible features as plasmons. It is somehow surprising that, using these concepts, one gets an accurate description of the complicated microscopic time evolution described by the equations of quantum mechanics and quantum electrodynamics.

However some of these simplified pictures fail in some situations. For example the description of the satellites discussed in Fig.~(\ref{fig:QP_renormalization}) requires a more elaborate representation then the single--particle picture.

\begin{figure}[t] 
 \centering
 \includegraphics[width=0.8\textwidth]{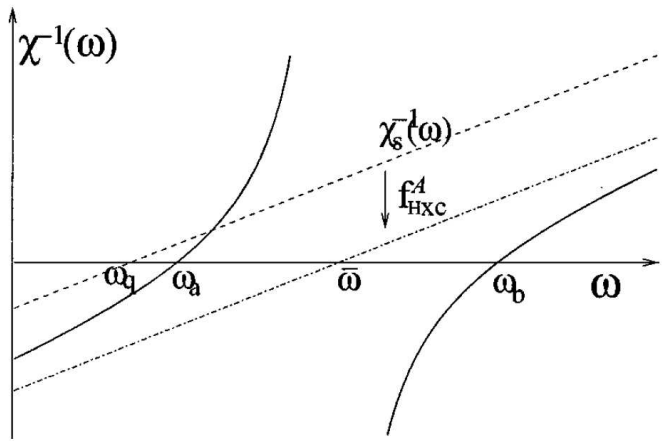}
 \caption{The frequency--dependent kernel of the Dyson equation for the response function splits
          the nn--interacting pole $\omega_q$ in two poles $\omega_a$ and $\omega_b$.
          The pole $\bar{\omega}$ is obtained using a static kernel. From Ref.~\cite{Maitra1}.}
 \label{fig:doubles_maitra_model}
\end{figure}

The existence of double excitations\index{double excitations} (DEs) is, in this sense, another situation where an intuitive picture, like the one of quasiparticle or the one of collective phenomena of the many body problem is problematic. It is still possible to describe the process as a result of the interaction among electrons: the light source hits an electron which is excited to an unoccupied state leaving a hole in the system. Due to the presence of both the hole and the extra electron the system can react by adjusting its configuration and exciting a second electron to an unoccupied state. This description is, however, not completely correct as it does not describe correctly the real evolution of the system. The alternative mathematical view is pictured in Fig.~(\ref{fig:doubles_maitra_model}). 
The frequency--dependent kernel describes the correction to an independent particle picture due to correlation effects. The structure is very close to the one described in Fig.~(\ref{fig:QP_renormalization}). The frequency dependence of the kernel reflects the fictitious time evolution used in MBPT to make the system evolve from a nn--interacting to an interacting eigenstate. DE\index{double excitations}s are then virtual processes needed to describe some features of absorption experiments.

It is important to observe, however, that Fig.~(\ref{fig:doubles_maitra_model}) describes a different process from the one of Fig.~(\ref{fig:QP_renormalization}). In Fig.~(\ref{fig:QP_renormalization}) we are moving from an independent particle picture to a QP picture with a finite lifetime which, in the sense shown in Fig.~(\ref{fig:Lehmann_Vs_QP}), contains more physics than the simple representation in terms of real poles. The common concept, however, is the failure of the static approximation which does not give the correct number of solutions. This happens when the kernel of the Dyson equation has a pole at an energy close to an independent particle solution.

The most common and widely used approach to the problem of DE\index{double excitations}s is given by post--HF methods based on the HF starting point which do not include correlation\footnote{Correlation is due to the effects of the interaction in a quantum many--particles system. In the present thesis I use the following ``language'': if it's possible to save the HF interpretation as a first order approximation the correlation is considered small, if not the correlation is considered big. When DE\index{double excitations}s are important, the interpretation in terms of single particles (or HF orbitals) starts to break but not completely and indeed diagrams one order beyond the HF (or TDHF) are enough to include this effect.}. However, while this is often reasonable for small molecules, in long 1D molecular chains the effects of correlation are crucial~\cite{1Dpolymers}. On the other hand the currently used approximations to electron correlation in the state--of--the--art approaches to the description of optical excitations, TDDFT and MBPT, fail to capture the physics of DE\index{double excitations}s~\cite{Casida1,Maitra2}. As post HF methods are designed to work in a small correlation regime, their extension to realistic nano--structured materials is very demanding, if not practically impossible and solutions within the TDDFT and MBPT framework are needed. 

In the TDDFT approach, the excitation energies of a system are obtained from the nn--interacting Kohn-Sham (KS) eigenenergies solving Eq.~(\ref{Dyson-tddft}) where the xc--effects are cast in the unknown xc--kernel $f_{xc}[\rho](r,r',t-t')$. Most of the success of the scheme is due to the success of the ALDA, which, despite being extremely simple, is surprisingly accurate in the case of many isolated systems. Nevertheless the ALDA suffers from some deficiencies that cause TDDFT to fail in some cases, such as in the description of  excited states with multiple--excitation character. The source of this failure has been traced back in the literature to the adiabatic approximation~\cite{Casida1,Maitra1}, which neglects the frequency dependence of the true xc kernel: it turns out that it is precisely this frequency dependence of the kernel that takes into account all the many--electron excitation effects. 

In MBPT the neutral excitations of the system are obtained by solving Eq.~(\ref{Bethe-Salpeter}). Similarly to TDDFT, xc effects in the BSE are cast in the four-point kernel $\Xi(1,2;3,4)$, which, unlike in TDDFT, can be written as a perturbative expansion. In the most common and widely used approximation to this kernel, introduced in Eq.~(\ref{common_BSE_kernel}), the xc effects in the BSE are described by the screened Coulomb interaction $W$, which is considered static, thus ruling out the possibility of describing DE\index{double excitations}s. 

In the TDDFT literature several solutions to the double-excitation problem have been proposed~\cite{Maitra1,Casida1,Li,WZ1,Casida_spinflip}. Wang and Zeigler used a non collinear representation of the xc kernel~\cite{WZ1}, which could be used to describe double-excitations, but only starting from the appropriate reference excited states, some of the weaknesses of this method are discussed in Ref.~\cite{Casida_spinflip}. Casida proposed a xc kernel which goes beyond the adiabatic approximation constructed from a superoperator formalism~\cite{Casida1} that contains as a special case the ``dressed-TDDFT'' recipe derived by Maitra et al.~\cite{Maitra1}. The dressed-TDDFT approach however is not predictive since the existence of DE\index{double excitations}s must be defined ``a priori''; only very recently Huix--Rotllant and Casida~\cite{Casida2} proposed an extension of the dressed TDDFT method, which clarifies the relation between Polarization Propagator (PP) approaches and the BSE methods and is presently being tested on an extensive set of molecules~\cite{Casida2}.

The most commonly used state--of--the--art approaches suffer, hence, from different types of pathologies. On one side, post-HF methods, based on the uncorrelated HF scheme, are in general designed to describe isolated systems with the idea that the interactions among particles can be treated perturbatively. In these, approximations are obtained by truncating the perturbative expansion to some finite order but always respecting key principles of quantum mechanics such as quantum statistics and Pauli exclusion principle. On the other side, in the BSE and TDDFT approaches correlation is treated to all orders of perturbation theory, but a well established method to include DE\index{double excitations}s does not exists yet.

In the present chapter we provide a detailed description of the phenomena where DE\index{double excitations}s play a role and we illustrate which direction can be followed to include DE\index{double excitations}s in TDDFT and BSE. In particular we write the equations of both approaches in a common formalism (see Eq.~(\ref{generalized_Dyson})) and we show that DE\index{double excitations}s are not described in the standard approximations because, in both, the kernel of Eq.~(\ref{generalized_Dyson}) is taken static. We focus on the BSE scheme, because within the MBPT it is more straightforward to look for practical approximations, in order to introduce a frequency--dependent kernel. First we solve the mathematical problems which arise trying to construct the dynamical version, i.e. with a dynamical kernel, of Eq.~(\ref{generalized_Dyson}) and then we show that indeed the frequency--dependent kernel obtained relaxing the static approximation in the standard BSE scheme capture, at least in part, the effects of DE\index{double excitations}s. The inclusion of this result within TDDFT is obtained thanks to the common language established, following an alternative, but similar, procedure to the one of Ref.~\cite{Romaniello1}.

Only in the following chapter, we will finally construct a fully consistent approximation to the problem of DE\index{double excitations}s within the BSE scheme. The explicit connection with TDDFT is not explored anymore in Ch. \ref{chap:A new approach to describe Double Excitations}; the extension of the results obtained to TDDFT, following the method illustrated in the present chapter, could be a possible development.

\section{Double excitations in quantum chemistry}
The concept of DE\index{double excitations}s is well known in the quantum chemistry literature. The absorption spectrum of a system is described with methods that, in contrast to DFT and MBPT, are based on the many body wave--function. The starting point is usually the Hartree-Fock (HF) approach, where the many body ground state is approximated with a single Slater determinant of one particles wave--functions:
\begin{equation}
\Psi_0^{HF}=
\left| \begin{array}{cccc}
\psi_1(\mathbf{x_1}) & \psi_1(\mathbf{x_2}) & ... & \psi_1(\mathbf{x_n})  \\
\psi_2(\mathbf{x_1}) & \psi_2(\mathbf{x_2}) & ... & \psi_2(\mathbf{x_n})  \\
\vdots               & \vdots               &     & \vdots                \\
\psi_n(\mathbf{x_1}) & \psi_n(\mathbf{x_2}) & ... & \psi_n(\mathbf{x_n})  
\end{array} \right|
\text{.}
\end{equation}
The excited state wave--functions can be expressed as a linear combination of Slater determinants, each of which is related to the HF ground-state through some excitation operator. For example, considering only single-particle excitatins one can write:
\begin{equation} \label{excited_HF}
\Psi_I\simeq\sum_{ij} c^I_{ij} \hat{a}^\dag_i\hat{a}_j \Psi_0^{HF}
\text{,}
\end{equation}
which we refer here as $\Psi_I^{HF}$, with $I$ labeling a particular excited--state. In the language of MBPT the HF approximation is equivalent to approximating the self--energy to its first order in the bare interaction. The approximation introduced by Eq.~(\ref{excited_HF}), instead, is equivalent, in the linear response regime, to approximating the BSE kernel to first order in the bare interaction within the Tamm--Dancoff Approximation (TDA) which neglects the coupling of the excitation space with the de--excitations one. The scheme obtained relaxing the TDA is known as Time Dependent HF (TDHF).

It is then natural to relax the introduced approximations to include single and higher order excitations in both the ground state and the excited state wave--function. This is the so called configuration interaction (CI) expansion:
\begin{equation} \label{CI_expansion}
\begin{split}
\Psi_0^{CI}=&\Psi_0^{HF} + \sum_{ij} c^0_{ij} \hat{a}^\dag_i\hat{a}_j \Psi_0^{HF}+
            \sum_{ijhk} d^0_{ijhk} \hat{a}^\dag_i\hat{a}^\dag_j\hat{a}_h\hat{a}_k \Psi_0^{HF} + ...  \\
\Psi_I^{CI}=&\Psi_I^{HF} + \sum_{ijhk} d^I_{ijhk} \hat{a}^\dag_i\hat{a}^\dag_j\hat{a}_h\hat{a}_k \Psi_0^{HF} + ... 
\hspace{1cm}\text{.}
\end{split}
\end{equation}

\begin{figure}[t] 
\begin{center}
\subfigure[states with $\Sigma_u$ symmetry]{\includegraphics[width=0.45\textwidth]{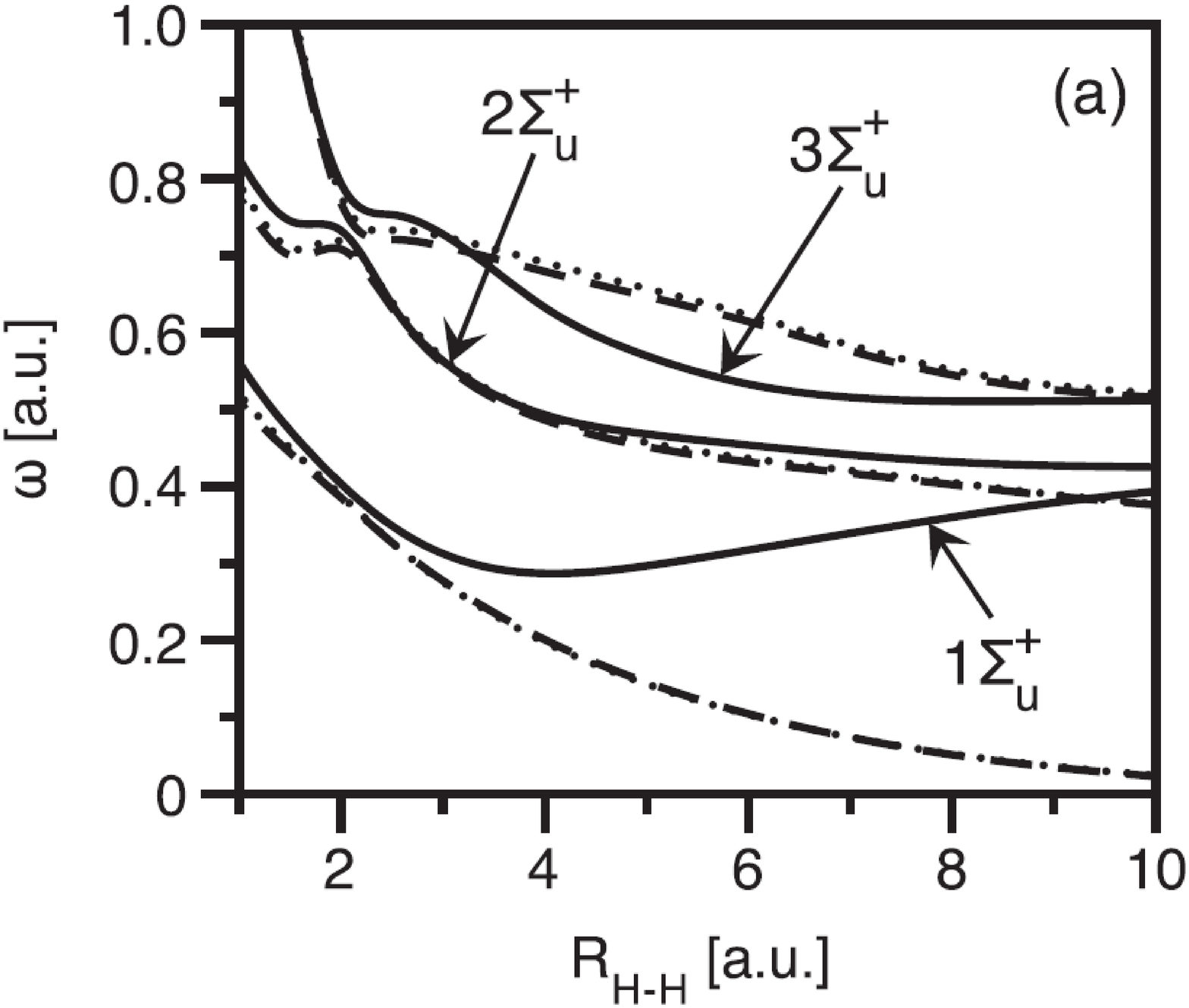}}
\subfigure[states with $\Sigma_g$ symmetry]{\includegraphics[width=0.45\textwidth]{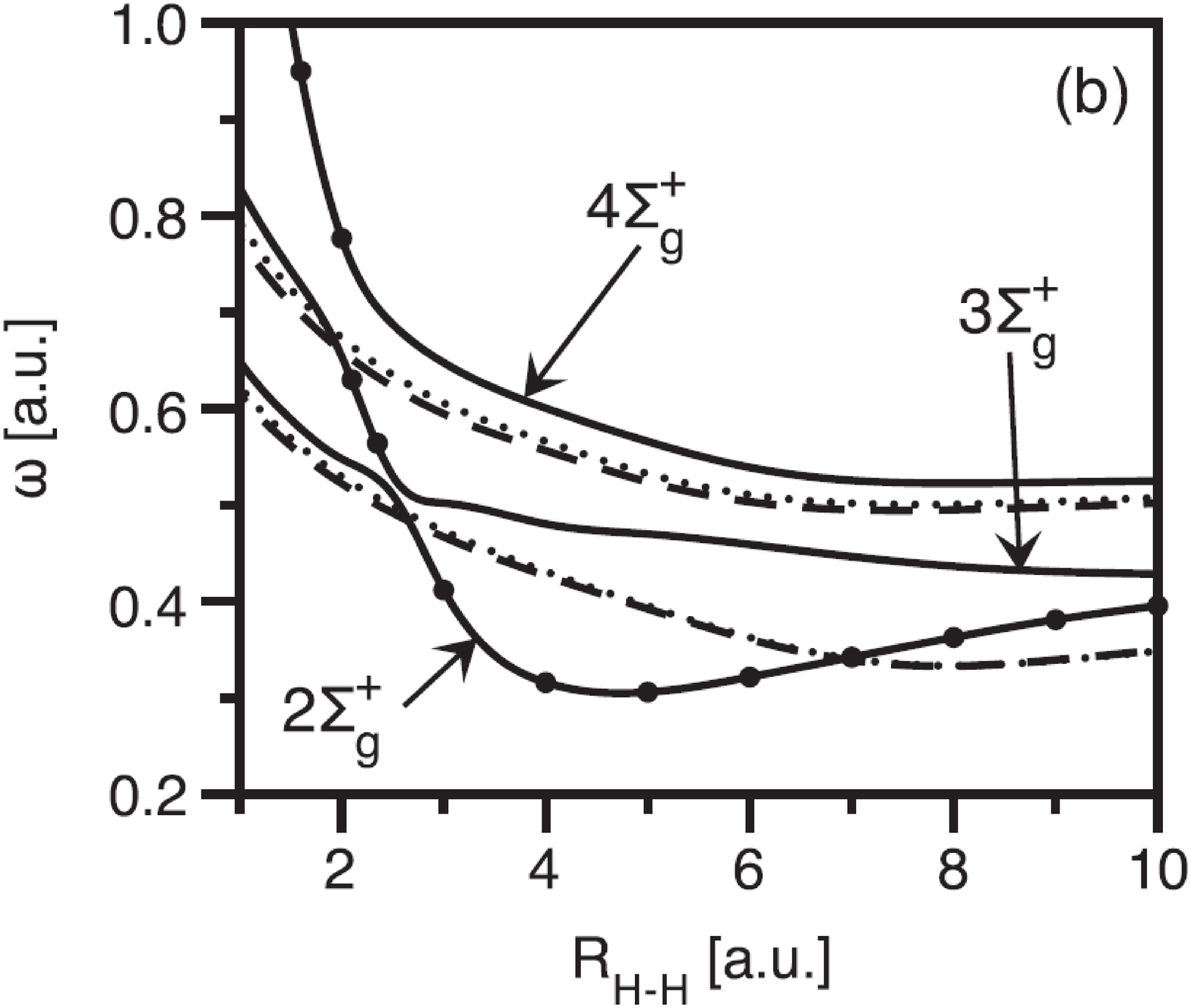}}
\caption{The lowest energies excitations energies of the $H_2$ molecule as a function of the distance among the nuclei. Solid lines are $CI$ results, while dotted and dashed lines are TDDFT calculation with two different functionals which does not include DE\index{double excitations}s. The excited state $^2\Sigma^+_g$ mainly composed by a DE\index{double excitations} is marked with dots. From Ref.~\cite{Baerdens_TDDM_polyenes}.}
\label{fig:H2_dissociation}
\end{center}
\end{figure}

In practice the full CI approach can be hardly used for a number of electrons larger than $10$\footnote{CI in the ``singles and doubles'' approximation (CISD) has recentrly been performed for systems up to almost 100 electrons, though under some approximations\cite{CI_manyel}.}, because the exponents grown at the number terms needed. However CI offers a clear mathematical description of multiple excitations, including DE\index{double excitations}s.

In the next chapter we will present an alternative Post-HF method, called second Random Phase Approximation (sRPA) which we will use to investigate more in detail the connection between the MBPT and the DE\index{double excitations}s concept. Here we want to start by using the CI scheme to understand how DE\index{double excitations}s are identified. We will also present two examples where the DE\index{double excitations}s have been identified to play an important role using CI calculations. The first example is constituted by small molecules, like the $H_2$ molecule considered here, for which DE\index{double excitations}s are known to be important in the description of molecular dissociation; the second is constituted by polyenic carbon chains saturated with hydrogen atoms.

$H_2$ is a very simple molecule and DE\index{double excitations}s play a crucial role already for such a small system. In Fig.~(\ref{fig:H2_dissociation}) we can see that during the dissociation process even the qualitative behavior of the excitation energies is wrong within TDDFT, if compared with a virtually exact full CI approach. In particular for the states with $g$ symmetry, panel $b$, we see that within TDDFT we have one excited state less than within CI. The double excited CI solution start at high energy at the equilibrium distance but then decrease in energy during the dissociation process becoming the lowest energy solution.

\begin{figure}[t] 
\begin{center}
\subfigure[polyenes structure]{\includegraphics[width=0.45\textwidth]{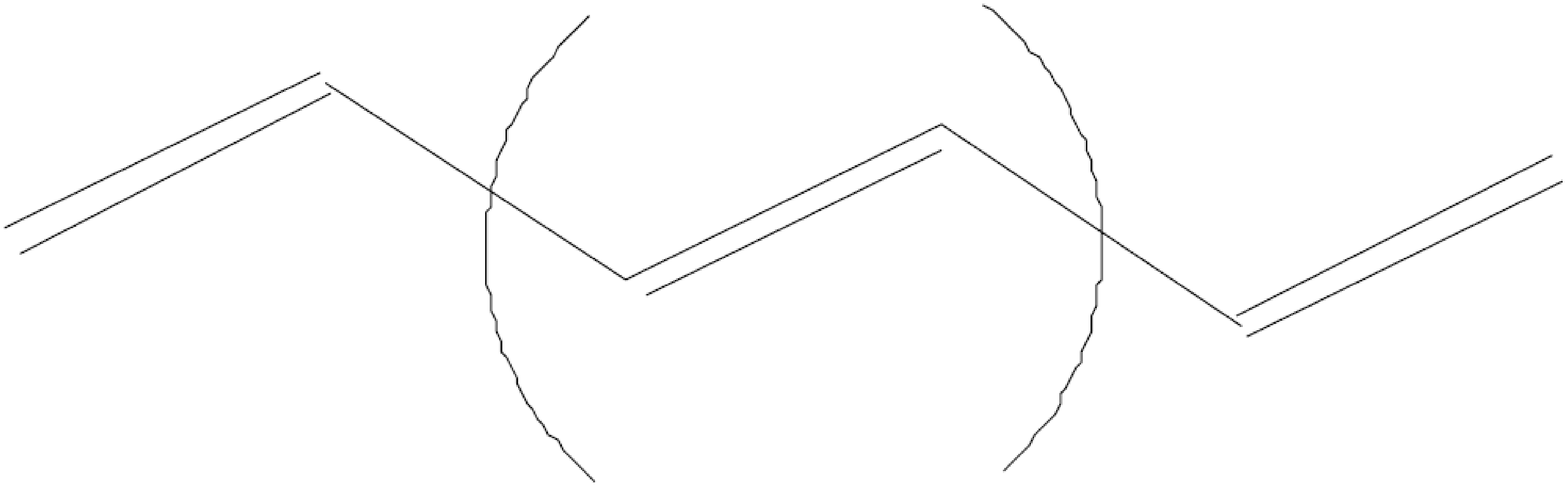}}
\subfigure[Double excitations]{\includegraphics[width=0.4\textwidth]{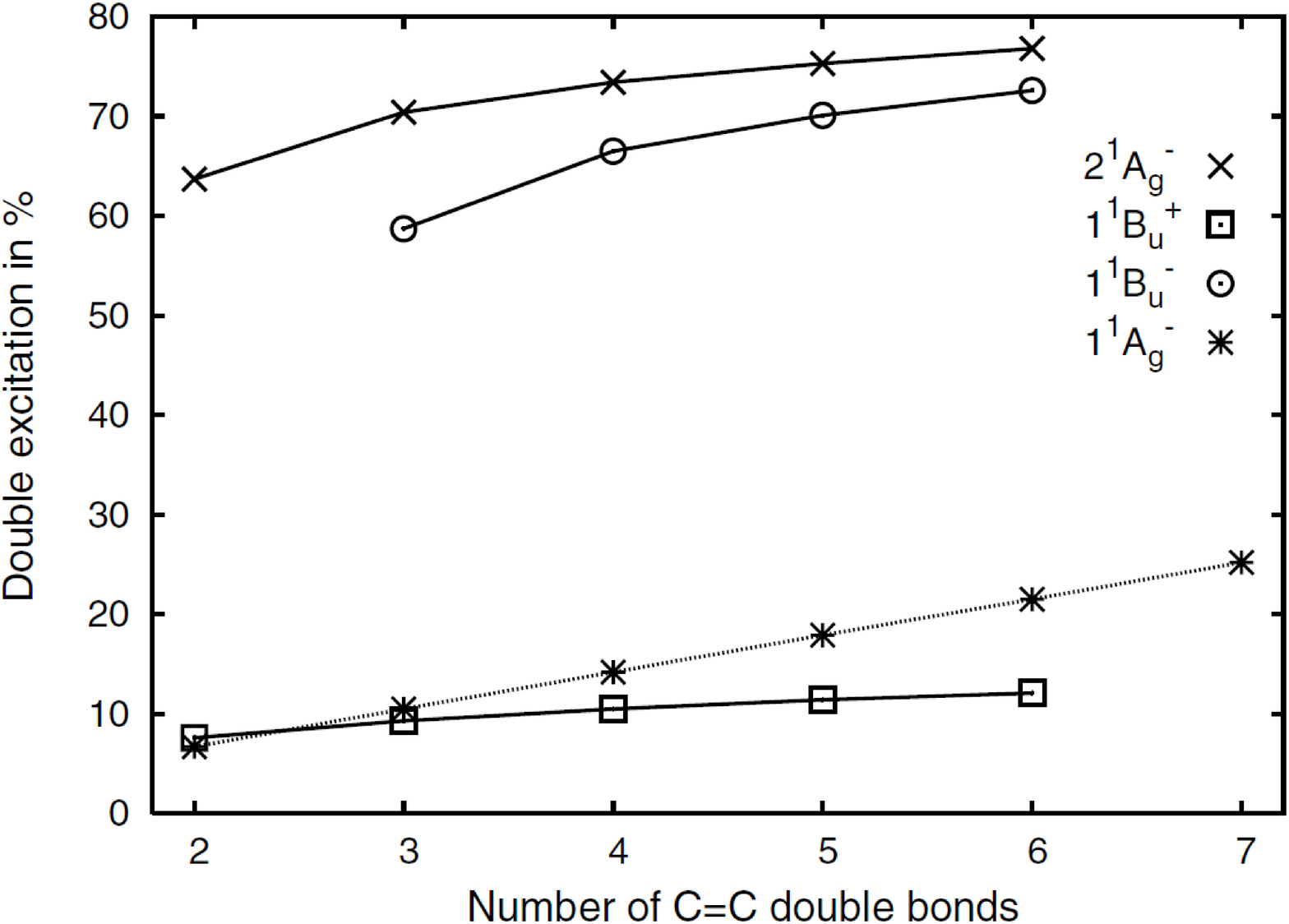}}
\caption{Structure of a polyene chain, $(a)$. The fundamental element of a chain is represented between brackets. At each vertex a carbon atom saturated with hydorgen is present. The weight of doubly excited determinants in the wave--functions of the electronic ground state $1^1 A_g^-$  and the energetically lowest three excited states $1^1 A_g^-$, $1^1 A_g^-$ and $1^1 A_g^-$ are given in percent. From Ref.~\cite{Polyenes_ADC}.}
\label{fig:polyenes}
\end{center}
\end{figure}

For polyenes, on the other hand, the description of DE\index{double excitations}s is known to be important already at the equilibrium geometry. The structure of the chains is represented in Fig.~(\ref{fig:polyenes}), together with a plot of the contribution of DE\index{double excitations}s to the ground state and the first three excited states energies as a function of the chain length.
The contribution to the ground state increases with the dimension of the chain from less then $10\%$ to around $25\%$ while for two of the three excited states is above $50\%$ (reaching the $75\%$) even for the shortest chains.

\begin{figure}[t] 
\begin{center}
\includegraphics[width=0.7\textwidth]{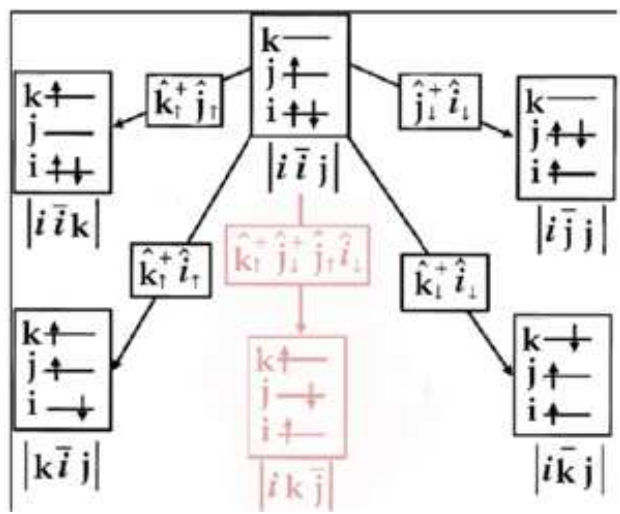}
\caption{Excited states in a spin polarized model with three electrons and three levels. The ground state configuration is plotted on the top, while the doubly excited configuration is highlighted with respect to the others. From Ref.~\cite{Casida1}}
\label{fig:Casida_doubles}
\end{center}
\end{figure}

Theoretically, DE\index{double excitations}s are predicted to be important for any open-shell molecule and in general for any system where the energy of a DE\index{double excitations} can be degenerate, or almost degenerate, with the energy of a single excitation, as shown in Fig.~(\ref{fig:doubles_maitra_model}). For open shell systems, in particular, the inclusion of DE\index{double excitations}s is imposed by spin symmetry requirements which forces single excited configuration to mix with double excited ones. This is shown in Fig.~(\ref{fig:Casida_doubles}) for a very simple three electrons system. The four single excited configuration represented must be mixed among themselves to obtain an eigenstate of the total spin operator $\hat{S}^2$.
However such operator contains a term of the form $\hat{a}^\dag_{i\uparrow}\hat{a}^\dag_{j\downarrow}\hat{a}_{j\uparrow}\hat{a}_{i\downarrow}$ which flips the spin of two particles. Applying these terms to the configurations $\psi_3=\hat{a}^\dag_{k\uparrow}\hat{a}_{i\uparrow}\psi_0$ or $\psi_4=\hat{a}^\dag_{k\downarrow}\hat{a}_{i\downarrow}\psi_0$ a doubly excited configuration $\psi_5=\hat{a}^\dag_{k\uparrow}\hat{a}^\dag_{j\downarrow}\hat{a}_{j\uparrow}\hat{a}_{i\downarrow}\psi_0$ is obtained. This is needed to construct the excited state wave--function with the correct spin symmetry~\cite{spin_adapted_TDDFT,Casida1}.

\subsection*{Satellites in the absorption spectrum}
We have already drawn a connection between the concept of satellites in the photoemission spectrum, illustrated in Ch.~\ref{chap:Green's Function approach} and the mathematical definition of DE\index{double excitations}s. In fact DE\index{double excitations}s can be seen as satellites in the absorption spectra: satellites in photoemission spectra are generated by the frequency dependence of the self--energy, whereas DE\index{double excitations}s by that of the BSE kernel. However, as opposed to photoemission spectra, the concept of satellites (due to pure many body effects) is not used for absorption spectra in the visible/UV range. The main reason being probably related to the different structure of the self--energy and of of the BSE kernel. The Self Energy is in fact usually almost diagonal and as a result the independent particle peaks are only shifted without involving a mixing of the independent particle transitions. On the other hand the kernel of the BSE is strongly non--diagonal and an exciton is composed by different independent particle transitions. As a consequence, usually\footnote{In strongly correlated materials the QP peak can be difficult to isolate as well.}, in the QP spectrum it is easy to isolate a peak which is not related to any independent--particle (IP) transition and so to trace it back to a satellite. On the other hand in the absorption spectrum many IP transitions usually mix in different ways to give well--defined excitonic peaks together with other less intense and dark peaks. Many features are already present in the spectrum. The inclusion of DE\index{double excitations}s can shift the excitonic peaks and increase the richness of the spectrum, however it is difficult to isolate a DE\index{double excitations} satellite without a CI--like analysis of the peak composition.

\begin{figure}[t] 
\begin{center}
\subfigure[Theory: TDDFT and LDA+U]{\includegraphics[width=0.5\textwidth]{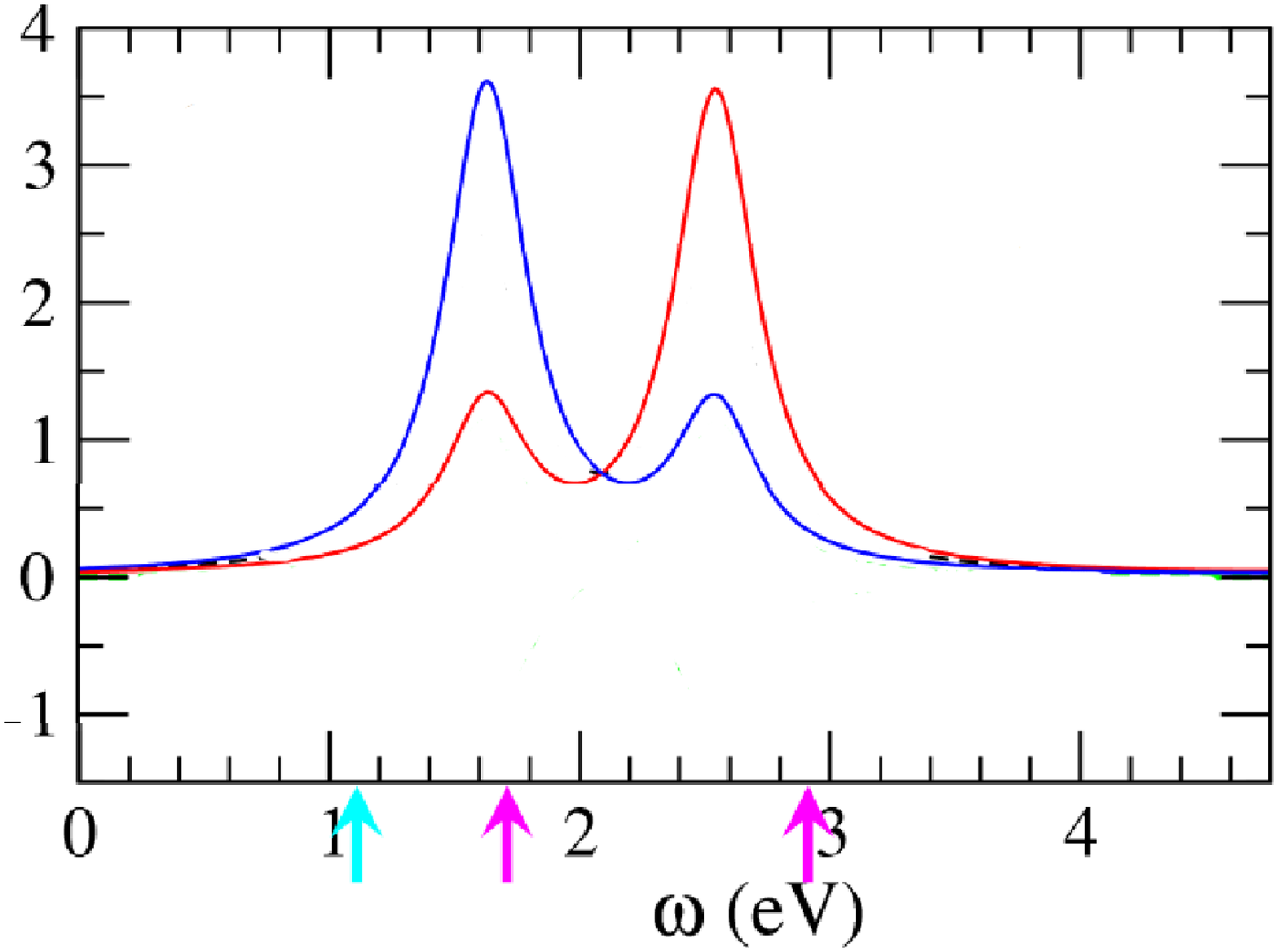}}
\subfigure[Experiment: NIXS]{\includegraphics[width=0.4\textwidth]{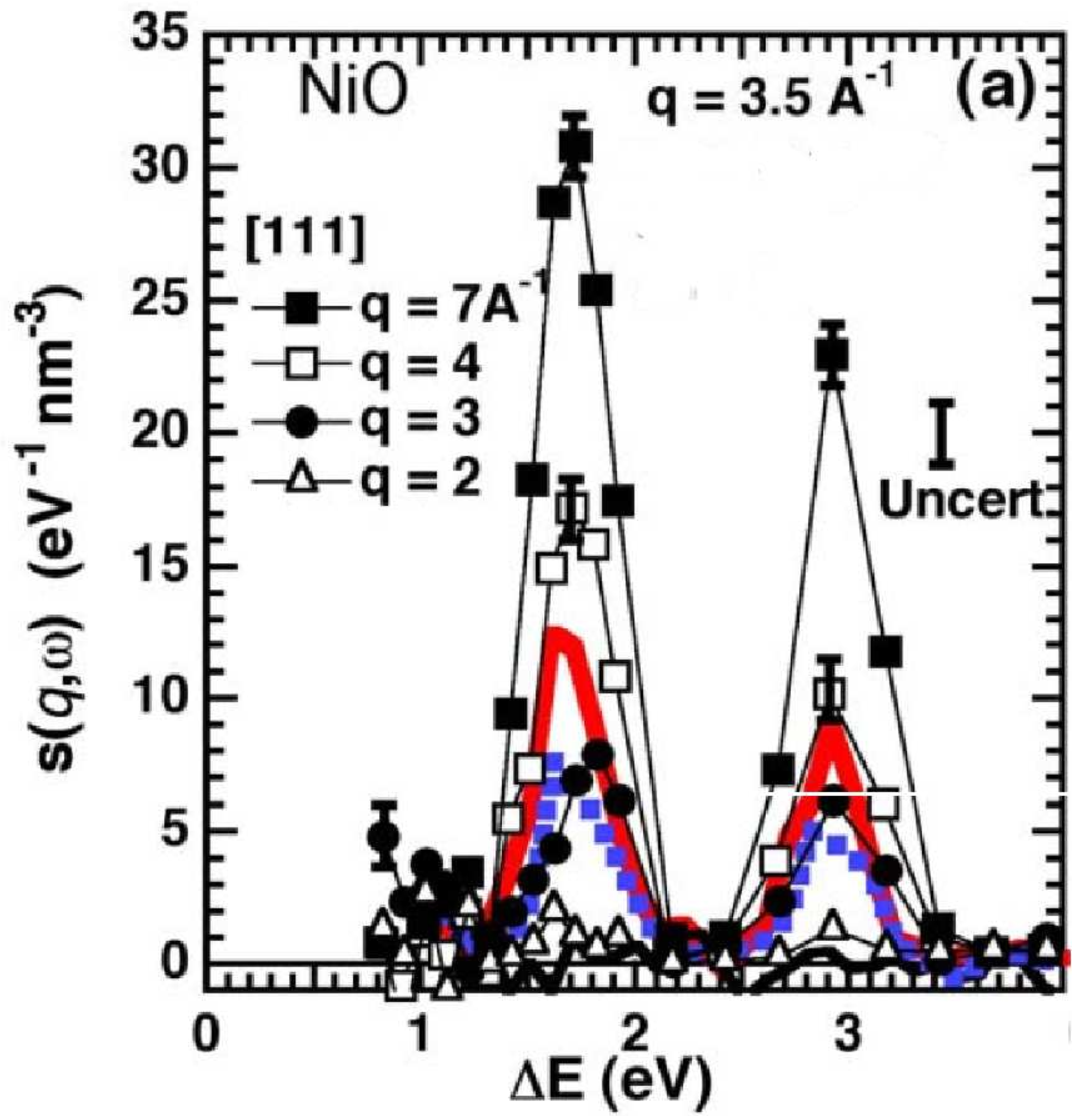}}
\caption{Non-Resonant Inelastic Xray Spectrum (NIXS) of NiO solving a series of tightly bound Frenkel excitons. The experimental spectrum~\cite{NiO_exp} shows more peaks than theoretically predicted~\cite{NiO_theory}. In the work on Lee et al.~\cite{NiO_theory} a frequency--dependent kernel is proposed as a possible solution to describe the experimental peaks (indicated with three arrows in $(a)$). Within the adiabatic description only two peaks are obtained.}
\label{fig:spectrum_NiO}
\end{center}
\end{figure}
$NiO$ has been shown to be a good candidate system where DE\index{double excitations}s can be found in the absorption spectrum. In two recent works the Non-Resonant Inelastic Xray Spectrum (NIXS) of NiO has been investigated both theoretically and experimentally~\cite{NiO_theory,NiO_exp}. At low energy NiO presents tightly bound Frenkel excitons (see Fig.~(\ref{fig:spectrum_NiO})). However the theoretical description within adiabatic TDDFT is not able to reproduce the detailed structure of the spectrum (the theory reproduces only 2 of the 3 peaks observed experimentally). In their work, Lee et al.~\cite{NiO_theory} indicate the need to go beyond the adiabatic approximation for a correct description of the spectrum: $<<$\emph{...to allow any fine (``multi--plets'') structure in strongly interacting systems, a non-adiabatic kernel is absolutely necessary (...). Obviously, this is one key aspect that almost all the existing approximate functionals lack and presents an essential and necessary step toward a proper description of local excitations in strongly interacting systems, within all the existing theoretical frameworks.\ }$>>$ \footnote{However alternative explanations to the extra peak observed in NiO, which do not make use of the concept of DE\index{double excitations}s, already exist in the literature~\cite{NiO_theory_2}.}.

\subsection*{Double excitations, in which basis set?}
Previously in this chapter we have defined  DE\index{double excitations}s from the CI expansion in terms of the HF wave--functions. However some of the experimental evidence we have illustrated has been compared against TDDFT calculations. It is then legitimate to question if we can use the same definition of DE\index{double excitations}s using the KS wave--functions, which, unlike the HF ones, do not have necessarily any physical meaning\footnote{There are several articles indicating that KS orbitals are typically good approximations to Dyson orbitals~\cite{Dyson_orb1,Dyson_orb2,Dyson_orb3}. However this is not true for any system. In correlated materials the overlap between KS wave--function and QP wave--function has been proven~\cite{PhD_Thesis_Gatti}.}.

As we stated in the introduction, when a feature of the spectrum need to be described in terms of DE\index{double excitations}s, the single particle description is breaks down. Then the HF wave--functions lose their physical meaning and have to be regarded only as a basis-set for the expansion of the Many-Body wave--function, and the same holds for KS wave--functions. The open question is if a double excitation in a basis set can be described as a single excitation in another basis set. In fact the single particle transitions, either KS or HF, do not form a complete basis--set in the space of the many--body excitation operators. Indeed the two span two different spaces.

As a partial answer it is possible to consider the number of excited states: the inclusion of DE\index{double excitations}s gives more solutions than the number of single particle excitations initially considered. However the space of single particle transitions is infinite and so it is the number of solutions.

\section{Double excitations in the many body approach} \label{BSE_and_TDDFT_common}
Both TDDFT and the BSE provides exact equations for the description of absorption spectra, therefore they can, in principle, describe DE\index{double excitations}s. In practice, however, approximations to the many--body effects of the system are needed, and the currently used ones, namely ALDA for TDDFT and the statically screened interaction within BSE, fail to reproduce DE\index{double excitations}s. This is why in this work we go beyond the standard approximations and we derive  a new kernel  for BSE  and TDDFT that can properly take into account DE\index{double excitations}s.

In the following we will introduce the single--particle transition space and we will show why a static kernel cannot describe DE\index{double excitations}s. By relaxing the static approximation to the kernel one can get DE\index{double excitations}s. We will illustrate this using the BSE, where the kernel has a clear physical meaning and includes, in a natural way, the many--electron excitations of the system. From the BSE kernel one can then obtain the TDDFT kernel using the technique of Ref. \cite{PhD_Thesis_Gatti}.
\subsection*{The Dyson equations in the transition space}
The main advantage of writing the equations for the response function in the space of single--particle transitions is that it offers a clear interpretation in terms of the single--particle wave--function. In the present work we will assume that the differences among the QP and the KS wave--functions are small and we will project the TDDFT and BS equations in the KS basis set. This assumption is usually done in the implementation of the BSE scheme within many ab--initio codes and it is justified in many materials though exceptions have been found~\cite{PhD_Thesis_Gatti}. In particular the starting point is usually a DFT--LDA calculation with self--energy effects introduced according to the scheme outlined in Ch.~\ref{chap:Green's Function approach}. Then the excitonic spectra can be computed within TDDFT starting from the $\psi_i^{KS}$ and $\epsilon_i^{LDA}$ or from the MBPT-$GW$ approach where the BSE is solved starting from $\psi_i^{KS}$ and $\epsilon_i^{GW}$.

We start from a Dyson equation for a generalized four--points response function in the frequency domain
\begin{multline} \label{generalized_Dyson}
\tilde{L}(\mathbf{1,2;3,4}|\omega)=\tilde{L}_0(\mathbf{1,2;3,4}|\omega)+  \\
\tilde{L}_0(\mathbf{1,2;1',2'}|\omega)K(\mathbf{1',2';3',4'}|0)\tilde{L}(\mathbf{3',4';3,4}|\omega)
\text{,}
\end{multline}
where we have already introduced the static approximation for the kernel $K$. Eq.~(\ref{generalized_Dyson}) can be obtained from Eq.~(\ref{Bethe-Salpeter}) by using
\begin{equation}
\begin{split}
&\tilde{L}(\mathbf{1,2;3,4}|\omega)=\int d(t_2-t_1) e^{-i\omega (t_2-t_1)}
             L(\mathbf{1}t_1,\mathbf{2}t_1;\mathbf{3}t_2,\mathbf{4}t_2)   \text{,}     \\
&\tilde{L}_0(\mathbf{1,2;3,4}|\omega)=\int d(t_2-t_1) e^{-i\omega (t_2-t_1)}
             L_0(\mathbf{1}t_1,\mathbf{2}t_1;\mathbf{3}t_2,\mathbf{4}t_2) \text{,}     \\
&K(\mathbf{1,2;3,4}|0)=\int d(t_2-t_1) e^{-i\omega (t_2-t_1)}
             \bigg(\delta(1,2)\delta(3,4)w(1,3)                       \\
             &\phantom{K(\mathbf{1,2;3,4}|0)=}\phantom{\int d\omega e^{-i\omega (t_2-t_1)}\delta(1,2)\delta(3)}
                   -\delta(1,3)\delta(2,4)W(\mathbf{1}t_1,\mathbf{2}t_2)\delta(t_2-t_1) \bigg)
\text{;}
\end{split}
\nonumber
\end{equation}
we have introduced $\tilde{L}$ to be distinguished from $L$. $\tilde{L}$ is a function of four space variables as $L$ but two time variables only. An equation of the form of Eq.~(\ref{generalized_Dyson}) can be obtained from Eq.~(\ref{Dyson-tddft}) using
\begin{equation}
\begin{split}
&\tilde{L}(\mathbf{1,2;3,4}|\omega)=\int d\omega e^{-i\omega (t_2-t_1)}
             \chi(\mathbf{1}t_1;\mathbf{3}t_2)\delta(1,2)\delta(3,4)     \text{,}    \\
&\tilde{L}_0(\mathbf{1,2;3,4}|\omega)=\int d\omega e^{-i\omega (t_2-t_1)}
             \chi_{KS}(\mathbf{1}t_1;\mathbf{3}t_2)\delta(1,2)\delta(3,4)   \text{,}         \\
&K(\mathbf{1,2;3,4}|0)=\int d\omega e^{-i\omega (t_2-t_1)}
             \(w(\mathbf{1,3})+\frac{\delta v_{xc}^A[\rho_{t_1}](\mathbf{1})}{\delta \rho_{t_1}(\mathbf{3})}\)
             \delta(1,2)\delta(3,4)
\text{.}
\end{split}
\nonumber
\end{equation}
Notice that not only the kernel is different, but also the four--points functions $\tilde{L}$ and $\tilde{L}_0$ differ in the two cases.
In the space of single particle wave--functions the equation for $\tilde{L}$ reads~\footnote{In the TDDFT formalism, in order to obtain simpler equations, when moving to the single--particle wave--functions space, the delta functions are discarded in the definition of $\tilde{L_0}$ which is constructed from $L_{KS}(1,2;3,4)=G(1,3)G(4,2)$. Only in this way $\tilde{L}_0$.}:
\begin{equation} \label{generalized_Dyson_matrix}
\tilde{L}_{ij,hk}(\omega)=\tilde{L}^0_{ij,hk}(\omega)+\tilde{L}^0_{ij,i'j'}(\omega)K_{i'j',h'k'}(0)\tilde{L}_{h'k',hk}(\omega)
\text{,}
\end{equation}
where we used the following change of basis
\begin{equation}
\begin{split}
\tilde{L}_{ij,hk}(\omega)=&\int d^3\mathbf{x_1}...d^3\mathbf{x_4}\ 
                  \psi_i(\mathbf{x_1})\psi^*_j(\mathbf{x_2})
                  \tilde{L}(\mathbf{1,2;3,4}|\omega)
                  \psi^*_h(\mathbf{x_1})\psi_k(\mathbf{x_1})   \\
                 =&\ \langle ij |\ \tilde{L}(\mathbf{1,2;3,4}|\omega)\ | hk\rangle
\text{.}
\end{split}
\end{equation}
Here $i$ is a generalized index for the KS wave--function containing all quantum numbers, spin included. Within this basis--set the generalized response function $\tilde{L}_0$ is diagonal
\begin{equation} \label{Lo_transition}
\tilde{L}^0_{ij,hk}=\frac{\delta_{i,k}\delta_{j,h}(f_j-f_i)}{\omega-(\epsilon_i-\epsilon_j)}
\text{.}
\end{equation}
We have dropped the $i\eta$ factors in order to avoid having two different expressions for the retarded and the time--ordered response function. If needed one can restore them remembering that we deal with retarded quantities within TDDFT and with time--ordered quantities within MBPT \footnote{One have to take care of this difference if a connection among the two theories need to be established. The difficulties which can arise trying to combine theories with different time ordering can be overcome thanks to the Keldish contour techniques.}. Moreover we have introduced the occupation factors\footnote{The occupation factors can be used as Fermi distribution functions to introduce a numerical smearing to get a faster convergence. The same Fermi functions are sometimes used to introduce a temperature dependence in the equations.}, $f_i=1$ if $i$ is occupied and $f_i=0$ otherwise, to have a compact expression for $\tilde{L}_0$. Using Eq.~(\ref{Lo_transition}) and writing the matrix equation in the form $\tilde{L}^{-1}=\tilde{L}_0^{-1}-K$ one can find the zeros of $\tilde{L}^{-1}$ solving an eigenvalue problem:
\begin{equation} \label{generalized_Dyson_eigenvalue_simple}
H^{2p} A_I = \omega_I A_I
\text{,}
\end{equation}
with $H_{ij,hk}=(\epsilon_i-\epsilon_j)\delta_{i,k}\delta_{j,h}+\sqrt{f_j-f_i}\ K_{ij,hk}\ \sqrt{f_h-f_k}$. Taking explicitly into account the occupation factors the eigenvalue problem can be recast in the electron--hole ($eh$) and hole--electron ($he$) transitions (see \cite{Casida_TDDFT} for TDDFT and \cite{RMP_Onida} for BSE):
\begin{equation} \label{generalized_Dyson_eigenvalue}
\left( \begin{array}{cc}
H_{res} & H_{coup} \\
H^{*}_{coup} & H^*_{res}
\end{array} \right)
\left( \begin{array}{c}
X_I \\
Y_I \end{array} \right)
= \omega_I
\left( \begin{array}{cc}
1 & 0 \\
0 & -1
\end{array} \right)
\left( \begin{array}{c}
X_I \\
Y_I \end{array} \right)
\text{.}
\end{equation}
We defined the resonant term $H_{res}=H_{eh,e'h'}$ and the coupling term $H_{coup}=H_{eh,h'e'}$ and we have assumed that $K(\mathbf{1,2;3,4}|\omega)$ is real. Let us call $S$ the matrix defining the metric of the system:
\begin{equation}
S=
\left( \begin{array}{cc}
1 & 0 \\
0 & -1
\end{array} \right)
\end{equation}
The block form of Eq.~(\ref{generalized_Dyson_eigenvalue}) allows us to have a clear interpretation of the physics involved. In particular the resonant part of the Hamiltonian describes the neutral excitations of the system. In terms of a linear composition of single $eh$ transitions. This can be directly compared with $\Psi_I^{HF}$ defined in Eq.~(\ref{excited_HF}). The term $H^*_{res}$ is the anti--resonant part of the Hamiltonian and describes the de--excitations. Finally $H^*_{coup}$ describes the coupling among the $eh$ and the $he$ space. When these are different from zero the many--body excitations contains terms involving single--particle de--excitations. These latter processes are clearly forbidden if one consider the $HF$ ground state and for this reason the coupling terms are said to describe ground state correlation~\cite{Gambacurta}.

While the eigenvalues of the problem gives the excitations energies the eigenvector can be used to construct excitation operators. Since we are working within the DFT basis set, it is tempting to describe the excited state wave--function directly applying the excitations operator to the ground state, in analogy to HF--based methids. However one has to keep in mind that the TDDFT and MBPT linear response equations have been derived starting from a variation of the density with no assumption on the wave--function. For this reason the excitation operator should be applied not to the DFT wave--function but to the correlated Many-Body wave--function. Only the TDA, which assumes that de--excitations do not need to be considered, is consistent with the approximation of the ground state as a single slater determinant. Within this approximation excited states wave--functions can be constructed from the KS ground state and interpreted, as it is often done for KS the ground--state, as approximations to the real many body wave--function. However only within the TDHF scheme and starting from the HF ground--state the approximation is formally correct.

Moreover, with the exact kernel, Eq.~(\ref{generalized_Dyson_eigenvalue}) would give the exact excitations energies, but not the full excitation operators, since only the single--particle part is accessible by construction. This means that multi--particle transitions must be hidden in the kernel of the equation. It becomes now clear why the eigenvalue problem (\ref{generalized_Dyson_eigenvalue}) cannot give DE\index{double excitations}s, unless the kernel is frequency--dependent. In the next chapter we will clarify these points exploring the Second--RPA method where an equivalent eigenvalue problem will be derived exactly from the projection on the space of single particle transition of an exact equation for the excitation operators.

Eq.~(\ref{generalized_Dyson_eigenvalue}) is equivalent to a Dyson equation for the generalized response function. From its solution we can construct the response function $\chi(\mathbf{1,2}|\omega)=\tilde{L}(\mathbf{1,1;2,2}|\omega)$.
As explained in Ch.~\ref{chap:Many Body Systems} 
we can use $\chi$ to define the dielectric function $\langle\epsilon^{-1}(\omega)\rangle=\langle1+w\chi\rangle$ and the polarizability $\langle\alpha\rangle=\langle\mathbf{x}\chi\mathbf{x}\rangle$. As an example we write here the esplicit connection of the macroscopic measurable quantity $\alpha(\omega)$ with the Eq.~(\ref{generalized_Dyson_eigenvalue}).

The expression for the generalized response function in terms of the eigenvalues and the eigenvectors of Eq.~(\ref{generalized_Dyson_eigenvalue}) is
\begin{equation}\label{L_solution}
\tilde{L}_{ij,hk}(\omega)=\sum_{I,J}\frac{A^I_{ij}S_{I,J} A^J_{jk}}{(\omega-\omega_I)}
\text{.}
\end{equation}
Using Eq.~(\ref{L_solution}) the polarizability can be written as
\begin{equation} \label{polarizability_eh}
\begin{split}
\alpha_{x_a x_b}(\omega)=&\sum_{ij,hk}\langle x^a_{ij} \tilde{L}_{ij,hk}(\omega) x^b_{hk} \rangle  \\
                        =&\sum_{ij,hk}\langle i | x^a | j \rangle
                          \sum_{I,J}\frac{A^I_{ij}S_{I,J} A^J_{jk}}{(\omega-\omega_I)}
                                      \langle k | x^b | h \rangle
\end{split}
\end{equation} 
The direct expression for the dielectric function in the space of transitions, instead, can be used to describe the concept of local fields effect if the basis--set of the block wave--functions is used. This is done in App. \ref{App:Block wave functions}.

\section{The dynamical Bethe--Salpeter equation (step I)}

\begin{figure}[t]
 \centering
 \includegraphics[width=0.99\textwidth]{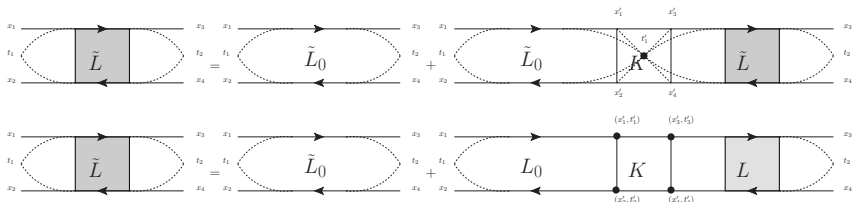}
 \caption{The BSE with a static kernel (upper diagrams) is compared with the BSE with an exact kernel (lower diagrams). The contraction of time variables to obtain $L\rightarrow\tilde{L}$ is represented with small dashed lines. The static approximation for the kernel is represented as a collapse of the time dependency to a single point highlighted in the figure. When using the exact kernel there is no time contraction and it is not possible to do the contraction $L\rightarrow\tilde{L}$ on the last term of the r.h.s.\ . The BSE does not reduce to a closed equation for $\tilde{L}$.}
\label{fig:BSE_time_contraction}
\end{figure}

In order to explore the effects of a dynamical kernel on the description of DE\index{double excitations}s as start we simply relax the static approximation to the screened interaction in the BSE~\footnote{We choose BSE as starting point because within TDDFT we do not have a straightforward way to insert a frequency--dependency in the $f_{xc}$ kernel}. However we immediately realize that, by using a frequency--dependent interaction $W(\omega)$, Eq.~(\ref{generalized_Dyson_matrix}) cannot be written any more as a simple matrix equation, as it involves a convolution in the frequency space. To analyze the problem we rewrite the exact BSE for $\tilde{L}(\omega)$ obtained by using a frequency--dependent kernel:
\begin{multline} \label{Exact-BSE}
\tilde{L}_{ij,hk}(\omega)=\tilde{L}_{ij,hk}^0(\omega)+\frac{1}{4\pi^2}\int d\omega' d\omega''
L_{ij,i'j'}^0(\omega,\omega')  \\ \Xi_{i'j',h'k'}(\omega,\omega',\omega'') L_{h'k',hk}(\omega,\omega'')
\text{,}
\end{multline}
The kernel $\Xi$ depends on four time variables making it impossible to contract the variables in the last term of the r.h.s. of Eq.~(\ref{Exact-BSE}) for $L$ and $L_0$, as shown in Fig.~(\ref{fig:BSE_time_contraction}) in order to obtain a closed equation for $\tilde{L}$. This problem does not appear within TDDFT where we have only two point (and so two times) quantities. For the Fourier transform we have adopted the following conventions~\cite{Strinati_book,Romaniello1}:
\begin{equation}
L(\omega,\omega',\omega'')=\int d\tau d\tau' d\tau'' e^{i\omega\tau} e^{i\omega'\tau'} e^{i\omega''\tau''}
                           L(t_1,t_1',t_2,t_2')
\end{equation}
with $\tau=(t_1+t_1')/2-(t_2+t_2')/2$, $\tau'=t_1-t_1'$ and $\tau'=t_2-t_2'$.

Commonly the BSE kernel follows from the GW approach to the self--energy and it reads
\begin{equation}
\Xi_{ij,hk}(\omega''-\omega')=w_{ij,hk}-W_{ij,hk}(\omega''-\omega')
\text{,}
\end{equation}
where we considered the full frequency dependence in the screened interaction.
To obtain a Dyson equation we insert the identities $L_0L_0^{-1}$ and $LL^{-1}$ on the left and of the right of the second term on the l.h.s. of Eq.~(\ref{Exact-BSE}) and define a new kernel
\begin{multline}
\(\Xi^d_2\)_{ij,hk}(\omega)\simeq w_{ij,hk} -L^{0}_{ij,ij}\phantom{)}^{-1}(\omega) \int d\omega' d\omega''   \\
  L^0_{ij,ij}(\omega,\omega') W_{ij,hk} (\omega''-\omega') L^0_{hk,hk}(\omega,\omega'')
 L^0_{hk,hk}\phantom{)}^{-1}(\omega)
\text{.}
\end{multline}
Here we have approximated $L\simeq L_0$ (linearization) in the kernel expression and we have used the fact that $L_0$ is diagonal in configuration space. In this way we have a closed equation, the dynamical BSE (DBSE), that we can project in the transition space in order to obtain an eigenvalue problem:
\begin{equation} \label{Freq_dep_BSE}
\tilde{L}_{ij,hk}(\omega)=\tilde{L}^0_{ij,hk}(\omega)+\tilde{L}^0_{ij,i'j'}(\omega)\tilde{K}_{i'j',h'k'}(\omega)\tilde{L}_{h'k',hk}(\omega)
\text{,}
\end{equation}
which is formally identical to SBSE except for the presence of a frequency--dependent kernel $\tilde{K}(\omega)=\Xi^d_2(\omega)$. Note that the if a static kernel is used the DBSE reduces exactly to the usual Static BSE (SBSE). 

From Eq. (\ref{Freq_dep_BSE}) we can construct an eigenvalue equation with frequency--dependent excitonic Hamiltonian
\begin{equation}\label{eigen_omega_dependent}
H^{2p}(\omega) A_I(\omega) = \omega_I(\omega) A_I(\omega)
\text{.}
\end{equation}
Note that Eq. (\ref{eigen_omega_dependent}) is similar to the frequency--dependent eigenvalue equation obtained by Strinati in Ref.~\cite{Strinati_book}.

\subsection*{A first analysis of the kernel}
\begin{figure}[t]
 \centering
 \includegraphics[width=0.9\textwidth]{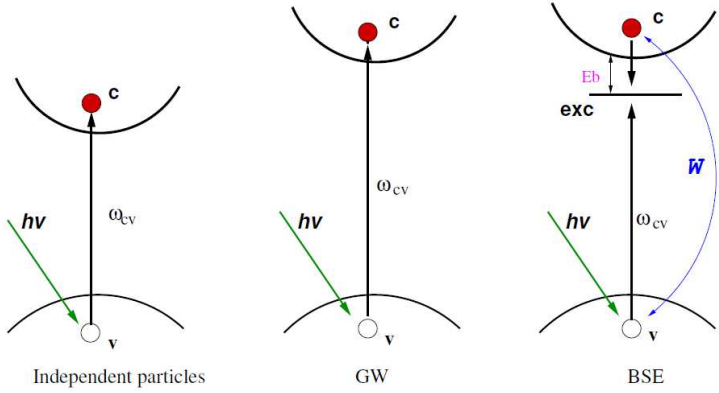}
 \caption{Self--energy corrections open the LDA gap due to a better description of the (screened) exchange effect which tends to push particles away. However the (screened) attraction between the electron and the hole gives a binding energy which partially compensates the self--energy effects and so the value of the optical gap is often closer to the LDA gap rather than to the QP one. From Ref. \cite{PhD_Thesis_Daniele}}
\label{fig:BSE_and_W_effects}
\end{figure}

The frequency integrals in the definition of $\tilde{K}(\omega)$ can be performed analytically as we know the frequency dependence of both $L_0(\omega)$ and $W(\omega)$. As for the static screening, we have here the problem that the kernel of the equation should, in principle, depend on the solution of the equation itself. However, in the SBSE the exact position of the poles in the construction of the kernel is not important and LDA eigenvalues are used to construct the screening instead of the exact poles of the response function or of the QP energies. QP eigenenergies are not chosen in order to prevent the larger QP gap to underestimate the screening\footnote{In a recent work \cite{GW_screening} the use of QP energies in the construction of the kernel has been proven to give better results for some materials. Other work on this point can be found in \cite{GW_screening2,GW_screening3}}. Indeed LDA gap is often comparable with the optical gap due to partial cancellation of self-energy and kernel effects; see Fig.~(\ref{fig:BSE_and_W_effects}).
However we will show in Ch.~(\ref{chap:A new approach to describe Double Excitations}) that the DBSE solutions are very sensitive to the exact position of the poles (see also Fig.~(\ref{fig:doubles_maitra_model}) ). Therefore a KS screening might not be accurate anymore. To overcome this problem one can use RPA or static BSE energies. From now on we will use the RPA energies $\Omega_\nu$, with eigenvectors $R^\nu$, which are solutions of Eq.~(\ref{generalized_Dyson_eigenvalue}) with $K(\mathbf{1,2,3,4}|0)=\delta(1,2)\delta(3,4)w(\mathbf{1,3})$. With this choice we can write the frequency--dependent kernel within TDA as
\begin{equation}
\(\Xi^d_2\)_{ij,hk}(\omega)= w_{ij,hk} + 2\sum_{\nu}\sum_{(nq)(n'q')}
                             v_{(jk)(nq)}\frac{R^\nu_{(nq)}R^{\nu\ *}_{(n'q')}}{\omega-(\Omega_\nu+\Delta\epsilon)}  v_{(n'q')(ih)}
\text{,}
\end{equation}
where the indexes $(nq)$ run over the possible $eh$ couples only and where the factor 2 is obtained using $R^\nu_{(nq)}=R^{-\nu}_{(qn)}$ which holds within TDA.

A frequency--dependent kernel imposes to solve Eq.~(\ref{generalized_Dyson_eigenvalue}) self-consistently with respect to the frequency. In practice we need the solution of the equation $E_I(\omega)=\omega$ with $E_I(\omega)$ the eigenenergies of a DBSE at a given frequency. This assumption resembles the QP concept as it relies on the assumption that
\begin{equation} \label{QP_exciton}
\begin{split}
\tilde{L}_{ij,hk}(\omega)&=\sum_{I,J}\frac{A^I_{ij}(\omega)S_{I,J} A^J_{jk}(\omega)}{(\omega-\omega_I(\omega))}  \\
                    &\simeq\sum_{I,J}\frac{A^I_{ij}(E_I)S_{I,J} A^J_{jk}(E_J)}{(\omega-E_I)}  
\text{.}
\end{split}
\end{equation}

For further details on this idea see App.~\ref{App:On the quasiparticle concept}

\subsection*{Connection with time dependent density functional theory}
In sec. \ref{BSE_and_TDDFT_common}, Eq.~(\ref{generalized_Dyson_matrix}) has established a common language between the SBSE and TDDFT. The DBSE, Eq. (\ref{Freq_dep_BSE}), is also the generalization of Eq. (\ref{generalized_Dyson_matrix}) to the frequency--dependent case. Indeed, with the choice $\tilde{K}(\omega)=f_{Hxc}(\omega)$ in Eq. (\ref{Freq_dep_BSE}), the TDDFT formalism is recovered.

A kernel for the TDDFT can be derived taking advantage of this common language. Under the assumption that the differences between the QP and the KS wave--functions can be neglected we obtain
\begin{equation}\label{tddft_kernel}
f_{Hxc}(\omega)=\chi_{KS}^{-1}(\omega)-\chi_0^{-1}(\omega)+\tilde{\Xi}^d_2(\omega)
\text{  ,}
\end{equation}
where the term $\chi_{KS}^{-1}(\omega)-\chi_0^{-1}(\omega)$ takes into account for the difference between the $QP$ and the $KS$ starting point. The term $\tilde{\Xi}^d_2(\omega)$ is the space contracted version of $\Xi^d_2(\omega)$, obtained changing the four points $L_0$ with the two points $\chi_0$ in the definition of the latter. Indeed, choosing the static approximation for the screened interaction in the definition of $\tilde{\Xi}^d_2(\omega)$, we obtain the so called Nanoquanta kernel~\cite{PhD_Thesis_Gatti}.

From now on we will work in the BSE framework, focusing our attention on the problem of DE\index{double excitations}s. However, under the assumptions of this section, a TDDFT kernel can be derived using Eq. (\ref{tddft_kernel}) and using a modified DBSE kernel (here $\tilde{\Xi}^d_2(\omega)$ ) which can be derived as long as the spatial contraction $L_0\rightarrow\chi_0$ is possible. This will be the case for the kernel proposed in the next chapter\footnote{However to define the contracted kernel the substitution  $L_s\rightarrow\chi_s$ instead of $L_0\rightarrow\chi_0$ will have to be considered. $L_s$ will be defined in Ch. \ref{chap:A new approach to describe Double Excitations} }. For a further study of the problem of the construction of a TDDFT kernel from a many--body formalism we address the reader to the references \cite{Casida2,PhD_Thesis_Gatti}. In particular the spatial contraction (or localization) is extensively discussed in \cite{Casida2}. As pointed out there however, it is not necessary if the only goal is to obtain excitations energy, since one can do a DFT based BSE calculation. This simply because $\chi$ and $L$ have the same poles; on the other hand if one is interested in intrinsic TDDFT quantities, as for example the time evolution of the density, the localization process can be crucial~\cite{Casida2}.
     
\subsection*{Some preliminary tests on a model system}
The performance of the DBSE can be tested in a two electrons and two levels model, that is the simplest possible system where a DE\index{double excitations} can appear. We will work here within the Tamm-Dancoff approximation to keep the equations as simple as possible. We look for the solution of the eigenvalue problem solving the equation $det(H^{2p}(\omega)-1\omega)=0$. For our model we obtain:
\begin{equation}\label{model_Pina}
\left(\Delta\epsilon+V-\tilde{W}(\omega)-\omega\right)^2-V^2=0
\text{,}
\end{equation}
where we defined $V=w_{vc,vc}$\ , $\tilde{W}(\omega)=\(\Xi^d_2\)_{vc,vc}(\omega)$\ , $\Delta\epsilon=\epsilon_c-\epsilon_v$ and we used the fact that $\tilde{W}(\omega)$ is diagonal in the spin space. The RPA solutions $\Omega_{1,2}$ needed to construct the $\tilde{W}(\omega)$ term have eigenvectors $R_1=1/\sqrt{2}\ (1\ \ 1)^t$ and $R_2=1/\sqrt{2}\ (1\ -1)^t$, thus we get
\begin{equation}
\tilde{W}(\omega)=A+\frac{B}{\omega-\Omega_1-\Delta\epsilon}
\end{equation}
where with $A=w_{vv,cc}$ and $B=2Re[w_{ccvc}w_{vcvv}]$. Eq.~(\ref{model_Pina}) has four solutions, although one expects only three for this system, i.e., a singlet single excitation, a triplet single excitation, and a singlet DE\index{double excitations}. One of the four is an unphysical state. We argue that the occurrence of this extra pole is related to the self-screening interaction that the GW approximation to the self-energy suffers from. This is related to the fact that $\tilde{W}$ is the test charge--test charge screening, whereas the charges to be screened are fermions, not classical charges. This can be cured by introducing a vertex correction to the self-energy. Indeed, if one considers only one electron in this model system, then Eq.~(\ref{eigen_omega_dependent}) produces two poles, one corresponding to a single excitation and the other one, unphysical, corresponding to a DE\index{double excitations}. In this case, there are no dynamical self-energy effects involved, and the extra pole arises, indeed, from the fact that the electron screens itself. We can recognize the spurious solution by solving Eq.~(\ref{model_Pina}) independently of the dynamical structure of $\tilde{W}$. We then obtain two groups of solutions: one for singlet states, $\omega=\Delta\epsilon+2V-\tilde{W}$, and one for triplet states, $\omega=\Delta\epsilon+2V-\tilde{W}$. Since the excited state involving a DE\index{double excitations} is a singlet, the correct double-excitation energy is the one coming from the singlet-group solutions. The four solutions ($\omega_{1,2}$ the singlet solutions, and $\omega_{3,4}$ the triplet
solutions) are
\begin{equation}
\begin{split}
\omega_{1,2} =& \frac{2\Delta\epsilon+\Omega_1-A\mp\sqrt{(\Omega_1-2V+A)^2-4B}}{2}   \\
\omega_{3,4} =& \frac{2\Delta\epsilon+\Omega_1-A\mp\sqrt{(\Omega_1+A)^2-4B}}{2}
\end{split}
\text{.}
\end{equation}
The energy $\omega_4$ (the solution with the sign $+$) is a spurious pole. In the next chapter we will understand better the origin of spurious excitation energies and we will show how to derive an approximation to the DBSE kernel which does not suffer of this problem.

\chapter{A new approach}                                \label{chap:A new approach to describe Double Excitations}
As we have shown in Ch.~\ref{chap:Introduction to the problem}, DEs are essential for the description of the optically excited states in open-shell molecules~\cite{Casida1}; however they can play an important role also in closed-shell systems, such as in polyenes, where the lowest-lying singlet state is known to have a HOMO$^2$-LUMO$^2$ double-excitation character~\cite{Maitra2}. The theoretical description of Double Excitations (DE) in conjugated polymers constitutes an important challenge for the state-of-the-art approaches used in physics and physical chemistry.

On one side there are the Post--HF methods that descibes DEs in a natural way, but at the price of a very demanding description of correlation effects; on the other side there are methods as TDDFT and BSE, which treat better the correlation, but within the standard approxmiations cannot capture the physics of DEs. The limitation of the latter approaches lies in the adiabatic approximation to the exchange--correlation effects.

In Ch. \ref{chap:Introduction to the problem} we showed that simply relaxing this approximatin, DEs are in fact described; however, together with the desired excitations, non-physical excitations also appear. Spurious excitations have been interpreted as due to the self-screening error embodied in the $GW$ self-energy~\cite{Romaniello1,Romaniello2}. 

In this chapter we investigate more in details this problem showing that uncontrolled effects, such as unphysical excitations, can appear as the quantum statistics and Pauli exclusion principle are easely broken in simple approximations, like the one we introduced. For this reason here we propose a novel approach to describe DEs in correlated materials by embodying the mathematical properties of post-HF methods (here we will use as reference the second--RPA) in a coherent Many-Body framework. In order to achieve this we first define the conditions for a Number Conserving (NC) approach, which avoids the appearance of spurious excitations; we then embody the NC condition in an extension to the BSE that describes DEs in a consistent manner.

\section{The second random phase approximation}  \label{Sec:sRPA}
\subsection*{A number--conserving approach}
The second--RPA (sRPA) is a particular appealing starting point because the scheme is directly derived approximating the many body excitation operators to DEs~\cite{sRPA1}:
\begin{multline}\label{exc_op_double}
\hat{O}_\nu\simeq\sum_{ij}\[ X^{(1)}_{ij}(\omega_\nu)\hat{a}^\dag_i\hat{a}_j - Y^{(1)}_{ij}(\omega_\nu)\hat{a}^\dag_j\hat{a}_i \]+ \\
            \sum_{ijmn}\[ X^{(2)}_{ijmn}(\omega_\nu)\hat{a}^\dag_i\hat{a}^\dag_m\hat{a}_j\hat{a}_n -
                          Y^{(2)}_{ijmn}(\omega_\nu)\hat{a}^\dag_j\hat{a}^\dag_n\hat{a}_i\hat{a}_m \]
\text{,}
\end{multline}
where $\hat{a}^\dag_i$ / $\hat{a}_i$ are creator / annichilation operators in a single particle wave--function basis set, the HF wave--functions are used in the original derivation of the sRPA equations, and $\omega_I$ are the excitation energies. The scheme can be constructed inserting Eq.~(\ref{exc_op_double}) in a double commutator equation which is satisfied by the operator $\hat{O}_\nu$~\cite{sRPA1}:
\begin{equation} \label{EOM}
\langle HF | \[ \hat{R}, \[\hat{H}, \hat{O}^\dag_\nu  \] \] | HF \rangle = 
    \omega_\nu\  \langle HF | \[ \hat{R}, \hat{O}^\dag_\nu  \] | HF \rangle
\text{.}
\end{equation}
Here $|HF\rangle$ is the HF ground state, $\hat{H}$ is the many body Hamiltonian and $\hat{R}$ is an operator in the same space of the excitation operators $\hat{O}_\nu$. The result can be written in the form of an eigenvalue equation:
\begin{equation}\label{EOM_eigen}
\left( \begin{array}{cc}
 \mathcal{A}      &  \mathcal{B}      \\
-\mathcal{B}^*    & -\mathcal{A}^* \end{array}\right)
\left( \begin{array}{c} \mathcal{X}(\omega_\nu) \\ \mathcal{Y}(\omega_\nu) \end{array}\right)
= \omega_\nu \left( \begin{array}{c} \mathcal{X}(\omega_\nu) \\ \mathcal{Y}(\omega_\nu) \end{array} \right)
\text{,}
\end{equation}
where
\begin{equation}
\mathcal{A} =
\left( \begin{array}{cc}
 A_{ij,hk}      &  A_{ij,hkpq}      \\
 A_{ijmn,hk}    & A_{ijmn,hkpq}     \end{array}\right)
\ \ \ \ 
\mathcal{B} =
\left( \begin{array}{cc}
 A_{ij,hk}      &  A_{ij,hkpq}      \\
 A_{ijmn,hk}    & A_{ijmn,hkpq}     \end{array}\right)
\end{equation}
and 
\begin{equation}
\mathcal{X} =
\left( \begin{array}{c}
 X^{(1)}_{ij}\ \ \            \\
 X^{(2)}_{ijmn} \end{array}\right)
\ \ \ \ 
\mathcal{Y} =
\left( \begin{array}{c}
 Y^{(1)}_{ij}\ \ \           \\
 Y^{(2)}_{ijmn} \end{array}\right)
\end{equation}
The elements of $\mathcal{A}$ are obtained from Eq.~(\ref{EOM}) using Eq.~(\ref{exc_op_double}), for example
\begin{equation}
A_{ij,hkpq}=\langle HF | \[ \hat{a}^\dag_j\hat{a}_i , \[\hat{H},
           \hat{a}^\dag_h\hat{a}^\dag_p\hat{a}_k\hat{a}_q  \] \] | HF \rangle
\text{,}
\end{equation}
and similarly for the other components. The elements of $\mathcal{B}$ have a similar form as the elements of $\mathcal{A}$, the only differences being: (i) in the operators on the right of the Hamiltonian particle-hole indexes are inverted, (ii) there is a minus sign.

Eq.~(\ref{EOM_eigen}) has the same formal properties of the RPA-TDHF equations and this guarantees that spectral sum--rules are respected\footnote{While the spectral sum rules are respected, the excitation energies are well described but, often, the oscillator strengths are not~\cite{Casida2}.}. Moreover  measurable quantities can be constructed from the solution of the problem using the same equation of the RPA method~\cite{sRPA1}.

\subsection*{Second random phase approximation and the folding} \label{sec:sRPA folding}
The Hamiltonian associated to the sRPA equation of motion can then be written~\cite{sRPA1,sRPA2} in the Fock space of single and DEs
\begin{equation}
\label{Rn-problem}
\left( \begin{array}{cc}
S & C \\
C^\dag & D \end{array}\right)
\left( \begin{array}{c} \mathbf{e}_1 \\ \mathbf{e}_2 \end{array}\right)
= \omega_I \left( \begin{array}{c} \mathbf{e}_1 \\ \mathbf{e}_2 \end{array} \right).
\end{equation}
Here $S$ and $D$ represent, respectively, the Hamiltonian in the space of single excitations (dimension $N_s\times N_s$) and of DEs (dimension $N_d\times N_d$). $C$ represents the coupling between single and DEs. The number of eigenvalues of Eq.~(\ref{Rn-problem})  is, thus, $N_s+N_d$. $\mathbf{e}_1$ and $\mathbf{e}_2$ are the sRPA excitation operator components~\cite{sRPA1,sRPA2} in the singles and doubles subspaces, respectively. 

The question now is how to obtain these $N_d$ poles working only in the space of single excitations, without introducing explicitly the doubles subspace. This step is crucial to create a link between the sRPA, Eq.~(\ref{polarizability_eh}), and the BSE, which is strictly defined only in the singles subspace. To create this link we fold the total Hamiltonian matrix in the $R^{N_s}$ subspace~\cite{sRPA1,sRPA2}. This is done by expressing  $\mathbf{e}_2$ in terms of $\mathbf{e}_1$, and then solving the equation for $\mathbf{e}_1$:
\begin{equation}\label{Rn-projected}
\left(S+ \Xi(\omega) \right)
\mathbf{e}_1=\omega_I \mathbf{e}_1 \text{,}
\end{equation}
with $\Xi(\omega)=C(\omega_I-D)^{-1}C^\dag$. Eq.~(\ref{Rn-problem}) and Eq.~(\ref{Rn-projected}), then, have the same $N_s+N_d$ eigenvalues but Eq.~(\ref{Rn-projected}) is solved in the single-excitation subspace, and the frequency-dependent kernel $\Xi(\omega)$ takes into account the down--folding of the double-excitation space to the single-excitation space. The correct structure of the $\Xi$ kernel is thus crucial to get the correct number of solutions. In particular, if $D$ can be diagonalized, then Eq.~(\ref{Rn-projected}) can be written in terms of the diagonal matrix $D^\prime=U^\dagger D U$:
\begin{equation} \label{Eqn:K-structure-corr}
\left(S+\sum_{\xi=1}^{N_d}\frac{K^{(\xi)}}{(\omega_I-D^\prime_{\xi\xi})} \right)
\mathbf{e}_1=\omega_I \mathbf{e}_1,
\end{equation}
with $K=C^\prime C^{\prime\dag}$ and $C^\prime=CU$. 

The explicit expression for the $\Xi$ kernel of sRPA can be obtained~\cite{sRPA1,sRPA2}, within the TDA, starting from Eq.~(\ref{EOM}) and Eq.~(\ref{exc_op_double}) and constructing all the matrix elements $A$ and $B$:
\begin{equation}\label{Eqn:2nd_kernel}
\Xi_{(ij),(hk)}(\omega)=\sum_{(nq)(mp)} \frac{C_{(ij),(nm)(pq)}C^\dag_{(nm)(pq),(hk)}}
                           {\omega-(\epsilon_n-\epsilon_m+\epsilon_p-\epsilon_q)},
\end{equation}
with
\begin{multline}\label{half_2nd_order}
C_{(ij),(nm)(pq)}=\frac{1}{2}\bigg(v_{(in),(pq)}\delta_{j,m}+v_{(jm),(pq)}\delta_{i,n} \\
            -\{n\leftrightarrow p\}-\{m\leftrightarrow q\}+\{(nm)\leftrightarrow (pq)\}\bigg).
\end{multline}
Here, in the space of DEs, the matrix elements of the interaction term in the hamiltonin
has been neglected; $\epsilon_i$ are the poles of the HF one particle GF, $G_{HF}$, whereas
\begin{equation}
v_{(ij),(hk)}=\int dxdx^\prime
\phi^*_j(x)\phi_i(x)v(xx^\prime)  \phi_{k}(x^\prime )\phi^*_{h}(x^\prime)
\text{,}
\end{equation}
are the projections of the Coulomb interaction in the space of single--particle wave--functions. The structure of Eq.~(\ref{Eqn:2nd_kernel}) is the same of the kernel in Eq.~(\ref{Eqn:K-structure-corr}). A key property of the $\Xi$ kernel is that it is unchanged under  $-\{n\leftrightarrow p\}$, $-\{m\leftrightarrow q\}$ (Pauli exclusion principle) and $\{(nm)\leftrightarrow (pq)\}$ (particle indistinguishably) transformations due to the symmetry of the $C_{(ij),(nm)(pq)}$ factors. 

Therefore {\em the algebraic structure of Eq.~(\ref{Eqn:2nd_kernel}) ensures the respect of the particle indistinguishability and of the Pauli exclusion principle which constitute necessary conditions for a number--conserving\,(NC) theory of DEs.}

This can be shown in detail by solving the characteristic equation of the eigenvalue problem Eq.~(\ref{Rn-projected}), i.e. $det(\omega-S-\Xi(\omega))=0$. Using the non linearity of the determinant operator,
\begin{equation}
det(\frac{K^{(\xi)}}{\omega-D'_{\xi\xi}})=\frac{det(K^{(\xi)})}{(\omega-D'_{\xi\xi})^{N_s}}
\text{,}
\end{equation}
where $N_s$ is the dimension of the matrix $K$, and $\xi$ stands for the set of indexes $\{(nm)(pq)\}$, and exploiting the relation~\cite{det_prop}
\begin{equation}
det(A+B)= \sum_{P_R,P_C} minor(A)\ minor(B)
\text{,}
\end{equation}
with $P_R$ and $P_C$ partitions of the rows and the columns of $A$ and $B$~\footnote{We recall that a minor of a matrix $A$ is the determinant of a submatrix $M$ obtained from erasing a fixed number $n$ of columns and rows. The terms $n=0$, i.e. the determinant of the matrix $M=A$, is considered too.}, the eigenvalue equation can then be rewritten as
\begin{equation}\label{det_eq2}
\begin{split}
det(\omega-S-\Xi(\omega))&= \sum_{P_R,P_C} 
 minor(\omega-S)\ minor(\Xi(\omega))          \\
         &= det(\omega-S) + \sum_{\xi=1}^{N_d} \frac{det(K^{(\xi)})}{(\omega-D'_{\xi\xi})^{N_s}} +
            \text{...}.
\end{split}
\end{equation}
In the second line of Eq.\ (\ref{det_eq2}) we considered the two terms in the minor expansion that have the maximum and the minimum degree in $\omega$, respectively $N_s$ and $-N_s$. Thus, assuming a completely general structure for the $K^{(\xi)}$ terms, Eq.~(\ref{det_eq2}) is a polynomial equation of degree $N_s+N_dN_s$. Consequently the introduction of a frequency-dependent kernel yields, in general, more solutions then the single electron transitions ($N_s$), although {\em larger} then the correct number of poles, $N_s+N_d$.  However, in our case, the particular structure of the matrices $K_{ij,hk}^{(\xi)}= C_{ij,\xi}C_{hk,\xi}^*$ ensures that the determinant of any but the one-dimensional sub-block of $K^{(i)}$ is zero. This means that the second term on the r.h.s. of Eq.~(\ref{det_eq2}) is zero and in the minor expansion only $N_d$ terms of degree $-1$ survives, from which it follows that the total degree of the polynomial expression $det(\omega-S-\Xi(\omega))$ is $N_s+N_d$.

In the notation of Eq.~(\ref{polarizability_eh}) $\mathbf{e_1}=A_{\lambda,eh}$. By plugging the eigenvectors and eigenvalues of Eq.~(\ref{Rn-projected}) in Eq.~(\ref{polarizability_eh}) we see immediately that the $N_d$ DEs will appear as poles of $\overleftrightarrow{\alpha}(\omega)$. 

\subsection*{Feynman diagrams reteined in the second random phase approximation}
\begin{figure}[!h]
 \begin{center}
 \subfigure[bubble diagrams]{\includegraphics[width=0.95\textwidth]{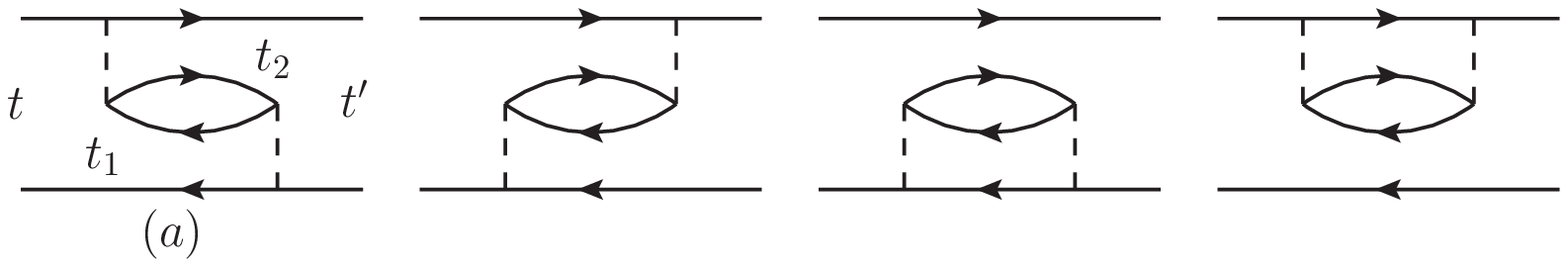}}
 \subfigure[e--h exchange diagrams]{\includegraphics[width=0.95\textwidth]{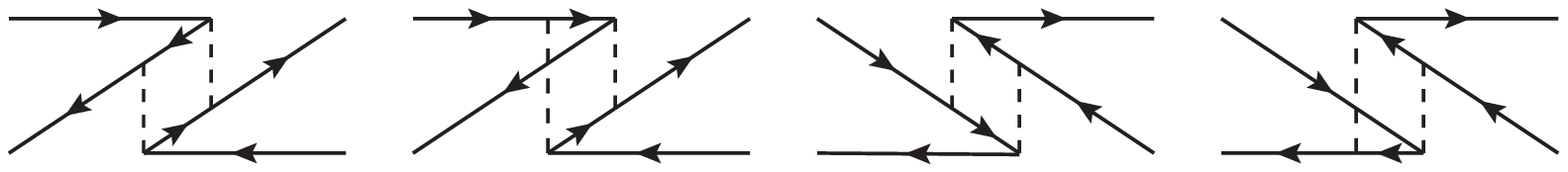}}
 \subfigure[one particle exchange diagrams]{\includegraphics[width=0.95\textwidth]{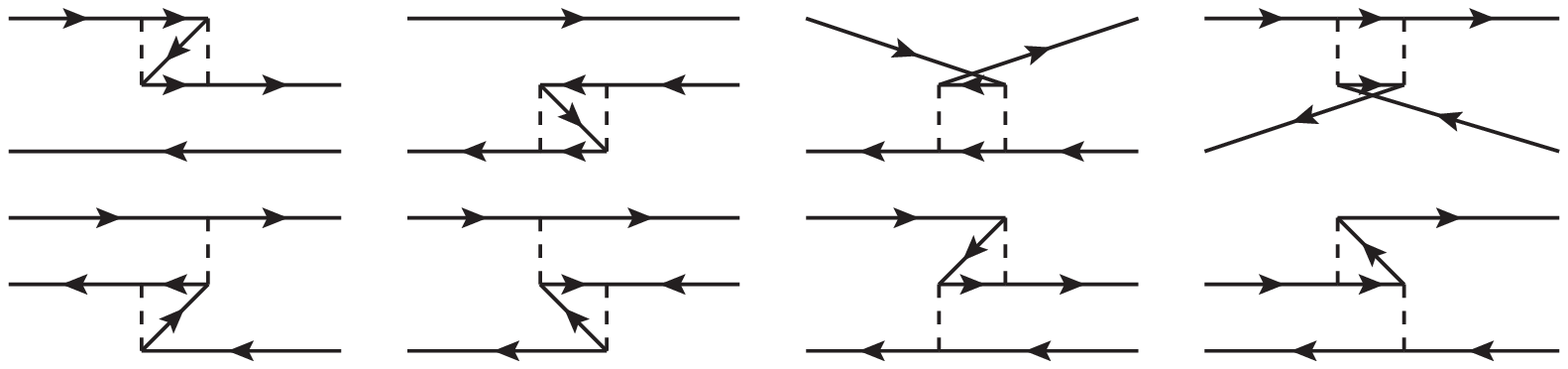}} 
 \caption{Basic Feynman diagrams included in the second--RPA approach~\cite{sRPA1,sRPA2} beyond the standard TDHF.
          The time flows from left to right respecting the Tamm--Dancoff approximation.
          The sRPA approach, when the TDA is relaxed, includes other 16 basic diagrams obtained by inverting
          the direction of all GF. The complete set of diagrams is obtained 
          by iterating the Dyson equation.}
 \label{fig:2nd_order_diagrams}
 \end{center}
\end{figure}

In order to create a common language between the sRPA and the DBSE approaches we start by noticing that, within TDA, the kernel of sRPA contains all Feynman diagrams up to second order. The $16$ second order diagrams included are represented in Fig.~(\ref{fig:2nd_order_diagrams}).

To understand why DEs are described within this approximation we focus on diagram $(a)$ of Fig.~(\ref{fig:2nd_order_diagrams}), drawn for a specific time ordering. The diagram describes a physical process where the electron-hole pair created at time $t$ emits a photon that generates another electron-hole pair at time $t_1$. The second e-h pair is annihilated at time $t_2$. Therefore this Feynman diagram is describing the coupling between a single-- and a double--excitation.

\subsection*{Second random phase approximation, correlation, and TDA: a closed end}
In extended systems the dressing up of bare particles induced by correlation effects is mediated by collective charge oscillations, i.e. by plasmons. Therefore a coherent approach to DEs in correlated materials should also describe the interaction with plasmons. The key problem in the description of plasmons is the possible breakdown of the TDA, as it occurs, for example, in nano--structures~\cite{Myrta}. Indeed, within the TDA neutral excitations are described as packets of electron-hole pairs propagating only forward in time, and, therefore, charge oscillations (plasmons) cannot be captured.

sRPA can, in principle, describe plasmons by going beyond the TDA. However, as a matter of fact, the complexity of the method imposes to retain only a few terms beyond TDA. Indeed, the sRPA, given by Eq.~(\ref{Rn-projected}), is equivalent to a Dyson equation for the response function that can be analysed by using the diagrammatic technique.
\begin{figure}
 \begin{center}
 \subfigure[\ kernel diagrams]{\includegraphics[width=0.95\textwidth]{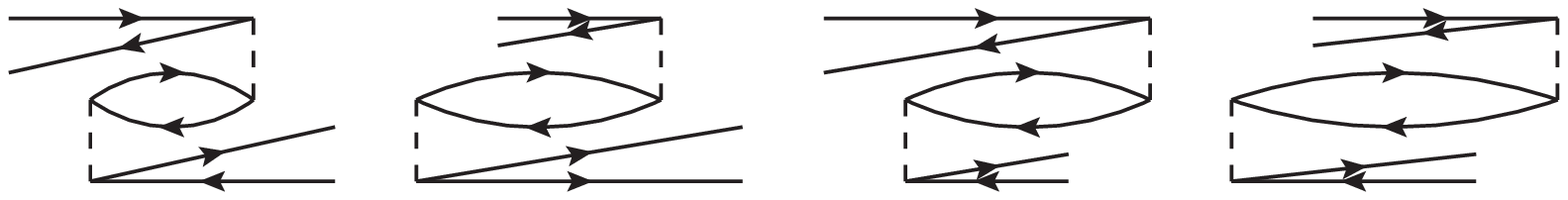}}
 \subfigure[\ self-energy diagrams]{\includegraphics[width=0.95\textwidth]{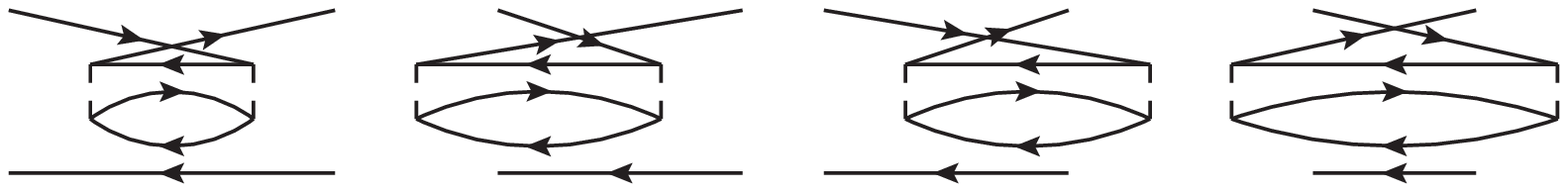}}
 \caption{Second order Feynman diagrams relevant to the description of collective excitations.
          Time flows from left to right. While kernel diagrams (a) are included in the sRPA, trough
          the iteration of the Dyson equation, self-energy ones (b) are not. This inconsistency prevent
          the sRPA to work in a correlated regime.}
 \label{fig:sRPA_failure}
 \end{center}
\end{figure}
It results that while kernel diagrams (see Fig.~(\ref{fig:sRPA_failure}), panel (a)) are included in the sRPA, self-energy diagrams (see Fig.~(\ref{fig:sRPA_failure}), panel (b)) are not. It has been shown that, starting from the HF approach, including only the kernel diagrams yields an incorrect description of the excitation energies~\cite{Cederbaum2}. In a recent paper by Gambacurta et al.~\cite{Gambacurta}, studying the spectrum of Sodium clusters, this problem is discussed and identified as lack of ground state correlation. The same problem is identified by Huix-Rotllant and Casida~\cite{Casida2}.

This is one of the major reasons why the sRPA approach is not very popular in the condensed matter field. Approaches like the Algebraic Diagrammatic Construction (ADC) are preferred~\cite{Polyenes_ADC,ADC}. However in the ADC approach kernel and self-energy diagrams beyond TDA are included only up to finite orders\footnote{Another limit of the ADC scheme, from our point of view, is that this explicitely includes all diagrams related by particle exchange and Pauli exclusion principle. For double excitations this forces to perform a matrix diagonalization in the space of $2p-2h$, when the electron hole interaction among virtal particles is considered (ADC(2)--x approximation~\cite{Polyenes_ADC}) }. While this is a reasonable approach for small systems, it is expected to fail in extended correlated ones. In extended system any order diagram in the bare interaction is relevant and kernel and self-energy diagrams must be included up to an infinite order.

\section{The dynamical Bethe--Salpeter equation (step II)}\label{Sec:DBSE}
It is now clear that a well defined approach to the description of DEs must be NC, i.e., it must not introduce spurious non-physical solutions. At the same time it must include diagrams up to infinite order and beyond the TDA in order to describe screening effect and collective excitations. The BSE approach is an alternative scheme which includes the infinite series of both kernel and self-energy diagrams, thus providing a suitable approach to achieve both goals in a coherent manner.

However in the DBSE presented in Ch.~\ref{chap:Introduction to the problem} the kernel $\Xi^d_2(\omega)$ includes the frequency dependency of kernel diagrams only, whereas the static self--energy effects are included as a rigid shift of the QP eigen--energies.
It has been already shown that, at linear order, dynamical effects have to be included in both the kernel and the self--energy~\cite{Marini1}. Following this input, in order to construct a consistent dynamical approximation, we improve the DBSE including dynamical self--energy effects in a term $\Xi^d_1(\omega)$ to be added to the kernel $\Xi^d_2(\omega)$. This extra term originates from the  $\tilde{L}_0(\omega)=-i\int d\omega'/(2\pi) G(\omega'+\omega/2)G(\omega'-\omega/2)$ as described in the following. We use the Dyson equation for the GF written in the form:
\begin{equation}
\begin{split}
&G^{-1}=g^{-1}-\Sigma_s-\Sigma_d(\omega)=G_s^{-1}-\Sigma_d(\omega)
\end{split}
\end{equation}
where we separated the Self--Energy in its static $\Sigma_s$ and dynamic $\Sigma_d$ parts, and we defined $G_s^{-1}=g^{-1}-\Sigma_s$. In this way we can write:
\begin{multline} \label{self_kernel_deriv}  
\tilde{L}_0(\omega)=\tilde{L}_s(\omega)\\
  -i\int \frac{d\omega'}{2\pi} G_s(\omega'+\omega/2)G_s(\omega'-\omega/2)\Sigma_d(\omega'-\omega/2)G(\omega'-\omega/2)\\
  -i\int \frac{d\omega'}{2\pi}G_s(\omega'+\omega/2)\Sigma_d(\omega'+\omega/2)G(\omega'+\omega/2)G_s(\omega'-\omega/2)\\
  -i\int \frac{d\omega'}{2\pi}G_s(\omega'+\omega/2)\Sigma_d(\omega'+\omega/2)G(\omega'+\omega/2) \\
  G_s(\omega'-\omega/2)\Sigma_d(\omega'-\omega/2)G(\omega'-\omega/2)
  \text{,}
\end{multline} 
where $\tilde{L}_s=-iG_sG_s$. Using the same trick adopted for the kernel, we multiply the second, third, and fourth term on the right-hand side of Eq.~(\ref{self_kernel_deriv}) by $\tilde{L}_s \tilde{L}_s^{-1}$ from the left and by $\tilde{L}_0^{-1} \tilde{L}_0$ from the right, and we obtain
\begin{equation}
\tilde{L}_0(\omega)=\tilde{L}_s(\omega)+\tilde{L}_s(\omega)\Xi_1^d(\omega)\tilde{L}_0(\omega)
\end{equation}
with 
\begin{multline}
\Xi_1^d(\omega)=-i\tilde{L}_s^{-1}(\omega)\int \frac{d\omega'}{2\pi} \\
    \big[ G_s(\omega'+\omega/2)G_s(\omega'-\omega/2)\Sigma_d(\omega'-\omega/2)G(\omega'-\omega/2)\\
         +G_s(\omega'+\omega/2)\Sigma_d(\omega'+\omega/2)G(\omega'+\omega/2)G_s(\omega'-\omega/2)\\
         +G_s(\omega'+\omega/2)\Sigma_d(\omega'+\omega/2)G(\omega'+\omega/2)\\
           G_s(\omega'-\omega/2)\Sigma_d(\omega'-\omega/2)G(\omega'-\omega/2)\big]
 \tilde{L}_0^{-1}(\omega).\nonumber
\end{multline}
For the description of DEs we set to zero the last term on the right-hand side of Eq.~(\ref{self_kernel_deriv}) as it describes a process where six Green-function lines appear in the same moment, so a triple excitation. We thus obtain a total kernel $\Xi^d(\omega)=\Xi^d_1(\omega)+\Xi^d_2(\omega)$ to be inserted in Eq.~(\ref{Freq_dep_BSE}), which we write here again for clarity
\begin{equation}\label{DBSE_configuration_all}
\tilde{L}_{ij,hk}(\omega)=\tilde{L}^s_{ij,hk}(\omega)+\tilde{L}^s_{ij,i'j'}(\omega)\tilde{K}_{i'j',h'k'}(\omega)\tilde{L}_{h'k',hk}(\omega)
\text{.}
\end{equation}
The zero order term is now called $L_s(\omega)$ to underline that it includes only the static effects of the self energy.

\begin{multline} \label{K1}
\Xi_1^d(\omega)=-i\tilde{L}_s^{-1}(\omega)\int \frac{d\omega'}{2\pi}
    \big[ G_s(\omega'+\omega/2)G_s(\omega'-\omega/2)\Sigma_d(\omega'-\omega/2)G(\omega'-\omega/2) \\
+G_s(\omega'+\omega/2)\Sigma_d(\omega'+\omega/2)G(\omega'+\omega/2)G_s(\omega'-\omega/2)
\text{,}
\end{multline}
\begin{multline} \label{K2}
\Xi_2^d(\omega)=\tilde{L}_0^{-1}(\omega)\frac{1}{(2\pi)^3} \int d\omega' d\omega'' d\omega'''L_0(\omega,\omega',\omega'') \\
    \Xi(\omega,\omega',\omega''') L(\omega,\omega''',\omega'') \tilde{L}^{-1}(\omega) \text{.}
\end{multline}
The complexity of the original Eq.~(\ref{Exact-BSE}) is thus transferred in the structure of the DBSE kernel $\Xi_d(\omega)$. We can, however, simplify the dependence on $G$ and $L$ in $\Xi_d(\omega)$ by starting from its linear limit where $G(\omega)\simeq G_s(\omega)$ in Eq.~(\ref{K1}) and $L(\omega,\omega',\omega'')\simeq L_0(\omega,\omega',\omega'')\simeq L_s(\omega,\omega',\omega'')$ in Eq.~(\ref{K2}). This limit is fully justified in the DBSE by the fact that it accounts for the simultaneous evolution of two e-h pairs, which represent the dominant channel in the description of DEs. Accordingly it is crucial that the static part of the self-energy is treated in a separate way. In contrast to $\Sigma^d$ in fact, the static part of the self energy $\Sigma_s$ cannot be treated through a linearized kernel, as this would lead to numerical instabilities~\cite{Bruneval_JCP}.

We need now to approximate the unknown quantities $\Sigma_s$, $\Sigma_d(\omega)$ and $\Xi(\omega,\omega',\omega'')$. One can verify that the DBSE is equivalent to sRPA if one chooses $\Sigma_s=\Sigma_{HF}$, $\Sigma_d=\Sigma_2(\omega)$, thus including all second order Feynman diagrams of sRPA, and $\Xi(\omega,\omega',\omega'')=\frac{\partial \(\Sigma_s+\Sigma_d(\omega)\)}{\partial G}$. Starting from this observation in the next section we derive a diagrammatic number conserving rule which we then use to construct an approximation able to properly describe correlated system. To do this we will consider the GW approximation to the Self--Energy evaluated at the QwP eigen--energies and we will derive the dynamical part $\Sigma_d$ and the kernel $\Xi(\omega,\omega',\omega'')$ starting from the screened Coulomb interaction, in order to include the static GW-BSE scheme in the $\omega\rightarrow 0$ limit. The derivation will be carried on within TDA in order to keep the discussion as simple as possible. 

\section{A number--conserving kernel for correlated systems}\label{Sec:Kernel}
The DBSE equation provides a powerful starting point to tackle the double-excitation problem, as the diagrammatic approach makes possible to introduce different levels of approximation that overcome the limits of the sRPA. We achieve this by following two essential steps: i) we use the sRPA to create a close link between the diagrams introduced in the DBSE kernel, the particle indistinguishability and the Pauli exclusion principle; ii) we use this link to define a number--conserving correlated kernel starting from the standard $GW$ approximation.

\subsection*{The diagrammatic number conserving rule}

\begin{figure}[t]
 \begin{center}
 \subfigure[e--h pairs exchange]{\includegraphics[width=0.4\textwidth]{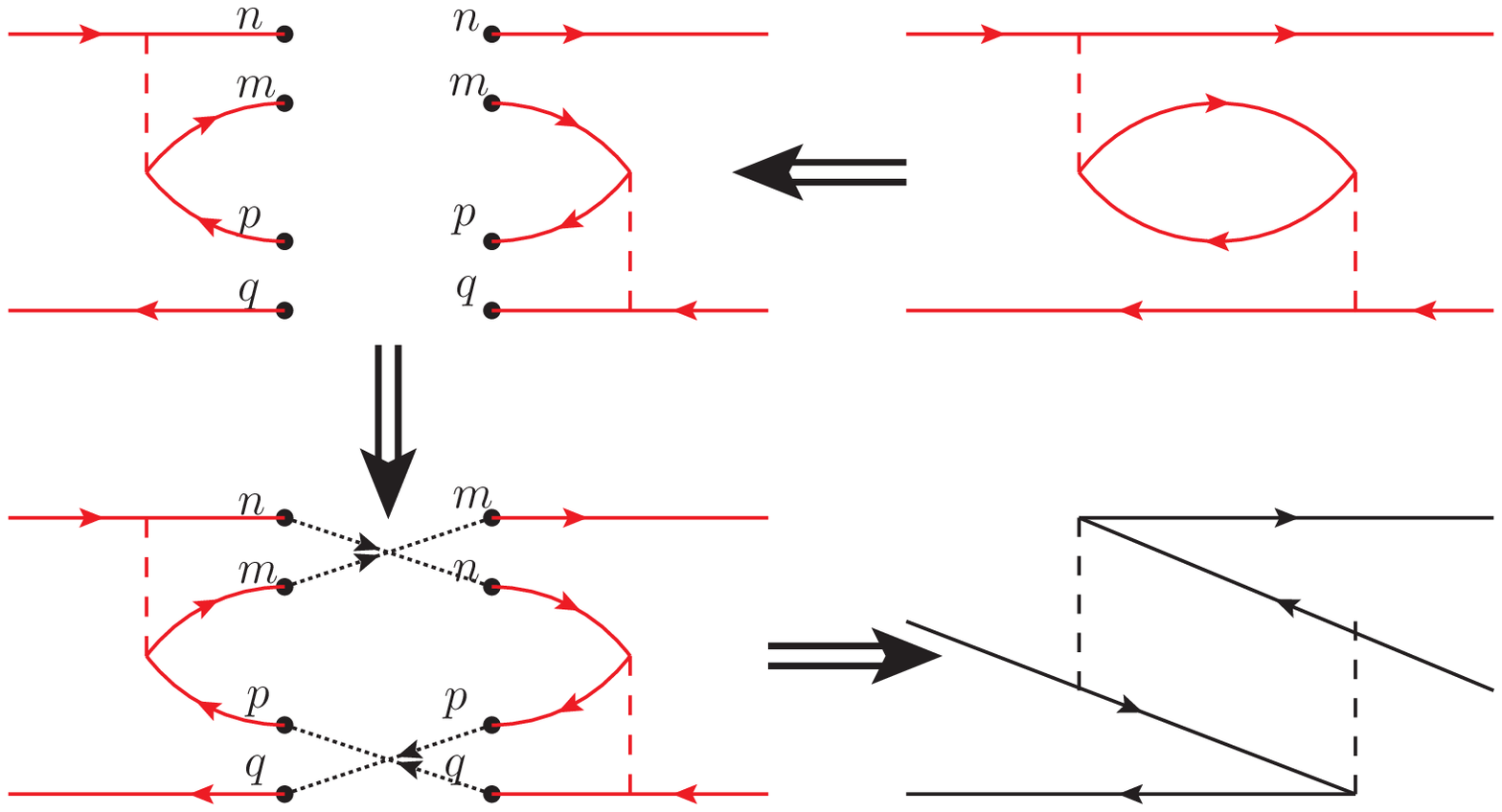}}
 \hspace{0.1\textwidth}
 \subfigure[Single particles exchange]{\includegraphics[width=0.4\textwidth]{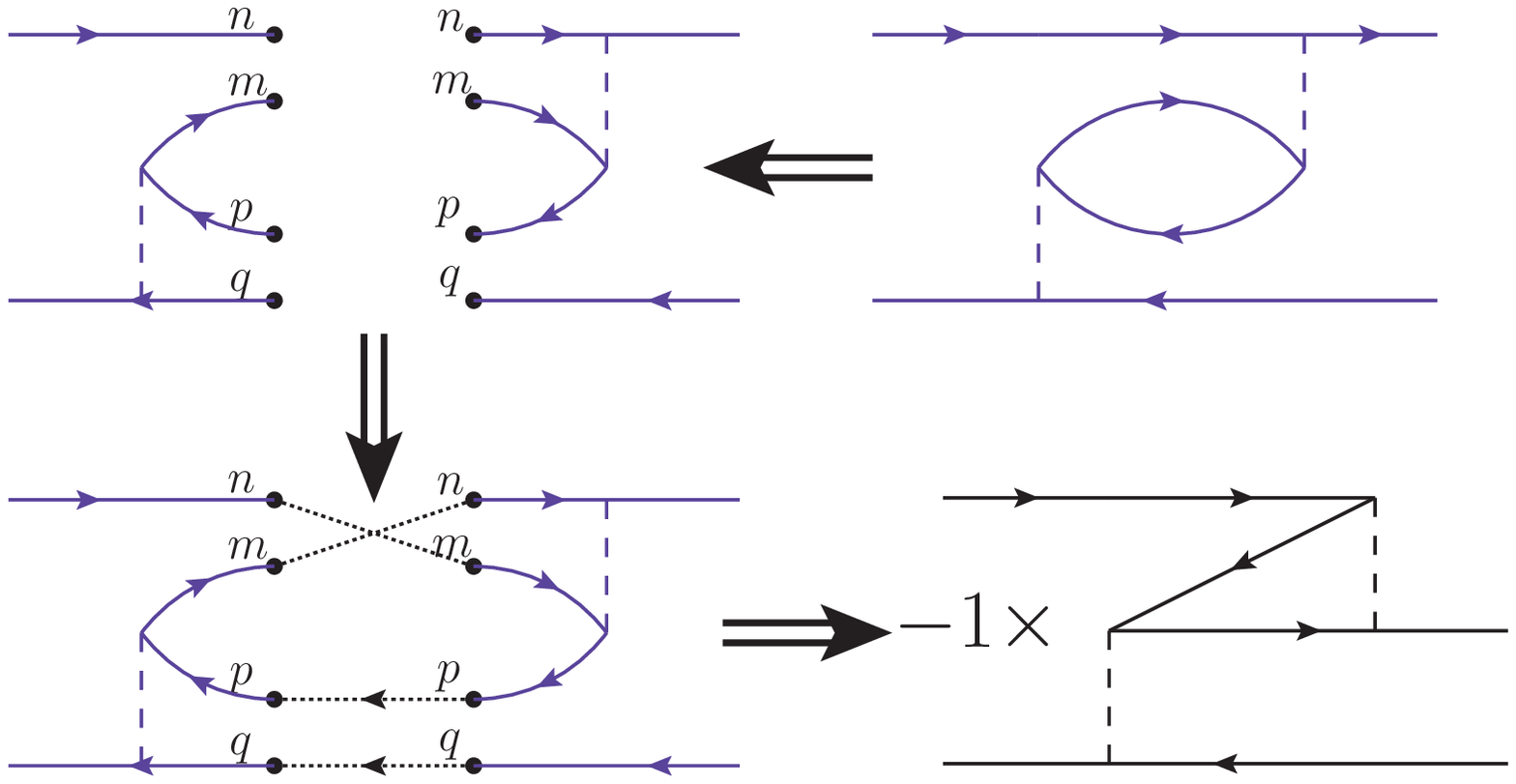}}
 \subfigure[Half diagrams recombination]{\includegraphics[width=0.7\textwidth]{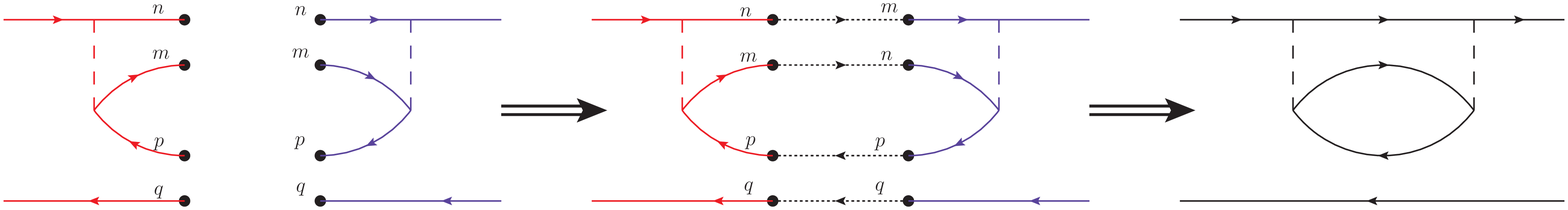}}
 \end{center}
 \caption{The DNCR in practice. We take as reference the two time orderings of the kernel bubble diagram,  corresponding to the two first diagrams of Fig.~(\ref{fig:2nd_order_diagrams}). The general procedure to get a NC kernel is to split each initial diagram in two half--diagrams. Then these half diagrams must be connected by exchanging in all possible ways all e--h pairs and all single  particles. This produces a new group of diagrams that must be processed using the same procedure. When no new diagrams appear the resulting kernel is NC.}
 \label{fig:Half_diag}
\end{figure}

By taking into account the 16 diagrams of Fig.~(\ref{fig:2nd_order_diagrams}) the DBSE (and consequently the sRPA) correctly describes {\em particle indistinguishability} and  {\em Pauli exclusion principle}. Here we illustrate how this can be deduced from the inspection of the Feynman diagrams. The 16 diagrams describe processes in which a DE appears from a photon emitted either from the electron or from the hole and then absorbed back (these two possibilities are the first two terms in the definition of $C_{(ij),(nq)(mp)}$, see Eq.~(\ref{half_2nd_order}) ). Therefore each double-excitation process can start and end in two ways so that there are 4 possible processes, which are the four bubble diagrams of Fig.~(\ref{fig:2nd_order_diagrams}). The other 12 diagrams reflect the particles indistinguishability that imposes the electron lines, as well as the hole lines, to be interchangeable among themselves. 

Following Fig.~(\ref{fig:Half_diag}) we can derive a graphical rule that any approximation has to respect in order to be $NC$, this is the proposed {\em Diagrammatic Number Conserving Rule}\,($DNCR$). First we consider an initial group of diagrams, chosen in such a way to describe the relevant physics we want to introduce in the theory (like plasmons and excitons). Then we split each diagram in two parts that, connected in all possible ways obtained by imposing particles exchange, lead to a new group of diagrams. When the same procedure applied to the resulting diagrams does not lead to any new diagram, then the approximation is, by definition, NC. As an illustration, the DNCR can be applied to the sRPA diagrams, shown in Fig.~(\ref{fig:2nd_order_diagrams}). It can be shown that {\em all} sRPA diagrams can be obtained from the first two by applying the proposed DNCR. 

\newpage

\subsection*{The diagrammatic number conserving rule applied to the Bethe-Salpeter equation}

\begin{figure}[t]
 \begin{center}
 \includegraphics[width=0.7\textwidth]{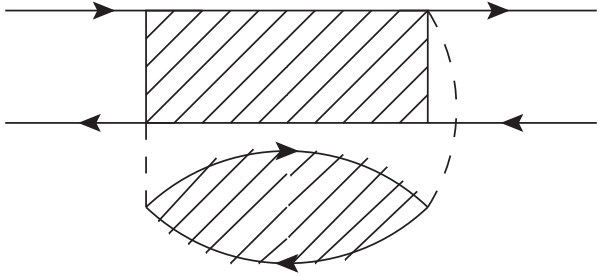} 
 \caption{Basic RPA diagram used as a starting point for the correlated kernel. All other diagrams are obtained by applying the DNCR, as discussed in the text.}
 \label{fig:RPA_basical}
 \end{center}
\end{figure}

A crucial consequence of the DNCR is that, as exemplified in Fig.~(\ref{fig:Half_diag}), a NC kernel must include all kind of diagrams.  Therefore, whatever initial approximation is chosen the repeated application of the DNCR will create a balanced mixture of diagrams in order to respect  particle indistinguishability. If the DNCR is not respected by selecting only a class of diagrams, then spurious solutions are expected to appear. This is the case of the kernel proposed in Ch. \ref{chap:Introduction to the problem} that was obtained from the standard $\Xi\simeq W\(\omega\)$ by simply relaxing the static approximation for $W$. This kernel introduces an infinite series of RPA diagrams only in the interaction $W$, neglecting all consequent diagrams imposed by the DNCR. As a consequence spurious poles in the polarizability are found as predicted by the DNCR.

\begin{figure}[t]
 \begin{center}
 \includegraphics[width=0.95\textwidth]{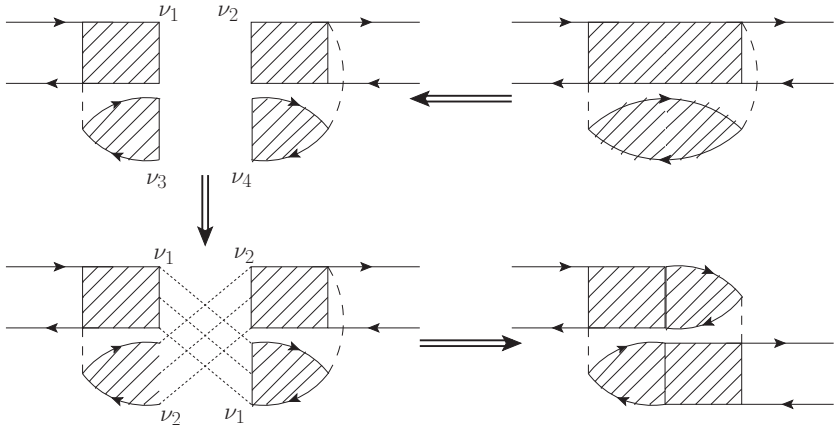} 
 \caption{Build up of correlated Feynman diagrams connecting two Feynman diagrams. The effect of exchange among two RPA excitations is shown using dotted lines.}
 \label{fig:Half_RPA_diag}
 \end{center}
\end{figure}

Nevertheless the kernel proposed in the previous chapter describes the interaction with plasmons, which is a desirable property which we want to retain, at the same time forcing the kernel to be NC. However, before applying the DNCR, we have to note that the $W(\omega)$ propagator describes the evolution of charge oscillations, composed by renormalized packets of e--h pairs. This clearly makes a distinction between the e-h pairs embodied in $W\(\omega\)$ and the real e--h pairs created by the scattering process leading to the breakdown of the particle indistinguishability. A better starting point is instead the basic diagram showed in Fig.~\ref{fig:RPA_basical}, where all e--h pairs are correctly renormalized.
In this diagram the filled bubble and the filled rectangle represent the RPA response function $\chi^{RPA}\(\omega\)$. By introducing the Lehman representation for $\chi^{RPA}$ in the same notation of the previous chapter we can write
\begin{equation}\label{polarizability_RPA}
\chi^{RPA}_{eh,e'h'}(\omega)=\sum_{\nu}\frac{R_{\nu,eh} R^{*}_{\nu,e'h'}}{\omega-\Omega_\nu}.
\text{.}
\end{equation}
We will call the poles of $\chi^{RPA}$ RPA excitations. Note that these capture the physics of the plasmonic oscillations.

The DNCR imposes to consider all possible diagrams obtained from Fig.~\ref{fig:RPA_basical} by exchanging the basic excitation propagators. The key point here is to rotate from the independent e--h pairs to the RPA basis, where e--h pairs are replaced by the RPA excitations. Therefore we proceed by splitting the $RPA$ propagators, using Eq.~(\ref{polarizability_RPA}), as sketched in Fig.~(\ref{fig:Half_RPA_diag}). Then we consider all diagrams where the RPA excitations are exchanged.

Mathematically the procedure sketched in  Fig.~(\ref{fig:Half_RPA_diag}) corresponds to rotate in the RPA excitation space the residuals and poles
of Eq.~(\ref{Eqn:2nd_kernel}). Each term in the rotated counterpart of Eq.~(\ref{half_2nd_order}) will correspond to a possible connection induced by the DNCR:
\begin{equation}
\label{Eqn:C-corr}
C^{RPA}_{ij,\nu_1\nu_2}=\frac{1}{2}\sum_{nq,mp}\bigg((v_{in,mp}\delta_{j,q}+v_{mp,jq}\delta_{i,n})R_{\nu_1,np}R_{\nu_2,mq}
          +\{\nu_1\leftrightarrow \nu_2\}\bigg).
\text{.}
\end{equation}
As a consequence the correlated version of Eq.~(\ref{Eqn:2nd_kernel}) will look like
\begin{equation}
\label{Eqn:k-structure-corr}
\left(\Xi_{RPA}^{d}\right)_{(ij),(hk)}= \sum_{\nu_1\neq\nu_2}
\frac{ C^{RPA}_{ij,\nu_1\nu_2} \[C^{RPA}_{hk,\nu_1\nu_2}\]^* } 
{\omega-(E_{\nu_1}+E_{\nu_2}+2i\eta)}.
\end{equation}
The symbol $\{\nu_1\leftrightarrow \nu_2\}$ in Eq.~(\ref{Eqn:C-corr}) imposes the invariance of the correlated kernel under exchange of RPA excitations. Consequently the kernel $\Xi_{RPA}^{d}$ is by definition invariant under exchange of two RPA excitations. However RPA excitations are bosons so that Pauli exclusion principle is not taken into account and the obtained kernel is not fully NC. To fix this problem it is sufficient to impose the condition $\nu_1\neq\nu_2$ in Eq.~(\ref{Eqn:k-structure-corr}).

\begin{figure}[t]
 \begin{center}
 \includegraphics[width=0.95\textwidth]{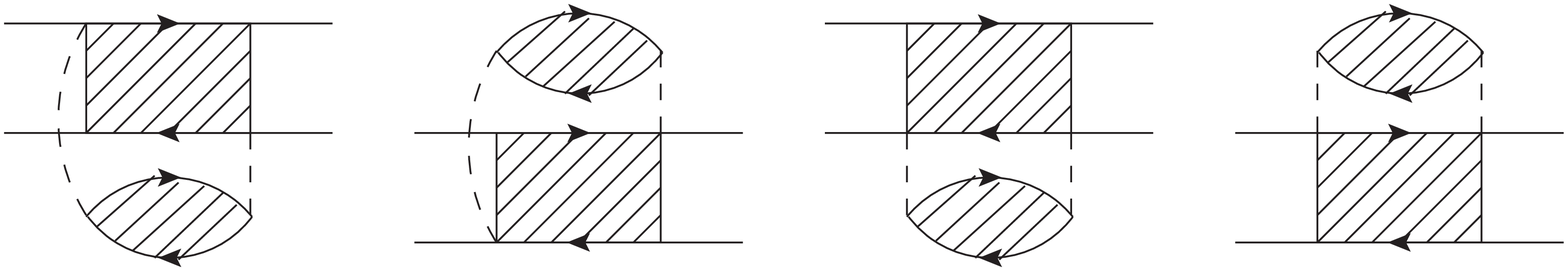} \\[.5cm]
 \includegraphics[width=0.95\textwidth]{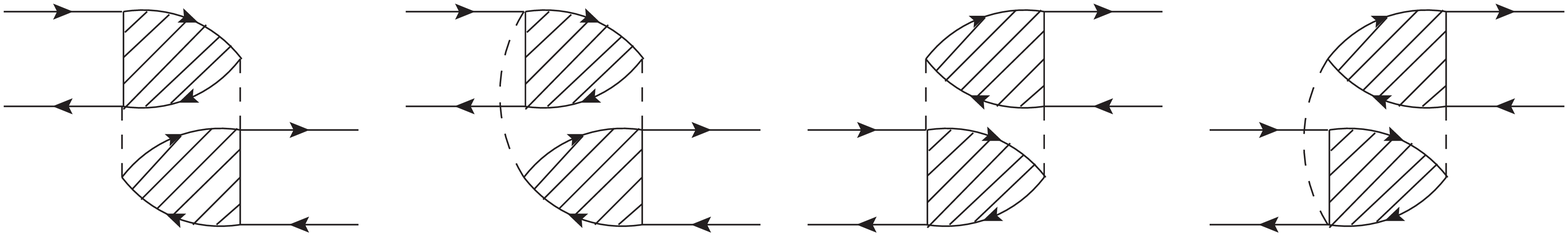} \
 \caption{The final correlated DBSE kernel. The filled 
 regions represent  the propagation of RPA excitations.}
 \label{fig:correlated_diagrams}
 \end{center}
\end{figure}

The DBSE obtained by using the $\Xi_{RPA}^{d}$ kernel includes all self--energy terms obtained from $\Sigma_d=GW$ and $\Xi(\omega,\omega',\omega'')=\delta(\omega-\omega')W(\omega'-\omega'')$. In addition extra terms appear in order to fulfil the $NC$ condition. Interestingly $\Xi_{RPA}^{d}$, within TDA, also embodies the full frequency dependent term $G\ \delta W / \delta G(\omega)$ which is usually neglected in the standard BSE approach. In the present case these second order diagrams in $W$ are indeed needed to correctly account for the particle indistinguishability. The resulting kernel, whose diagrammatic expression is sketched in Fig.~(\ref{fig:correlated_diagrams}), does have the right mathematical structure by construction, so that no spurious solutions are present.

\newpage

\section{Numerical results on model molecular systems}\label{Sec:Numerical results on DE}
In the following we will illustrate various conceptual and technical aspects of our approach using
two benchmark model systems, based on the $C_8H_2$ and the $C_4H_6$ molecules. These unsaturated
hydrocarbon chains are often chosen as benchmark systems to test theoretical methods aimed
to describe double excitations.
By calculating the polarizability of these systems we will show: i) the role played by subgroups of
diagrams in the description of double excitations; ii) the fact that the number conserving rule
not only applies to the total number of poles, but also to the number of optically active poles;
iii) the absence of spurious double excitation peaks that appear that appear in approaches~\cite{Romaniello1}
that violate the NC rule.

The calculations have been performed using the YAMBO code~\cite{Yambo}, where we implemented sRPA
for closed--shell systems, within the TDA. Furthermore we approximate both
QP and HF wave--functions with KS--LDA wave--functions.

sRPA produces results similar to the GW--BSE approach or to the DBSE when only
``bubble diagrams'' (first row of Fig.~\ref{fig:2nd_order_diagrams})
or bubble diagrams and ``eh exchange diagrams'' (second row of Fig.~\ref{fig:2nd_order_diagrams}),
respectively, are selected. Therefore this implementation allows us to explore the performances of
the various approaches by selecting specific subgroups of diagrams.

We first performed a ground-state calculation with the ABINIT code~\cite{Abinit}, within
DFT/LDA, with an energy plane--wave cut-off of 20 Hartree and a super--cell of 25 $\times$ 25 $\times$ 40
Bohr for the $C_8H_2$ (a linear molecule $\approx 21$ Bohr long) and a smaller super--cell of 25 $\times$
25 $\times$ 15 Bohr for the $C_4H_6$ (the molecule extends for $\approx 10$ Bohr both in the $x$ and $y$
directions). Then we performed excited-state calculations in the basis--set of KS--states,
considering only the states from HOMO-3 to LUMO+3. In this way our systems can be mapped into an
eight level model with 16 single and 240 double excitations. All the $C_8H_2$ eigenvalues are doubly
degenerate due to the symmetry of the molecule.

\begin{figure}[t]
 \begin{center}
 \includegraphics[width=0.95\textwidth]{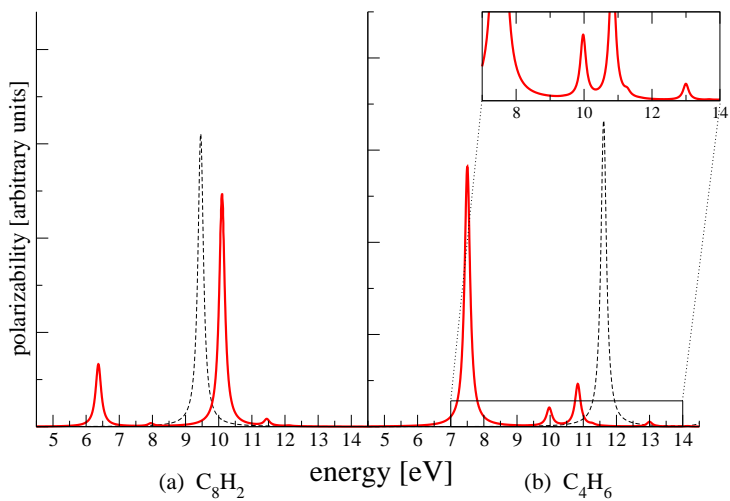} 
 \end{center}
 \caption{Second Random Phase Approximation spectra. For both model systems the frequency--dependent
          kernel produces extra
          peaks (red line) which cannot be described by a static kernel. The black thin dashed line
          is the Independent--Particle spectrum. The inset is present here as reference to detect
          spurious peaks in the insets of Fig.~\ref{fig:subgroups_of_diagrams} and \ref{fig:bubbles_and}.
          Alla spectra in this and in the following figures include an artificial broadening due to the
          use o an imaginary factor $i\eta=0.05 eV$ in the Green's function denominator.}
 \label{fig:eigenvalues}
\end{figure}

In the description of double excitations the kernel frequency dependence becomes crucial when
one or more poles fall in the absorption spectrum energy range.
In this case the static approximation fails, and extra peaks appear. In order to artificially
simulate this situation in our systems we use HF eigenvalues to construct $L_0$, while the kernel
 is built with KS--LDA ones. This choice gives us the possibility to
investigate more physical situations which could arise for correlated materials.

The results of these calculations are plotted in Fig.~\ref{fig:eigenvalues}. For both systems at
the HF Independent--Particle (IP) level there is a clear peak, which falls close to 9 eV
for the $C_8H_2$, and close to 12 eV for $C_4H_6$. As expected the kernel constructed with KS--LDA
eigenvalues has poles in these energy ranges, so that extra peaks appear in the spectrum.
The effect is visible in both model systems: in $C_8H_2$ the main
peak is essentially split in two (see Fig.~\ref{fig:eigenvalues} panel $(a)$);
for $C_4H_6$ several extra peaks appear as shown in the inset of Fig.~\ref{fig:eigenvalues} panel $(b)$.

\begin{figure}[t]
 \begin{center}
 \includegraphics[width=0.95\textwidth]{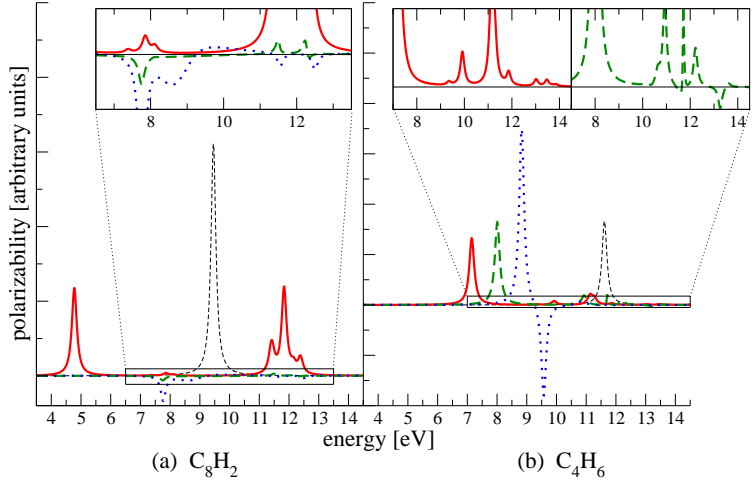}
 \end{center}
 \caption{The spectra obtained selecting only specific subgroup of diagrams. By selecting only ``eh exchange''
          (green dashed line) or ``one particle exchange" (blue dots) an unphysical
          (negative) polarizability is observed. Only the spectra obtained with the kernel constructed using the
          ``bubble diagrams'' (red line) is positive defined. However when only ``bubble diagrams'' are
          used, as proposed by Romaniello et. al~\cite{Romaniello1}, spurious peaks appear. These peaks 
          do not appear in
          the spectra obtained from the full sRPA kernel (see Fig.~\ref{fig:eigenvalues}).}
 \label{fig:subgroups_of_diagrams}
\end{figure}

We will now explore the role played by the various subgroups of
diagrams, namely (a) the ``bubble diagrams'' (first row of Fig.~\ref{fig:subgroups_of_diagrams}); (b) the
``$eh$ exchange diagrams'' (second row of Fig.~\ref{fig:subgroups_of_diagrams}), which are obtained from
the bubble diagrams via $eh$ exchange; (c) the ``particle exchange
diagrams'' (third and fourth row of Fig.~\ref{fig:subgroups_of_diagrams}), which are obtained from the
bubble diagrams via single--particle exchange.

Fig.~\ref{fig:subgroups_of_diagrams} shows
the spectra obtained taking into account, beyond the TDHF scheme, only selected families of
diagrams. By selecting only diagrams of type (b) or (c) the spectra are not 
positive defined. This unphysical property can be understood by noticing that
the frequency--dependent kernel constructed
from diagrams (b) and (c) does not have the mathematical structure of Eq.~(\ref{Eqn:2nd_kernel}).
On the contrary
the kernel constructed from the bubble diagrams (a) is positive, though particle indistinguishability and
Pauli exclusion principle are not respected as illustrated in previous sections.

Indeed the spectra constructed from bubble diagrams is positive defined, though
spurious peaks appear: in $C_8H_2$ (Fig.~\ref{fig:subgroups_of_diagrams} panel $(a)$) one has three peaks
at around $12$ eV, and in the $C_4H_6$ many peaks appear (see Fig.~\ref{fig:subgroups_of_diagrams} panel $(b)$,
the left inset) which are not present in the full sRPA spectra.

\begin{figure}[t]
 \begin{center}
 \includegraphics[width=0.95\textwidth]{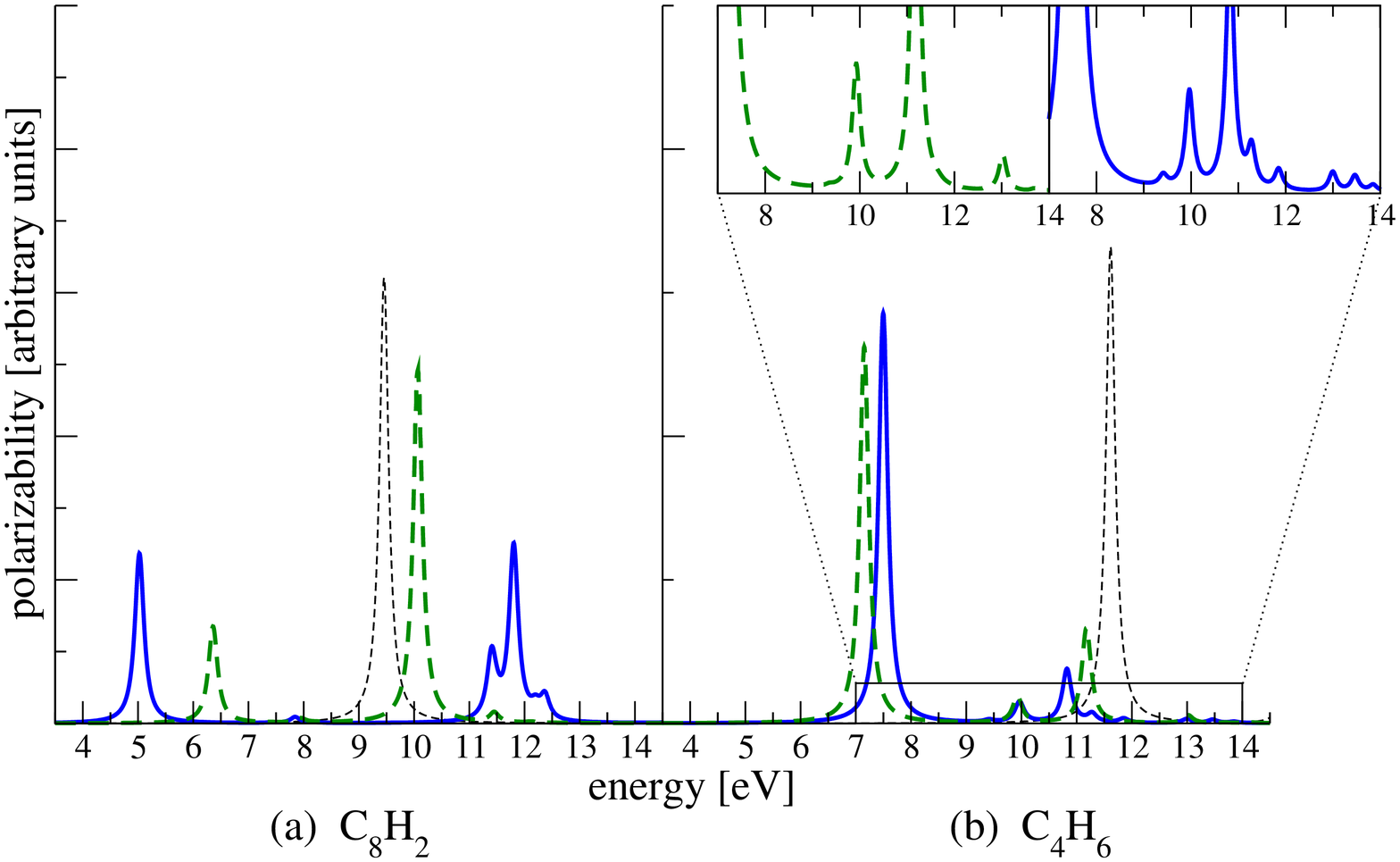}
 \end{center}
 \caption{Spectra obtained with the kernel constructed using the ``bubble+eh exchange'' diagrams
          (green dashed line) and the ``bubble+eh exchange'' diagrams (blue line).
          Both spectra are positive defined but only
          the combination ``bubbles + eh exchange'' gives the same number
          of poles of the full sRPA spectra (see Fig.~\ref{fig:eigenvalues}). On the contrary the
          combination "bubbles + one particle exchange" gives spurious solutions.}
 \label{fig:bubbles_and}
\end{figure}

Fig.~\ref{fig:bubbles_and} shows the spectra constructed taking into
account both subsets of diagrams (a) and (b) or (a) and (c) together.
The spectra are positive defined. However, only the former combination gives the right
number of peaks (i.e. the same number of the full sRPA spectrum) whereas the
latter produces spurious poles.

In this perspective it is interesting to compare the two cases. In the $C_8H_2$ model
the subset of diagrams (a) and (b) (green dashed line) gives a spectrum which is very close to
the full sRPA spectrum of Fig.~\ref{fig:eigenvalues} (red line) both in the structure and number
of poles. Diagrams of kind (c) are, instead, negligible.

In the $C_4H_6$ model, on the contrary, diagrams of kind (c) play an important role:
they shift the peak of the ``bubbles'' polarizability towards the results obtained
with the sRPA kernel.
Diagrams of kind (b), instead, in this case have a negligible effect on the position of the
peaks. However the choice (a) + (c) gives several spurious poles (see Fig.~\ref{fig:bubbles_and} panel $(b)$,
blue line in the right inset) and, as for $C_8H_2$, only the combination (a)
and (b) yields the correct number of poles (see Fig.~\ref{fig:bubbles_and} panel $(b)$,
green dashed line in the left).

The sum of diagrams (a) and (b) describes $eh$ pairs as indistinguishable bosons,
whereas the sum of diagrams (a) and (c) does not correspond to any defined statistic. We can then
conclude that diagrams of kind (c) are meaningful only if added to the other two classes of
diagrams in order to describe particle indistinguishability. However, the spectrum obtained
combining the diagrams (a) and (b) has indeed the same number of peaks of the spectrum
obtained using the complete kernel, thus supporting our recipe to construct a correlated
kernel discarding the subset of diagrams (c)\footnote{The subset of diagrams (c) could be included
in a correlated kernel only at the price of a direct diagonalization in the space $2p-2h$ as in
the ADC(2)--x scheme}. Another conclusion we draw from these results is that
our approach, by respecting the NC rule, ensures that the theory produces
not only the correct total number of poles, but also the correct
total number of {\em optically active} (and optically not active) poles.

\newpage
\section{Conclusions}
In this part of the thesis we presented a method to include double excitations in a consistent
manner within the GW+BSE approach. The main idea has been to correct the standard
BSE kernel in order to go beyond the static approximation fulfilling the Number Conserving (NC)
condition. The resulting scheme keeps all the advantages of the Many Body approach, that is
the ability to describe extended and correlated materials in a consistent manner, without
producing spurious excitations, provided the cndition $\nu_1\neq\nu_2$ in Eq.~(\ref{Eqn:k-structure-corr})
is imposed. This is not an exact condition of the kernel but is sufficient to give the correct numbering
of poles.

The NC condition results from an inspection of the similarities and the differences between
the BSE scheme, designed for solids, and the sRPA approach, designed for isolated
systems. The main character in the first is the screening, while in the second it is the role of
exchange. As pointed out in the very recent work by Huix-Rotllant and Casida~\cite{Casida2},
there is a
great interest in this direction in order to develop approximations at the nanoscale interface
between molecules and solids. As opposed to other works, however, we do not directly consider
all exchange diagrams related to the RPA screening re--summation, because we believe
that such an approach would be impractical, especially for nanostructured materials. Instead,
our method is aimed to capture the main feature related to the exchange principle
without requiring matrix diagonalizztion in the space of double excitations.

\part{Carbon nanotubes in magnetic fields}
\chapter{The Aharonov Bohm effect in carbon nanotubes}  \label{chap:The AB effect}
The Aharonov--Bohm (AB) effect~\cite{Aharonov1959} is a purely quantum mechanical effect which does not have a counterpart in classical mechanics. A magnetic field $\mathbf{B}$ confined in a closed region of space alter the kinematics of charged classical particles only if they move inside this region. Electron dynamics, instead, governed by the Schr\"{o}dinger equation, is influenced even if the particles move on paths that enclose the region where the magnetic field is confined, where the Lorentz force is strictly zero. If this closed region is the inner part of a nano--tube, electrons traveling around the cylinder are expected to manifest a shift of their phase. The mathematical interpretation of this effect is connected with the definition of the vector potential, which, in the case of confined magnetic fields, cannot be nullified everywhere.

This extraordinary effect, first predicted by Aharonov and Bohm~\cite{Aharonov1959}\,(AB) in 1960, was interpreted as a proof of the reality of the electromagnetic potentials. The idea that electrons could be affected by electromagnetic potentials without being in contact with the fields was skeptically received by the scientific community. At the same time the  AB paper spawned a flourishing of experiments and extension of the original idea. The first experiment aimed at proving (or disproving) the AB effect revealed a perfect agreement with the theoretical predictions~\cite{Chambers1960}. Nevertheless only some years later, in 1986, the experiment which can be considered as a definitive proof of the correct interpretation of the AB effect was realized. Tonomura et al.~\cite{Tonomura1986}, using superconducting niobium cladding, were in fact able to completely exclude the possibility of stray fields as alternative explanation of the predicted and observed AB oscillations.

Nowadays the AB effect can be used in a wide range of experiments, from the investigation of the properties of mesoscopic normal conductors to the experiments designed to reveal the structure of flux lines in superconductors. Growing interest is emerging in the field of nanostructured materials. One of the most well--known case is given by Carbon NanoTubes\,(CNTs) that, if immersed in a uniform magnetic field aligned with the tube axis, have been predicted to show peculiar oscillations of the electronic gap. These oscillations are characterized by a period given by the magnetic flux quantum $h/e$ and are commonly interpreted as caused by the change in the wave functions of the electrons localized on the tube surface induced by the Aharonov-Bohm effect.

The first experiment carried on CNTs, in 1999, described the oscillations in the electronic conductivity~\cite{Bachtold1999}, but with period of $h/2e$. This deviation from the predicted AB oscillation period has been explained in terms of the weak localization effect~\cite{Abrahams1979} induced by defects and dislocation by Al'tshuter, Aronov, Spivak~\cite{AAS1981}\,(AAS effect). Only in 2004 a clear proof of the existence of AB oscillations with an $h/e$ period have been given by Coskun et coll.~\cite{Coskun2004} by measuring the conductance oscillations in quantum dots. The dots were built using concentric Multi--Wall (MW) CNTs of different radii, short enough to prevent the appearance of weak localization. In the same year Zaric et al.~\cite{Zaric2004} observed modulation in the optical gap of pure Single Wall (SW) CNTs with oscillations of $h/e$ period. 

\section{What is the Aharanov--Bohm effect?}

\begin{figure}[t]
 \begin{center}
 \subfigure[AB picture]{\includegraphics[width=0.4\textwidth]{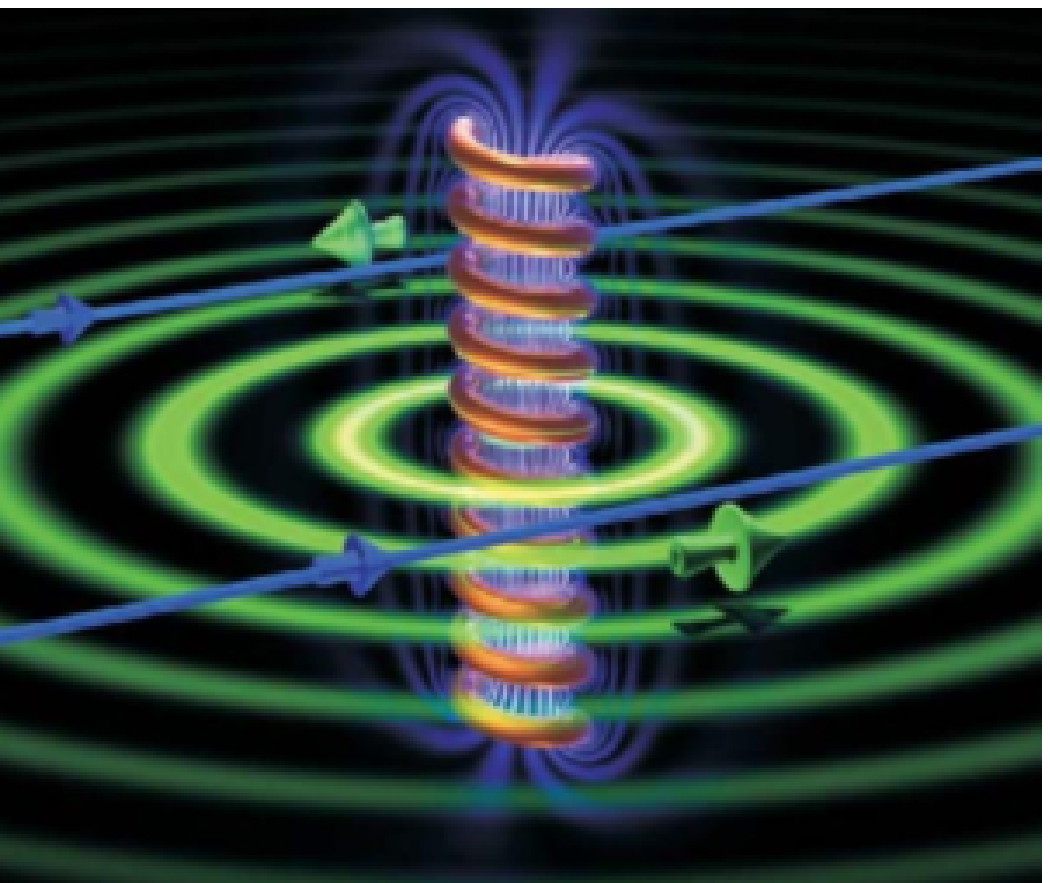}}
 \hspace{0.05\textwidth}
 \subfigure[AB experiment]{\includegraphics[width=0.46\textwidth]{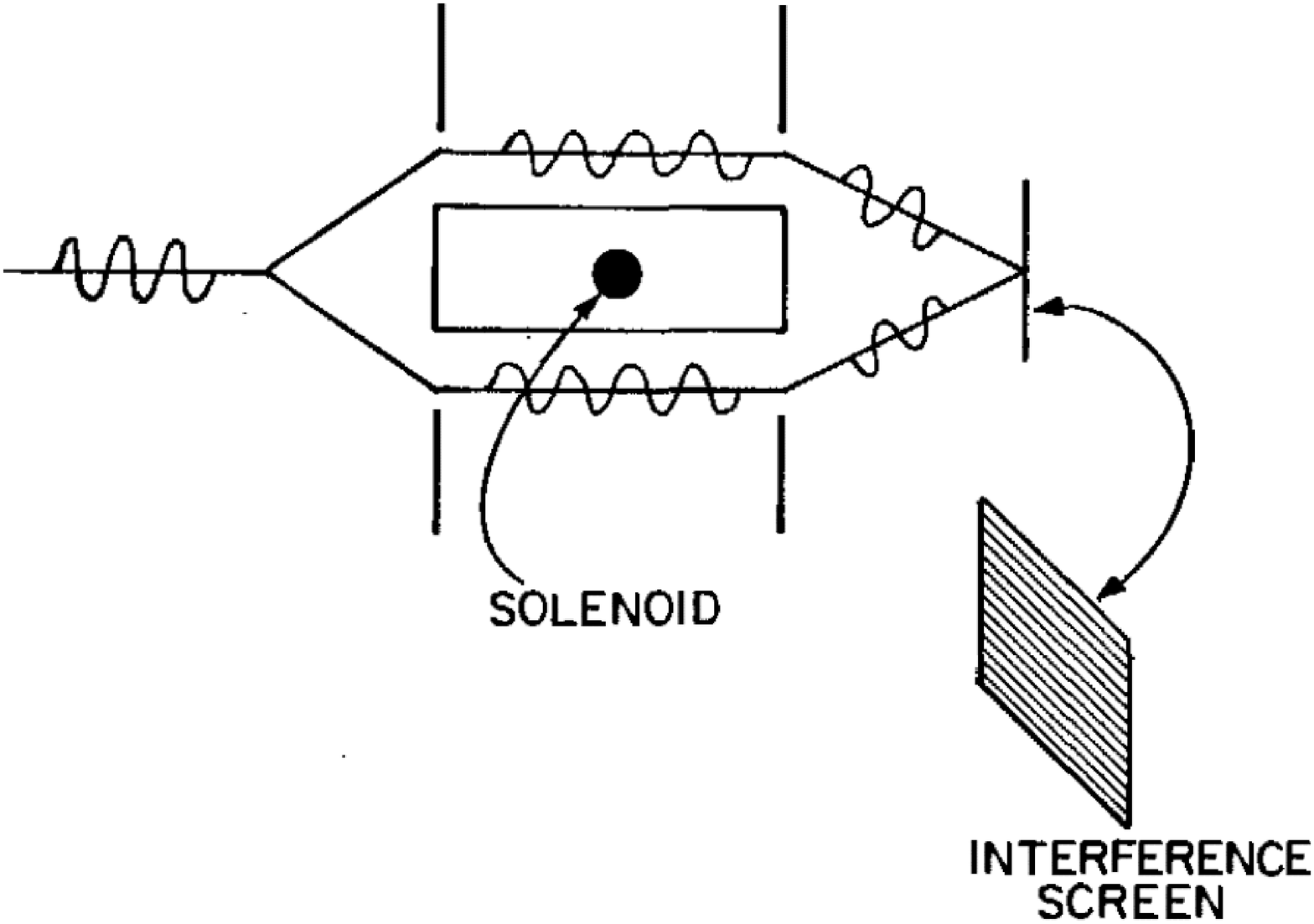}}
 \end{center}
 \caption{Representation of the AB effect. In the left panel, panel $(a)$, in green the vector potential generated from a solenoid in the region $\mathbf{B=0}$ in the symmetric gauge. From Ref.~\cite{Physics_today_AB}. On the the right the ideal experimental setup of the AB effect. From Ref.~\cite{Aharonov1959}.}
 \label{fig:AB_effect}
\end{figure}

The AB effect was introduced in~\cite{Aharonov1959} by considering the interference experiment described in Fig.~(\ref{fig:AB_effect}$.b$). An (ideally infinite) solenoid generates a magnetic field only inside the solenoid itself. In contrast to the magnetic field, the vector potential $\mathbf{A}$, which satisfy the condition $\mathbf{\nabla\times A=B_0}$ inside the solenoid, will not be zero outside. Indeed in the symmetric gauge the total vector potential written in cylindrical coordinates, $\mathbf{A}=(0,0,A_\phi)$ is
\begin{equation}
\begin{split}
&A_\phi= \frac{1}{2} B_0 r \ \ \ \ \ \text{for } r<r_0        \text{,}    \\
&A_\phi= \frac{1}{2} B_0 \frac{r_0^2}{r} \ \ \ \text{for } r>r_0  \text{,}
\end{split}
\end{equation}
where $r_0$ is the radius of the solenoid. In Fig.~(\ref{fig:AB_effect}$.a$), the vector potential around the solenoid is represented by the green circumferences. In real experiments, as the solenoid is not infinite, the return magnetic field must properly set--up in order to avoid the regions where the electrons are permitted. Electrons are injected in the experiment from a point outside and far from the solenoid (see Fig. \ref{fig:AB_effect}$.b$)

If electrons were classical particles their equation of motion would be completely determined by the magnetic field only and in the experiment of Fig.~(\ref{fig:AB_effect}$.a$) they would not be affected at all by the presence of the solenoid. Electrons, instead, are quantum particles and are governed by the Schr\"{o}dinger equation, where the potentials do enter. In~\cite{Aharonov1959} Aharonov and Bohm demonstrated that the electrons feel the presence of the solenoid by acquiring a phase shift between the two paths of Fig.~(\ref{fig:AB_effect}$.a$). This phase can be measured as change in the interference pattern on the screen. The phase shift, $S(\mathbf{x})$, between the two paths can be expressed in terms of the vector potential as 
\begin{equation}
S(x)=-\frac{e}{h}\int_{\gamma(\mathbf{x})} A(\mathbf{x'}) d\mathbf{x'}
\end{equation}
and, computing the integral along the closed line obtained from the path of the two wave--packets, we obtain $S(\mathbf{x})=e/h\ \Phi$ with $\Phi$ the magnetic flux. The electronic wave--function which describes the image on the screen is then $\Psi'=\Psi_0\ e^{-i\Phi/\Phi_0}$, with $\Psi_0$ the wave--function when the experiment is carried out without the solenoid and $\Phi_0=h/e$ the flux quantum.

A static version of the AB effect also exists. If we consider a free electron on a ring which encloses a solenoid we can compute its eigenvalues as~\cite{AB_review1989}
\begin{equation}
E_l=\frac{\hbar^2}{2mr^2}\(l_z-\frac{\Phi}{\Phi_0}\)^2
\text{,}
\end{equation}
where $l_z$ is the canonical angular momentum of the electron, $r$ is the radius of the ring and $\Phi$ the magnetic flux through the ring. We see that the presence of a magnetic flux modifies the eigenvalues splitting their $\pm l_z$ degeneracy.

The quantum mechanical nature of the AB effect is clearly shown by its proportionality to the magnetic flux $\Phi_0$, which goes to zero in the classical limit.

\subsection*{Interpretation of the Aharonov--Bohm effect}
When they first proposed the existence of the effect, Aharonov and Bohm claimed that it was a proof of the reality of the electromagnetic potentials. Their paper generated an intense debate in the scientific community which is not yet terminated. Here we offer some considerations which are inspired by the review by Peshkin and Tonomura~\cite{AB_review1989} on the AB and in particular on the \emph{``central role of the quantized angular momentum''}. 

Consider the static AB effect previously described. We will show in the following that, if the AB effect did not exist, than we would obtain, as a result, that the eigenstate of the system depend on the history of the system. This is in sharp contrast with the foundations of quantum mechanics which states that the Hamiltonian of a system is, at any time, a well defined operator with unique eigenvalues and eigenvectors.

The hypothesis that the AB does not exist means that and we can compute the electronic eigenstates in a (multi-connected) region of space from the sole knowledge of the magnetic field in that region. This means, in our static example, on the ring and nearby it where the electronic wave--function is different from zero.

Suppose that at an initial time $t=t_0$ there is no current flowing through the solenoid and one electron is in a steady state of the ring, $E_l=\frac{\hbar^2}{2mr^2}l_z^2$, with total kinetic angular momentum in the $z$-direction $K_z=\hbar l_z$. Then we turn on the current and, during the transient, a time dependent electromagnetic field is generated by the solenoid. This field will generate a torque on the electron $\Delta K_z=\hbar \(e\Phi/h\)$ (details of the calculation can be found in~\cite{AB_review1989}) so that the total angular momentum of the electron, which must then be conserved, becomes $K_z=\hbar (l_z-e\Phi/h)$.

Let's now consider a second possibility, where the electron on the ring originates from an $eh$ pair created long after the electromagnetic wave has been dissipated. In this case the electromagnetic field is zero on the ring and nearby. Then the steady states on the ring, accordingly to our hypothesis that the AB effect does not exist, have eigenvalues $E_l=\frac{\hbar^2}{2mr^2}l_z^2$ independently of the presence of the solenoid.

The natural conclusion would, then, be that the electronic eigenstates depend on the history which is clearly in contradiction with the principles of quantum mechanics. As a consequence the AB effect must exist in order to ensure that the eigenstate in presence of a magnetic field, possesses angular momentum $K_z=\hbar (l_z-e\Phi/h)$.  
Thus we can interpret the AB effect as a witness of the previous switching on of the solenoid, which modified the space around itself.

\subsection*{Persistent Currents}

\begin{figure}[t]
 \centering
 \includegraphics[width=0.8\textwidth]{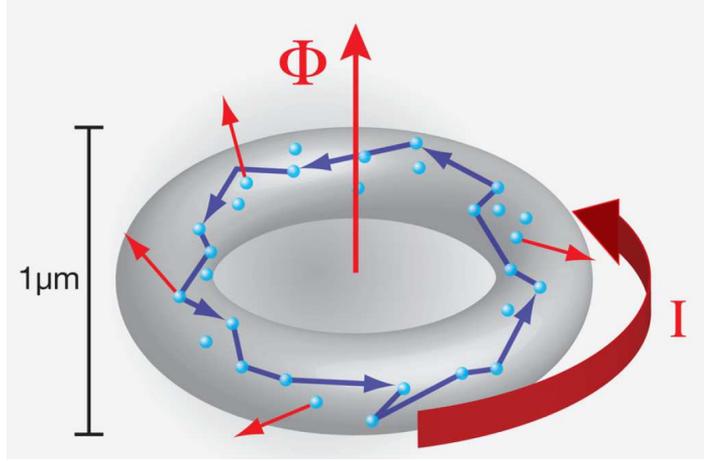}
 \caption{Schematic representation of a PC in a mesoscopic metal ring
          threaded by a magnetic flux quantum, $\Phi$. For rings $\sim 1\mu m$ in size
          and at a temperature $T \sim 1K$, the flux quantum induces a PC
          due to the AB effect. From Ref.~\cite{PC_picture}}
  \label{fig:persistent_currents}
\end{figure}

The existence of the AB effect is strongly related to the quantization of the angular momentum in quantum mechanics~\cite{AB_review1989}; to be precise, in the Hamiltonian formalism the canonical angular momentum $\mathbf{L}=\mathbf{r}\times\mathbf{p}$ is quantized. When a vector potential $\mathbf{A}$ exists, the mechanical angular momentum of the electron is $\mathbf{K}=\mathbf{r}\times\mathbf{(p-A)}$ which is, in general, non quantized. This observation has an important consequence: the existence of Persistent Currents (PCs) in quantum mechanics generated when elecrons move in some particular topologies, like rings or cylindrical shaped objects.

In classical mechanics if we move a metallic ring in a region of space where a magnetic field is present the change in the magnetic flux induces a transient voltage and a current, that will eventually disappear due to the existence of dissipation mechanisms. The appearance of an angular current proportional to the vector potential $\mathbf{j^{(A)}}=+e^2/m^2\ \mathbf{A} |\Psi|^2$ is in fact counterbalanced by an opposite current $\mathbf{j^{(p)}}=-e/m\ Re[\Psi^*\ \mathbf{p}\ \Psi]$ that relaxes the system to the lower energy configuration with $\mathbf{j}=0$. The current flowing in the angular direction is proportional to the angular momentum divided by the radius of the ring. However the term $\mathbf{r\times p}$ can assume only integer values and the total current can be nullified only when $\mathbf{r\times A}$ is an integer, while for any other value PCs exist. If we consider a constant magnetic field then, in the symmetric gauge,
\begin{equation}\label{j_psi}
j_\phi^{(A)}=\frac{e^2}{m^2} |\Psi|^2 A_\phi \simeq \frac{\hbar e\rho}{m^2R_{ring}}\frac{\Phi}{\Phi_0}
\end{equation}
can be nullified only when $\Phi/\Phi_0$ is an integer. Here $\rho$ is the electronic density and $R_{ring}$ the radius of the ring where electrons are trapped. The existence of a periodic lattice partially breaks the quantization of the anguar momentum, however in mesoscopic rings (or cylinders) the quantization is almost exact and even at the nanoscale the argument is correct in first approximation.

The possibility of observing PCs in non superconducting mesoscopic metallic rings was first proposed by B\"{u}ttiker et al. in 1983~\cite{Landauer1983}, while experimental confirmation was reported in 1990 by a research group at Bell Laboratories~\cite{Exp_pc_1990}.  Fig.~(\ref{fig:persistent_currents}) provides a schematic representation of PCs in a mesoscopic ring. PCs have been predicted to exist in  nanostructured materials, like carbon nanotubes~\cite{PC_CNT}. However PCs have been measured only in 2009 in gold and aluminum rings~\cite{Exp_pc_2009_A,Exp_pc_2009_B}. PCs in fact are easily destroyed increasing the temperature of the system by the smearing on the electronic occupations. In particular if the smearing is greater than the difference in energy between states with opposite angular momentum the currents vanish; for this reason PCs do not exist in macroscopic objects.

PCs have never been measured in CNTs. We will discuss this subject more in details in the next chapter.

\subsection*{The Al'tshuler, Aronov and Spikav effect}

In the next section we will describe how the AB effect influences the electronic properties of CNTs. We will provide some experimental evidences, like the resistivity oscillations observed in MW--CNTs with a period which is half the AB period. To understand this result we need to introduce here the concept of Weak Localization (WL), which was first proposed in the 1979~\cite{Abrahams1979}. WL is often seen as a precursor of strong localization in disordered materials and the detailed derivation of the phenomena is rather intricate while its phenomenological interpretation is quite intuitive. 

\begin{figure}[t]
 \centering
 \includegraphics[width=0.6\textwidth]{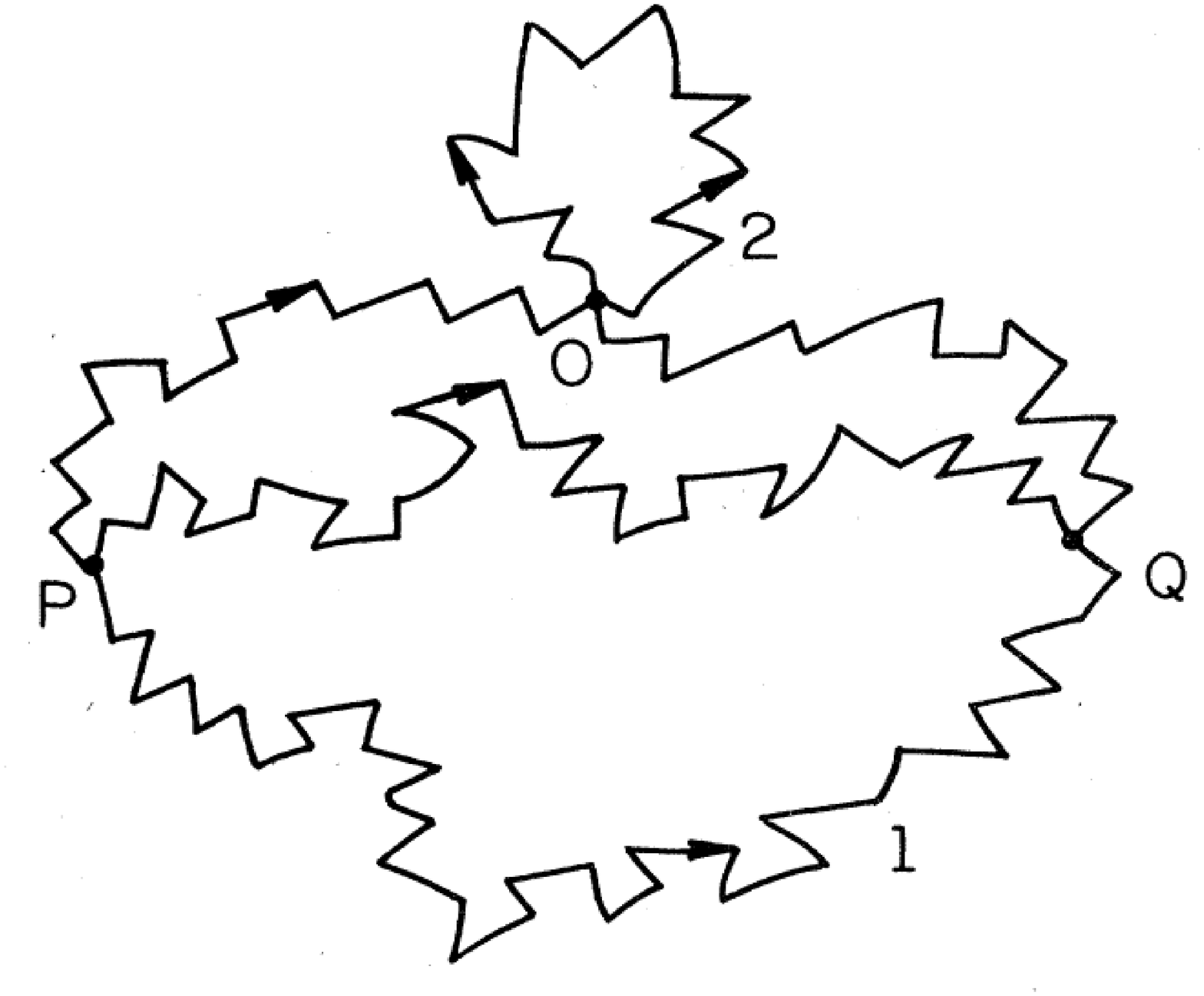}
 \caption{Different types of quasi--classical particle trajectories connecting P and Q. Point O is a self--crossing point of two possible trajectory. From Ref.~\cite{WL_picture}}
 \label{fig:weak_localization}
\end{figure}
In the classical theory of transport phenomena the total probability for a particle to transfer from point $P$ to point $Q$ (Fig.~(\ref{fig:weak_localization}) ) is the sum of probabilities of such a transfer over all possible trajectories. In quantum mechanics this result corresponds to neglect the interference of scattered electrons propagating along different paths and having approximately random phases under the quasi--classical condition $\lambda << l$, with $l$ the length of the propagation path and $\lambda$ the De Broglie wave--lenght of the wave--packet. There is, however, a specific class of trajectories, namely, self--crossing trajectories (trajectory 2 in Fig.~(\ref{fig:weak_localization}) ) for which the wave interference turns out to be essential. Indeed two waves propagating along such trajectories in two opposite directions (conjugated waves) accumulate the same phase difference. Therefore the contribution of these trajectories to the probability of coming to the same point (point $O$ in Fig.~(\ref{fig:weak_localization})) will be 
\begin{equation}
|A_1+A_2|^2 = |A_1|^2+|A_2|^2+2 Re[A_1^*A_2]
\end{equation}
which is twice the sum of the squared amplitude moduli. A higher probability of returning back to point $O$ means a lower probability of transfer from point $P$ to point $Q$. Thus (weak) localization is favourite and, hence, results in an increase of the resistivity.

If the sample is placed in a magnetic field then the probability amplitudes of completing the loop on contour 2 of Fig.~(\ref{fig:weak_localization}) acquire an additional phase
\begin{equation}
A_1 \rightarrow A_1 e^{ 2i\pi \frac{\Phi}{\Phi_0}}  \ \ \ \ \ \ \ \ \
A_2 \rightarrow A_2 e^{-2i\pi \frac{\Phi}{\Phi_0}}
\end{equation}
and then the phase difference will be $\Delta\varphi/(2\pi)=2\Phi/\Phi_0$. Indeed the combined WL and AB effects predict the existence of resistivity oscillations with a period of $\Phi_0/2$ and is known as AAS effect from the names of Al'tshuler, Aronov and Spikav who first proposed its existence in 1980~\cite{AAS1981}.

\section{An introduction to carbon nanotubes}\label{Sec:Intro_CNTs}

\begin{figure}[t]
 \centering
 \includegraphics[width=0.45\textwidth]{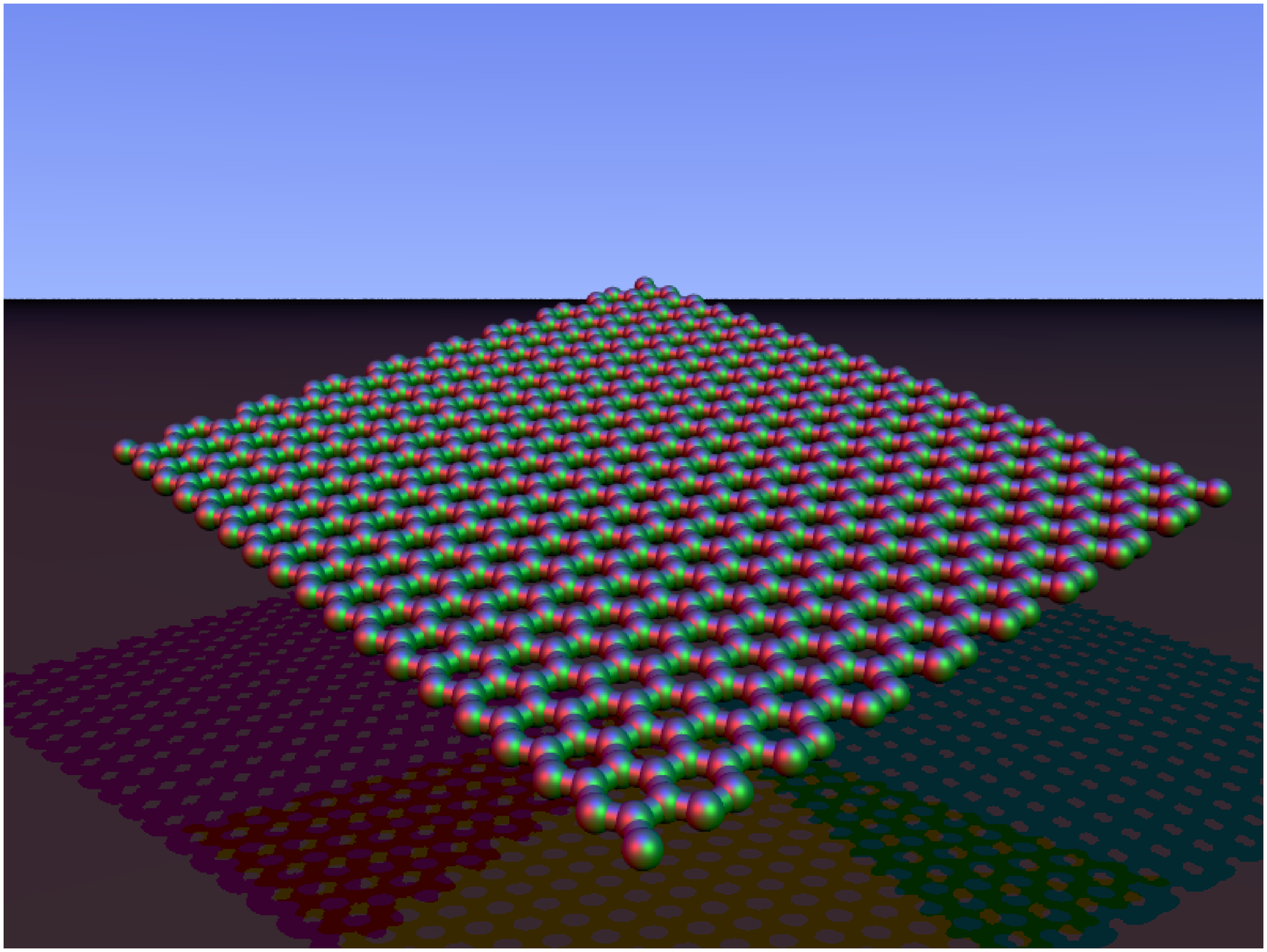}
 \includegraphics[width=0.465\textwidth]{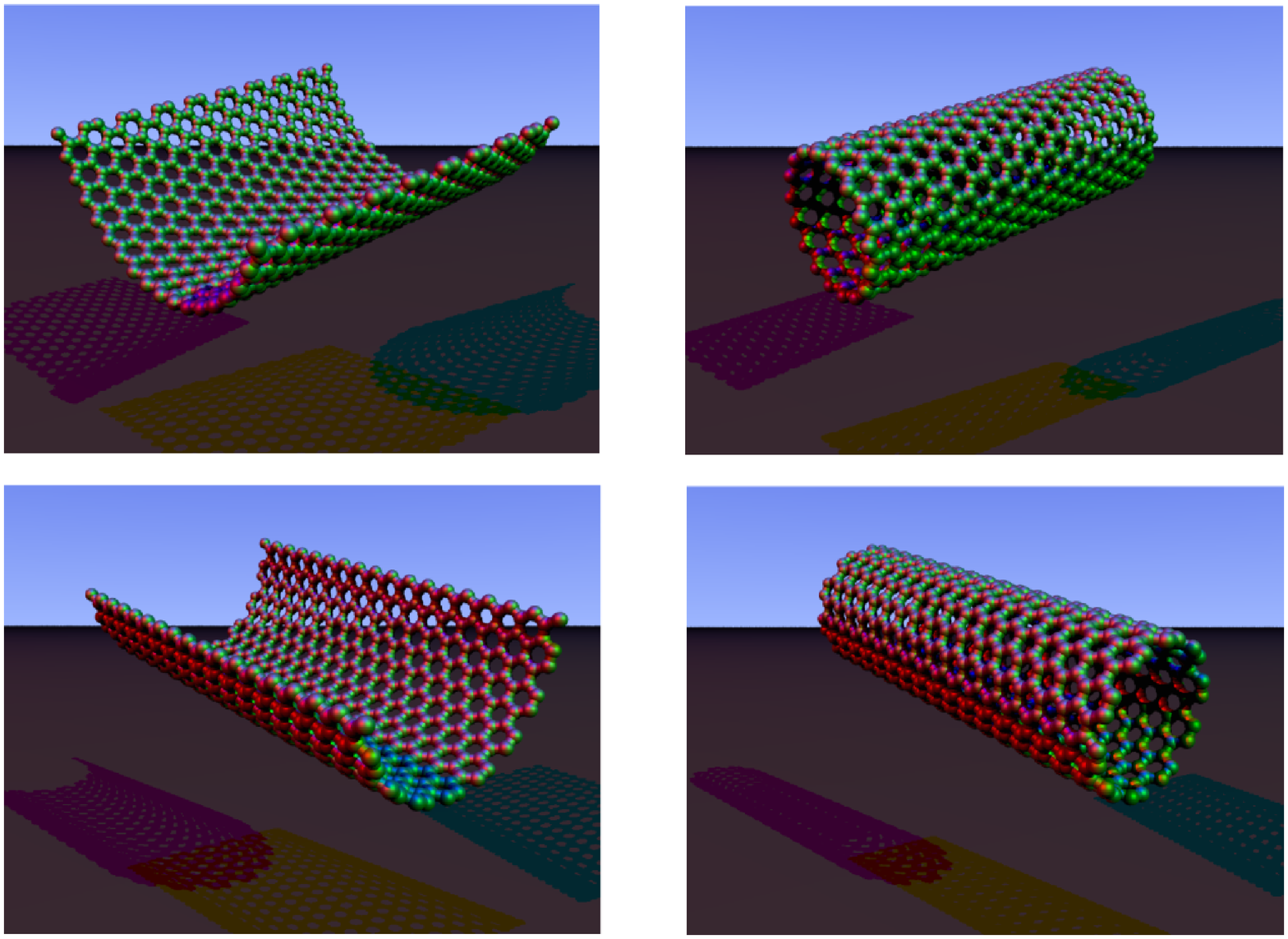}
 \caption{Rolling a Plane of graphene to obtain (n,0) and (n,n) CNTs. From Ref.~\cite{wikipedia}}
 \label{fig:rolling_graphene}
\end{figure}

A CNT is a honeycomb lattice rolled into a hollow cylinder with nano--metric diameter and $\mu m$ length. CNTs were discovered and first characterized in 1991 by Iijima from NEC laboratories (Japan)~\cite{Iijima_1991}. The first CNTs discovered were made of several concentric cylindrical--like shells regularly spaced by an amount of about 3.4 $A$ as in conventional graphite materials. These Multi--Wall CNTs (MWCNTs) were first synthesized with diameters ranging from a few nanometers to several hundred nanometers for the inner and outer shells, respectively. As for the length, MWCNTs extending over several microns are currently synthesized. Shortly after the discovery of MWCNTs, Single--Wall CNTs (SWCNTs) were synthesized in abundance using arc--discharge methods with transition--metal catalysts~\cite{Bethune_1993,Iijima_1993}. These tubes have quite small and uniform diameter, on the order of $1 nm= 10^{-9} m$. This unprecedentedly small diameter, combined with the crystalline perfection of the atomic network, explains why these objects were quickly considered as the ultimate carbon--based 1D systems.    Crystalline ropes or bundles of SWNTs, with each rope containing tens to hundreds of tubes of similar diameter, closely packed in a hexagonal configuration, have also been synthesized using a laser vaporization method~\cite{Guo_1995} and other methods.

Depending on the community, specific interests, and targeted applications, nanotubes are regarded as either single molecules or quasi-one-dimensional crystals with translational periodicity along the tube axis. As there are an infinite number of ways of rolling a sheet into a cylinder (two of them are represented in Fig.~(\ref{fig:rolling_graphene}) ) the large variety of possible helical geometries, defining the tube chirality, provides a family of CNTs with different diameters and microscopic structures. Some properties of these nanotubes, such as the elastic ones, can be explained within a macroscopic model of a homogeneous cylinder. Others depend crucially on the atomic configuration. For instance, the electronic and transport properties, are certainly among the most significant physical properties of CNTs, and crucially depend on the diameter and chirality. This dependence on the atomic configuration is quite unique in solid-state physics.

CNTs can be either semi--metallic or semi--conducting, with a band gap varying from zero to a few tenths of an $eV$, depending on their diameter and chirality. Further, the band gap of semi--conducting tubes, or the energy difference between the peaks in the electronic density of states, the so--called Van Hove singularities, can be shown to first order to be simply related to the tube diameter. Such remarkable results can be obtained from a variety of considerations, starting from the so-called Zone Folding Approach (ZFA), based on knowledge of the electronic properties of the graphene (a single sheet of graphite), to the direct study of nanotubes using semi--empirical Tight--Binding (TB) approaches. The comparison with more sophisticated Ab--initio calculations, and with available experimental results, permits to find the limits of these simple approaches.

\subsection*{The Zone Folding Approach}

In the ZFA the wave--functions $\Psi_{n,k_x,k_z}(\mathbf{x})=e^{i (k_x x+k_z z)}u_{n,k_x,k_z}(\mathbf{x})$ of the graphene sheet are used to describe electrons in CNTs assuming that the curvature of the sheet gives negligible effects if the tube radius is large enough. Then the only difference the electrons feel, with respect to graphene, is the quantization of the angular momentum $l_z=k_x R$ (we are assuming here that the tube direction is the $z$-axis). Accordingly only the wave--functions $e^{i (l_z/R \phi+k_z z)}u_{n,l_z/R,k_y}(\mathbf{x})$ are considered in the model.

\begin{figure}[t]
 \centering
 \includegraphics[width=0.610\textwidth]{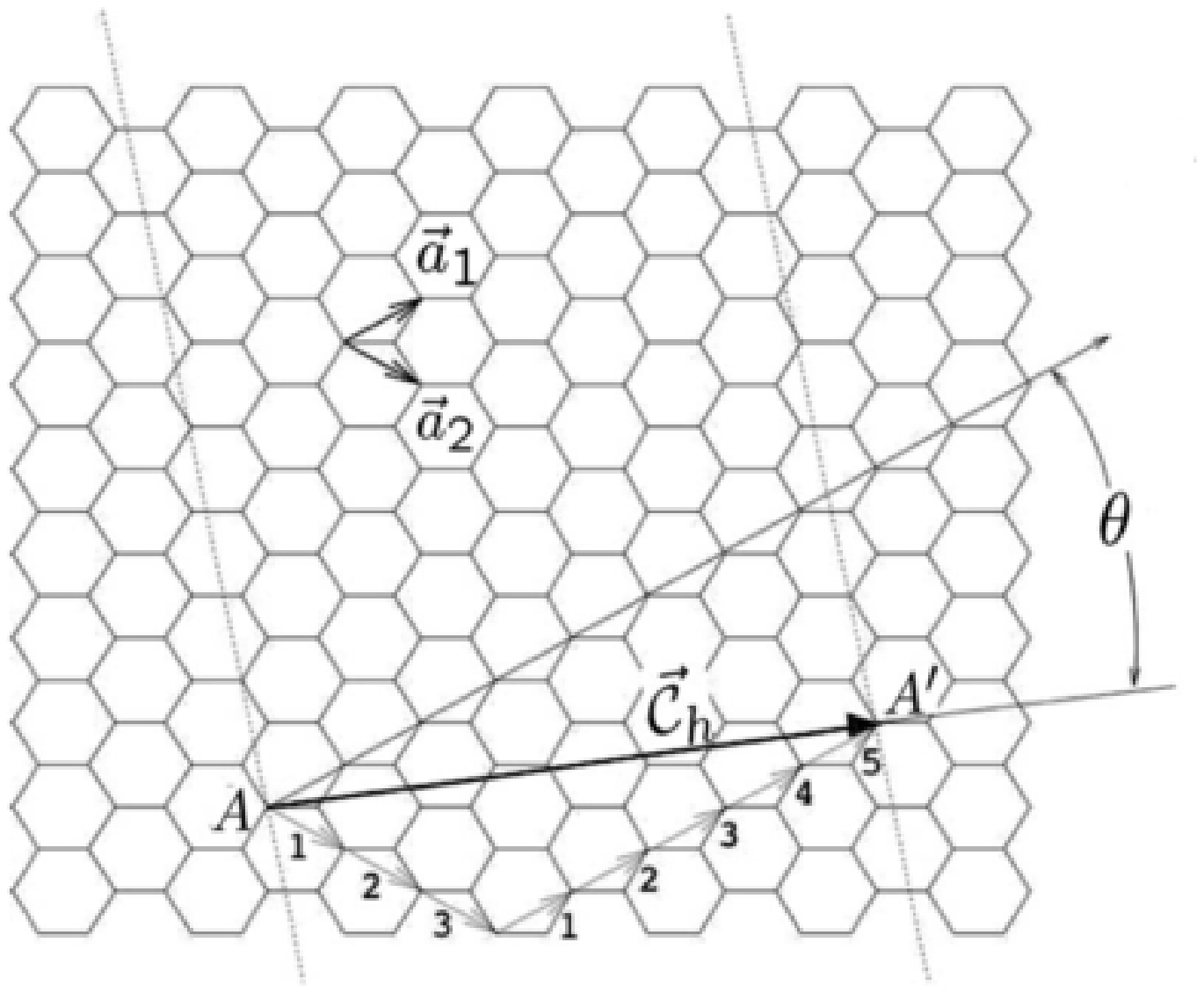}
 \includegraphics[width=0.315\textwidth]{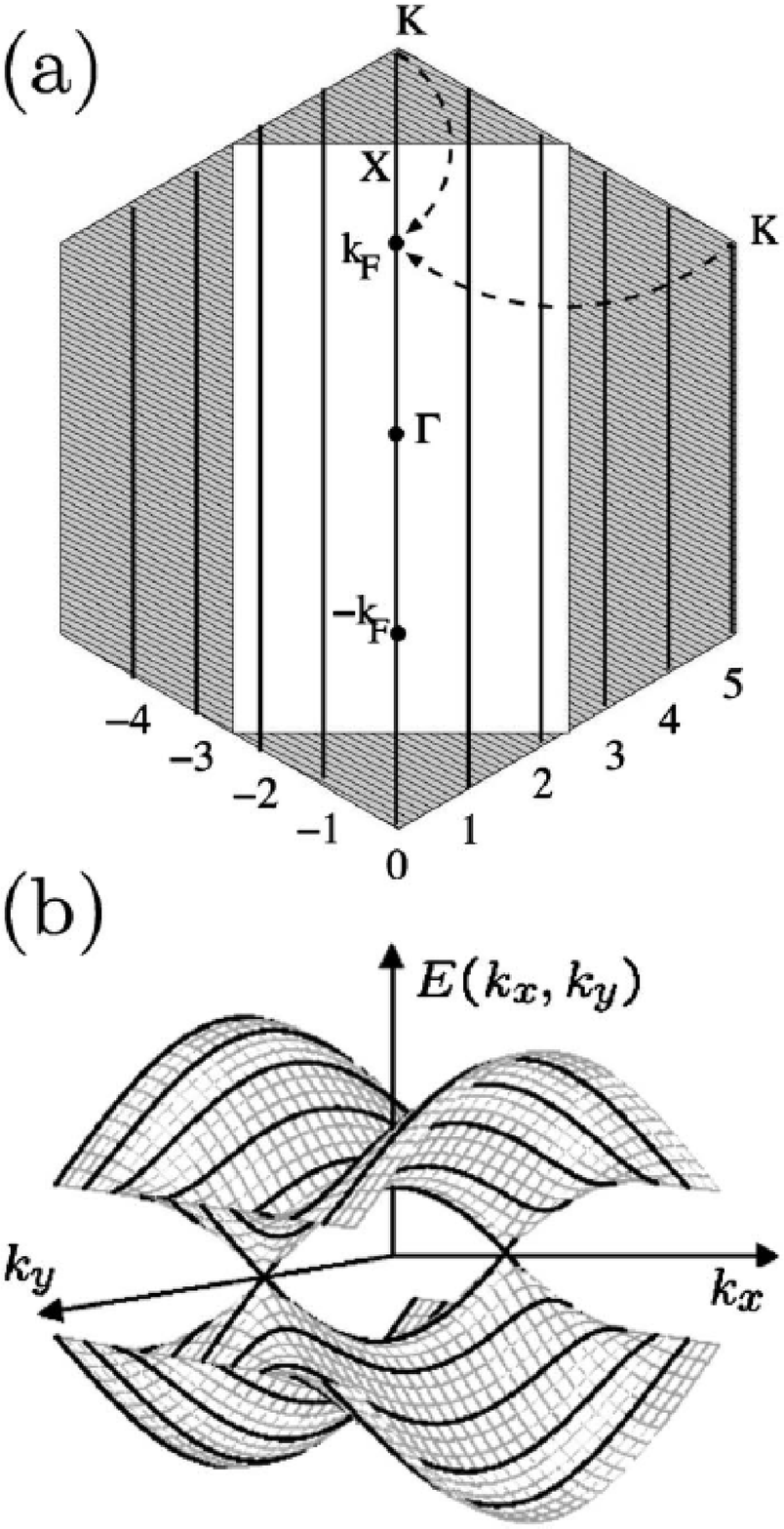}
 \caption{The Zone Folding Approximation. On the left panel the section of graphene which describes the $(3,5)$ CNT within the ZFA. Periodic boundary conditions are imposed on the dotted lines. On the right panel the Brillouin Zone of graphene with the $k$--points lines which respect the the rolling condition of a $(5,5)$ CNT. From Ref.~\cite{Charlier2007}.}
 \label{fig:ZF_graphene}
\end{figure}

In Fig.~(\ref{fig:ZF_graphene}), on the left $(a)$ panel, a CNT is represented as a stripe in the plane of graphene. In the ZFA each CNT can be identified by two numbers, which represent the circumference vector in the basis of the direct lattice vectors $\mathbf{a_1,a_2}$ of graphene. Boundary conditions are then imposed which results in selecting specific $k$--points in the reciprocal space. In Fig.~(\ref{fig:ZF_graphene}), on the right top panel, the Brillouin Zone (BZ) of graphene is represented and the lines correspond to the $k$ points which respects the boundary conditions of a $(5,5)$ CNT. The right bottom panel, instead, shows the energy surface of the $\pi-\pi^*$ bands of graphene, cut by the allowed $k$--points lines which can be used to construct the CNT band structure. 

The predictions of the ZFA give a good description of the properties of CNTs but some corrections have to be considered both for SWCNTs, to include the curvature of the tubes, and for MWCNTS, to include the effect of the interaction among different tubes. For example the position of the so called Dirac points (see the next subsection for the definition of the Dirac points) has to be shifted in the Brillouine Zone due to curvature effects.  Finally the ZFA results depend on the method used to compute the band structure of graphene. The energy surfaces in the Fig.~(\ref{fig:ZF_graphene}) have been calculated within the TB model, for example.

\subsection*{The Tight Binding Model for graphene}\label{Subsection:TB}

The graphene plane is an hexagonal lattice with two atoms per unit cell (A and B) and a basis defined by the vectors $(a_1,a_2)$, as in Fig.~\ref{fig:ZF_graphene}. The condition $\mathbf{a}_i\cdot \mathbf{b}_j = 2\delta_{i,j}$ allows one to obtain the reciprocal lattice vectors $(b_1,b_2)$. Every carbon atom possesses four valence electrons (two $2s$ and two $2p$ electrons). When the atoms are placed onto the graphene hexagonal lattice the electronic wave functions from different atoms overlap. However, such an overlap between the $p_z$ orbitals and the $s$ or $p_x$ and $p_y$ electrons is strictly zero by symmetry. Consequently, the $p_z$ electrons, which form the $\pi$ bonds in graphene, can be treated independently of other valence electrons. Within this $\pi$-band approximation, the A atom (/ B atom) is uniquely defined by one orbital per atom site $p_z(r-r_A)$ [/ $p_z(r-r_B)$]. To derive the electronic spectrum of the total Hamiltonian, the corresponding Schr\"odinger equation has to be solved, and by applying the Bloch theorem, the wave functions can be written as follows
\begin{equation}
\Psi_\mathbf{k}(\mathbf{r})=c_A(\mathbf{r})\tilde{p}_{z,\mathbf{k}}^A(\mathbf{r})+
                            c_B(\mathbf{r})\tilde{p}_{z,\mathbf{k}}^B(\mathbf{r})
\text{,}
\end{equation}
where
\begin{equation}
\tilde{p}_{z,\mathbf{k}}^J(\mathbf{r})=\frac{1}{\sqrt{N}} \sum_\mathbf{l} 
                                       e^{i\mathbf{kl}} p_z(\mathbf{r-r_J-l}) \ \ \ \ \ \ \ J=A,B\text{\ .}
\end{equation}
$\mathbf{k}$ is the crystal momentum, $N$ is the number of unit cells in the graphene sheet, and $\mathbf{l}$ is the cell position index. 

\begin{figure}[t]
 \centering
 \includegraphics[width=0.7\textwidth]{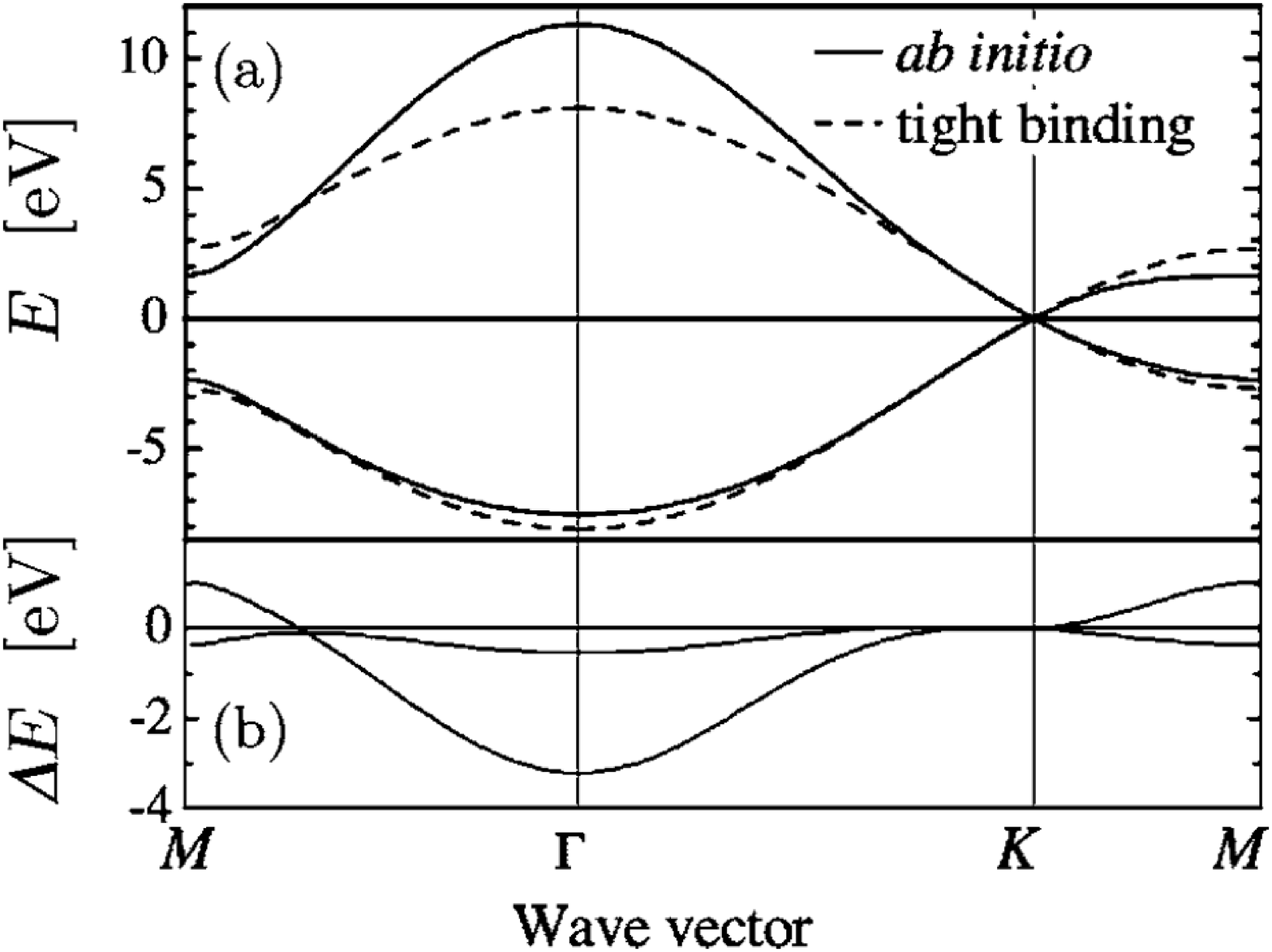}
 \caption{Band structure of graphene evaluated with the tight binding model compared with the
          result of ab--initio calculations. From Ref.~\cite{Charlier2007}.}
 \label{fig:ZF_Vs_AB}
\end{figure}

The spectrum is derived by solving the Scr\"odinger equation which reduces to the diagonalization of
a $2\times 2$ matrix
\begin{equation}
\left( \begin{array}{cc}
H_{AA}-E & H_{AB} \\
H_{BA} & H_{BB}-E
\end{array} \right)
\text{,}
\end{equation}
in the space defined by the $|J\rangle=\tilde{p}_{z,\mathbf{k}}^J(\mathbf{r})$ wave--functions. Neglecting the overlap $\langle A | B \rangle$, restricting interactions to nearest neighbors only and setting $H_{AA}=H_{BB}=0$ as energy reference the dispersion relation are then:
\begin{equation}
E^{\pm}(\mathbf{k})=\pm\gamma_0\sqrt{3+2cos(\mathbf{ka_1})+2cos(\mathbf{ka_2})+2cos(\mathbf{k[a_2-a_1]})}
\text{.}
\end{equation}
These are the $\pi$ and the $\pi^*$ bands in the TB model. One of the two bands, which represent the valence and the conduction bands, is completely filled and the other completely empty. Moreover they intersect only in two points, known as Dirac points, in the BZ, as shown in Fig.~(\ref{fig:ZF_graphene}). For this reason graphene is a semi--metal with a one--dimensional Fermi surface. For the same reason CNTs are predicted to be either metallic or semi--conducting according to whether or not the Dirac points belong to the set of $k$--points allowed in the ZFA.

\begin{figure}[t]
 \centering
 \subfigure[$(5,5)$ CNT]{\includegraphics[width=0.44\textwidth]{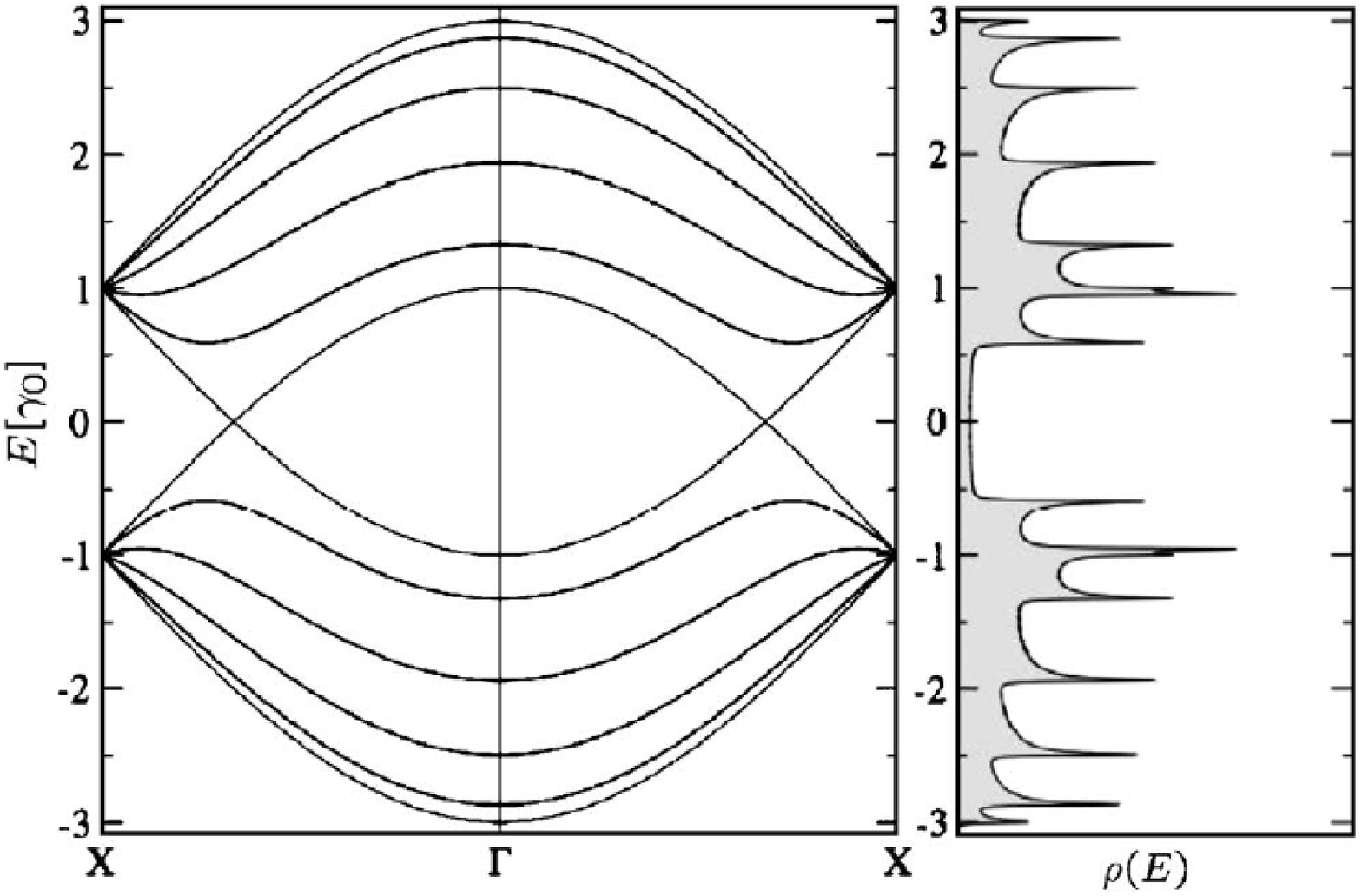}\label{subfig:ZF_55_bands}}
 \hspace{0.05cm}
 \subfigure[$(9,0)$ CNT]{\includegraphics[width=0.43\textwidth]{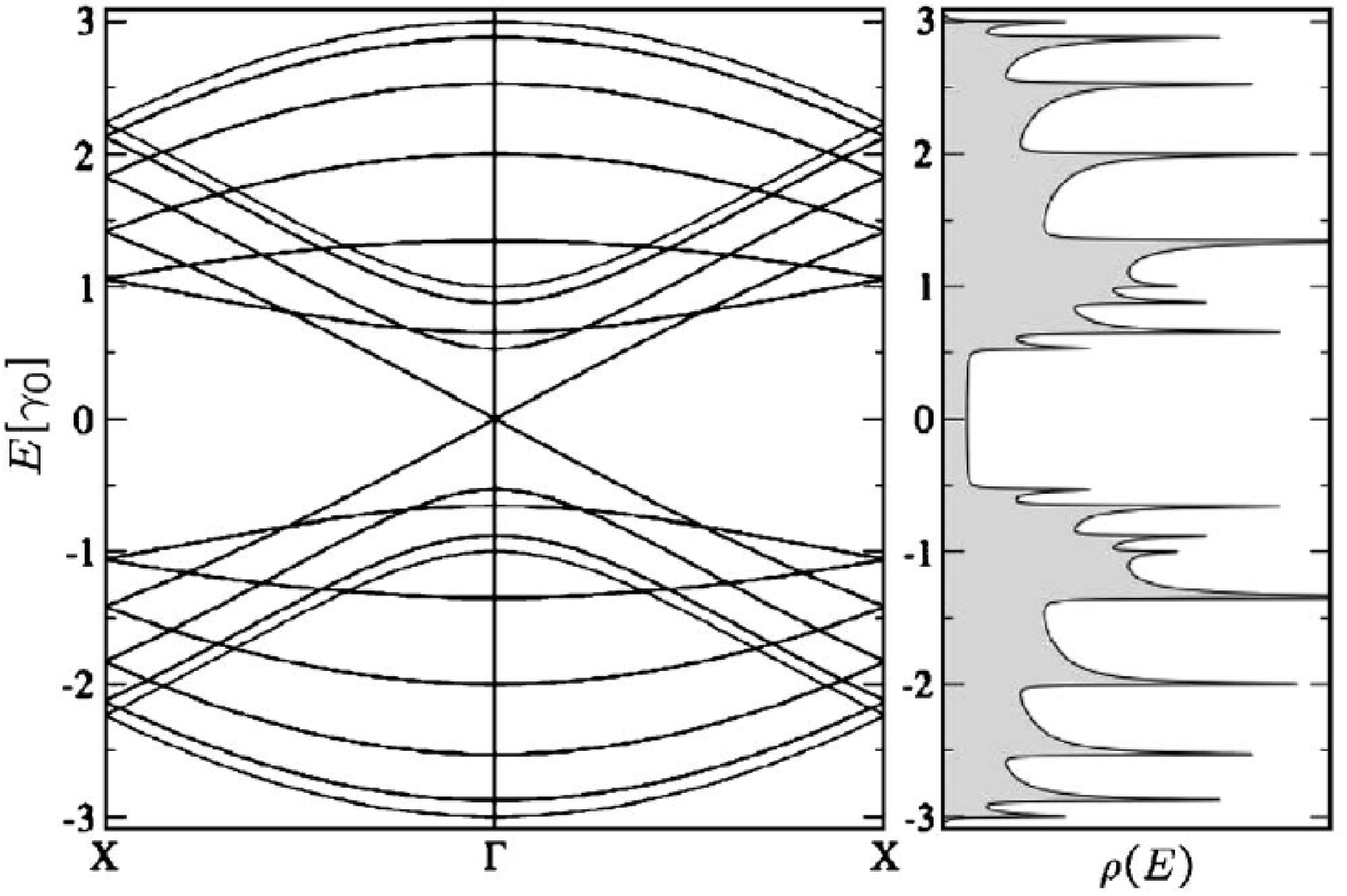}}
 \subfigure[$(8,2)$ CNT]{\includegraphics[width=0.43\textwidth]{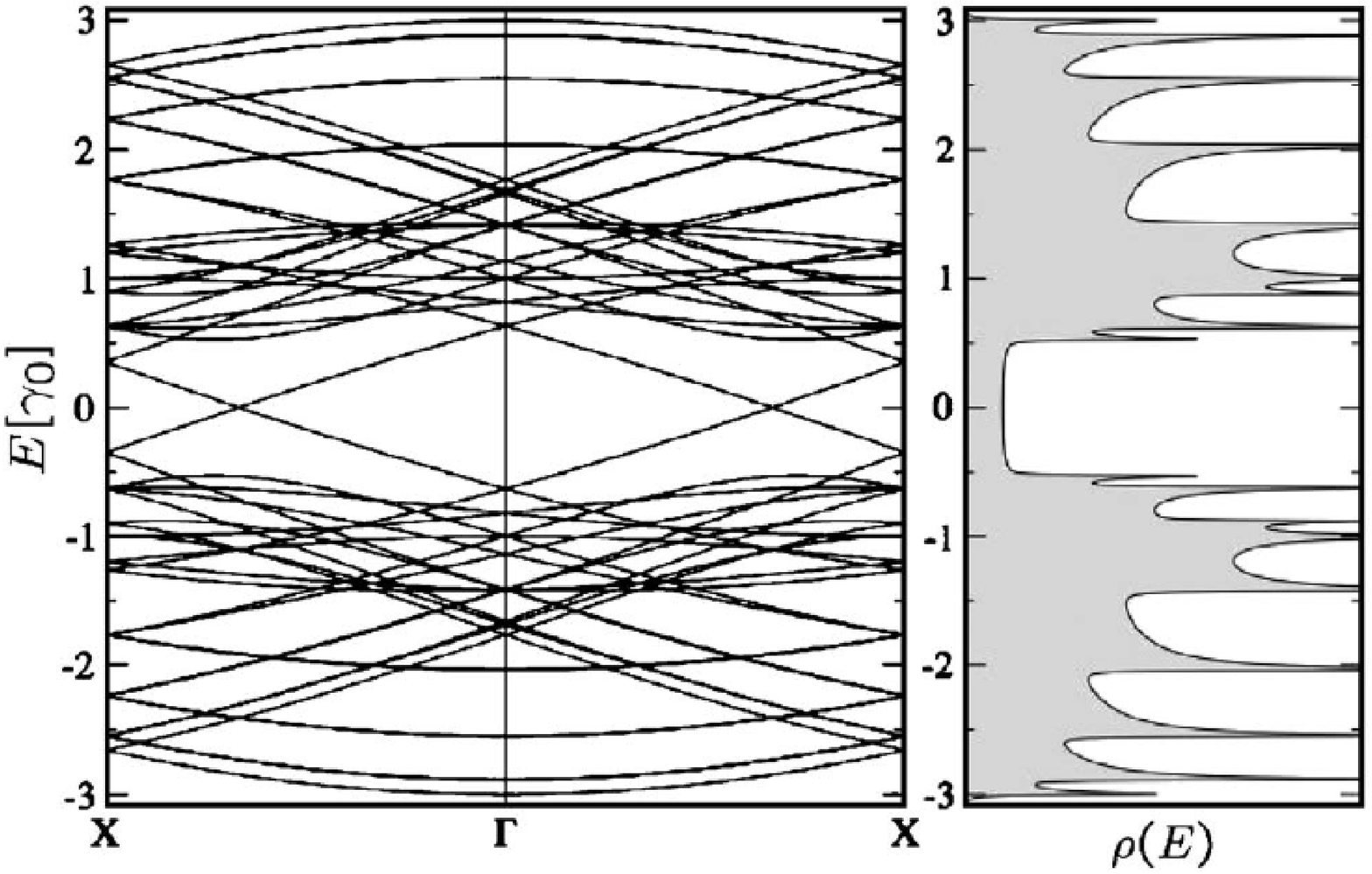}}
 \hspace{0.05cm}
 \subfigure[$(10,0)$ CNT]{\includegraphics[width=0.43\textwidth]{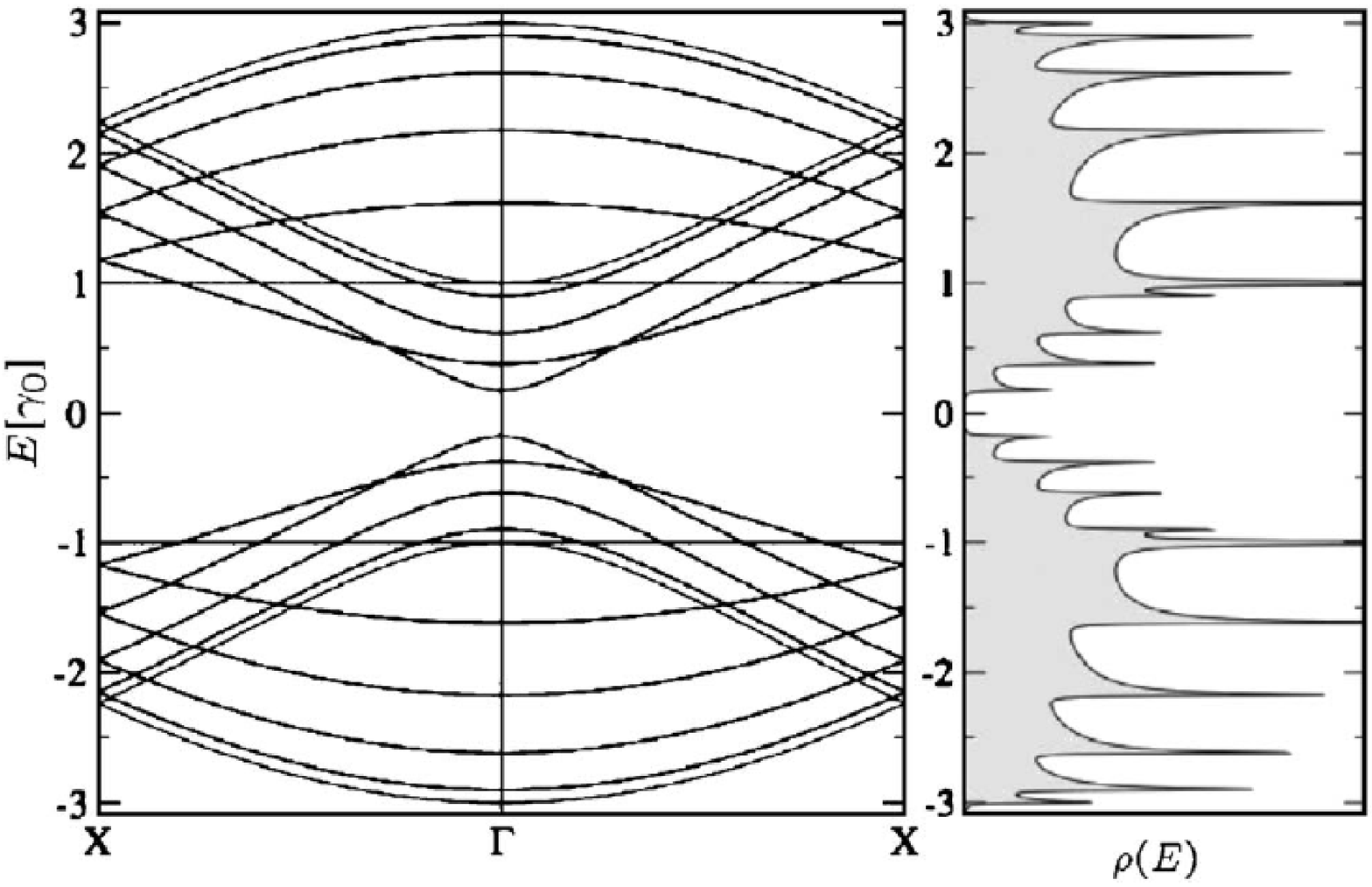}}
 \caption{Band structures and DOS of several CNTs calculated within the ZFA. From Ref.~\cite{Charlier2007}.}
 \label{fig:ZF_on_TB}
\end{figure}

The predictions of the TB model are partially confirmed by Ab--initio calculations even if some differences appear. For example the DFT band structure of graphene is not symmetric with respect to the chemical potential, as the TB one (Fig. (\ref{fig:ZF_Vs_AB}) ). In the rest of this chapter we will work in the ZFA using as starting point the TB band structure. Ab--initio corrections will be discussed later.

Within the TB+ZFA scheme, it can be show that for example all $(n,n)$ and $(3n,0)$ CNTs are metallic, with $n$ any integer, while all the remaining $(n,0)$ tubes are semi--conducting. CNTs of different kinds, $(n,m)$ CNTs, are said to be chiral and can be either metallic or semi--conducting. The general rule is that a CNT is metallic if $n-m$ is a multiple of $3$~\cite{Charlier2007}. In Fig.~(\ref{fig:ZF_on_TB}) the bands structures for the $(5,5)$, the $(9,0)$, the $(10,0)$ and the $(8,2)$ CNTs are shown. 

\section{Theoretical predictions and experimental results}
\subsection*{The Zone Folding Approach}
The state--of--the--art theoretical approach to the AB oscillations in CNTs is based on the ZFA.

\begin{figure}[t]
 \centering
 \includegraphics[width=0.45\textwidth]{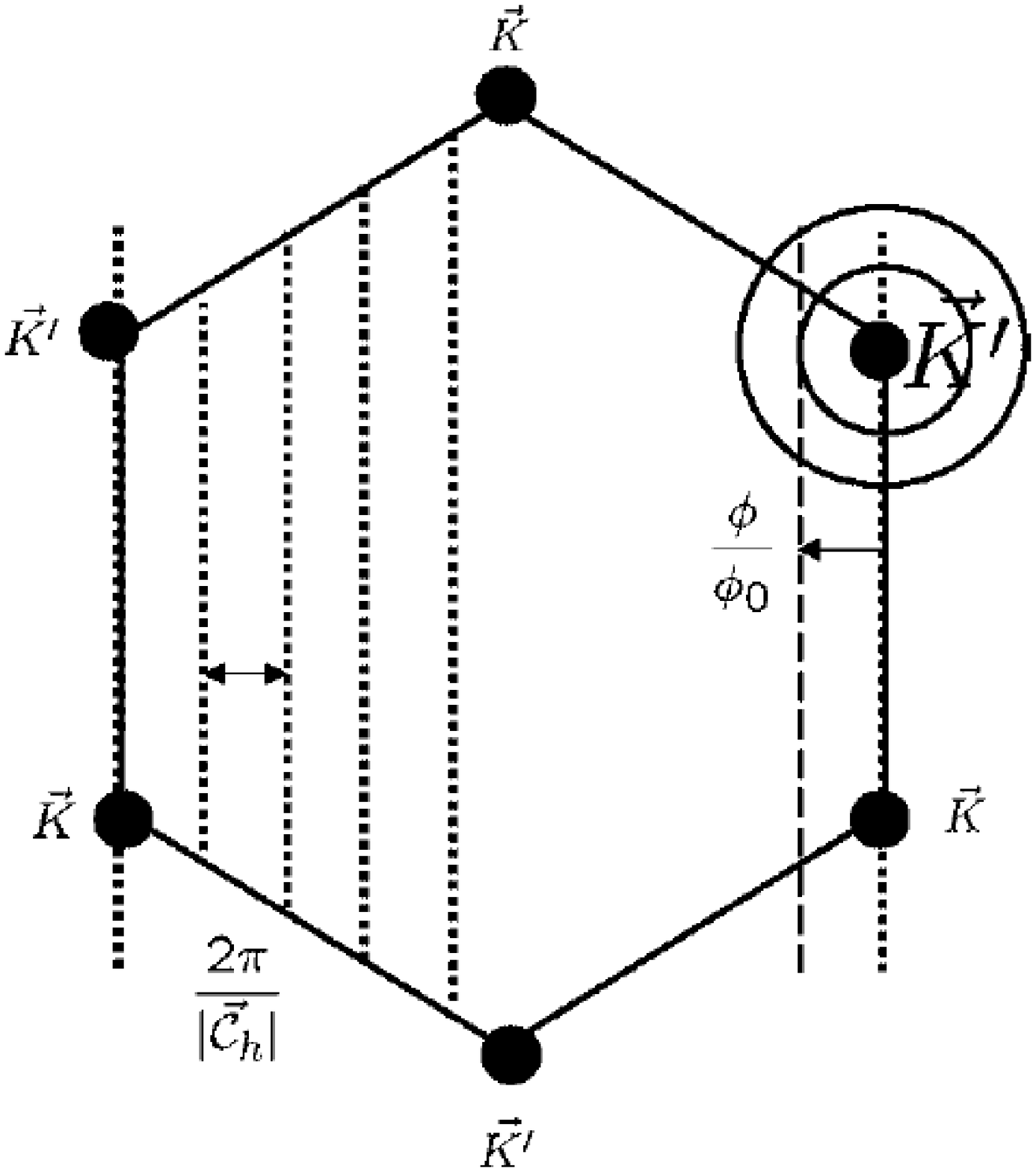}
 \caption{Aharonov--Bohm effect within the Zone Folding Approach. From Ref.~\cite{Charlier2007}.}
 \label{fig:AB_within_ZF}
\end{figure}

The ZFA is introduced by observing that the Hamiltonian of a CNT can be expressed in cylindrical coordinates as
\begin{equation}\label{H_cylindrical}
H=-\frac{\hbar^2}{2m}\left[ \frac{\partial^2}{\partial r^2}+\frac{1}{r}\frac{\partial}{\partial r}+
                            \frac{1}{r^2}\left(i\frac{\partial}{\partial\varphi}-\frac{\Phi}{\Phi_0}\right)+
                            \frac{\partial^2}{\partial z^2}
                    \right] + V(r,\varphi,z)
\text{.}
\end{equation}
Then in the ZFA the following map is applied
\begin{eqnarray}
&&\phi \rightarrow x/R_{CNT} \ \ \text{,}  \\
&&r \rightarrow y  \ \ \ \ \ \ \ \ \ \ \ \text{,}            \\
&&z \rightarrow z  \ \ \ \ \ \ \ \ \ \ \ \text{,}
\end{eqnarray}
where $(x,y,z)$ are cartesian coordinates with the graphene sheet oriented in the $xz$--plane.
The ZFA Hamiltonian, when a magnetic field is present, is then
\begin{equation}\label{H_ZFA}
H=-\frac{\hbar^2}{2m}\left[ \frac{\partial^2}{\partial y^2}+
                            \left(i\frac{\partial}{\partial x}-\frac{\Phi}{\Phi_0 R_{CNT}}\right)^2+
                            \frac{\partial^2}{\partial z^2}
                    \right] + V(x,y,z)
\text{.}
\end{equation}
As the ZFA approximates the CNT with a planar graphene sheet in Eq. (\ref{H_ZFA}) the $2^{nd}$ term of Eq. (\ref{H_cylindrical}) is set to zero and $r \approx R_{CNT}$. Applying the resulting Hamiltonian to the block wave--function $\Psi_{k_x,k_z}(x,y,z)=e^{i (k_x x+k_z z)} u_{k_x,k_z}(\mathbf{x})$ and defining
\begin{equation}
k'_x=k_x-\frac{\Phi}{\Phi_0 R_{CNT}}
\end{equation}
we obtain, for the periodic part of the wave--function, an Hamiltonian identical to the case without magnetic field a part from a shift of $k_x$.

\begin{figure}[t]
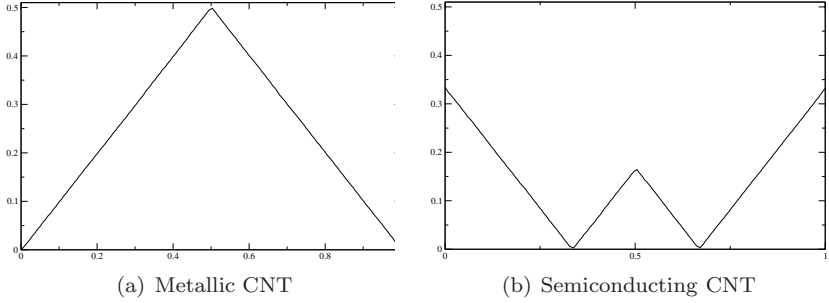

 \centering
 \subfigure[Metallic CNT]{\includegraphics[width=0.45\textwidth]{images/CNTs/Metallic_oscillations_ZF.eps}}
 \hspace{0.1cm}
 \subfigure[Semiconducting CNT]{\includegraphics[width=0.45\textwidth]{images/CNTs/Semiconducting_oscillations_ZF.eps}}
 \caption{Aharonov--Bohm gap oscillations in carbon nanotubes according to the zone folding approach.}
 \label{fig:gap_oscillations_ZF}
\end{figure}

This means that the eigen--functions of the Hamiltonian with $\mathbf{B\neq 0}$ can be written in term of the eigen--function of the Hamiltonian with $\mathbf{B=0}$:
\begin{eqnarray}
\Psi^{\mathbf{B}}_{n,k_x,k_z}(x,y,z)&=&e^{i (k_x x+k_z z)}u^{\mathbf{B}}_{n,k_x,k_z}(\mathbf{x})     \\
                                    &=&e^{i (k'_x x+k_z z)}u^{\mathbf{B=0}}_{n,k'_x,k_z}(\mathbf{x}) e^{i (\Phi/\Phi_0)(x/R)}
\text{.}
\end{eqnarray}
Consequently we can obtain the eigen--functions and the corresponding shift of the energies of the allowed $k$-point grid as shown in Fig.~(\ref{fig:AB_within_ZF}). This means that, increasing the magnetic field, all the electronic properties of the CNTs will oscillate with period $\Phi_0$ as, each times $\Phi=n\Phi_0$ the allowed $k$--points will coincide with the ones at $\mathbf{B=0}$. In particular the electronic gap is predicted to oscillate because periodically $k_x'$ match a Dirac point. For $(n,0)$ or $(n,n)$ CNTs, two kinds of oscillations can exist according to whether the CNT is metallic or semi--conducting as shown in Fig. (\ref{fig:gap_oscillations_ZF})

\subsection*{The experimental evidences}

\begin{figure}[t]
 \centering
 \subfigure[Experimental setup]{\includegraphics[width=0.43\textwidth]{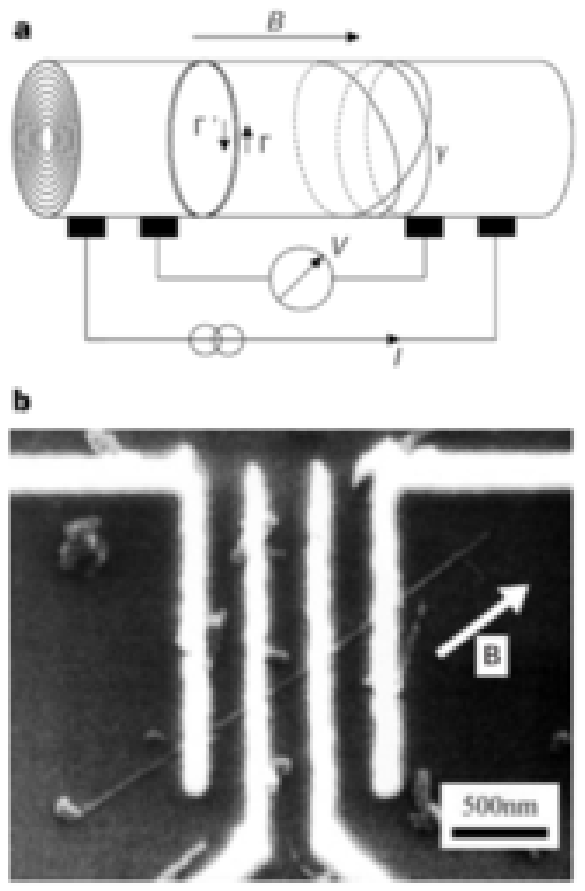}}
 \subfigure[Resistivity oscillations]{\includegraphics[width=0.5\textwidth]{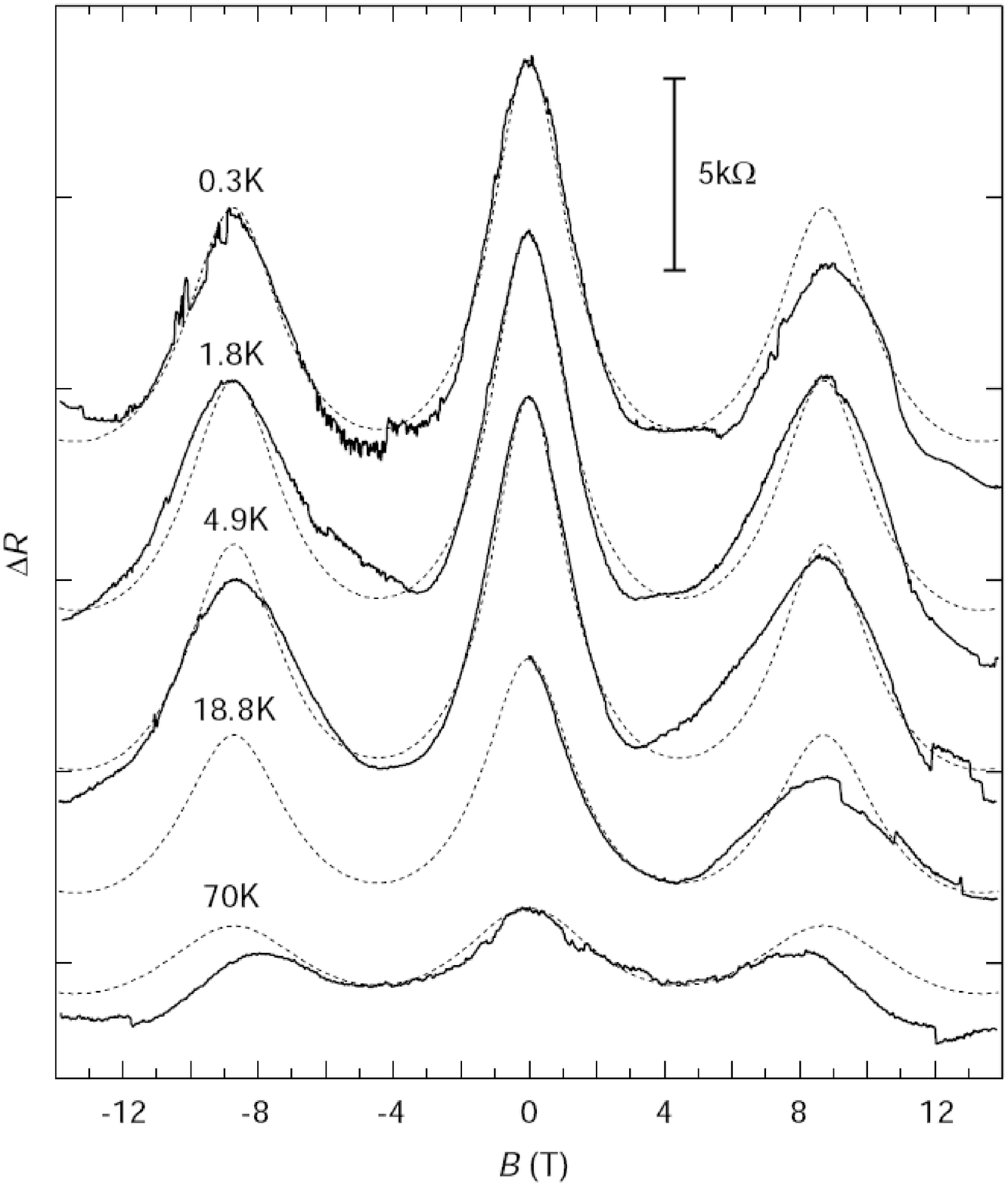}}
 \caption{Description of the experimental setup and results of the experiment carried out in~\cite{Bachtold1999}.  In the right panel resistivity oscillations with a period $\Phi_0/2$ can be observed.}
 \label{fig:resistivity_oscillations}
\end{figure}

The first experimental evidences of the AB effect in CNTs is reported in Fig. (\ref{fig:resistivity_oscillations}). In their work A. Bachtold et al.~\cite{Bachtold1999} showed that the resistivity of a MWCNTs oscillates when an increasing magnetic field is applied. They measured oscillations with a period of $h/2e$ in agreement with the prediction of the AAS theory. Assuming that the current carriers where localized, in the radial direction, on a single shell, they estimated their average radius from the relation  $\pi r^2 B=\Phi_0 /2 = h/2e$, valid after a complete oscillation. Knowing the external magnetic field they obtained an average radius for the current carriers of $r=8.6\pm0.1\ nm$, in excellent agreement with Atomic Force Microscopy measurements of the most external tube in the MWCNT: $r=8\pm0.8\ nm$. That is, assuming that the carriers were mainly localized on the external tube, the AAS oscillations were confirmed.

The results obtained by A. Bachtold et al.~\cite{Bachtold1999} however where not an experimental evidence of the gap oscillations existence in CNTs under the effect of an increasing magnetic field. Indeed the AAS effect is independent on the electronic gap and AAS oscillations can be measured in mesoscopic metallic rings too, contrary to pure AB gap oscillations which are a peculiar characteristic of CNTs.

\begin{figure}[t]
 \centering
 \subfigure[Experimental setup]{\includegraphics[width=0.55\textwidth]{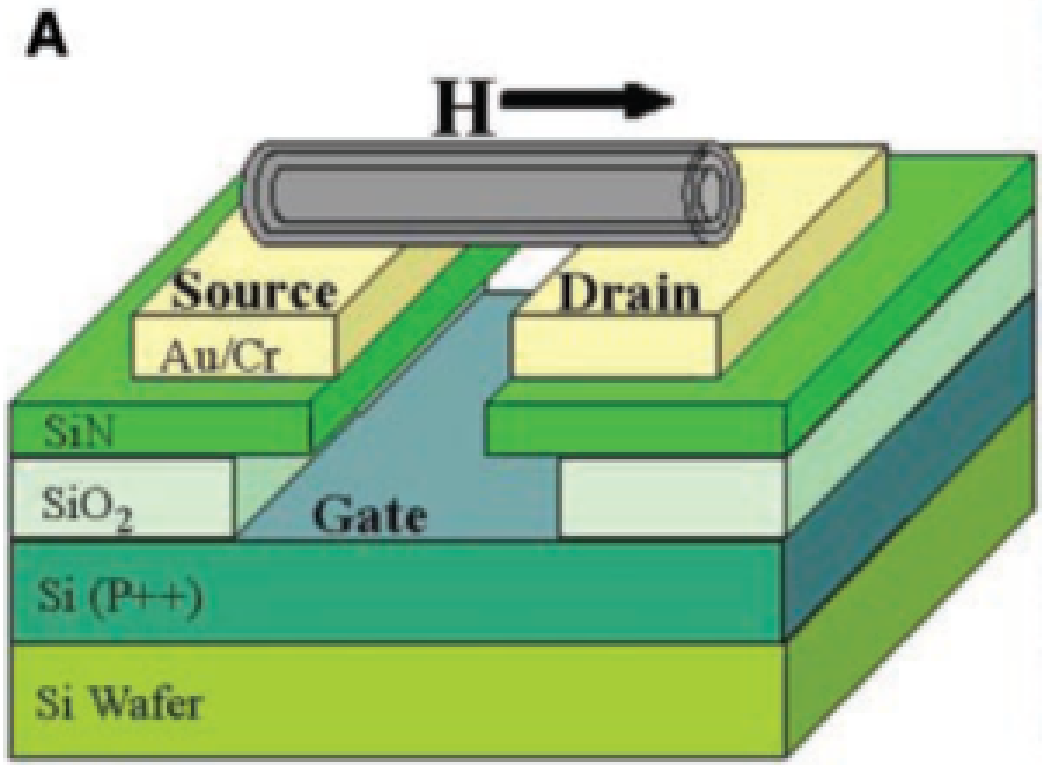}}
 \subfigure[Conductance]{\includegraphics[width=0.35\textwidth]{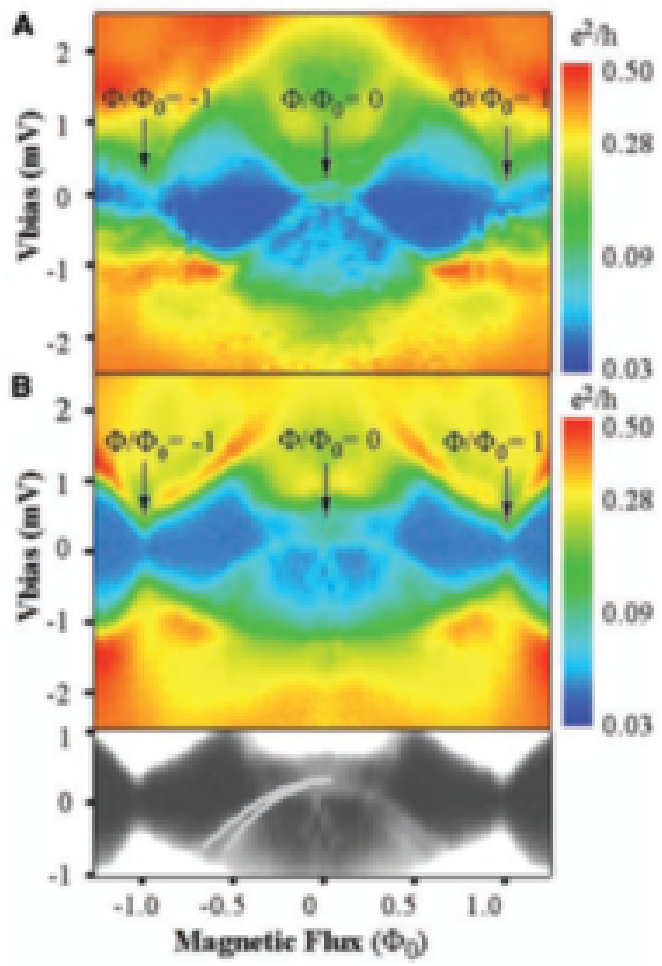}}
 \subfigure[Photoluminescencie spectrum of a SW--CNT]{\includegraphics[width=0.90\textwidth]{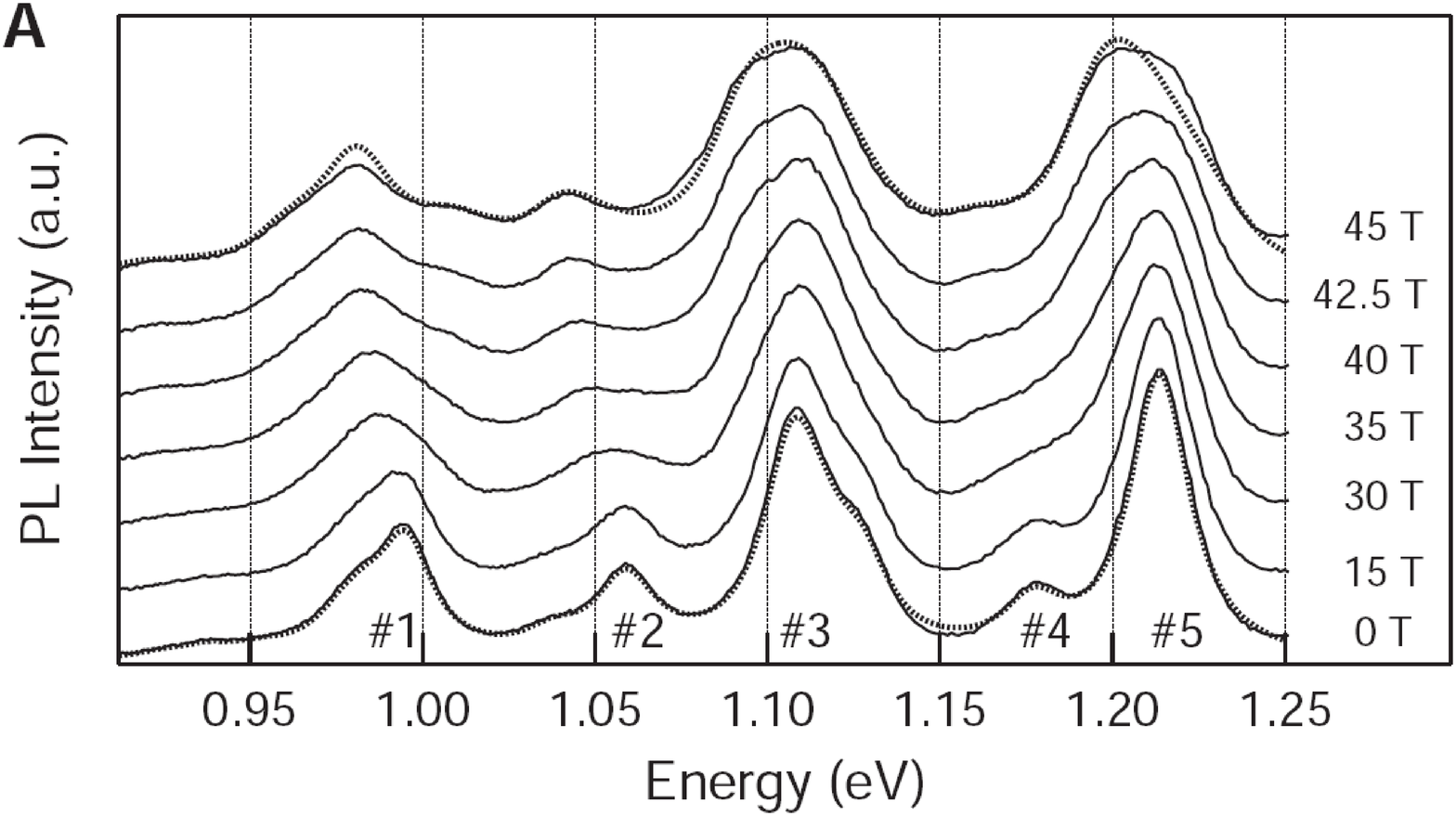}}
 \caption{Top panels: description of the experimental setup and results of the experiment carried out in~\cite{Coskun2004}. In panel $(b)$ the differential conductance oscillates with the imposed magnetic field in a standard circuit with an applied bias potential at two different values of the gate potential. Low panel: photoluminescencie spectrum of a SW--CNT immersed in a static magnetic field from Ref.~\cite{Zaric2004}. The beginning of a gap oscillation is displayed by the behavior of peak $\#1$}
 \label{fig:CNT_exp_gap_AB}
\end{figure}

The first experiments able to test the predictions of the ZFA for the gap oscillations were performed in 2004. Coskum et al.~\cite{Coskun2004} prepared an experimental setup (see Fig. (\ref{fig:CNT_exp_gap_AB}$.a$)) similar to the one of Ref.~\cite{Bachtold1999} but using MWCNTs shorter than the dephasing length to probe a qualitatively different phenomenon, namely, the electronic energy spectrum modulation by a coaxial magnetic field. In particular they measured the differential conductance at finite bias on a single--electron tunneling transistor formed by a MWCNT acting as coherent Coulomb island. In this way they were able to observe the interconversion of semi--conducting and metallic nanotubes. The results are shown in Fig. (\ref{fig:CNT_exp_gap_AB}$.b$), where the differential conductance is plotted in units of $e^2/h$. Oscillations of the conductance with the correct AB period, $\Phi_0=h/e$ are clearly shown in the figure. Moreover they were able to observe the splitting due to the interaction of the external field with the electron spin $\mathbf{B}\cdot\boldsymbol{\sigma}$.

Zaric et al. instead \cite{Zaric2004} measured the photoluminescencie spectrum of a SWCNT immersed in a magnetic field. They showed that the lowest energy excitonic peak (peak $\#1$ in Fig. (\ref{fig:CNT_exp_gap_AB}$.c$)) moves to lower energy when the magnetic field increase. This result is in agreement with the prediction of the ZFA and in particular the gap closing in semi--conducting CNTs which, at low magnetic fluxes, is predicted to follow the rule $E_g=E_0(1-3\Phi/\Phi_0)$.

\chapter{Numerical results}                             \label{chap:numerical results CNT}
\section{Details of the implementation}

In order to describe magnetic field effects ``ab--inito'' we used the Yambo~\cite{Yambo} code.
Yambo starts from a previous SCF computation of the ground state of the system at zero
magnetic field, taking as input the KS (LDA) wave--functions $\psi_i$ and energies $\epsilon_i$.
Then it constructs a new Hamiltonian,
$H=H_{DFT}+H_{magn}$, where $H_{DFT}$ is the DFT Hamiltonian with no external fields, and
$H_{magn}=\mathbf{A}^{ext}\cdot\mathbf{j}$~\footnote{Other considerations 
on the Hamiltonian in presence of a magnetic field can be found in Appendix
\ref{App:DFT and magnetic fields}.}. The total Hamiltonian is 
constructed in the space of the KS wave--functions as
\begin{equation}
H_{ij}=\delta_{i,j}\epsilon_i +V^H_{ij}[\delta\rho]+\(V^{xc,new}_{ij}[\rho]-V^{xc,old}_{ij}[\rho_0]\)
                              +H^{magn}_{ij}
\text{,}
\end{equation}
with $\rho_0$ the density at $\mathbf{B=0}$, $\rho$ the self--consistent density
and $\delta\rho=\rho-\rho_0$. At a first step $\rho=\rho_0$, then the Hamiltonian is diagonalized
and a new set of wave--functions is obtained, together with a new $\rho$. A new Hamiltonian is then
constructed from the new density and projected in the space of the new set of wave--functions.
The process is carried on until the convergence is reached. A convergence threshold
for both the density end the eigen--energies is defined within the code.

The implementation is done assuming that a small number of wave--functions in the KS basis--set
are needed to have a good description of the system. Otherwise this implementation would be highly
impractical as the code needs to store all the wave--functions of the basis set in memory and
to compute all integrals numerically. 
Indeed, as long as the perturbation is small,
the density of the system is not expected to change much and so a small basis--set is required.

Among other reasons, our motivation to investigate ab--initio the effects of magnetic fields in
CNTs is to test the correctness of the assumptions embodied in the ZFA.
In the previous chapter, indeed, we have shown that the pure AB effect has to be
measured with an experimental setup where a confined magnetic field is present and electrons move
in a region where the magnetic field is strictly zero. Indeed the prediction of the AB effect 
had a strong impact on the scientific community because it introduced the possibility of an effect which
is non local in the magnetic field.
However the experimental setup in which CNTs are always studied is in sharp contrast with this assumption,
since CNTs are \emph{fully immersed} in a constant magnetic field.

We have designed the code to simulate this two different geometries.
First, we simulate the {\it pure} AB effect, where electrons travel in a space where {\bf B}$=${\bf 0}.
Second, we use a uniform field. We will refer to the first implementation as {\it confined geometry},
because the magnetic field is confined inside the CNT and null outside. This setup will be compared with the
standard {\it experimental setup} where the magnetic field is uniformly distributed (we will refer to this case
as {\it extended geometry}).

\section{Gap oscillations} \label{Sec:gap oscillations}

We consider five CNTs: two metallic, the $(5,5)$ and the $(8,8)$, two semi--conductive, the $(8,0)$ and the $(14,0)$ and one Multi--Walled, the $(5,5)@(10,10)$. CNTs ground state have first been computed with the Abinit~\cite{Abinit} code using a super--cell with dimensions $(10$ \AA$+2 R_{CNT})\times (10$ \AA$+2 R_{CNT})\times h_{CNT} $ and an angle of $120^\circ$ between the vectors defining the super--cell in the $xy$ plane in order to maximaze the distance between CNTs in the periodic array of super--cells. Here $R_{CNT}$ is the radius of the CNT while $h_{CNT}$ is the dimension of the supercell in the periodic direction, $z$. $h_{CNT}=2.46$ for metallic $(n,n)$ CNTs where we used a reciprocal space grid of $20 k$--points in the Brillouin zone, and $h_{CNT}=4.26$ for semi--conducting $(n,0)$ CNTs where we used a grid of $10 k$--points. However finer grids and bigger cells in the $xy$ directions have been tested when needed. Self Consistent calculations have then been performed with the Yambo code using a basis--set of about 40 states per $k$--point, above the last occupied state.

\subsection*{Metallic single wall carbon nanotubes}

\begin{figure}[t]
 \begin{center}
  \epsfig{figure=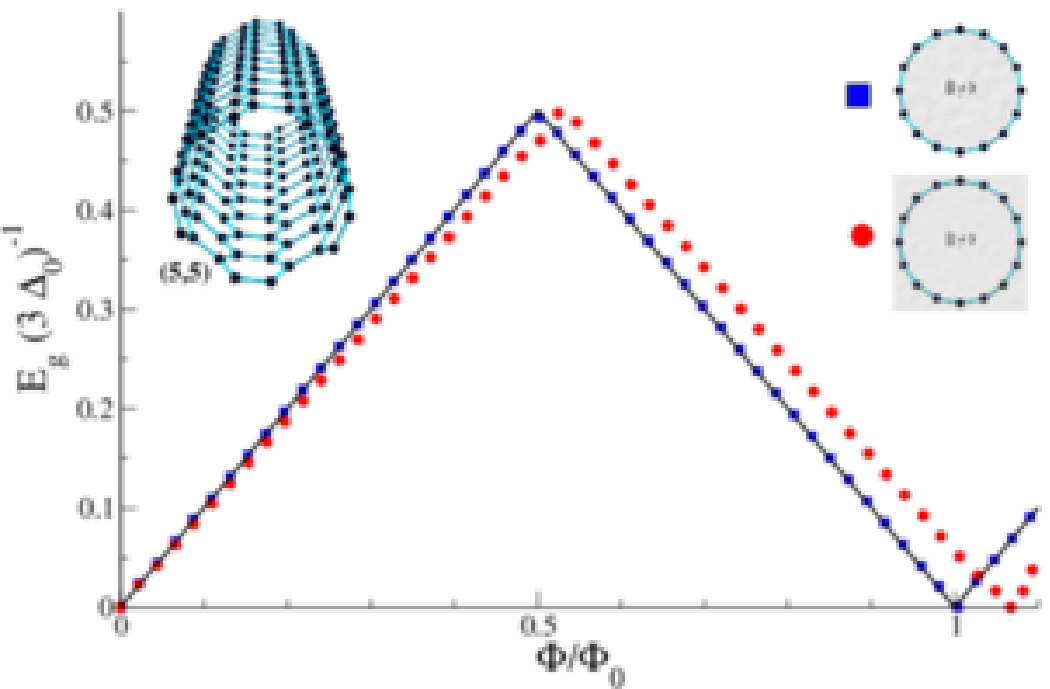,width=0.45\textwidth}
  \epsfig{figure=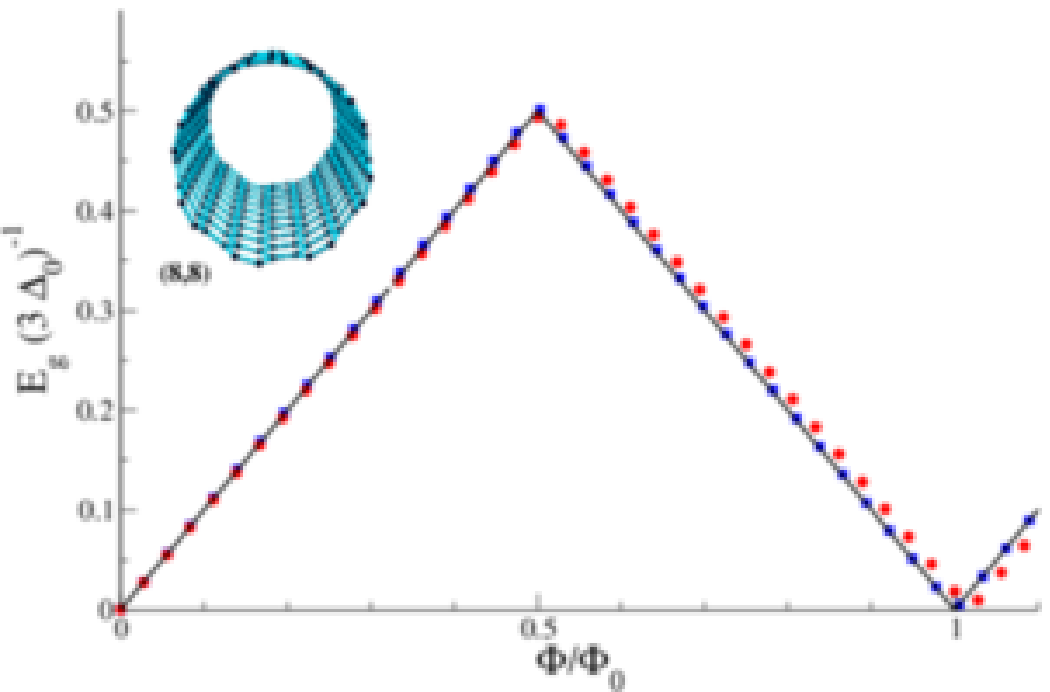,width=0.45\textwidth}
 \caption{\footnotesize{Gap oscillations in the metallic (5,5) and (8,8) CNTs. Two different geometries
are considered. In the extended geometry the magnetic field is applied uniformly in all the space (blue boxes). In the confined geometry, instead, the magnetic field is confined inside the CNT (red spheres). We compare the ab--initio calculations with the ZFA results (black line). We see that in the extended geometry, which represents the standard experimental setup, the Lorentz correction (see text) induces an overestimation of the elementary magnetic flux $\Phi_0$.}}
 \label{fig:CNT_gap_metallic}
 \end{center}
\end{figure}

First we consider two metallic CNTs: a (5,5) tube with radius 3.39 \AA \ and a (8,8) tube with radius 5.41 \AA. In Fig.~(\ref{fig:CNT_gap_metallic}) we compare the gap dependence on the applied magnetic flux in the two  geometries with the result of the ZFA. In the case of the smaller (5,5) tube we immediately see a first important difference between the extended geometry and the confined geometry. The extended geometry, which represents the standard experimental setup, overestimates by $\sim$7\% the elementary flux $\Phi_0$ which defines the periodicity of the gap oscillations.

To explain the different gap dependence obtained in the two geometries we introduce, in
a formal manner, the Hamiltonian which governs the AB effect in the specific case of a CNT:
\begin{equation}
\label{eq:H}
H=-\frac{\hbar^2}{2m}\[ \frac{\partial^2}{\partial r^2}+\frac{1}{r}\frac{\partial}{\partial r}+
                            \frac{1}{r^2}\(i\frac{\partial}{\partial\varphi}-\frac{er}{\hbar}A\(r\)\)^2+
                            \frac{\partial^2}{\partial z^2}
                     \] + V(r,\varphi,z)
\text{.}
\end{equation}
Eq.~(\ref{eq:H}) describes the electronic dynamics under the action of 
a static magnetic field, written in cylindrical coordinates centred in the center of the CNT.
$A\(r\)$ is the vector potential which, in the symmetric gauge, describes a static magnetic
field along the $z$ direction and $V(r,\varphi,z)$ is the local DFT potential, which includes the 
ionic potential plus the Hartree and exchange--correlation terms.

\begin{figure}[t]
 \begin{center}
  \epsfig{figure=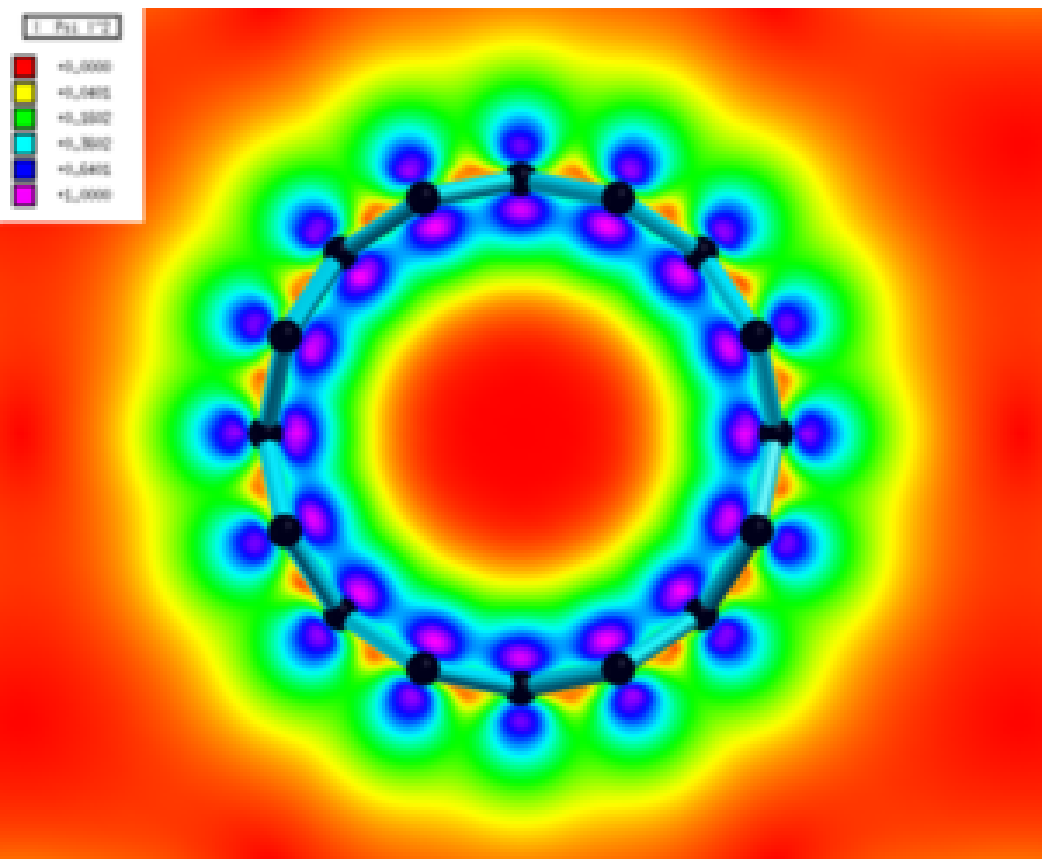,width=0.45\textwidth}
  \epsfig{figure=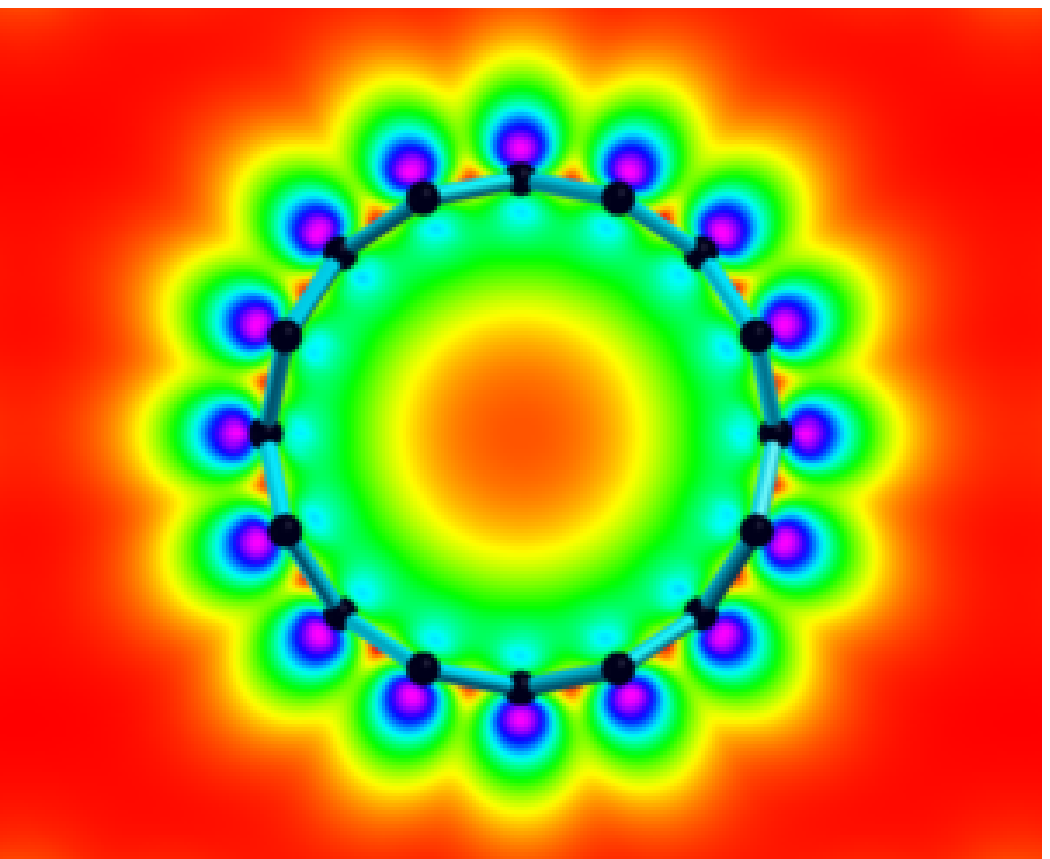,width=0.45\textwidth}
  \caption{\footnotesize{Two dimensional plot of the last occupied and the first unoccupied band
          wave--function at the $\Gamma$ point for the $(8,0)$ CNT. The $\pi$ orbitals,
          deformed by the curvature of the CNT, have a larger amplitude in the outer part of the CNT surface.
          As a consequence we have a finite difference between the Hamiltonians corresponding to the extended
          geometry compared to the confined geometry.
          }}
 \label{fig:wf_plot}
 \end{center}
\end{figure}

The only term of Eq.~(\ref{eq:H}) which reflects the different geometry ({\it extended} or {\it confined}) is
$A\(r\)$. In the extended geometry
\begin{equation}\label{ext_pot}
A^{Extended}\(r\)=\frac{1}{2}B_0 r,
\end{equation}
with $B_0=\left|\mathbf{B}\right|$. In the confined geometry, instead, we
have~\footnote{The potential in Eq.~(\ref{conf_pot}) is not defined at $r=0$. However this problem can be easily overcome in the implementation, setting the magnetic potential $A=1/2 (\Phi/r_0^2)r$ for $r<r_0$ with $r_0$ a tunable parameter. If $r_0$ is small enough then the vector potential is different from the AB potential only in a small region where the wave--function is almost zero and so the results are independent of $r_0$. In our calculation we checked $r_0=0.5 Bohr$ is enough.}
\begin{equation}\label{conf_pot}
A^{Confined}\(r\)=\frac{1}{2}\frac{\Phi}{\pi r},
\end{equation}
with $\Phi=\pi B_0 R_{CNT}^2$ and $R_{CNT}$ the CNT radius. From Eqs.~(\ref{ext_pot}) and (\ref{conf_pot}) we see that
$A^{Extended}\(R_{CNT}\)=A^{Confined}\(R_{CNT}\)$, which implies that, if the electrons would {\bf exactly} move on the
tube surface the extended geometry and the confined geometry would lead to the same gap oscillations. 
The different gap dependence observed in Fig.~(\ref{fig:CNT_gap_metallic}), is then due to the different correction
induced in the total Hamiltonian by $A^{Extended}$ and $A^{Confined}$.
If we plug the two different expressions for $A\(r\)$ into Eq.~(\ref{eq:H}) we get two different
Hamiltonians, $H^{confined}$ and $H^{extended}$, whose difference is
\begin{equation}
\label{eq:H_corr}
\Delta H=H^{extended}-H^{confined}=-\frac{e\hbar B_0}{4 m}
          \left( 2i\frac{\partial}{\partial\varphi}- \frac{\Phi}{\Phi_0}\right)
          \(1- \frac{R_{CNT}^2}{r^2}\)
\text{,}
\end{equation}
with $\Phi_0=h/e$. This term is zero when $r=R_{CNT}$, while near the 
tube surface behaves like $\sim B_0\(1-\frac{R_{CNT}^2}{r^2}\)$. 
Now, as shown in Fig.~(\ref{fig:wf_plot}), the electrons are localized near, but not exactly {\it on} the tube surface.
Consequently $1-\langle R^2_{CNT}/r^2 \rangle\neq 0$ and, $\langle H^{extended}-H^{confined} \rangle \neq 0$.

We will refer to the correction defined by Eq. (\ref{eq:H_corr}) as Lorentz Correction (LC) as it 
introduces a magnetic term which depends on the electronic trajectory (through the term
$rA^{extended}_\phi$).
The LC appears in the extended geometry as an effective different radius 
of the electronic orbitals, as the correction would be zero defining the flux with respect
to the effective radius which satisfy the equation $1-\langle R^2_{eff}/r^2 \rangle= 0$.
In the confined geometry this correction is zero by definition; physically this is related to the fact that
the AB effect does not depend on the specific electronic trajectory.

The LC goes to zero in the limit of SW--CNTs with infinitely large 
radius. Nevertheless, even in large CNTs, impurities or defects can alter the electronic
trajectory creating deviations from a perfect circle of radius $R_{CNT}$. In all this cases we
predict the LC to induce deviations from the AB oscillations when uniform $\mathbf{B}$ field is
applied. 

From Fig.~(\ref{fig:CNT_gap_metallic}) we see that the ZFA matches the ab--initio simulation of the pure AB effect
corresponding to the {\it confined geometry} setup. This agreement is due to the fact that in the ZFA
the LC is strictly zero as the electrons are assumed to move exactly on the graphite sheet, i.e. on 
the CNT surface. Consequently, in the ZFA  the electronic gap is function of the flux only.



\subsection*{Semi--conducting single wall carbon nanotubes}

\begin{figure}[t]
 \begin{center}
 \subfigure{\includegraphics[width=0.45\textwidth]{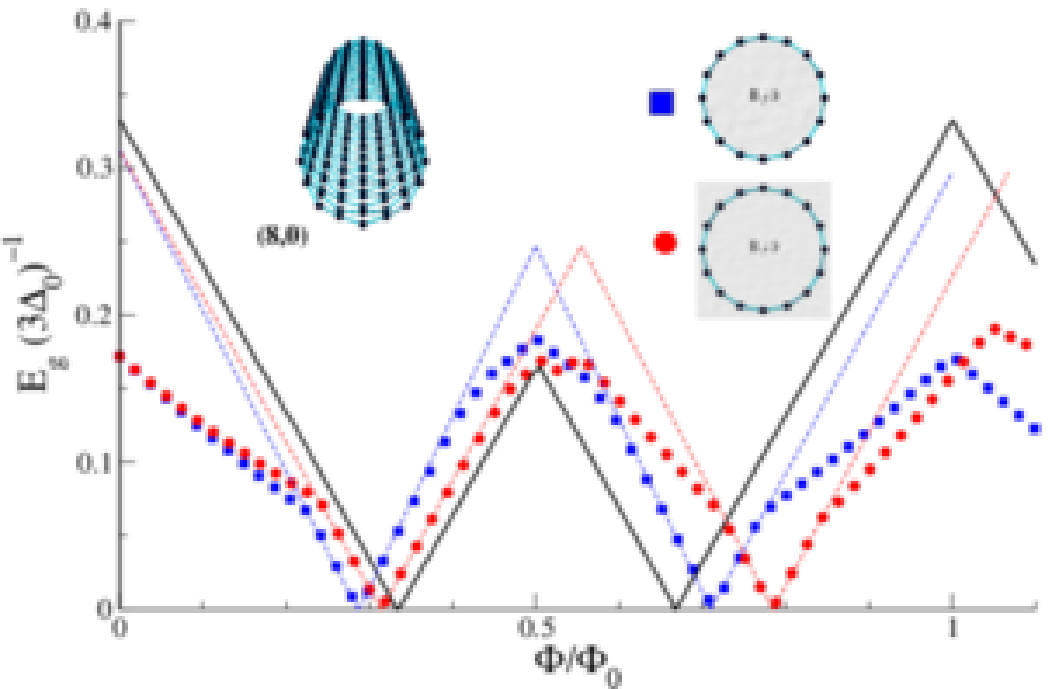} }
 \subfigure{\includegraphics[width=0.45\textwidth]{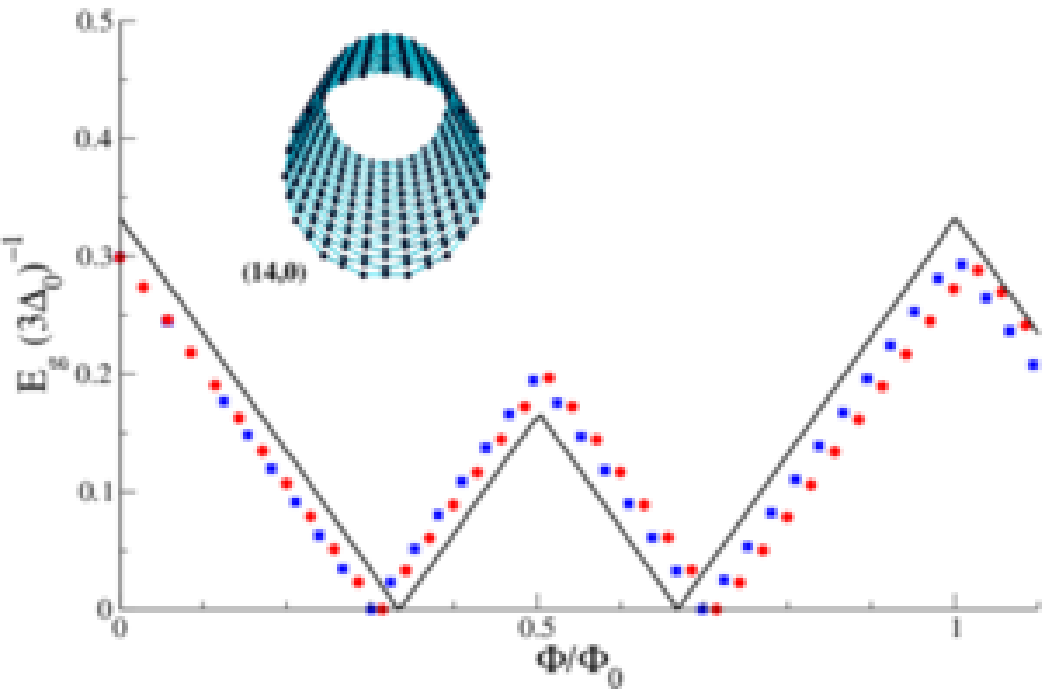} }
 \caption{Gap oscillations in a semiconducting CNTs: (8,0) and (14,0). In contrast to the
metallic case the curvature effects induce more evident differences betweem {\it confined} and
the {\it extended} geometries.  The same conventions of Fig.~\ref{fig:CNT_gap_metallic} are used here.}
 \label{fig:CNT_gap_semiconducting}
 \end{center}
\end{figure}

We now consider two semi-conducting CNTs: the (8,0) and the (14,0). 
The flux dependent electronic gap is shown in Fig.~(\ref{fig:CNT_gap_semiconducting}).
Similarly to the metallic case, LC makes the extended geometry to oscillate with a period
greater than $h/e$.
In contrast to the metallic case, the gap vanishes at two values of $\Phi$, which the
ZFA predicts to be at $\Phi_0/3$ and $2\Phi_0/3$, when the Dirac points become allowed
$k$ points~\cite{Charlier2007}. Noticeably both points are renormalized 
in the ab--initio simulation by curvature effects. 
It is well known, indeed, that, compared to graphene, curvature effects shift the Dirac points~\cite{Charlier2007}
$K$ (see also Fig.~(\ref{fig:bands_55})) at a position $|K|<2 \pi / 3a$, with $2 \pi / 3a$ being the Dirac point position in graphene.
Accordingly a lower magnetic field is needed
to bring the Dirac point in coincidence with the set of the allowed $k$--points, and
semi--conducting CNTs become metallic at $\Phi<\Phi_0 / 3$.
Being the oscillations symmetric, the second metalization
point is reached at $\Phi>2\Phi_0 /3$.

The deformation of the oscillations in the $(n,0)$ CNTs is smaller in 
bigger tubes. However it goes to zero slowly, because both the shift and the magnetic period depend on the size
of the tube. For this reason the effect is still not negligible in the large $(14,0)$ tube, as shown in
Fig.~(\ref{fig:CNT_gap_semiconducting}).

From Fig.~(\ref{fig:CNT_gap_semiconducting}) we see that the $(8,0)$ gap
oscillations strongly deviate from the ZFA that does not reproduce, even qualitatively, the 
full ab--initio results. The reason for this large discrepancy traces back to the presence of a 
metallic--like band located near the Fermi surface.

\begin{figure}[t]
 \centering
 \includegraphics[width=0.9\textwidth]{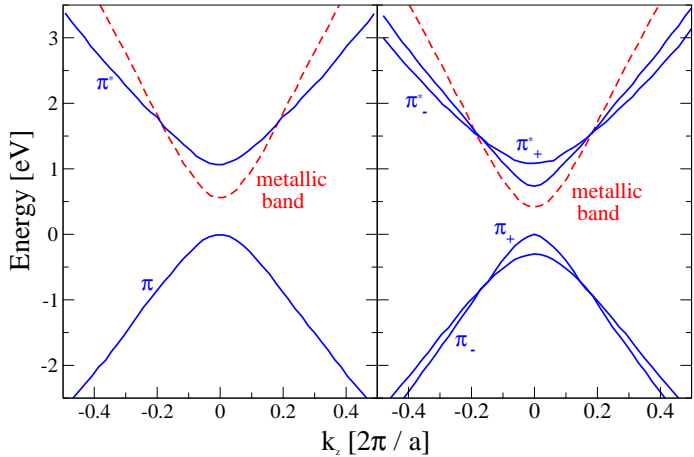}
 \caption{Metallic--like (in red) and $\pi$ / $\pi^*$ (in blue) bands of the $(8,0)$ CNT.
          In presence of a magnetic field electrons spinning clockwise around the CNT have a different
          energy with respect to electrons spinning counter--clockwise.
          On the left panel $\Phi=0$, while on the right $\Phi>0$. We notice that the upper metallic--like
          band lays in the gap without acquiring a splitting due to the breaking of the time reversal symmetry,
          i.e. the energy of electrons spinning around the tube in opposite direction is different when a 
          magnetic field is present.}
 \label{fig:CNT_80_metallic_band}
\end{figure}

This band is shown in Fig.~(\ref{fig:CNT_80_metallic_band}) together with the $\pi / \pi^*$ bands closest
to the Fermi level. When the $\mathbf{B}$ field is increased we see that, in contrast to the
$\pi / \pi^*$ bands, the metallic--like band does not shift, but moves inside the gap produced by
the $\pi / \pi^*$ states. Consequently by changing the flux intensity the gap is defined by
transitions between the $\pi / \pi^*$ states or between the $\pi$ and the metallic--like band.
This explains the anomalous dependence of $E_g$ by $\Phi$ in Fig.~(\ref{fig:CNT_gap_semiconducting}).

\subsection*{Multi wall carbon nanotubes}

\begin{figure}[t]
 \begin{center}
 \includegraphics[width=0.9\textwidth]{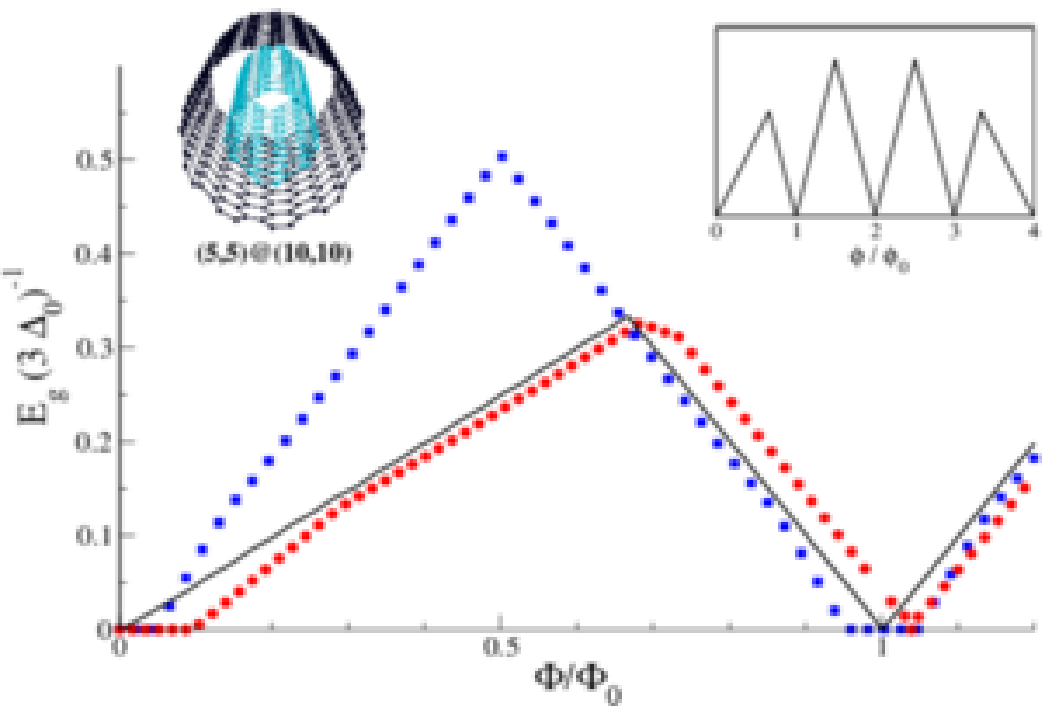} 
 \caption{Gap oscillations in a Multi--Walled CNT: (5,5)@(10,10). In contrast to the SW--CNTs here the confined and the extended geometry display a qualitative different behavior. In the inset a complete period of oscillation according to the ZFA which describes the MW--CNT as two not interacting SW--CNTs.  The same convention of Fig.~\ref{fig:CNT_gap_metallic} is used here.}
 \label{fig:MWCNT_gap}
 \end{center}
\end{figure}

Although SW--CNTs are routinely syntetized, MW--CNTs still constitute the majority of cases
used in the experiments. In Fig.~(\ref{fig:MWCNT_gap}) we consider the case of a $(5,5)@(10,10)$ CNT
with radii 3.39 and 6.78 \AA. In this case the confined geometry is implemented considering 
a flux $\Phi=\pi B_0 R_{(10,10)}^2$. This flux is roughly the same experienced, in the extended geometry,
by the electrons of the $(10,10)$ CNT. In the case of MWCNT the ZFA  is used only by considering
two different Hamiltonians, one for the $(5,5)$ and one for the $(10,10)$ and by neglecting the
tube--tube interaction. Our ab--initio results, instead, reveal a quite different picture.

The gap calculated in the extended geometry follows the ZFA prediction except in the very low field regime and near 
the first inversion point $(\Phi\approx \Phi_0)$. While the LC causes the shift of the inversion point,
the metallic regime observed for $\Phi<\Phi_0/10$ is not described at all by the ZFA. It is, indeed,
a consequence of the different chemical potential felt by the electrons moving on the $(5,5)$ and the
$(10,10)$ surfaces. Here, moreover the LC is enhanced because the presence of the inner CNT tends
to attract electrons and accordingly to modify the effective radius defined by the LC. Indeed the flux
renormalization is $\approx 4\%$ here, where $R_{(10,10)}=6.78$ \AA, while in the SW configuration the
renormalization is $\approx 7\%$ and $\approx 1\%$ for the $(5,5)$ and $(8,8)$, where $R_{(5,5)}=3.39$
and $R_{(8,8)}=5.42$ \AA.

\begin{figure}[t]
 \begin{center}
 \includegraphics[width=0.85\textwidth]{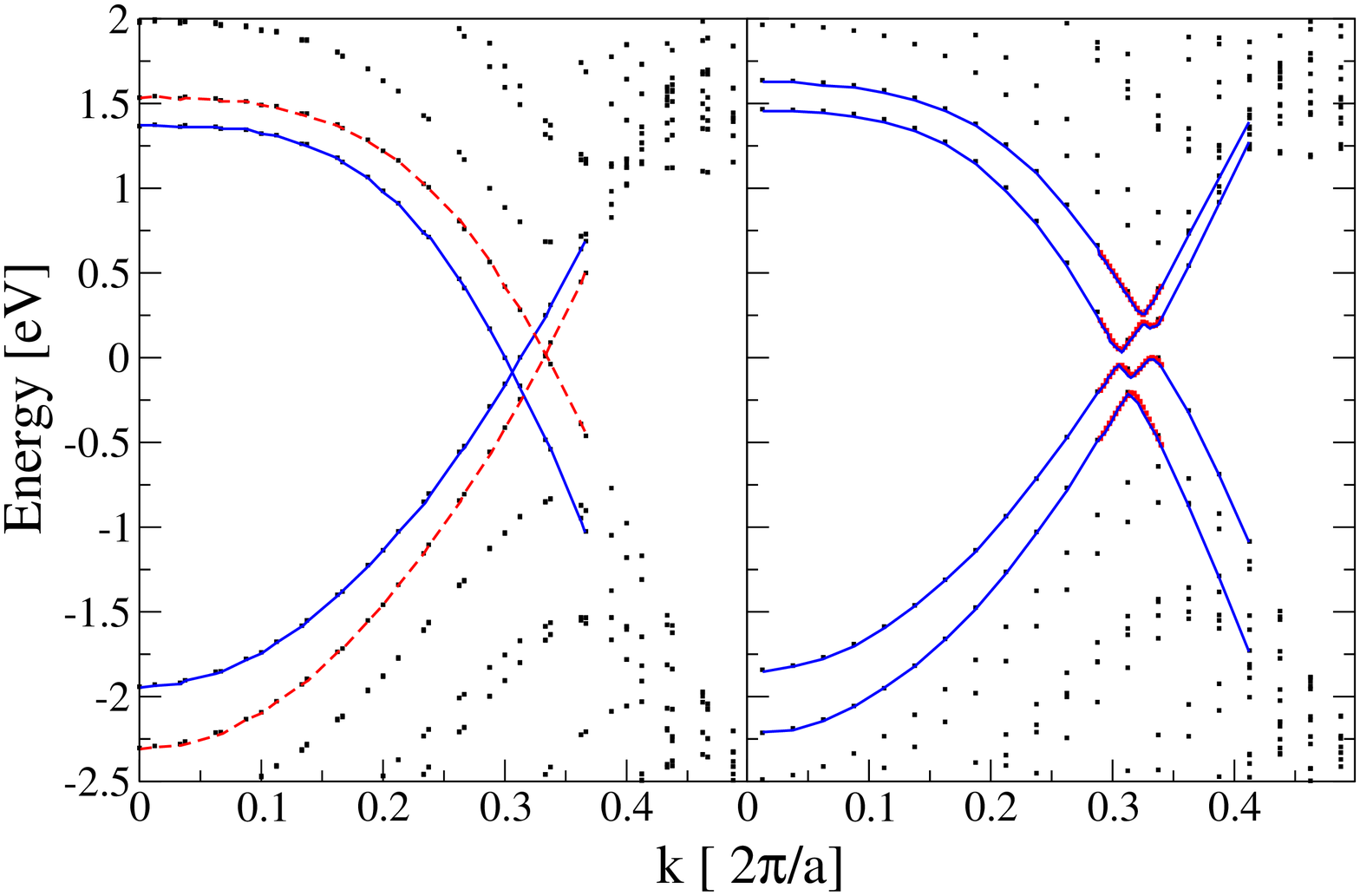}
 \end{center}
 \caption{Band structure of the metallic $(5,5)@(10,10)$ CNT at zero magnetic flux (left panel) and at low magnetic flux, extended geometry (right panel). In the left panel in blue the bands, near the Fermi level, occupied from electrons on the inner shell, while in red the ones occupied from electrons on the outer shell. When the magnetic field is present it is not possible anymore to distinguish to which shell belongs each band. Due to the shift in energy between the point where the $\pi$--$\pi^*$ bands of the two CNTs cross, at low flux, though two small direct gaps open, the system remains metallic until an indirect gap opens (right panel). Only at higher fluxes the system becomes again direct gap semi--conductor.}
 \label{fig:MWCNT_bands}
\end{figure}

The ab--initio calculations show that there is a $\approx 0.09 eV$ shift between the two
chemical potentials, as shown in Fig.~(\ref{fig:MWCNT_bands}).
Here the band structure of the double walled CNT is plotted at $\Phi=0$,
on the left, and at small flux on the right. At zero magnetic flux there are two pairs of crossing bands
which can be identified as the $\pi-\pi^*$ bands of the $(5,5)$ and the $(10,10)$ CNTs respectively. The two 
crossing points however are not aligned in energy so that when a small magnetic flux is present two
small direct gap opens but the CNT remains metallic as long as the tip of the $\pi^*$ band of the 
$(5,5)$ is lower in energy than the one of the $\pi$ band of the $(10,10)$ CNT.
Even when a gap opens the system remains an indirect gap semiconductor for certain range of values
of the applied magnetic flux. 

In the confined geometry the gap dependence is deeply modified. In the low $\Phi$ regime a metallic region appear
for the same reason outlined above. At difference with the extended geometry case, however, the gap increases with
increasing $\Phi$ with a much larger slope, dominated by the $(10,10)$ CNT. The slower slope of the 
extended geometry, instead,  is dictated by the gap of the $(5,5)$ CNT.

Another drastic difference with the extended geometry case is a second metallic phase near
$\Phi\approx \Phi_0$. This phase is due to the fact that, in the confined geometry, the electrons on the two
CNTs feel the same magnetic flux. As a consequence the gaps of the 
two CNTs move coherently as $\Phi$ approaches $\Phi_0$ and the same situation of
$\Phi \approx 0$ occurs. The coherent variation of the gaps of the $(5,5)$ and the
$(10,10)$ tube result, for the gap oscillations, in a period which is $1/4$ the
period of the extended geometry.

\subsection*{Numerical instabilities: a ``gauge fixing'' solution}

The results from the previous section have been obtained with an ``improved
implementation'' which takes advantage of the
gauge freedom (see Appendix \ref{App:Gauge transformations}) to describe a magnetic
field. 

We observe that the AB effect can be described (as we did in the confined geometry)
by a vector potential
\begin{equation}\label{A_pure_AB}
A_\psi=\frac{\hbar}{er} \frac{\Phi}{\Phi_0}
\end{equation}
which is a pure gauge field when $\Phi=n \Phi_0$ with $n$ any integer. A potential is
pure gauge if a function $\Lambda(r,\psi,z)$ exist, which plugged in
Eqs. (\ref{gauge_transformations}) nullifies it
everywhere\footnote{The scalar potential $V$ must be zero. This condition is
automatically satisfied in the static case, as a time independent gauge generating function
can be used.}.
In this situation we need a function which satisfies the relation
$1/r\ \partial \Lambda / \partial\phi=-A_\phi(r)$ with $A_{\phi}$ defined
by Eq. (\ref{A_pure_AB}).

Indeed such function can be constructed for any value of the magnetic flux $\Phi$:
\begin{equation}\label{gauge_generator}
\Lambda(r,\phi,z)=-\frac{\hbar}{e} \frac{\Phi}{\Phi_0} \phi
\text{.}
\end{equation}
However in quantum mechanics a gauge transformation in the electro--magnetic
potentials has always to be realized together 
with the corresponding transformation of the wave--function.
The new wave--function obtained using Eqs. (\ref{gauge_generator}) is well defined only if 
$\Phi=n \Phi_0$, while for any other value of $\Phi$ it is a multi--valued wave--function.
For this reason any effect induced by the AB effect is periodic
with period $\Phi_0$.

When a vector potential defined by Eq. (\ref{A_pure_AB}) with $\Phi=n \Phi_0$ is applied, 
all the physical quantities of the system have to remain unchanged,
while all wave--functions acquire a phase factor. However the KS basis--set is not a convenient
choice to describe this change in the phase of the wave--functions.

This can be understood if we try to construct the wave--functions at $\Phi=\Phi_0$ as a linear
combination of the KS wave--functions at $\Phi=0$.
The wave function of any single particle state must be expressed as
\begin{equation}
\psi_{n,k_z,l_z}^{\Phi=\Phi_0}=\psi_{n,k_z,l_z}^{\Phi=0}e^{i\phi}=
     \sum_{n,l_z} c_{n,k_z} \psi_{n,k_z,l_z}^{\Phi=0}
\text{.}
\end{equation}
If few elements of the basis--set are used to build $\psi_{n,k_z,l_z}^{\Phi=0}e^{i\phi}$
using a basis set which contains $\psi_{n,k_z,l_z}^{\Phi=0}$ and few other orthogonal wave--functions
orthogonal, this leads to severe numerical instabilities that we first encountered when we tryed to 
compute the CNTs gap oscillations.

\begin{figure}[t]
 \centering
 \includegraphics[width=0.9\textwidth]{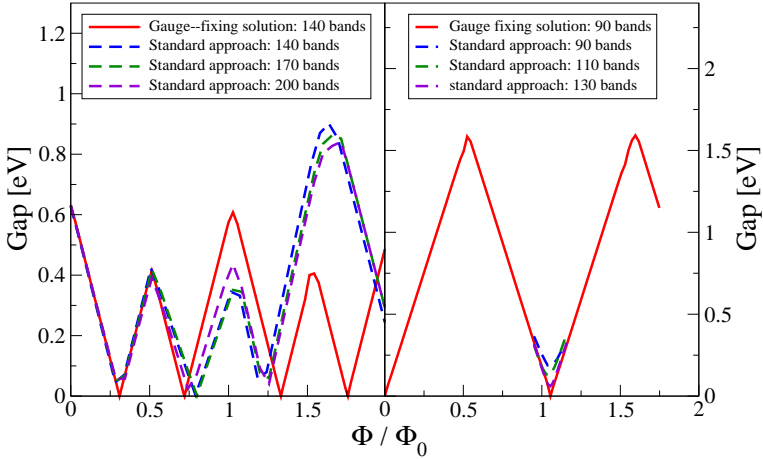}
 \caption{The gap oscillations of the (14,0) CNT (the left panel) and of the (5,5) CNT (right panel). The results of the gauge--fixing implementation are compared with the results obtained with the standard implementation. Increasing the number of KS wave--function in the basis--set the two methods converge to the same result.}
 \label{fig:check_phase_fixing}
\end{figure}

The metallic CNTs were not returning metallic
when $\Phi=\Phi_0$. Moreover for the semiconducting CNTs we found that using a small
basis--set the gap oscillations, as well as other physical quantities, were not periodic
in the flux. 

To solve this problem we observed that at $\Phi=n\Phi_0$ it is possible to impose a
phase--matching solution. That is, instead of projecting the Hamiltonian, at $\Phi=n\Phi_0$,
on the starting KS basis--set we can
project it in a basis--set modified adding the needed phase factor.
In this new basis--set the Hamiltonian will be diagonal. Then
for intermediate values of the magnetic flux we can select the most appropriate basis--set
according to which
integer values of the flux we are closer. This numerical procedure avoids the self--consistent
cycle to remain trapped in the initial gauge. 

In Fig.~(\ref{fig:check_phase_fixing}) the gap oscillations of the $(14,0)$ CNT with and without the 
phase fixing implementation are shown. Without fixing the gauge the gap oscillations seem to converge towards
the gauge fixing solution increasing the number of bands. However the convergence is very slow as very few
wave--function from the basis--set can be used for computational limits. Indeed if in the plane--wave
basis--set thousands of states are needed to reproduce the KS wave--functions, we expect that thousands
of states would be needed to correctly reproduce a phase factor.

The same is true for metallic CNTs. For these the convergence problem in the standard implementation appears
for field values corresponding to the metallic phase, as shown in Fig. (\ref{fig:check_phase_fixing}).

\section{The band structure}\label{Sec:Band structure}
The electronic gap of a CNT depends only on the behaviour of the last occupied and the first unoccupied band.
In particular the gap depend of their behaviour near a specific $k$--point, that is near the Dirac
point (see Sec.~\ref{Sec:Intro_CNTs}).

We have already seen how some deviation of the gap oscillations from the ZF predictions, can
be understood in terms of the bands structure, as in the case of the (8,0) CNT or of the multi--walled
$(5,5)@(10,10)$ CNT.

\begin{figure}[t]
 \centering
 \includegraphics[width=0.9\textwidth]{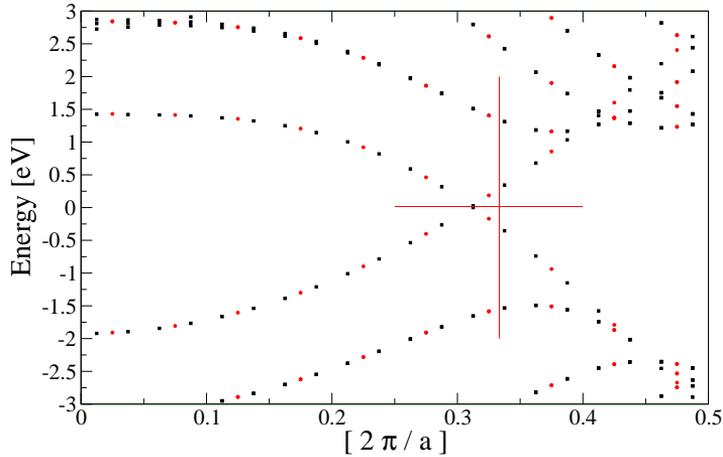}
 \caption{Band structure of the (5,5) CNT at zero magnetic field near the Fermi level. The Dirac point,
          according to the prediction of the ZF model, should be placed where the two red
          lines cross. Due to curvature effects, not included in the ZF model, the
          Dirac point, i.e. the point where the $\pi$ and the $\pi^*$ bands cross, is shifted
          in the DFT approach.}
 \label{fig:bands_55}
\end{figure}

In this section we will explore more in details the effects of the magnetic field on the bands structure CNTs.
As a reference we will use the prediction of the ZF model. In Fig.~(\ref{fig:bands_55})
for example we see that the band structure of the (5,5) CNT computed with Yambo is similar to the prediction
of the ZF model (see Fig.~(\ref{subfig:ZF_55_bands}) but with the shift of the Dirac K point. This is
a result known from the literature.

\begin{figure}[t]
 \centering
 \includegraphics[width=0.9\textwidth]{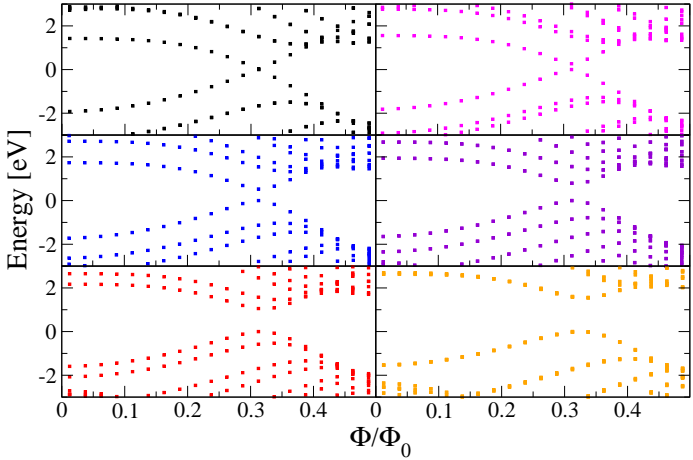}
 \caption{Band structure of the (5,5) CNT near the fermi level at increasing values
         (black--pink--blue--violet--red--orange dots) of the magnetic field. Here
          the magnetic flux increase from $\Phi=0$ (black dots) to $\Phi=\Phi_0/2$ (orange dots).}
 \label{fig:bands_55_landau}
\end{figure}

When a magnetic field is present, then the band structure follow, in first approximation, the predictions
of the ZF model (see Fig.~(\ref{fig:bands_55_landau}) ) and the deviation are the one we have already 
explored in the previous section describing the gap oscillations of different CNTs.

In this section we will show (i) the role of the
non--local part of the pseudo--potential for a correct description of the more bounds valence electrons and (ii)
the effect of a not perfect alignment of the magnetic field with the CNT on the band structure.

\subsection*{The non local part of the pseudo--potential}

When the term $H_{magn}=\mathbf{A \cdot j}$ is included in the Hamiltonian the non--local part of the pseudo
potential must be changed accordingly (see App.~\ref{App:DFT and magnetic fields}). In the present work we
followed the strategy of Ref.~\cite{Mauri_nl_psp} where the non local pseudo--potential reads:
\begin{equation}\label{Eq:mauri_nl_psp}
V_{NL}^{\mathbf{A}}(\mathbf{r,r'})=\sum_{\mathbf{R}}V_{NL}^{\mathbf{A=0}}(\mathbf{r,r'})
                                   e^{i/c\int_{\mathbf{r}\rightarrow\mathbf{R}\rightarrow\mathbf{r'}}\mathbf{A}\cdot d\mathbf{l} }
\text{.}
\end{equation}

In order to explore the effect of the expression appearing in Eq.~\ref{Eq:mauri_nl_psp} we considered the
behaviour of the KS eigenvalue at the Dirac point, 
not only for the last occupied and the first unoccupied band, but for all the eigenvalues. As for the electronic gap,
all eigenvalues must be periodic with period $\Phi=\Phi_0$ and the oscillation symmetric with respect to
$\Phi=\Phi_0/2$.

\begin{figure}[t]
 \centering
 \includegraphics[width=0.9\textwidth]{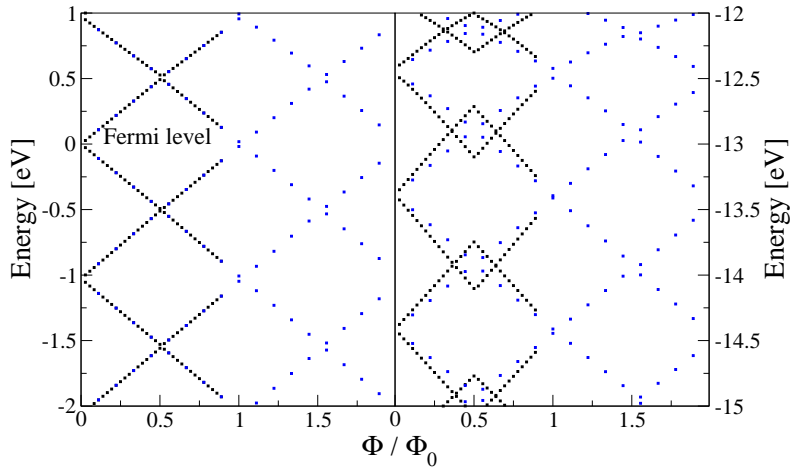}
 \caption{KS eigenvalues at the Dirac point for the (8,8) CNT as a function of the magnetic flux.
          The results obtained using the standard pseudo--potential $V_{NL}^{\mathbf{A=0}}(\mathbf{r,r'})$
          (black squares) and the corrected pseudo--potential $V_{NL}^{\mathbf{A}}(\mathbf{r,r'})$
          (blue squares) are compared. Near the Fermi level (left panel) there are not appreciable differences,
          while for the more bound states (right panel) there is a difference of about $20 \%$.
          }
 \label{fig:KS_88_landau}
\end{figure}

In Fig.~(\ref{fig:KS_88_landau}) we can see the different behaviour of the KS eigenvalues induced by
the correction to the non local part of the pseudo--potential, Eq.~(\ref{Eq:mauri_nl_psp}).
In particular for the deeper valence states the change of the gauge at $\Phi=\Phi_0$
(see Sec.~\ref{Sec:gap oscillations}, subsection on the gauge--fixing solution)
induces a discontinuity of the derivative when $V_{NL}^{\mathbf{A=0}}(\mathbf{r,r'})$ only is
used. This is an indication that the KS eigenvalues are wrong in this case. The corrected pseudo--poential
fixes the problem.

\subsection*{The effect of a magnetic field not aligned to the tube axis}

Yambo~\cite{Yambo} is a plane--waves based code, devised to treat periodic systems.
The applied uniform magnetic field
is described with a vector potential of the form $A_\phi=B_0 r$ (with $r$ the radius in cylindrical coordinates)
in the symmetric gauge. However such a vector potential is not periodic in the $xy$ plane
and numerically it is replaced by a saw--like dependence,
which induces jumps at the super--cell borders in order to keep the potential periodic.
CNTs are isolated systems in the $xy$ plane and the electronic wave--function is almost zero
on the borders, consequently the $A_\phi$ jumps do not affect the results.

\begin{figure}[t]
 \centering
 \includegraphics[width=0.9\textwidth]{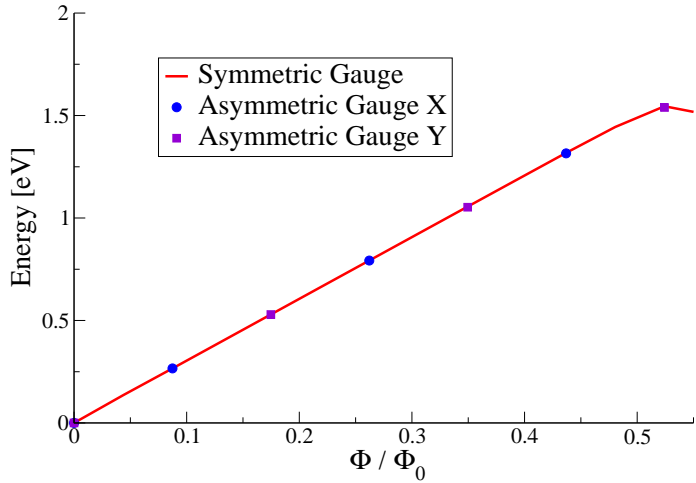}
 \caption{Convergence checks. Three different gauges describing the same magnetic field
          with no appreciable differences.
          The test are performed on the $(5,5)$ CNT at low magnetic field in order to
          avoid wave--function phases problems.}
 \label{fig:check_asymmetric_gauges}
\end{figure}

However, if the CNT is not perfectly aligned with the magnetic field, i.e. in the $z$ direction
a saw--like vector potential cannot be used.
To overcome this problem it is possible to use an asymmetric
gauge. Suppose for example that the CNT axis lays on the $yz$ plane, then the vector potential
\begin{equation}
A_y = B_0 x
\end{equation}
in the $y$-asymmetric gauge can be used. The generating function needed to switch from the symmetric to
the $y$--asymmetric (/ $x$--asymmetric)
gauge is $\Lambda=\pm 1/2\ B_0 xy $ . We have used this possibility in order to verify the effects of 
a possibly not perfect alignment between the magnetic field and the CNT in the experimental setup. First we checked
that the different gauges gave the same results on a CNT in the $z$ direction, Fig. (\ref{fig:check_asymmetric_gauges})).
As expected there are not appreciable differences between the three gauges.

\begin{figure}[t]
 \centering
 \includegraphics[width=0.9\textwidth]{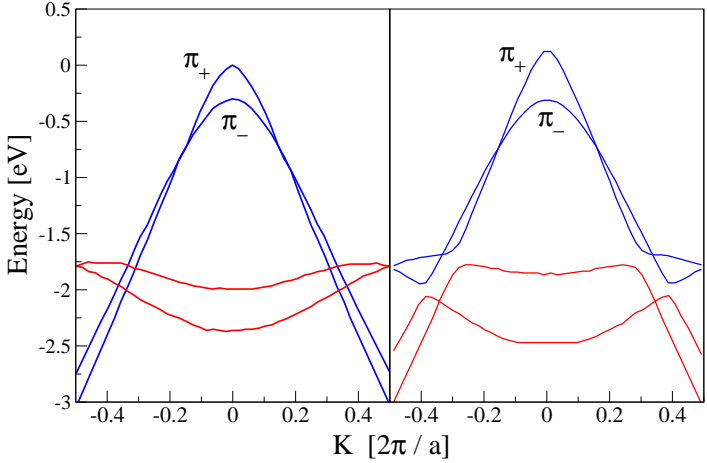}
 \caption{Low magnetic flux. A CNT perfectly aligned with the magnetic field in the left panel and
          the same CNT oriented with an angle of $15^\circ$ at higher magnetic field on the right.
          The component of the magnetic field perpendicular to the CNT breaks the accidental
          degeneracies in the band structure. Here as example we show the effect on the last occupied bands.}
 \label{fig:not_aligned_field}
\end{figure}

Then we considered a setup with a small misalignment between the magnetic field and the CNT ($\theta=15^\circ$),
Fig. (\ref{fig:not_aligned_field}). In this configuration the small component of the magnetic field perpendicular
to the axis of the tube has, in general, a negligible effect on the properties of the CNT. 
The band structure of the systems and the gap oscillations are dominated by the component of the magnetic field
aligned with the CNT, that breaks the time--reversal symmetry and give the usual gap oscillation.
However the perpendicular component is not always negligible, when two degenerate bands
are present it breaks the accidental degeneracies in the band structure as shown
in Fig. (\ref{fig:not_aligned_field}).

\section{Persistent currents} \label{Sec:PC}
In the previous chapter we have introduced the concept of PCs induces by the AB effect. These currents have never been measured experimentally in SWCNTs because they are too small. However PC have been measured in mesoscopic rings~\cite{Exp_pc_1990,Exp_pc_2009_A,Exp_pc_2009_B} and have been predicted to be measurable in toroidal shaped CNTs~\cite{PC_toroidal_CNT} within the TB model.

We have compared the predictions of our ab--initio approach against TB results in order to estimate the reliability of the TB methods. Our scheme in fact has the advantage of including many--body effects and to describe the $\sigma$ states which are not in the TB model.

To compute PCs we've implemented two different schemes: we evaluated the expectation value of the current--density operator $\hat{\mathbf{j}}$ in order to construct the angular current $I_\phi$ and we also computed the angular current as the derivative of the total energy with respect to the applied magnetic flux. The second scheme includes many--body effect, that is the contribution due to the Hartree and the xc energies. The first scheme, instead, has been used to validate the method as the average on the KS wave--functions of the current operator can be compared with the contribution from the bands--energy, i.e. the sum of the KS eigenvalues.

We will proceed as follow: (i) we will show how the total energy, and the different contributions to the energy, depend on applied the magnetic flux and how the limitations of our approach appear around $\Phi=\Phi_0/2$. (ii) We will then show the results obtained for the total current using the two approaches. The preliminary results shown here are for the metallic (8,8) CNT.

\subsection*{The energy of the system}

\begin{figure}[t]
 \centering
 \includegraphics[width=0.9\textwidth]{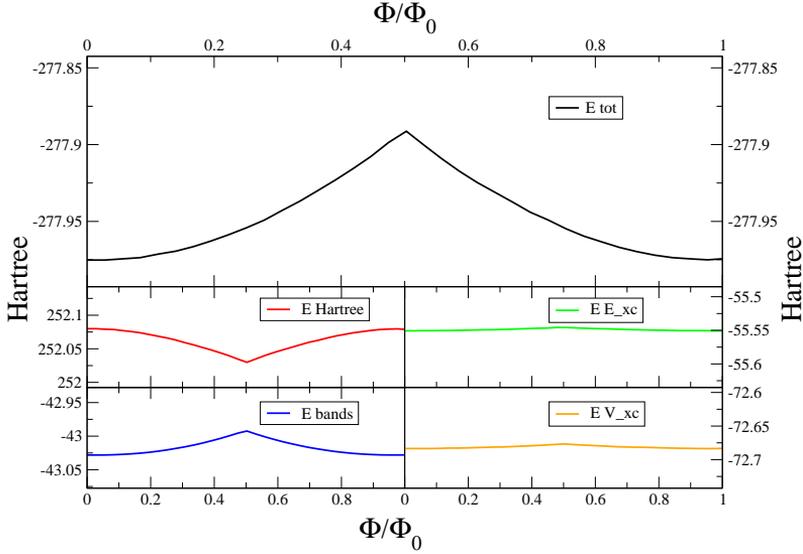}
 \caption{Energy components of the (8,8) CNT as a function of the applied flux in the confined geometry.
          From the top to the bottom: the total energy (black line), the Hartree energy (red line), the
          xc energy (green line), the sum of the KS eigenvalues (blue line), the expectation
          value of the xc potential (orange line).}
 \label{fig:88_energy_components}
\end{figure}

The total energy of the system within DFT is expressed as~\cite{DFT_book}
\begin{equation}
E_{tot}[\rho]=E_{bands}-E_H[\rho]-\langle V_{xc}[\rho] \rangle + E_{xc}[\rho]
\text{,}
\end{equation}
where $E_{bands}=\sum_{i\mathbf{k}}\epsilon^{KS}_i(\mathbf{k})f_i(\mathbf{k})w(\mathbf{k})$ with $\epsilon^{KS}_i(\mathbf{k})$ the KS eigenvalues, $f_i(\mathbf{k})$ the occupation factors and $w(\mathbf{k})$ the weights of the k--points in the BZ. The Hartree energy, $E_H[\rho]$, is double--counted in the $E_{bands}$ term and is subtracted while the xc energy is incorrectly described by the $V_{xc}[\rho]$ term which is subtracted.

The different components of $E_{tot}[\rho]$ are shown in Fig.~\ref{fig:88_energy_components}, that shows how the total energy (and also all other components of the energy) goes quadratically with $\Phi$ and present a derivative discontinuity at $\Phi=\Phi_0/2$. The quadratic behaviour can be understood, in first approximation, from the structure of the Hamiltonian
\begin{equation}
H_{magn}=\mathbf{A\cdot j}= \mathbf{B\cdot L} +q \rho A^2
\text{.}
\end{equation}

At low magnetic fields the contribution from the term $\mathbf{B\cdot L}$ is almost zero because all the contribution $+L_z$ and $-Lz$ cancel almost exactly\footnote{The wave--functions are not exact eigenstates of $\hat{L}_z$ because the rotation invariance is broken by the presence of the carbon atoms.} while the quadratic term grows. This behaviour, which is exact at low magnetic flux within the DFT scheme~\cite{CDFT_book}, should be modified at higher magnetic fields. However in our approach the high fluxes correction to the quadratic behaviour is not correctly described, as witnessed by the discontinuity at $\Phi=\Phi_0/2$. As a consequence the gauge--fixing solution induces a unphysical jump in the current in order to restore the correct periodicity.

\subsection*{The current}
\begin{figure}[t]
 \centering
 \includegraphics[width=0.48\textwidth]{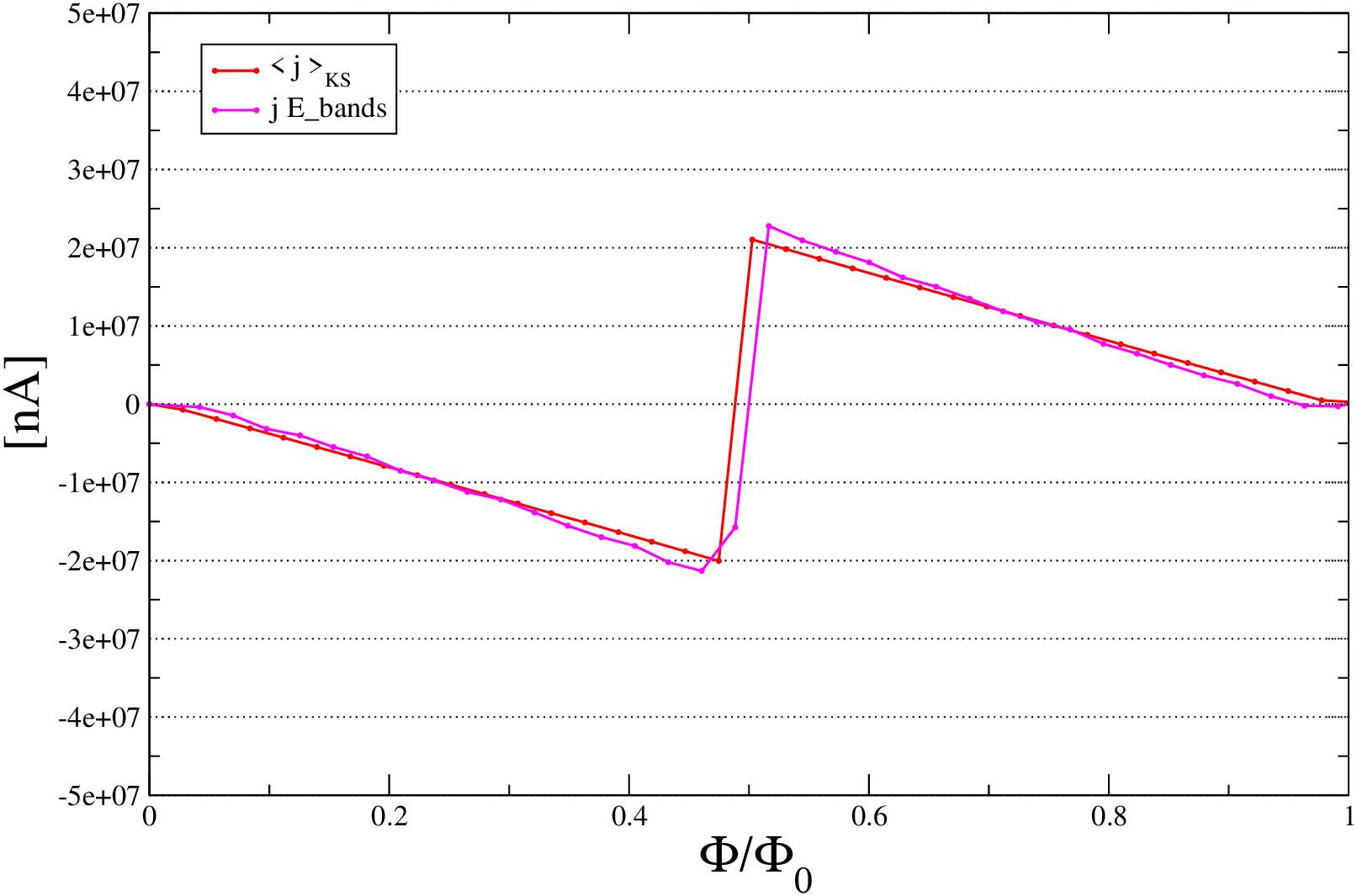}
 \includegraphics[width=0.48\textwidth]{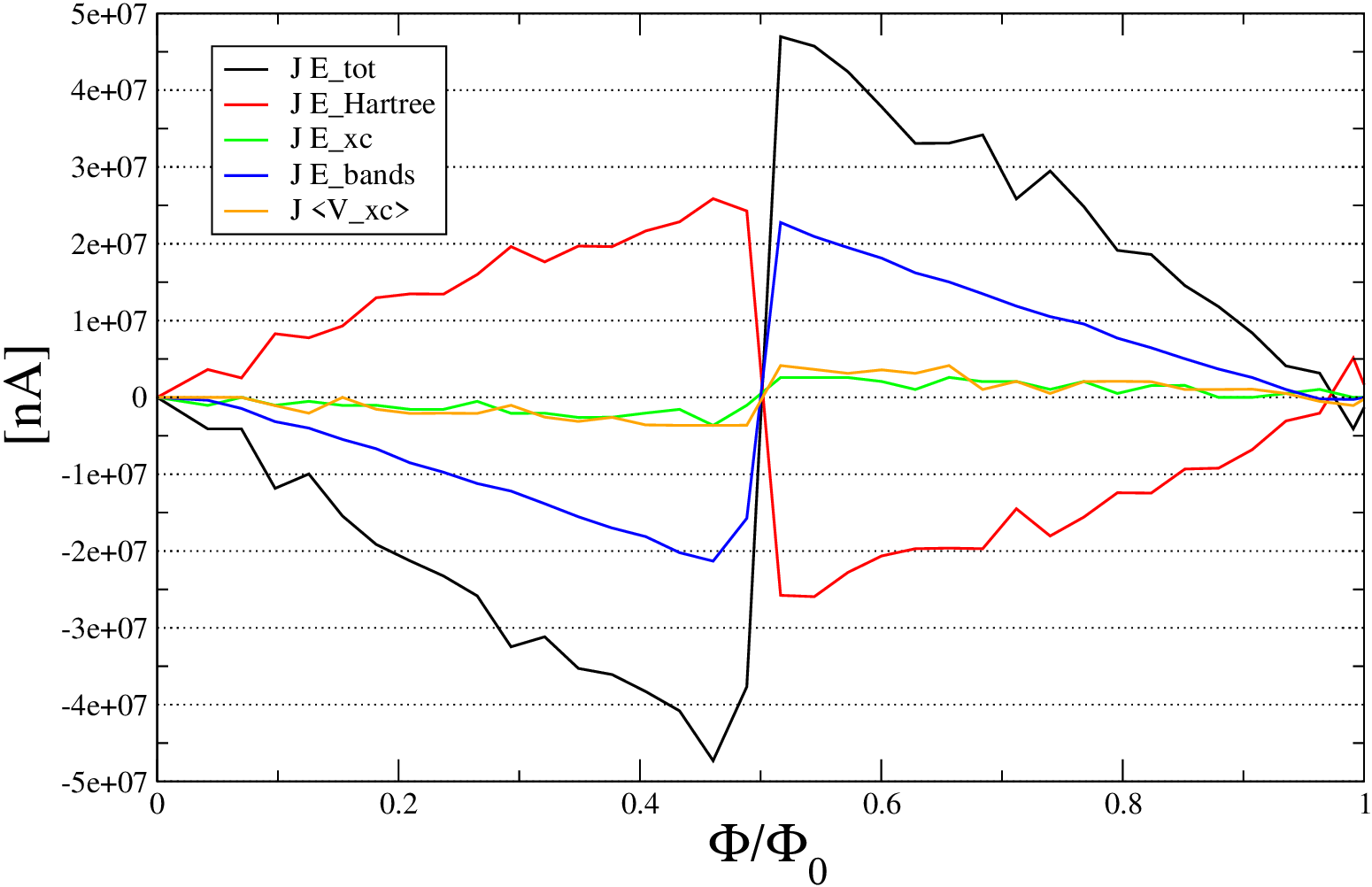}
 \caption{Current components of the (8,8) CNT as a function of the applied flux in the confined geometry.
          The expectation value of the $\hat{j}$ operator on the KS wavefunctions is compared
          against $j=\partial E_{bands} / \partial \Phi$ (left panel). The different components
          of the current are then computing deriving the different contributions to the total energy
          (right panel), cfr. Fig.~\ref{fig:88_energy_components}.}
 \label{fig:88_current_components}
\end{figure}

The quadratic behaviour of the energy is reflected in the linear behaviour of the persistent current as a function of the magnetic flux (see Fig.~\ref{fig:88_current_components} and the main components of the total current are the bands contribution and the Hartree contribution. First we can see that the two methods gives the same results for the bands term of the current\footnote{Here it is plotted the total current in the unit--cell, that is the current for a length of 2.46 \AA of the infinite CNT.}.

The result obtained both for the bands current and for the total current are much bigger than the prediction within the TB model~\cite{PC_toroidal_CNT,PC_CNT}. In particular the self--induced flux, which within the TB scheme is around $10^{-3}\Phi$, is here close to $0.15\Phi$. This strong discrepancy can be due to many factors: in the TB scheme the many--body effect are not included and only the $\pi$ states are used to describe the total current. On the contrary in our scheme the many--body effect are, at least in part, included and the sigma states are fully described, though we do not have an accurate description of the current in the region near $\Phi_0/2$. We are presently working on an improved implementation able to describe the current in the whole range of applied fluxes.

\section{Conclusions}
In this part of the thesis we have studied the effects of a magnetic field on
the electronic properties of Carbon Nano--Tubes (CNTs) aligned with the field.
In particular we focused
our attention on the oscillations
in the electronic gap induced by the Aharonov--Bohm (AB) effect.
By using parameter--free approach we made 
a direct comparison between the pure AB interpretation (confined geometry)
and the common experimental setup (extended geometry).

In the extended geometry we confirmed the general behaviour of the pure AB interpretation for
ideally perfect Single Wall (SW) CNTs. However we revealed the existence of corrections due to
a trajectory dependent term (the Lorentz Correction (LC)).
This effect decreases increasing the dimension of the CNT in
ideally defect--free CNTs, but it is likely to be
enhanced if vacancies or impurities, which can alter the electronic trajectory, are present.

In the literature the pure AB interpretation is usually used to describe magnetic field
effects within the Zone Folding Approach (ZFA). It is known, however, that 
the ZFA does not take into account curvature effects that can modify the band structure
of CNTs. We revealed these effects as corrections to the behaviour of the gap
oscillations, both in the extended and in the confined geometry.
The shift in the position of the metallization points
in semi--conducting CNTs or the completely different shape of the gap
oscillations in the $(8,0)$ CNT are examples of curvature effects.

We also discussed how the interaction among different shells in a Multi Wall (MW)
CNT can alter the gap oscillations obtained in the SW configuration. Indeed in the MW configuration
there is a drastic difference between the extended and the confined geometry. The AB interpretation
in the extended geometry can be recovered only at the price of describing the MWCNT as two not
interacting SW--CNTs. This is the standard approach used in the literature and again our approach 
has revealed important corrections like the existence of a metallic phase at low magnetic flux followed
by an indirect gap phase.

In the last two sections we have shown some preliminary results regarding the
case of a not perfect aligned magnetic field and the
existence of persistent currents in CNTs.
Further work is planned in this direction in the near future.

In conclusion we have described, for the first time at our knowledge, the AB effect in CNTs
with a first principles approach. We showed how curvature effects
modify the properties of CNTs under the effect of a magnetic field. Our results are grounded to
well--known facts in the ab--initio community.
For example the 
shift of the metalization points we observed in
semi--conducting CNTs can be related to the shift of the Dirac points in the band structure of 
small metallic CNTs. In the same way the peculiar behavior of gap oscillations
we observed in the $(8,0)$ CNT is related to the existence of a metallic--like valence
band, not predicted by the ZFA.

On the other hand the existence of LCs, which can effect the electronic properties of
CNTs is a completely new effect, which to our knowledge has never been addressed in the literature.
Moreover in the case of the $(5,5)@(10,10)$ CNT we pointed out how the pure AB interpretation
of magnetic field effects in MWCNTs is not free of ambiguities. Indeed the pure AB interpretation
gives different predictions with respect to what is measured experimentally in the extended geometry.
The pure AB effect can be recovered, as a first order approximation, only under the assumption 
that electrons can be distinguished according to weather they orbit on the inner or on the outer
shell of the MWCNT.

\newpage
\thispagestyle{empty}
\vspace{10cm}
\begin{flushright}
\large
\emph{
Nanos gigantium     \\
humeris insidentes  \\
}
\rule{\linewidth}{0.2mm}
\emph{
Dwarfs standing on      \\
the shoulders of giants
}
\end{flushright}
\newpage
\thispagestyle{empty}

\chapter*{Conclusions}
\fancyhead[RE]{\textbf{Conclusions}}
\fancyhead[LO]{\textbf{Conclusions}}
\addcontentsline{toc}{chapter}{Conclusions}
When I started to work on the subjects of the present thesis, more than
three years ago, my idea was to carry on a project which would have been
concluded at the end of the PhD.  Now I realize that a research
project, often, opens more question than the number of answers and solutions
it founds. The work presented in this thesis is the result of 
a long learning process that is far from being closed.

The initial plan of the thesis was to
developed a new method able to describe double excitations,
within a many body approach. This purpose required a deep study of 
the different approaches available in the literature:
from the Bethe--Salpeter Equation and
the time dependent density functional theory for extended systems, to the
Hartree--Fock and Post Hartree--Fock approaches to isolated systems.
From the Configuration Interaction expansion and 
other wave--functions based approaches adopted in quantum chemistry to the random
phase approximation and second random phase approximation used to describe nuclear systems.

Our resulting idea has been to merge some aspects of
techniques used for extended systems with the key ingredients used in quantum
chemistry and in nuclear physics. Indeed the two relevant aspects
we focused on are the idea of screening and the mathematical properties
imposed by exchange effects. Following both physical intuition and mathematical
rigor we proposed the resulting scheme as a possible choice to describe
double excitations in correlated materials.

An interesting result is that the inclusion of double (and even higher order)
excited configurations naturally emerge as a frequency dependent Bethe--Salpeter equation kernel in the
space of single particle transition. This result is similar to the one obtained in
other works~\cite{PhD_Thesis_Gatti} were the frequency dependency, i.e. temporal non 
locality, appears from the contraction of an higher
to a lower dimensional space. This projection implies that the system cannot be
regarded anymore as closed, and so its evolution depends
from its past history. 

The time--dependent density--functional theory and the Bethe--Salpeter equation
kernels are frequency dependent because these schemes,
instead of the many body wave--function, are based on 
the time dependent density and on the two particles Green's function respectively.
Here we rediscovered the frequency dependence of the kernel as a consequence
of the projection of the excited states operator in the space of
single particles transitions. 
\emph{``There is a pleasure in recognizing old things from a new point of view.
Also there are problems for which the new point of view offers a distinct
advantage''}~\cite{Feynmann}. 

Only systematic tests on realistic materials will reveal if the 
proposed approximation will work on realistic systems.
This is an open question which will likely need much more time
then the one available for a PhD thesis to find a definitive answer. 
When we arrived to formulate our final expression for the kernel of the
Bethe--Salpeter equation we felt we had reached a satisfying point of our investigation.

In the second part of the present thesis we tackled the description of
magnetic field effects in carbon nanotubes aligned in the field
direction, within a first principles approach.

The state--of--the--art on the subject describes these effects in terms of the
Aharonov--Bohm effect within the zone--folding 
Approach. Many of the properties are obtained starting from a tight binding 
calculation on a graphene sheet, which can be performed analytically.
We have shown how the first principles description confirms the qualitative
results of model. However corrections need to be considered. The
first--principles approach includes, in a consistent manner,
many effects beyond the zone--folding approach, such as the curvature of
the graphene sheet in the carbon nanotubes or the correction (Lorentz correction) to the
pure Aharonov--Bohm description of the magnetic field.

In this part of the thesis the strategy adopted is, for certain aspects,
opposite to the one adopted in the first part. Instead of mathematical rigor
(the approximations used are, sometimes, not fully justified 
from a theoretical point of view) the key aspect here have been to focus
on the physical behavior of the system. The description of a
physical system often involves many different aspects and it is crucial to 
find out which of these are the most important.

Moreover mathematical inspection has been used a posteriori
to check and improve the approximation involved. 
This is the example of the ``gauge fixing'' solution which we have developed to
overcome numerical instabilities, observing that the Aharonov--Bohm effect has to be
a pure gauge effect for some values of the vector potential.

The resulting approach enabled us to give a much accurate description of the
system compared to what can be obtained using models such as the the zone--folding
approach.

This second part of the work has been an occasion
to study a new subject and learn a different approach to a research project.
The concept of double excitations had a clear mathematical
definition though at the beginning we had no clue on how to incorporate them in a
many body approach.

On the other hand we had a clear idea of the experimental
setup used to simulate carbon nanotubes immersed in a magnetic field, but it
took some time to understand how the concept of the Aharonov--Bohm effect can be used
to describe such setup. Indeed according to Aharonov and Bohm the Aharonov--Bohm effect
arise because \emph{``there exist effects of potentials on charged particles,
even in the region where all the fields (and therefore the forces on the particles)
vanish''}~\cite{Aharonov1959}.
However the experimental setup considered for carbon nanotubes is in sharp contrast with the
situation suggested by Aharonov and Bohm themselves,
since the carbon nanotubes are fully immersed in the magnetic field.
Only understanding this difference we have been able to address specific
questions and recognize, for example, the renormalization of the gap
oscillations due to the Lorentz correction effect.

Indeed when, at first, we decided to work on the description of magnetic field effects
in carbon nanotubes our idea was to tackle the description of either magneto--optical 
spectra, and the ``dark to bright transition'' of the lowest energy exciton~\cite{Rivastava2008}
due to the Aharonov--Bohm effect, or the resistivity oscillations, related to the combined effect
of Aharonov--Bohm and weak localization~\cite{AAS1981,Bachtold1999}. However, the description of
the Aharonov--Bohm effect within a full ab--initio approach turned out to be an almost
unexplored field, and we were forced to first address the many open questions
which in the end became the core of the present work.

As I stated at the beginning of this conclusions, many questions have been opened.
I hope that the present thesis will be used as a starting point by someone, maybe
myself, to look for new answers. Many possible paths have been highlighted, which could 
possibly inspire new projects. Started from the results presented in part II of the thesis
a project apply the approach proposed to describe double excitations on realistic materials.
From part III, on the other hand, it could be interesting to look for experimental
configurations where the Lorentz correction is more pronounced than in ideally perfect carbon nanotubes.
Similarly a new project could improve the ab--initio description of persistent currents, in order
to say the last word on their real intensity and on the role played by many--body effects.

\appendixpage
\fancyhead[RE]{\bfseries\leftmark}
\fancyhead[LO]{\bfseries\rightmark}
\begin{appendices}
\addcontentsline{toc}{part}{Appendices}
\chapter{Connection to the experiments: extended systems} \label{App:Block wave functions}
\section{The Dielectric constant}

In Ch.~\ref{chap:Many Body Systems} we draw the connection between the microscopic quantities and the observable measured in an absorption experiment, that is the dielectric constant $\epsilon(\omega)$ for extended systems and the polarizability $\alpha(\omega)$ for isolated systems. Then in Ch.~\ref{chap:Introduction to the problem} we specialized the description to isolated systems and we wrote the expression of the polarizability in the space of single--particle wave--functions starting from the solution of Eq.~(\ref{generalized_Dyson_eigenvalue_simple}). In sec.~\ref{Sec:Numerical results on DE} we used this result to describe the spectra of two molecules ($C_8H_2$ and $C_4H_6$) and to test different aspects of the kernel proposed in Ch.~\ref{chap:A new approach to describe Double Excitations}.

However the kernel we propose, including the concept of screening, could in principle be used to describe extended systems too. For this reason we introduce here the direct expression for the dielectric function obtained from the solution Eq.~(\ref{generalized_Dyson_eigenvalue_simple}). It is convenient for this purpose to work in the space of the Block wave--functions:
\begin{equation}
\Psi_{n\mathbf{k}}(\mathbf{x})=u_{n\mathbf{k}}(\mathbf{x})e^{\mathbf{kx}}=1/\sqrt{V}\sum_{\mathbf{G}} c_{n}(\mathbf{k+G})e^{\mathbf{(k+G)x}}
\text{.} 
\end{equation}
With this choice the structure $\mathbf{k+G}$ is introduced in the reciprocal space, and the four point response function can be expressed as
\begin{multline}
\tilde{L}_{G_1,G_2,G_3,G_4}(\mathbf{k_1,k_2,k_3,k_4}|\omega)=\sum_{nm,st} 
       c_{n}(\mathbf{k_1+G_1})c^*_{m}(\mathbf{k_2+G_2})  \times                                        \\
       \tilde{L}_{nm,st}(\mathbf{k_1,k_2,k_3,k_4}|\omega)
       c_{s}(\mathbf{k_3+G_3})c^*_{t}(\mathbf{k_4+G_4})
\text{.}
\end{multline} 
Thanks to translation invariance we then reduce the dependence to three momenta $\mathbf{k}=(\mathbf{k_2}-\mathbf{k_1})$, $\mathbf{k'}=(\mathbf{k_4}-\mathbf{k_3})$ and $\mathbf{q}=1/2\ [(\mathbf{k_1}+\mathbf{k_2})-(\mathbf{k_4}+\mathbf{k_3})]$. The contraction in real space to obtain the response function $\chi(1,2)=L(1,1;2,2)$ is equivalent to set $(\mathbf{k_1,G_1})=(\mathbf{k_2,G_2})$ and $(\mathbf{k_3,G_3})=(\mathbf{k_4,G_4})$. So we obtain the function
\begin{equation}
\chi_{G_1,G_2}(\mathbf{q},\omega)=L_{G_1,G_1,G_2,G_2}(\mathbf{q},\omega)
\text{.}
\end{equation}
We recall that the macroscopic dielectric function (see Ch.~\ref{chap:Many Body Systems}) is obtained by averaging over the unit--cell, that is, in this notation, taking the the component $(\mathbf{G_1=G_2}=0)$ of the microscopic dielectric function. By using the expression
\begin{equation}
c_{i}(\mathbf{k_1+G_1})=\int d^3\mathbf{x} e^{i(\mathbf{k_1+G_1})}\Psi_{i\mathbf{k_1}}(\mathbf{x})
\text{,}
\end{equation}
 and rewriting the interaction in the space $(\mathbf{k+G})$, $w_\mathbf{G,G'}(\mathbf{q})=\delta_\mathbf{G,G'}\ 4\pi/|\mathbf{k+G}|^2$ we obtain
\begin{equation}\label{average_epsilon}
\begin{split}
\epsilon^{-1}_\mathbf{0,0}(\mathbf{q},\omega)&=
      1+w_\mathbf{0}(\mathbf{q})\chi_\mathbf{0,0}(\mathbf{q},\omega) \\
                                             &=
      1+\frac{w_\mathbf{0}(\mathbf{q})}{V}\sum_{nm,st} \langle n|e^{i\mathbf{qx}}|m\rangle
      \sum_{I,J}\frac{A^I_{nm}(\mathbf{q})S_{I,J} A^J_{st}(\mathbf{q})}{(\omega-\omega_I(\mathbf{q}))}
      \langle t|e^{-i\mathbf{qx}}|s\rangle
\text{.}
\end{split}
\end{equation}
The absorption spectrum is described by the $q\rightarrow 0$ limit of Eq.~(\ref{average_epsilon}). This is because the photon, at the range of energy of few $eV$, has negligible momentum $p_f=E/ c\simeq 10^{-27}\ [kg\ m/s]$, if compared to the momentum of the electron at the same energy $p_e=\sqrt(2m_eE)\simeq 5\cdot 10^{-25}\ [kg\ m/s]$, that is $p_f / p_e \simeq 2 \cdot 10^{-3}$.

\section{Electron energy loss spectroscopy and absorption}     \label{App:EELS vs absorption}
Once we computed the macroscopic dielectric function in extended systems electron energy loss spectroscopy (EELS) experiments can be described. In contrast to absorption experiments where the quantity $Img[\epsilon]$ is measured, EELS experiments measure the quantity $Img[\epsilon^{-1}]$ \cite{Schafer_Wegener}. Here we show that this difference is related to how the long range term of the Coulomb interaction $w$ enters in the equation to construct the macroscopic dielectric function. Indeed, using the microscopic relations $\epsilon^{-1}=1+w\chi$, Eq. (\ref{epsilon_micro}), and $\epsilon_M^{-1}=\langle \epsilon^{-1} \rangle$, Eq. (\ref{epsilon_Mac}), we will derive for EELS 
\begin{equation}
\epsilon^{-1}_M(\mathbf{q},\omega)=1+w_{0}(\mathbf{q})\chi_{00}(\mathbf{q},\omega)  \label{mac_Abs}  
\text{ ,}
\end{equation}
as opposite to the equation
\begin{equation}
\epsilon_M(\mathbf{q},\omega)     =1-w_{0}(\mathbf{q})\bar{\chi}_{00}(\mathbf{q},\omega)  \label{mac_EELS}  
\text{,}
\end{equation}
which holds for absorption experiments.

Here $\bar{\chi}$ is the response function obtained from the Dyson equation $\bar{\chi}=\chi^0+\chi^0 (\bar{w}+f_{xc} )\bar{\chi}$ where, with respect to Eq. (\ref{Dyson-tddft}), the $\mathbf{G=0}$ term of the interaction is not included in the Hartree part of the kernel, i.e. $\bar{w}_0(\mathbf{q})=0$.

The derivation is a mathematical exercise, but the result has an intuitive physical interpretation. While in absorption experiments the photons probe the system locally, so they do not feel the effect of the long range contribution of the interaction, in EELS experiments the electrons travel through the medium for long distances and the system reacts in a different manner. This difference disappear when isolated system are considered, as long as the dimension of the system is smaller than the wave--length of the photons (i.e. around $10^2 - 10^3 nm$).

From Eq.~(\ref{mac_Abs}) it is possible to understand why, sometimes, independent particle spectra are referred to as RPA spectra in the literature. At the RPA level $f_{xc}=0$, the term $w$ of the kernel can then be divided in two parts: the long range contribution $w_0(\mathbf{q})$, which account for the long range part of the electron--electron interaction, and the other terms $w_{G}(\mathbf{q})$ with $\mathbf{G}\neq 0$, the so called Local Fields (LFs), whose average on the unit cell is zero. If the LFs effects, which in extended system can be negligible, are not considered then $\bar{\chi}=\chi^0$ at the RPA level. For this reason the IP approximation is sometimes referred to as RPA without LFs effects.

\subsection*{Mathematical Derivation}
In order to obtain Eq.~(\ref{mac_EELS}) we observe that the microscopic dielectric function can be obtained inverting Eq.~(\ref{epsilon_micro}):
\begin{equation}\label{direct_epsilon_micro}
\epsilon(1,2)=\ \frac{\delta \varphi(1)}{\delta V(2)}=\ \delta(1,2)- \ w(1,1')\Pi^{\star}(1',2)
\text{.}
\end{equation}
We use here $\tilde{\chi}$, the retarded version of the T--ordered $\Pi^{\star}$, to write in reciprocal space
\begin{equation}\label{direct_epsilon_G}
\epsilon_{\mathbf{G,G'}}=\delta_{\mathbf{G,G'}}-w_G\tilde{\chi}_{\mathbf{G,G'}}
\text{.}
\end{equation}
We use now the general rule for a matrix of dimension $N\times N$
\begin{equation}
M= \left(
\begin{array}{c c}
m   & C_1 \\
C_2 & A  
\end{array}
\right)
\text{,}
\end{equation}
where $m=M_{1,1}$, $A$ is a matrix of dimension $(N-1)\times (N-1)$ and $C_1$ and $C_2$ are matrix of dimension $ 1 \times (N-1)$ and $(N-1)\times 1$ respectively,
\begin{equation}
M^{-1}=
\frac{1}{m-C_1 A^{-1} C_2}
\left(
\begin{array}{c c}
1          & -C_1 A^{-1} \\
A^{-1} C_2 & A^{-1}  
\end{array}
\right)
\text{,}
\end{equation}
to express $\epsilon^{M}=1/\epsilon^{-1}_{\mathbf{0,0}}$ starting from Eq.~(\ref{direct_epsilon_G}):
\begin{equation}
\epsilon^{M}=1-w_{\mathbf{0}}\tilde{\chi}_{\mathbf{0,0}}-
               \sum_{\mathbf{G,G'\neq 0}} w_{\mathbf{0}}\tilde{\chi}_{\mathbf{0,G}}\
                     \epsilon^{-1}_{\mathbf{G,G'}}\
                        w_{\mathbf{G'}}\tilde{\chi}_{\mathbf{G',0}}
\text{.}
\end{equation}
Finally we recognize that 
\begin{equation}
\tilde{\chi}_{\mathbf{0,0}}+\sum_{\mathbf{G,G'\neq 0}}\tilde{\chi}_{\mathbf{0,G}}\ \epsilon^{-1}_{\mathbf{G,G'}}\ w_{\mathbf{G'}}\tilde{\chi}_{\mathbf{G',0}} 
\end{equation}
is the zero component of a modified Dyson--like equation where the $\mathbf{G=0}$ component of the kernel is zero. Using the relation $\epsilon^{-1}=(1-w\tilde{\chi})^{-1}$ in fact we can define
\begin{equation}
\bar{\chi}=\tilde{\chi}+\tilde{\chi}\bar{w}\bar{\chi}
\text{,}
\end{equation}
and so express the macroscopic dielectric function as
\begin{equation}
\epsilon^{M}=1-w_{\mathbf{0}}\bar{\chi}_{\mathbf{0,0}}
\text{.}
\end{equation}

\chapter{On the quasiparticle concept} \label{App:On the quasiparticle concept}
To clarify the quasiparticle (QP) concept here we consider the poles of the one--particle Green's function (GF), that represent the description of photo--emission excitations.

The QP concept introduced by Landau derives from the physical intuition that some excited states of an interacting many--electron system reassemble a one particle resonance in experiments carried out in accelerators. The main difference is that real particles do not interact with the background (the vacuum) while in the interacting systems QP excitations have a (long) finite life--time due to the interaction with the many--electrons sea. This means that QPs are not exact eigenstates of the Hamiltonian, but are ``quasi--eigenstates''.

In the GF formalism the finite lifetime is described by the imaginary part of the self--energy which, evaluated at a QP pole, is not real. This concept is strongly related to the existence of a continuum of poles in an infinite system, that is a branch cut in the complex plane. 

First we give the mathematical construction which connects the Lehmann representation to the QP representation of a GF. Most of the concepts are in Ref.~\cite{RMP_Onida}.

Consider the function (here $x$ is in the complex plane)
\begin{equation}
f_{ik}(x)=\frac{a_i(k)}{x-\epsilon_i(k)+i\eta}
\text{,}
\end{equation}
which has a pole on the real axis at $x=\epsilon_i(k)$ (the small $i\eta$ off-set is there, as usual, for mathematical convenience). Now let's take the related function
\begin{equation}\label{Lehmann}
g(x)=\sum_{i,k} \frac{a_i(k)}{x-\epsilon_i(k)+i\eta}
\text{,}
\end{equation}
which has a series of poles on the real axis. Suppose that, in some limit (the thermodynamic limit in a physical system), the variable $k$ becomes a continuum variable and the sum over $k$ becomes an integral\footnote{In calculations on realistic system we have a discrete grid of $k$ points but we think this as an approximation to the whole Brillouin--zone}. Supposing for simplicity $\epsilon_i(k)=k\ \delta_{i,1}$ and that $a_i(k)= (1/\pi) (1/[(k-E_1)^2+E_2^2]$, the integrand has a branch cut in the lower complex plane for $k=x-i\eta$ and a simple complex pole at $k=E_1+iE_2$. We can perform a contour integral in the upper plane (avoiding the branch cut) and using the residue theorem we obtain
\begin{equation}\label{QP}
g(x)= \frac{2\pi i}{\pi}\ \frac{1}{\omega-(E_1+iE_2)}
\text{,}
\end{equation}
where I let the $\eta$ to go to zero as we do not need it any more here.

The poles of $g(x)$ can be either the branch cut, i.e. the poles of the Lehmann representation (Eq.~(\ref{Lehmann})), or the complex pole, i.e. the QP pole (Eq.~(\ref{QP})). This connects the Lehmann representation with the QP picture.

We have a connection with the definition I gave in my thesis with $E_1=\epsilon_i+Re[\Sigma_{ii}(E_1+iE_2)]$ and $E_2=Img[\Sigma_{ii}(E_1+iE_2)]$. This last connection can be obtained writing the analytic continuation of
\begin{equation}
G_{ii}(k,\omega)=\frac{1}{\omega-(\epsilon_i(k)+\Sigma_{ii}(k,\omega))}
\end{equation}
The GF has a QP pole when $Img(\Sigma_{ii}(k,\omega))$ is small (we need the QP to be almost an eigen--state of the Hamiltonian according to the request that we have a quasi--eigen--state).

In the analytic continuation we find the complex poles and not the branch cuts on the real axis because the GF is defined on the Riemann surface where the branch cuts are the connections of two sheets.

When we do not have a continuum of states, then all this procedure makes no sense and we remain with the simple poles on the real axis. Physically this is related to the fact that there is not a macroscopic number of states which can interact and make the QP poles to have a finite life--time.

The definition of QP as ``dressed'' one particle excitations, which never consider satellites as QP, is not exactly the same as the idea of a quasi--eigenstate of the Hamiltonian. If a satellite is a clear and well definite peak in the spectrum, then it can be considered a QP. On the other hand if the satellite appear, for example, as a shoulder of a QP peak in the spectrum it cannot be considered a QP.

However when, in practice, we do $GW$ calculations we assume that the QP wave--functions are well described by the KS wave--functions. This means that we do not have the QP wave--function which describes the satellite. Maybe in this sense just ``dressed'' one particle excitations are QPs. 

The procedure I've described in this section is usually not carried on in the BSE scheme where instead the Lehmann representation is used\footnote{Only in some works the imaginary--part of the self--energy is used to evaluate the life--time of neutral excitations}. It could be interesting to keep in mind this possibility, especially once one goes beyond the static approximation for the kernel of the BSE and so can think to an analytic continuation and to a kernel which can be non-real.

\chapter{Gauge transformations} \label{App:Gauge transformations}

In quantum mechanics a transformation is a gauge transformation if it leaves unchanged
all physical predictable quantities.
The most well known kind of gauge transformation are those related to
the electromagnetic potentials (or more in general to gauge potentials). 
The Schr\"odinger equation, for example, is invariant under the transformations
\begin{eqnarray}\label{gauge_transformations}
\mathbf{A} &\rightarrow&  \mathbf{A+\nabla} \Lambda             \text{ ,}             \\
        V &\rightarrow&  V-\frac{\partial\Lambda}{\partial t}   \text{ ,}             \\
     \Psi &\rightarrow&  \Psi e^{\frac{i\hbar}{e}\Lambda}
\text{ .}
\end{eqnarray}
$V$ and $\mathbf{A}$ are the scalar and the vector electro--magnetic potential 
respectively, $\Psi$ is the wave--function of the system, and $\Lambda$ is a generic function.

In Ch.~\ref{chap:numerical results CNT}, for example, we used this freedom to fix numerical
instabilities in the solution of the self--consistent problem. The concept of AB
effect itself is deeply related to the concept of gauge transformation. Indeed the existence of the AB effect
is due to the fact that, in quantum mechanics, the vector potential outside a solenoid can
be nullified only at integer values of the magnetic flux trapped by the solenoid, and accordingly
is not a pure gauge potential for non--integer values of the flux.

The many--body Hamiltonian introduced in Ch.~\ref{chap:Many Body Systems} is written approximating
the $H_{int}$ term in the non-relativistic limit, i.e. neglecting terms proportional to $1/c$ (with
$c$ the speed of the light). The retained term, $w$, is the interaction in the ``Coulomb gauge''
or transverse gauge obtained setting $\mathbf{\nabla \cdot A=0}$ and it is instantaneous.

Other possible gauge transformations exist. The response function introduced in Ch.~\ref{chap:Many Body Systems}
in the dipole approximation is proportional to the expectation value $\langle \mathbf{\hat{x}} \rangle$ of
the position operator. However the relation $[\hat{H},\hat{x}]=\hat{p}$ between the position
operator to the momentum operator can be used to write the response function in terms of either the
position (lenght gauge) or the momentum (velocity gauge).

\chapter{DFT and magnetic fields} \label{App:DFT and magnetic fields}

In Ch. \ref{chap:numerical results CNT} we have shown ab--initio results of the effect of a static magnetic field on CNTs. The implementation in the code has been done adding the term $H_{magn}=\mathbf{A^{ext}\cdot j} $. While this term enters in the many--body Hamiltonian, it is not the only term which enters be used in the DFT Hamiltonian. Our implementation is then approximated.

\subsection*{DFT vs CDFT}
When the coupling term $\mathbf{A\cdot j}$ is considered the HK theorem does not hold anymore and CDFT have to be used, as we stated in Ch. \ref{chap:Density Functional Theory}, where an extension of the HK theorem can be proved~\cite{CDFT_book,CDFT}. Within (non relativistic) CDFT the total energy is a functional of the density and the paramagnetic current $E[\rho,\mathbf{j}^{(p)}]$. DFT can be seen as an approximation to CDFT with $E[\rho,\mathbf{j}^{(p)}=0]$; accordingly $V_{xc}[\rho,j^{(p)}]\approx V_{xc}^{DFT}$ and $A_{xc}\approx 0$. This approximation is used in the ZF model and in the present thesis too.

\subsection*{Diamagnetic screening}
Beyond the $xc$ vector potential, the magnetic current generates a diamagnetic screening potential, that is the equivalent of the Hartree term which originates from the density:
\begin{equation}
A^{diam}(\mathbf{r})=\frac{\mu_0}{4\pi}\int d^3\mathbf{r}' \frac{\mathbf{j(r')}}{|\mathbf{r-r'}|}
\end{equation}
The diamagnetic screening term is usually negligible as the ``diamagnetic field'' is usually much smaller then the external applied field. In CNTs too is small, however, due to its cylindrical geometry it could be measurable. Indeed the ``diamagnetic field'' is, in first approximation, proportional to the radius of the electronic orbits. In simple geometries electrons spin around the atoms at a radius close to 1 Bohr. In CNTs and in mesoscopic rings however the radius is much bigger. Indeed in mesoscopic rings the field can be strong enough to have stationary states with self--sustaining currents\footnote{When the radius of the cylinder becomes very large and the classical limit is approached the diamagnetic current must be balanced by the paramagnetic current and the total effect is not always diamagnetic.}. 

In the present work we neglected this term as a first approximation, though the description of persistent currents in the ground state could be used to estimate the value of it.

\subsection*{Non--local term of the pseudo--potential}
When a vector potential is considered, the non local part of the pseudo--potential has to be modified. This can be understood observing that a non local operator $V_{nl}(\mathbf{r,r'})$ can be expressed as $V_{nl}(\mathbf{r,p})$~\cite{Louie_nl_psp,Mauri_nl_psp}. Then when a vector potential is present the substitution $\mathbf{p}\rightarrow\mathbf{(p-eA)}$ has to be performed in order to ensure that $\langle V_{nl}(\mathbf{r,p}) \rangle$ is gauge invariant and so that the total energy of the system is gauge invariant. 

We explored the effect of the non--local term of the pseudo--potential in Sec.~\ref{Sec:Band structure}.

The non--local term in the pseudo--potential enters even in the definition of the current operator. Indeed the current is proportional to the velocity and so to the commutator of the Hamiltonian with the position operator
\begin{eqnarray}
\mathbf{j}&=&q\ \[H,\mathbf{r}\] \text{,}  \\
          &=&q\ (p-e\mathbf{A}+\[V_{nl},\mathbf{r}\])
\text{.}
\end{eqnarray}
In the present work, when we compute the PCs we considered the effect due to the presence of the non--local term on the pseudo--potential. However we verified that this term is negligible with respect to the paramagnetic and the diamagnetic term of the total current.

\end{appendices}
\backmatter
\bibliographystyle{unsrt}
\bibliography{Biblio_sorted}

\begin{thebibliography}{10}

\bibitem{Iijima_1991}
{ S. Iijima, Nature \textbf{354}, 56 (1991) }.

\bibitem{Charlier2007}
{J.-C. Charlier, X. Blase and S. Roche, Rev. Mod. Phys. \textbf{79}, 677
  (2007)}.

\bibitem{CNT_discovery}
{ Carbon \textbf{44}, 1621 (2006), Guest Editorial }.

\bibitem{PhD_Thesis_Fabien}
{Fabien Bruneval, PhD thesis, \emph{Exchange and Correlation in the Electronic
  Structure of Solids, from Silicon to Cuprous Oxide: GW Approximation and
  beyond} defended on $7^{th}$ September 2005, \'Ecole Polytechnique,
  Palaiseau. \\
  http://theory.polytechnique.fr/people/bruneval/bruneval\_these.pdf }.

\bibitem{Fetter_Walecka}
{A.L. Fetter and J.D. Walecka, \\ \emph{Quantum Theory of Many-Particles
  Systems}, \\ Dover edition, New York (2003)}.

\bibitem{PhD_Thesis_Sottile}
{Francesco Sottile, PhD thesis, \emph{Response functions of semiconductors and
  insulators: from the Bethe--Salpeter equation to time-dependent density
  functional theory}, defended on $29^{th}$ September 2003, \'Ecole
  Polytechnique, Palaiseau. \\
  http://etsf.polytechnique.fr/system/files/users/francesco/Tesi\_dot.pdf }.

\bibitem{Schafer_Wegener}
{W. Sch\"{a}fer, M. Wegener, \\ \emph{Semiconductor Optics and Transport
  Phenomena} \\ Springer, New York (2001)}.

\bibitem{Hedin}
{L. Hedin, Phys. Rev. {\bf 139}, A796 (1965)}.

\bibitem{Landau_Fermi_Liquid}
{L. D. Landau, Zh Eksperim. i Theor Fiz. \textbf{30} 1058 (1956). \\ ENGLISH
  TRANSLATION: Soviet. Phys. JEPT \textbf{3}, 920 (1956) }.

\bibitem{Cederbaum1}
{L. S. Cederbaum, and J. Schirmer, Z. Physik \textbf{271}, 221-227 (1974)}.

\bibitem{PhD_Thesis_Gatti}
{Matteo Gatti, PhD thesis, \emph{Correlation effects in valence--electron
  spectroscopy of transition--metal oxides : many-body perturbation theory and
  alternative approaches}, defended on $4^{th}$ December 2007, \'Ecole
  Polytechnique, Palaiseau. \\
  http://etsf.polytechnique.fr/sites/default/files/users/matteo/matteo\_thesis%
.pdf }.

\bibitem{Thomas_1927}
{ L.H. Thomas, Proc. Camb. Philos. Soc. \textbf{23}, 542 (1927) }.

\bibitem{Fermi_DFT}
{ E. Fermi, Accad. Naz. Lincei, Rend. \textbf{6}, 602 (1927) }.

\bibitem{Dirac_DFT}
{ P. A. M. Dirac, Proc. Camb. Philos. Soc. \textbf{26}, 376 (1930) }.

\bibitem{HK}
{ P. Hohenberg, and W. Kohn Phys. Rev. \textbf{136}, B864 (1964) }.

\bibitem{KS}
{ W. Kohn, and L. J. Sham Phys. Rev. \textbf{140}, A1133 (1965) }.

\bibitem{DFT_book}
{R.M. Dreizler and E.K.U. Gross, \\ \emph{Density Functional Theory}, \\
  Springer-Verlag, Berlin (1990)}.

\bibitem{CDFT}
{ G. Vignale, and M. Rasolt Phys. Rev. Lett. \textbf{59}, 2360 (1987) }.

\bibitem{CDFT_book}
{G. F. Giuliani, and G. Vignale, \\ \emph{Quantum theory of the electron
  liquid}, \\ Cambridge University Press, Cambridgen (2005)}.

\bibitem{LDA_figure}
{O. Gunnarsson, M. Jonson, B. I. Lundqvist, Phis. Rev. B \textbf{20}, 3136
  (1979) }.

\bibitem{RG}
{ E. Runge and E.K.U. Gross Phys. Rev. Lett. \textbf{52}, 997 (1984) }.

\bibitem{TDDFT_Leeuwen}
{ R. van Leeuwen, Phys. Rev. Lett. \textbf{80}, 1280 (1998) }.

\bibitem{GGA_worse}
{ C. Loschen,J. Carrasco, K. M. Neyman, and F. Illas Phys. Rev. B \textbf{75},
  035115 (2007) }.

\bibitem{GGA_worse2}
{ C. Loschen,J. Carrasco, K. M. Neyman, and F. Illas Phys. Rev. Lett.
  \textbf{82}, 2544 (1999) }.

\bibitem{Maitra1}
{N. T. Maitra, F. Zhang, R. J. Cave, and K. Burke, J. Chem. Phys. \textbf{120},
  5932 (2004) }.

\bibitem{1Dpolymers}
{D. Varsano, A. Marini, and A. Rubio, Phys. Rev. Lett. \textbf{101}, 133002
  (2008) }.

\bibitem{Casida1}
{M. E. Casida, J. Chem. Phys. \textbf{122}, 054111 (2005) }.

\bibitem{Maitra2}
{R. J. Cave, F. Zhang, N. T. Maitra, and K. Burke, Chem. Phys. Lett.
  \textbf{389}, 39 (2004) }.

\bibitem{Li}
{ C. M. Isborn and X. Li, J. Chem. Phys. \textbf{129}, 204107 (2008) }.

\bibitem{WZ1}
{F. Wang and T. Ziegler, J. Chem. Phys. \textbf{121}, 12191 (2004) }.

\bibitem{Casida_spinflip}
{M. Huix--Rotllant, B. Natarajan, A. Ipatov, C. M. Wawire, T. Deutsch, and M.
  E. Casida, Phys. Chem. Chem. Phys. \textbf{12}, 12811 (2010) }.

\bibitem{Casida2}
{M. Huix--Rotllant, and M. E. Casida, \emph{Formal Foundations of Dressed
  Time-Dependent Density-Functional Theory for Many-Electron Excitations}, \\
  arXiv:1008.1478v1 [cond--mat. mes--hall] 9 Aug 2010 }.

\bibitem{Romaniello1}
{P. Romaniello, D. Sangalli, J. A. Berger, F. Sottile, L. G. Molinari, L.
  Reining, and G. Onida, J. Chem. Phys. \textbf{130}, 044108 (2009) }.

\bibitem{Baerdens_TDDM_polyenes}
{K. J. H. Giesberts, E.J. Baerdens, and O.V. Gritsenko Phys. Rev. Lett.
  \textbf{101}, 033004 (2008)}.

\bibitem{CI_manyel}
{T.S. Chwee and E.A. Carter, J. Chem. Phys. \textbf{132}, 074104 (2010) }.

\bibitem{Polyenes_ADC}
{J. H. Starcke, M. Wormit, J. Schirmer, A. Dreuw Chem. Phys. \textbf{329},
  39-49 (2006)}.

\bibitem{spin_adapted_TDDFT}
{Z.D. Li and W.J. Liu, J. Chem. Phys. \textbf{133}, 064106 (2010) }.

\bibitem{NiO_exp}
{B. C. Larson, Wei Ku, J. Z. Tischler, Chi-Cheng Lee, O. D. Restrepo, G.
  Eguiluz, P. Zschack, and K. D. Finkelstein, Phys. Rev. Lett. \textbf{99},
  026401 (2007) }.

\bibitem{NiO_theory}
{Chi-Cheng Lee, H. C. Hsueh, and Wei Ku, \\ Phys. Rev. B \textbf{82}, 081106(R)
  (2010) }.

\bibitem{NiO_theory_2}
{Chi-Cheng Lee, H. C. Hsueh, and Wei Ku, Phys. Rev. Lett. \textbf{99}, 257401
  (2007) }.

\bibitem{Dyson_orb1}
{P. Duffy, D. Chong, M. E. Casida, and D. R. Salahub, Phys. Rev. A \textbf{50},
  4707 (1994) }.

\bibitem{Dyson_orb2}
{M. E. Casida, Phys. Rev. A \textbf{51}, 2505 (1995) }.

\bibitem{Dyson_orb3}
{S. Hamel, P. Duffy, M.E. Casida, and D.R. Salahub, J. Electr. Spectr. and
  Related Phenomena \textbf{123}, 345 (2002) }.

\bibitem{Casida_TDDFT}
{ M.E. Casida, \emph{Recent Advances in Density Functional Methods, Part I}
  edited by D.P. Chong (World Scientific, Singapore, 1995) p.155 }.

\bibitem{RMP_Onida}
{G. Onida, L. Reining, and A. Rubio, Rev. Mod. Phys. \textbf{74}, 601 (2002) }.

\bibitem{Gambacurta}
{D. Gambacurta, and F. Catara, Phys. Rev \textbf{B 79}, 085403 (2009) }.

\bibitem{Strinati_book}
{G. Strinati, Rev. Nuovo Cimento \textbf{11}, 1, (1988) }.

\bibitem{GW_screening}
{L. Chiodo, J. M. Garcia-Lastra, A. Iacomino, S. Ossicini, J. Zhao, H. Petek,
  and A. Rubio, Phys. Rev. B \textbf{82}, 045207 (2010) }.

\bibitem{GW_screening2}
{ P.H. Hahn, W.G. Schmidt, and F. Bechstedt, Phys. Rev. B \textbf{72}, 245425
  (2005) }.

\bibitem{GW_screening3}
{Y. Shigeta, A.M. Ferreira, V.G. Zakrzewski, and J.V. Ortiz, Int. J. Quant.
  Chem. \textbf{85}, 411 (2001) }.

\bibitem{PhD_Thesis_Daniele}
{Daniele Varsano, PhD thesis, \emph{First principles description of response
  functions in low dimensional systems}, defended on $13^th$ July 2006, at
  University of the Basque Country, San Sebasti\'an (Spain). \\
  http://nano-bio.ehu.es/files/arenal\_phd.pdf }.

\bibitem{Romaniello2}
{ P. Romaniello, S. Guyot, and L. Reining, J. Chem. Phys. \textbf{131}, 154111
  (2009) }.

\bibitem{sRPA1}
{ C. Yannouleas, Phys. Rev. C {\bf 35}, 1159 (1987) }.

\bibitem{sRPA2}
{ J. Wambach, Rep. Prog. Phys. {\bf 51}, 989 (1988) }.

\bibitem{det_prop}
{V.E.Korepin, N.M.Bogoliubov, and A.G.Izergin, \\ \emph{Quantum Inverse
  Scattering Method and Correlation Functions}, \\ CAMBRIDGE MONOGRAPHS ON
  MATHEMATICAL PHYSICS. In particular Appendix IX.1 }.

\bibitem{Myrta}
{M. Gruning, A. Marini, and X. Gonze, Nano Letters \textbf{9}, 2820 (2009) }.

\bibitem{Cederbaum2}
{J. Brand, and L. S. Cederbaum, Phys. Rev. \textbf{A 97}, 4311 (1998) }.

\bibitem{ADC}
{J. Schirmer, Phys. Rev. \textbf{A 26}, 2395 (1982) }.

\bibitem{Marini1}
{A. Marini and R. Del Sole, Phys.\ Rev.\ Lett.\ \textbf{91}, 176402 (2003) }.

\bibitem{Bruneval_JCP}
{F. Bruneval, F. Sottile, V. Olevano, and L. Reining, J. Chem. Phys.
  \textbf{124}, 144113 (2006) }.

\bibitem{Yambo}
{A. Marini, C. Hogan, M. Grüning and D. Varsano, Comp. Phys. Comm.
  \textbf{180}, 1392 (2009). \\ http://www.yambo-code.org/ }.

\bibitem{Abinit}
{X. Gonze et al. Comp. Phys. Comm. \textbf{180}, 2582 (2009). \\
  http://www.abinit.org/ }.

\bibitem{Aharonov1959}
{Y. Aharonov, and D. Bohm, Phys. Rev. \textbf{115}, 485 (1959)}.

\bibitem{Chambers1960}
{R. G. Chambers, Phys. Rev. Lett. \textbf{5}, 3 (1960)}.

\bibitem{Tonomura1986}
{A. Tonomura, N. Osakabe, T. Matsuda, T. Kawasaki, and J. Endo, Phys. Rev.
  Lett. \textbf{56}, 792 (1986)}.

\bibitem{Bachtold1999}
{A. Bachtold, C. Strunk, J. P. Salvetat, J. M. Bonard, L. Forro, T. Nussbaumer,
  and C. Schonenberger, Nature \textbf{397}, 673 (1999)}.

\bibitem{Abrahams1979}
{E. Abrahams, P. W. Anderson, D. C. Licciardello, and T. V. Ramakrishnan, Phys.
  Rev. Lett. \textbf{42}, 673 (1979)}.

\bibitem{AAS1981}
{B. L. Al'tshuler, A.G. Aronov, and B. Z. Spivak, Pis'ma Zh. Eksp. Teor. Fiz.
  \textbf{33}, 101 (1981) }.

\bibitem{Coskun2004}
{U. C. Coskun, T. Wei, S. Vishveshwara, P. M. Goldbart, and A. Bezryadin,
  Science \textbf{304}, 1132 (2004)}.

\bibitem{Zaric2004}
{S. Zaric, G. N. Ostojic, J. Kono, J. Shaver, V. C. Moore, M. S. Strano, R. H.
  Hauge, R. E. Smalley, X. Wei, Science \textbf{304}, 1129 (2004)}.

\bibitem{Physics_today_AB}
{H. Bateelan, and A. Tonomura, Physics Today, September (2009), pag. 38 }.

\bibitem{AB_review1989}
{M. Peshkin, and A. Tonomura, \emph{The Aharonov-Bohm effect}, \\ Lectures
  Notes in Physics, 340 (1989)}.

\bibitem{PC_picture}
{H. Bouchiat, Physics \textbf{1}, 7 (2008). Viewpoint on Phys. Rev. Lett.
  \textbf{101}, 057001 (2008) }.

\bibitem{Landauer1983}
{M. B\"{u}ttiker, Y. Imry, and R. Landauer, Phys. Lett. A \textbf{96}, 365
  (1983)}.

\bibitem{Exp_pc_1990}
{L. P. Levy, G. Dolan, J. Dunsmuir, and H. Bouchiat, Phys. Rev. Lett.
  \textbf{64}, 2074 (1990) }.

\bibitem{PC_CNT}
{M. Szopa, M. Marganska, and E. Zipper, Phys. Lett. A \textbf{229}, 593 (2002)
  }.

\bibitem{Exp_pc_2009_A}
{A. C. Bleszynski-Jayich, W. E. Shanks, B. Peaudecerf, E. Ginossar, F. von
  Oppen, L. Glazman, and J. G. E. Harris, Science \textbf{326}, 272 (2009) }.

\bibitem{Exp_pc_2009_B}
{H. Bluhm, N. Koshnick, J. Bert, M. Huber, and K. Moler, Phys. Rev. Lett.
  \textbf{120}, 136802 (2009) }.

\bibitem{WL_picture}
{A. G. Aronov, and Yu. V. Sharvin, Rev. Mod. Phys. \textbf{59}, 775 (1987) }.

\bibitem{wikipedia}
{http://en.wikipedia.org/wiki/Carbon\_nanotubes }.

\bibitem{Bethune_1993}
{D. S. Bethune, C. H. Kiang, M. S. de Vries, G. Gorman, R. Savoy, J. Vazquez,
  and R. Beyers, Nature \textbf{363}, 605 (1993) }.

\bibitem{Iijima_1993}
{S. Iijima, and T. Ichihashi, Nature \textbf{363}, 603 (1993) }.

\bibitem{Guo_1995}
{T. Guo, C.-M. Jin, and R. E. Smalley, Chem. Phys. Lett. \textbf{243}, 49
  (1995) }.

\bibitem{Mauri_nl_psp}
{Chris J. Pickard, and Francesco Mauri, Phys. Rev. Lett. \textbf{91}, 196401
  (2003) }.

\bibitem{PC_toroidal_CNT}
{M. F. Lin , and D. S. Chuu, Phys. Rev. B \textbf{57}, 6731 (1998) }.

\bibitem{Feynmann}
{R. P. Feynmann, Rev. Mod. Phys. \textbf{20}, 367 (1948) }.

\bibitem{Rivastava2008}
{A. Srivastava, H. Htoon, V. I. Klimov, and J. Kono, Phys. Rev. Lett.
  \textbf{101}, 087402 (2008) }.

\bibitem{Louie_nl_psp}
{S. Ismail-Beigi, E. K. Chang, and S. G. Louie, Phys. Rev. Lett. \textbf{87},
  087402 (2001) }.

\end{thebibliography}

\pagestyle{fancy}
\fancyhf{}
\fancyhead[LE,RO]{\bfseries\thepage}
\renewcommand{\headrulewidth}{0.5pt}
\renewcommand{\footrulewidth}{0pt}
\addtolength{\headheight}{0.5pt}
\fancypagestyle{plain}{%
              \fancyhead{}
              \renewcommand{\headrulewidth}{0pt}}

\chapter{Acknowledgments}
Here we go! After three long years I'm at the end of my PhD thesis. There are so many people I feel to thank for helping and supporting me during this period.

First of all my supervisor prof. Giovanni Onida. I know I have not been the easiest student you met, often doing things in my own way, ready to quit science and to look for something different to work on. I have to thank you for the freedom I had to follow my ideas and to interact with many other scientists. For the possibility I had to get in touch with an international community and to do good science. This is something that in Italy seems to be always more and more difficult to do.

A special thank you to my co-supervisor, dott. Andrea Marini for the endless discussions, skype calls and email exchanges. For the help with coding, writing my thesis, papers and abstract, and preparing the presentations (thank you in advance even for the corrections to the present acknowledgements). From you I've learned and appreciated a completely different approach to the way of doing physics which, you know, will never be completely mine.

A special thank you also to dott. Pina Romaniello with whom I worked for the first year of my thesis and who I consider, especially for part II of the thesis, my second co--supervisor. Thank you for the help I received in many aspects of my research.

I would like to thank prof. Mark Casida with his careful reading of the thesis, providing me with comments, suggestions and criticism on the first version of the manuscript.

There are many scientist I have to thank for useful discussions and suggestions on physics. Prof. Gianluca Colo, prof. Luca Molinari, prof. Mark Casida and his student Huix--Rotllant, prof. Angel Rubio, Claudio Attaccalite, Matteo Gatti, Francesco Sottile and Adriano Moscaconte and many others.

Then all the researchers, professors, and not only, who I met during this three years in Milano, Roma, San Sebastian and at many conferences. You are too many, I'm sure I would forget someone trying to remember all names...thank you all! A special thank you to Ali Akbari with whom I lived for some months in Donostia (San Sebastian) for the great Iranian food I could taste and for the biathlon. To Diego Galli with whom I lived for some month in Rome having the opportunity to know a ``romano de core''. To Elena Cannuccia for many discussions on our PhD and on our common co--supervisor (sorry Andrea) and not only; Marco Genoni with whom I did the test for the CNISM PhD in Rome and with whom I've been many times both in Rome and in Trieste. Then thank you to the ``ssptg'' in Milan, to the ETSF nodes of Rome and San Sebastian and again to all my lunch--time friends in Milan.

Si, perch\'e gli ultimi tre anni sono stati un lungo interminabile viaggio in Italia ed in Europa, ed ogni volta ho trovato nuove motivazioni anche sapendo che a Gessate c'era un posto dove sarei potuto tornare, dove i miei genitori mi stavano aspettando. Un grazie speciale a loro per avermi sempre supportato anche se ero lontano e senza capire bene a cosa stessi lavorando. Un grazie a mio fratello che, ogni volta che tornavo, era il primo compagno per un'uscita in bici, una giornata a sciare o qualche palleggio a tennis. Un grazie anche a tutti i miei familiari che mi hanno visto andare e venire e con cui spesso non mi facevo sentire per mesi.

Ad ogni rientro a casa ritrovavo non solo parenti ma anche amici. Un grazie a Marco J. (MJ) che fin dal liceo \`e il mio compagno pi\`u costante per le escursioni in montagna d'estate ed a sciare d'inverno, nonstante l'inguaribile neo di restare un Berlusconiano convinto. Un grazie a lui e a tutti gli amici di Gessate che ogni volta reincontro con piacere. Un grazie anche al gruppo del tennis che in questi tre anni di dottorato ho sempre ritrovato con piacere a Gessate.

Un saluto ed un grazie ai miei ex compagni di corso e di Liceo, da dove tutto \`e cominciato, ed in particolare a coloro che con me hanno diviso anche se da lontano, l'esperienza di un dottorato in giro per l'Europa. Ogni volta che ci si rivede \`e sempre bello.

E per concludere, perch\'e le cose pi\`u importanti si lasciano sempre per ultime, un grazie a Veronica perch\'e lei pi\`u di tutti ha dovuto aspettare, sopportare e pazientare. Un grazie perch\`e pi\`u di ogni altra cosa \`e stata (e tuttora \`e) la mia certezza ovunque fossi e qualunque cosa facessi.

In questa tesi c'\`e anche un po' di noi.

%


\end{document}